\documentclass[10pt]{iopart}

\expandafter\let\csname equation*\endcsname\relax
\expandafter\let\csname endequation*\endcsname\relax
\usepackage{amsmath}
\usepackage{iopams}
\usepackage{amssymb}
\usepackage{color}
\usepackage[usenames,dvipsnames,svgnames,table]{xcolor}
\usepackage[colorlinks=true,linkcolor=blue,urlcolor=blue,citecolor=blue]{hyperref}
\usepackage{diagbox}
\usepackage{siunitx}
\usepackage{graphicx}
\usepackage{array}
\usepackage{lipsum}
\usepackage{bm}
\usepackage{subfigure}
\usepackage{tabularx}
\usepackage{braket}
\usepackage{ulem}
\usepackage{verbatim}
\usepackage{orcidlink}
\usepackage{comment}
\usepackage{tabularray}
\usepackage{lipsum}
\usepackage{mathrsfs}
\usepackage{MnSymbol}

\begin{document}

\title[WDM Roadmap]{Roadmap for warm dense matter physics}


\author{%
Jan Vorberger\,\orcidlink{0000-0001-5926-9192}$^{1}$, 
Frank Graziani\,\orcidlink{0000-0002-7204-0509}$^{2}$, 
David Riley\,\orcidlink{0000-0002-6212-3212}$^{3}$, 
Andrew D. Baczewski\,\orcidlink{0000-0001-8553-9934}$^{4}$,
Isabelle Baraffe\,\orcidlink{0000-0001-8365-5982}$^{5,6}$,
Mandy Bethkenhagen\,\orcidlink{0000-0002-1838-2129}$^{7}$,
Simon Blouin\,\orcidlink{0000-0002-9632-1436}$^{8}$,
Maximilian P. B\"ohme\,\orcidlink{0000-0003-0290-3628}$^{2}$,
Michael Bonitz\,\orcidlink{0000-0001-7911-0656}$^{9,10}$,
Michael Bussmann\,\orcidlink{0000-0002-8258-3881}$^{1}$,
Alexis Casner\,\orcidlink{0000-0003-2176-1389}$^{11}$,
Witold Cayzac\,\orcidlink{0000-0001-9682-1636}$^{11}$,
Peter Celliers\,\orcidlink{0000-0001-7600-1227}$^{2}$,
Gilles Chabrier\,\orcidlink{0000-0002-8342-9149}$^{6,5}$,
Nicolas Chamel\,\orcidlink{0000-0003-3222-0115}$^{12}$,
Dave Chapman\,\orcidlink{0000-0001-8741-1424}$^{13}$,
Mohan Chen\,\orcidlink{0000-0002-8071-5633}$^{14}$,
Jean Cl\'erouin\,\orcidlink{0000-0003-2144-2759}$^{15}$,
Gilbert Collins\,\orcidlink{0000-0002-4883-1087}$^{16,17,18}$, 
Federica Coppari\,\orcidlink{0000-0003-1592-3898}$^{2}$, 
Tilo D\"oppner\,\orcidlink{0000-0002-5703-7125}$^{2}$,
Tobias Dornheim\,\orcidlink{0000-0001-7293-6615}$^{1,19}$,
Luke B. Fletcher\,\orcidlink{0000-0002-7120-7194}$^{20}$,
Dirk O. Gericke\,\orcidlink{0000-0001-6797-9004}$^{21}$,
Siegfried Glenzer\,\orcidlink{0000-0001-9112-0558}$^{20}$,
Alexander F. Goncharov\,\orcidlink{0000-0002-6422-8819}$^{22}$,
Gianluca Gregori\,\orcidlink{0000-0002-4153-0628}$^{23}$,
Sebastien Hamel,\orcidlink{0000-0003-4246-0892}$^{2}$,
Stephanie B. Hansen\,\orcidlink{0000-0002-1886-9770}$^{24}$,
Nicholas J. Hartley\,\orcidlink{0000-0002-6268-2436}$^{20}$,
Suxing Hu\,\orcidlink{0000-0003-2465-3818}$^{16,17,18}$,
Omar A. Hurricane\,\orcidlink{0000-0002-8600-5448}$^{2}$,
Valentin V. Karasiev\,\orcidlink{0000-0003-3445-6797}$^{16}$,
Joshua J. Kas\,\orcidlink{0000-0002-6103-0356}$^{25}$,
Brendan Kettle\,\orcidlink{0000-0003-4424-4448}$^{26}$,
Thomas Kluge\,\orcidlink{0000-0003-4861-5584}$^{1}$,
Marcus D. Knudson\,\orcidlink{0000-0002-5916-4662}$^{24}$,
Alina Kononov\,\orcidlink{0000-0002-6600-224X}$^{4}$,
Zuzana Kon\^{o}pkov\'{a}\,\orcidlink{0000-0001-8905-6307}$^{27}$
Dominik Kraus\,\orcidlink{0000-0002-6350-4180}$^{28,1}$,
Andrea Kritcher\,\orcidlink{0000-0002-9694-1088}$^{2}$,  
Sophia Malko\,\orcidlink{0000-0003-1041-1092}$^{29}$,
G{\'e}rard Massacrier\,\orcidlink{0000-0002-1021-2950}$^{30}$,
Burkhard Militzer\,\orcidlink{0000-0002-7092-5629}$^{31}$,
Zhandos A. Moldabekov\,\orcidlink{0000-0002-9725-9208}$^{1,19}$,
Michael S. Murillo\,\orcidlink{0000-0002-4365-929X}$^{32}$,
Bob Nagler\,\orcidlink{0009-0002-5736-7842}$^{20}$,
Nadine Nettelmann\,\orcidlink{0000-0002-1608-7185}$^{28}$,
Paul Neumayer\,\orcidlink{0000-0002-5156-8830}$^{33}$,
Benjamin K. Ofori-Okai\,\orcidlink{0000-0002-0737-6786}$^{20}$,
Ivan I. Oleynik\,\orcidlink{0000-0002-5348-6484}$^{34}$,
Martin Preising\,\orcidlink{0000-0001-7689-2736}$^{28}$,
Aurora Pribram-Jones\,\orcidlink{0000-0003-0244-1814}$^{35}$,
Tlekkabul Ramazanov\,\orcidlink{0009-0009-2301-1152}$^{36}$,
Alessandra Ravasio\,\orcidlink{0000-0002-2077-6493}$^{7}$, 
Ronald Redmer\,\orcidlink{0000-0003-3440-863X}$^{28}$,
Baerbel Rethfeld\,\orcidlink{0009-0008-9921-4127}$^{37}$,
Alex P. L. Robinson\,\orcidlink{0000-0002-3967-7647}$^{38}$,
Gerd R{\"o}pke\,\orcidlink{0000-0001-9319-5959}$^{28}$,
Fran\c{c}ois Soubiran\,\orcidlink{0000-0002-0982-5178}$^{15}$,
Charles E. Starrett\,\orcidlink{0000-0002-1582-8148}$^{39}$,
Gerd Steinle-Neumann\,\orcidlink{0000-0001-7455-6149}$^{40}$,
Phillip A. Sterne\,\orcidlink{0000-0002-6398-3185}$^{2}$,
Shigenori Tanaka\,\orcidlink{0000-0002-6659-2788}$^{41}$,
Aidan P. Thompson\,\orcidlink{0000-0002-0324-9114}$^{4}$,
Samuel B. Trickey\,\orcidlink{0000-0001-9224-6304}$^{42}$,
Tommaso Vinci\,\orcidlink{0000-0002-1595-1752}$^{7}$,
Sam M. Vinko\,\orcidlink{0000-0003-1016-0975}$^{23}$,
Lei Wang\,\orcidlink{0009-0004-1348-8583}$^{43}$,
Alexander J. White\,\orcidlink{0000-0002-7771-3899}$^{39}$
Thomas G. White\,\orcidlink{0000-0002-3865-4240}$^{44}$,
Ulf Zastrau\,\orcidlink{0000-0002-3575-4449}$^{27}$,
Eva Zurek\,\orcidlink{0000-0003-0738-867X}$^{45}$,
Panagiotis Tolias\,\orcidlink{0000-0001-9632-8104}$^{46}$
}

\address{%
$^{1}$Helmholtz-Zentrum Dresden Rossendorf, 01328 Dresden, Germany\\
$^{2}$Lawrence Livermore National Laboratory, Livermore, CA 94550, USA\\
$^{3}$School of Mathematics and Physics, Queen's University of Belfast, BT7 1NN, United Kingdom\\
$^{4}$Center for Computing Research, Sandia National Laboratories, Albuquerque, 87185 New Mexico, United States \\
$^{5}$University of Exeter, Exeter, EX4 4QL, UK \\
$^{6}$Ecole Normale Supérieure de Lyon, CRAL, CNRS UMR 5574, 69364, Lyon Cedex 07, France \\
$^{7}$LULI, CNRS, CEA, Ecole Polytechnique—Institut Polytechnique de Paris, 91128 Palaiseau cedex, France \\
$^{8}$Department of Physics and Astronomy, University of Victoria, Victoria, BC V8W 2Y2, Canada\\
$^{9}$Institut für Theoretische Physik und Astrophysik, Christian-Albrechts-Universit\"at zu Kiel, Leibnizstraße 15, 24098 Kiel, Germany\\
$^{10}$Kiel Nano, Surface and Interface Science KiNSIS\\
$^{11}$CEA, DAM, DIF, F-91297 Arpajon, France\\
$^{12}$Institut d'Astronomie et d'Astrophysique, Universit\'e Libre de Bruxelles - CP226, 1050 Brussels,  Belgium\\
$^{13}$First Light Fusion Ltd., Pioneer Park, Oxford OX5 1QU, United Kingdom\\
$^{14}$HEDPS, CAPT, College of Engineering and School of Physics, Peking University, Beijing 100871, People’s Republic of China\\
$^{15}$Universit\'e Paris-Saclay, CEA, Laboratoire Mati\`ere sous conditions extr\^emes, 91680 Bruy\`eres-le-Ch\^atel, France\\
$^{16}$Laboratory for Laser Energetics, University of Rochester, Rochester, 14623 New York, United States \\
$^{17}$Department of Physics and Astronomy, University of Rochester, Rochester, 14627 New York, United States \\
$^{18}$Department of Mechanical Engineering, University of Rochester, Rochester, 14627 New York, United States \\
$^{19}$Center for Advanced Systems Understanding (CASUS), 02826 G\"orlitz, Germany \\
$^{20}$SLAC National Accelerator Laboratory, Menlo Park, CA 94025, USA \\
$^{21}$Centre for Fusion, Space and Astrophysics, Department of Physics, University of Warwick, Coventry CV4 7AL, United Kingdom\\
$^{22}$Carnegie Science, Earth and Planets Laboratory, 5241 Broad Branch Road, NW, Washington, DC 20015, USA\\
$^{23}$Department of Physics, Clarendon Laboratory, University of Oxford, Parks Road, Oxford OX1 3PU, UK\\
$^{24}$Pulsed Power Sciences Center, Sandia National Laboratories, Albuquerque, 87185 New Mexico, United States \\
$^{25}$University of Washington, Seattle, WA 98195, United States \\
$^{26}$The John Adams Institute for Accelerator Science, Blackett Laboratory, Imperial College London, London, SW7 2AZ, United Kingdom\\
$^{27}$European XFEL, Holzkoppel 4, 22869 Schenefeld, Germany \\
$^{28}$Universit{\"a}t Rostock, Institut f{\"u}r Physik, Albert-Einstein-Str.~23-24, D-18059 Rostock, Germany \\
$^{29}$Princeton Plasma Physics Laboratory, 100 Stellarator Rd, 08540, Princeton, NJ, USA \\
$^{30}$Univ Lyon, Ens de Lyon, CNRS, Centre de Recherche Astrophysique de Lyon UMR5574, F-69230, Saint-Genis-Laval, France\\
$^{31}$Department of Earth and Planetary Science, University of California, Berkeley, California 94720, USA\\
$^{32}$Michigan State University, East Lansing, Michigan USA 48824\\
$^{33}$GSI Helmholtzzentrum f\"ur Schwerionenforschung GmbH, 64291 Darmstadt, Germany \\
$^{34}$Department of Physics,  University of South Florida, Tampa,  FL 33620, USA \\
$^{35}$Department of Chemistry and Biochemistry, University of California, Merced, California 95340, USA\\
$^{36}$Al-Farabi Kazakh National University, Almaty, Kazakhstan \\
$^{37}$RPTU University Kaiserslautern-Landau, Department of Physics and State Research Center OPTIMAS, Kaiserslautern, Germany\\
$^{38}$STFC Rutherford-Appleton Laboratory, Didcot OX11 0QX, United Kingdom \\
$^{39}$Los Alamos National Laboratory, P.O. Box 1663, Los Alamos, NM 87545, U.S.A.\\
$^{40}$Bayerisches Geoinstitut, Universit\"at Bayreuth, 95440 Bayreuth, Germany \\
$^{41}$Graduate School of System Informatics, Kobe University, Kobe 657-8501, Japan \\
$^{42}$Department of Physics and Quantum Theory Project, University of Florida, Gainesville FL 32611, USA \\
%
%
$^{43}$Institute of Physics, Chinese Academy of Sciences, Beijing 100190, China \\
$^{44}$Department of Physics, University of Nevada, Reno, Nevada 89557, USA\\
$^{45}$Department of Chemistry, University at Buffalo, Buffalo, NY 14260, USA\\
$^{46}$Space and Plasma Physics, Royal Institute of Technology (KTH), SE-10044 Stockholm, Sweden \\
}

\ead{%
j.vorberger@hzdr.de,graziani1@llnl.gov,D.Riley@qub.ac.uk,tolias@kth.se
}

\vspace{10pt}

\begin{abstract}
This roadmap presents the state-of-the-art, current challenges and near future developments anticipated in the thriving field of warm dense matter physics. Originating from strongly coupled plasma physics, high pressure physics and high energy density science, the warm dense matter physics community has recently taken a giant leap forward. This is due to spectacular developments in laser technology, diagnostic capabilities, and computer simulation techniques. Only in the last decade has it become possible to perform accurate enough simulations \& experiments to truly verify theoretical results as well as to reliably design experiments based on predictions. Consequently, this roadmap discusses recent developments of and contemporary challenges for theoretical methods and experimental techniques needed to describe, create and diagnose warm dense matter. A large part of this roadmap is dedicated to specific warm dense matter systems and applications in astrophysics, inertial confinement fusion and novel material synthesis.
\end{abstract}

%
%
%
\maketitle
%
%

\tableofcontents
\ioptwocol

\newpage
\clearpage
\section*{Preamble}\label{section_01}
\author{Jan Vorberger$^1$, Frank Graziani$^2$, David Riley$^{3}$}
\address{
$^1$Helmholtz-Zentrum Dresden-Rossendorf (HZDR), D-01328 Dresden, Germany \\
$^2$Lawrence Livermore National Laboratory, Livermore, CA 94550, United States \\
$^3$Queen's University of Belfast, Belfast BT7 1NN, United Kingdom\\
}

\subsection*{Introduction}

Among all known states of ordinary baryonic matter, warm and dense matter, or in short warm dense matter (WDM), exists at the crossroads of condensed matter physics, dense liquid theory, and plasma physics. Since it doesn't occur easily in our laboratories or on our desks, is not the brightest in the (night) sky, nor does it feature the lowest or highest temperatures or energies, it has not attracted the same attention as, e.g., high energy particle physics or solid state physics.

Still, much of what we strive to learn about energy production, the origin of life and the universe as a whole, is strongly connected to our understanding of WDM. This roadmap serves as a collection of available knowledge and expert guidance on fruitful new paths for future research to venture into new territories of WDM physics where no one has gone before.

\subsection*{Importance and Occurence of WDM}

The deep interior of most planets and exoplanets is made up of WDM~\cite{chabrier_jpa_06,MilitzerJGR2016,Helled2020b,saumon2021current}. These states are expected to contain pure elements (e.g. hydrogen or carbon), but mostly are mixtures or alloys of light and heavy elements. We distinguish giant gas planets which are mainly hydrogen-helium mixtures; ice planets mainly featuring a mix of hydrogen, oxygen, carbon, nitrogen; rocky planets primarily composed of silicon, oxygen, magnesium, and iron; but also exotic planets made of carbon. Hydrogen-helium mixtures are also the main constituents of brown dwarfs, white dwarf envelopes as well as neutron star crusts. These materials exist at a wide variety of pressure-temperature combinations which cause phase transitions among high pressure phases, miscibility changes, and drastic shifts between different transport regimes, e.g., convection barriers or insulator-metal transitions. The modeling of planetary interiors and planetary property evolution, e.g. cooling rates, demands a solid understanding of the equation of state (EOS) of the required non-ideal mixtures of elements. Transport properties of WDM are required for magneto-hydrodynamic (MHD) simulations of dynamos and magnetic fields of planets.

In addition, the WDM state is an integral part of any inertial fusion experiment and must be understood for efficient inertial fusion energy production~\cite{Hurricane_RMP_2023,Hurricane_PPCF_2025}. Almost the entire compression path of the initially cryogenic hydrogen pellet lies within the WDM regime and only close to and after ignition is there enough energy in the system to speak of a hot plasma state. The cause of any hydrodynamic motion, as well as the seed of any instability or asymmetry in the compression must be understood and in order to do so, we need reliable knowledge of the EOS, electrical resistivity, thermal conductivity and opacities.

Furthermore, techniques from WDM experiments can be employed to synthesize novel materials and available theoretical tools can be utilized to predict the conditions where desirable or exotic properties emerge. In particular, materials that feature a high activation threshold or are native to high pressure regimes (but are metastable at ambient conditions) can be produced and investigated in this way~\cite{Kraus2016b,Kraus2017}. The study of laser-matter interaction, almost always the first step in any WDM experiment, lends itself to technological applications as well, as rescaled setups are used frequently to modify surfaces, e.g., in the semiconductor industry~\cite{Rethfeld_2017}.

The unique challenges of WDM physics have inspired advances in experimental and theoretical research~\cite{graziani2014frontiers}. Use of modern high energy or high power optical lasers in conjunction with x-ray free electron lasers, all of them capable of high repetition rates nowadays, has manifoldly increased the quantity of reliable experimental results~\cite{falk_wdm,Riley_2021}. X-ray scattering diagnostics, in particular x-ray Thomson scattering (XRTS), have become ever more powerful~\cite{Glenzer_revmodphys_2009}. The predictive power of theoretical methods in WDM research has been found wanting for a long time. Only in the last 15 years has substantial progress been made so that there is a certain reliability to simulation results now~\cite{bonitz_pop_19,Bonitz_POP_2020}. Naturally, the theoretical description of hydrogen has been high on the agenda~\cite{Bonitz_POP_2024}. Difficulties are still being encountered when it comes to transport properties~\cite{grabowski2020,Stanek_PoP_2024} and non-equilibrium~\cite{white_dynamic_2023}. Moreover, the results of WDM research are considered useful in improving density functional theory (DFT) and time-dependent density functional theory (TD-DFT) regardless of the conditions~\cite{Dornheim_PhysReports_2018,Moldabekov_PPNP_2024}.

\subsection*{Characterization of WDM}

WDM physics is always many-particle physics, always quantum physics (sometimes even relativistic), and nearly exclusively staged in a 3-dimensional space (not surfaces or dots). 
We describe it using the Hamiltonian
\begin{equation}\label{eq:general-hamiltonian}
    \hat{H}=\sum_{i=1}^N\frac{\hbar^2\nabla_i^2}{2m_i}+\sum_{i<j}^N\frac{e_ie_j}{|{\bf r}_i-{\bf r}_j|}\,,
\end{equation}
where a particle $i$ has a charge $e_i$ and a mass $m_i$ [$1/(4\pi\varepsilon_0)=1$] and the sum runs over all particles.
WDM is characterized by moderate-to-strong coupling, and structural order emerges among its electrons and ions, but long range order or crystal symmetry as in solids is not prominent. WDM is unlike gases or the traditional concept of plasmas, but more akin to fluids. However, fluids are generally assumed to consist of neutral particles, whereas the ionization degree in WDM can be considerable due to the elevated temperatures and increased densities. Ionization potential depression (IPD) and pressure ionization are among many quantum effects that emerge in WDM. All occurring scattering processes are inherently quantum, as are, of course, frequent bound-free, bound-bound, and free-bound transitions. The spin of the particles is important such that the proper quantum statistics can be observed -- most WDM states are partially degenerate meaning that there will be temperature dependent Pauli-blocking, a thermal occupation of states, and temperature dependent exchange--correlation (XC). 

For a rough estimation of the importance of all these effects, simple parameters and scaling laws have been introduced. We estimate the strength of the coupling between particle species $a$ and $b$ using the coupling parameter
\begin{equation}\label{g_general}
    \Gamma_{ab}=\frac{|\langle V_{ab}\rangle |}{\langle K_{a}\rangle}\,,
\end{equation}
with the mean potential energy $\langle V_{ab}\rangle$ and the mean kinetic energy $\langle K_{a}\rangle$. In the non-degenerate limit, Eq.~(\ref{g_general}) simplifies to the well-known $\Gamma_{ab}=|e_ae_b|/k_BTd$, with $k_B$ the Boltzmann constant, $T$ the temperature, $d=(3/4\pi n)^{1/3}$ the mean particle distance belonging to density $n$. Typical values of the coupling in WDM, irrespective of species, are above unity, $\Gamma_{ab}=1\ldots \sim 50$. The curve $\Gamma=1$ encircles the area of strong coupling that is bounded by temperature for low densities and limited at high density irrespective of temperature when degeneracy reduces the coupling. At high densities, coupling is also described by the Brueckner parameter $r_s=d/a_B$, which is simply the ratio of the mean particle distance $d$ and the Bohr radius $a_B$. Typically, perturbation expansions work for very low and very high densities ($r_s\ll 1$), but diverge for WDM densities ($r_s\sim 1$). 

The effect of quantum degeneracy may be estimated by either 
\begin{equation}
    \Theta_a=\frac{T}{T^F_a}\,\,\mbox{, or}\quad
    \chi_a=n_a\Lambda_a^3=n_a\left(\frac{h^2}{2\pi m_ak_BT}\right)^{3/2}\,.
\end{equation}
Here, the Fermi temperature is $T^F_a=(\hbar q_F)^2/2m_ak_B$ with the Fermi wavenumber $q_F=(3\pi^2 n)^{1/3}=\alpha/r_s$, $[\alpha=(9\pi/4)^{1/3}]$ that depends on the density. The thermal deBroglie wavelength $\Lambda_a$ has also been defined. Degenerate systems occur for $\Theta_a<1$ ($\chi_a>1$), Boltzmann statistics appear for $\Theta_a\gg 1$ ($\chi_a\ll 1$). Typical WDM conditions have both parameters around unity, thus again prohibiting the use of semi-classical or ground state simplifications.

The typical length scales of screening $\lambda$ are given by the inverse of the screening parameter $\kappa_a^{-1}=\lambda_a$
\begin{equation}
    \kappa_a^{2}=\frac{4\pi e_a^2 m_a}{\pi\hbar^3}\int\limits_0^{\infty}f_a(p)dp\,,
\end{equation}
where the integral is over the Wigner distribution function $f_a$. The time/energy scales of screening are given by the plasma frequency
\begin{equation}
    \omega_{\mathrm{pl}}^2=\frac{4\pi e_a^2n_a}{m_a}\,.
\end{equation}
In partially ionized matter, the ionization degree is 
\begin{equation}
    \alpha=\frac{n_e^{\mathrm{free}}}{n_e^{\mathrm{free}}+n_e^{\mathrm{bound}}}\,.
\end{equation}
The degree of ionization depends on temperature and density and in particular in a self-consistent way on the effective ionization energies that are described via the concept of ionisation potential depression (IPD)
\begin{equation}
    E_i^{\mathrm{eff}}=E_i-\Delta_{\mathrm{IPD}}\,.
\end{equation}
It is thus possible that electrons become free by either elevating their energy out of existing bound states or by the bound state vanishing into the continuum.

\subsection*{Structure of the roadmap}

The present roadmap consists of $26$ sections. The first $7$ sections describe theoretical tools that are available, employed, and developed to describe various WDM properties and WDM systems. They include first principle methods that are confined to small system sizes as well as methods that require external input but can describe larger spatiotemporal domains. The following $7$ sections are dedicated to the experimental generation of WDM states and their diagnosis. This includes the WDM creation by various laser types as well as mechanical devices. The diagnostic techniques that are available nowadays are equally wide-ranging based on either photons or particle beams. The next $6$ sections cover the physics properties of specific WDM systems. Starting from the archetypal uniform electron gas, specific occurrences of hydrogen and different mixtures in giant gas planets, ice planets, rocky planets and other astrophysical objects are discussed. Then, $3$ sections are devoted to the technological applications of WDM methods and insights concerning the discovery of novel materials, surface modification, and inertial confinement fusion. The final $3$ sections cover relevant developments in computing, data management and experimental infrastructure.

\newpage

\newpage 
\clearpage

\section{Path integral Monte Carlo}\label{section_02}
\author{Michael Bonitz$^{1,2}$, Tobias Dornheim$^{3,4}$, Burkhard Militzer$^{5}$}
\address{
$^1$Institut für Theoretische Physik und Astrophysik, Christian Albrechts Universität zu Kiel, 24089 Kiel, Germany \\
$^2$Kiel Nano and Interface Science (KiNSIS)\\
$^3$Center for Advanced Systems Understanding (CASUS), D-02826 G\"orlitz, Germany \\
$^4$Helmholtz-Zentrum Dresden-Rossendorf (HZDR), D-01328 Dresden, Germany \\
$^{5}$Department of Earth and Planetary Science, University of California, Berkeley, California 94720, USA
}

\subsection*{Introduction}

The simultaneous manifestation of strong coupling, quantum diffraction as well as exchange effects poses fundamental challenges for the theoretical description of WDM because standard perturbation theory cannot be applied. Moreover, the high temperature requires one to go beyond ground state calculations and often to include a challengingly large number of excited electronic states. Therefore, even in thermodynamic equilibrium, the physical processes are very complex and require \emph{ab initio} methods that take into account all these effects in a comprehensive way. 
In the regime where the electrons are moderately or highly excited, path integral Monte Carlo (PIMC) simulations~\cite{cep} have emerged as a successful theory and a practical computation tool. The PIMC method is based on Feynman's path integral picture of quantum statistical mechanics~\cite{Feynman_Hibbs} and is in principle capable of providing exact (within Monte Carlo error bars) results for a wide range of parameters without any empirical input. This is key for the construction of accurate input data tables and parameterizations for other methods~\cite{Dornheim_PhysReports_2018,FPEOS,KSDT2014PRL}, for the rigorous assessment of less accurate but computationally cheaper approaches~\cite{Bonitz_POP_2024,Dornheim_PhysReports_2018,moldabekov2025applyingliouvillelanczosmethodtimedependent,Moldabekov_JCTC_2024,dharmawardana2025xraythomsonscatteringstudies,dornheim2024modelfreerayleighweightxray},
and for the study of quantum many-body effects in WDM in their own right~\cite{Dornheim_ComPhys_2022,Dornheim_PRL_2018,Boehme2022}.

\begin{figure*}[t!]
    \centering
    \includegraphics[width=0.31\textwidth]{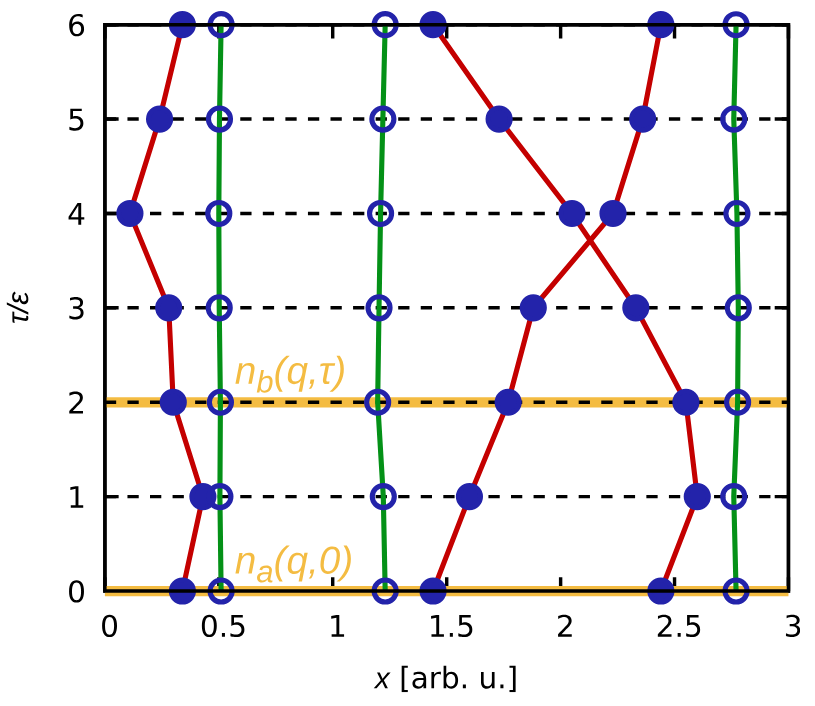}
    \hspace{0.01\textwidth}
    \includegraphics[width=0.31\textwidth]{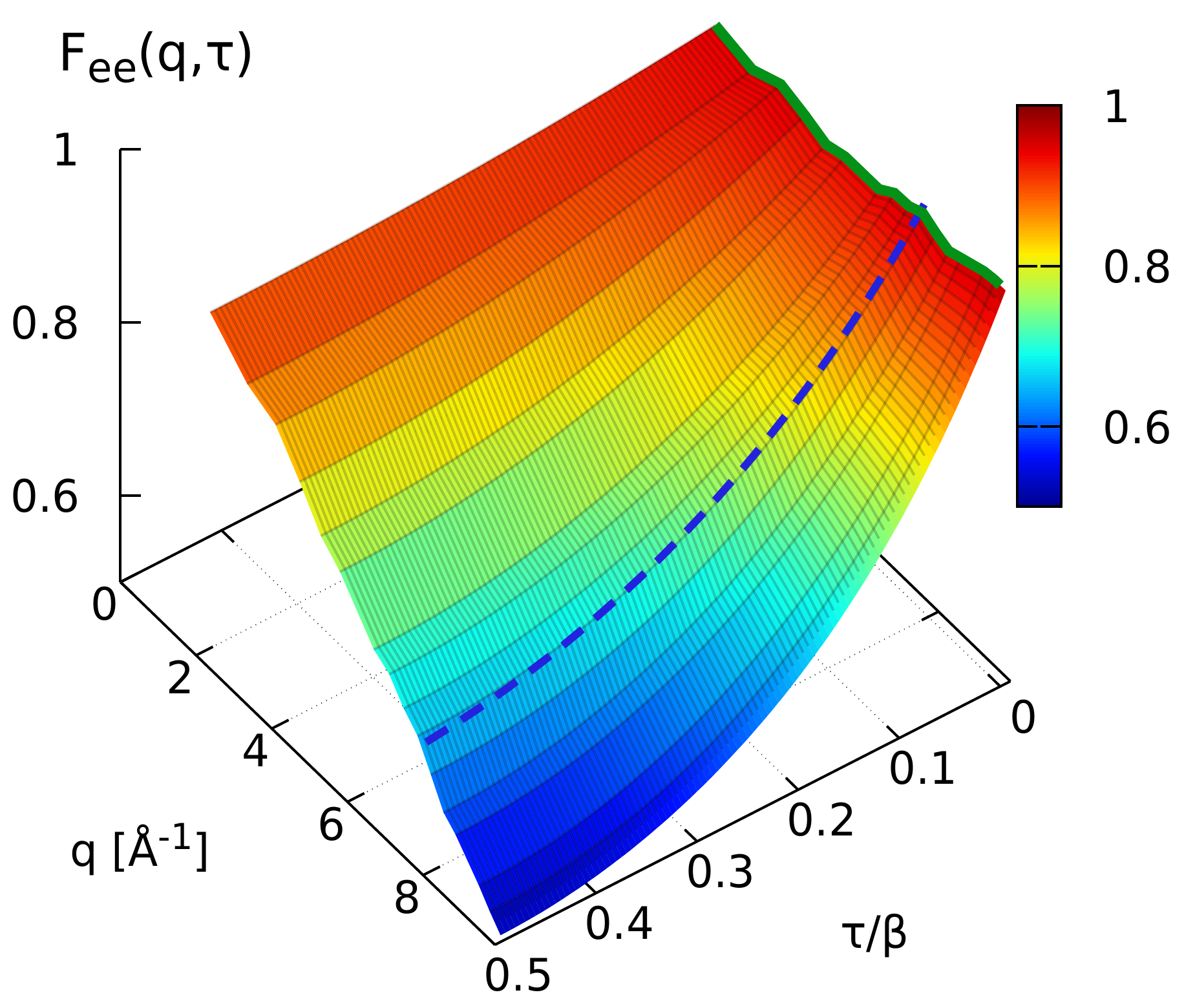} 
    \hspace{0.01\textwidth}
    \includegraphics[width=0.32\textwidth]{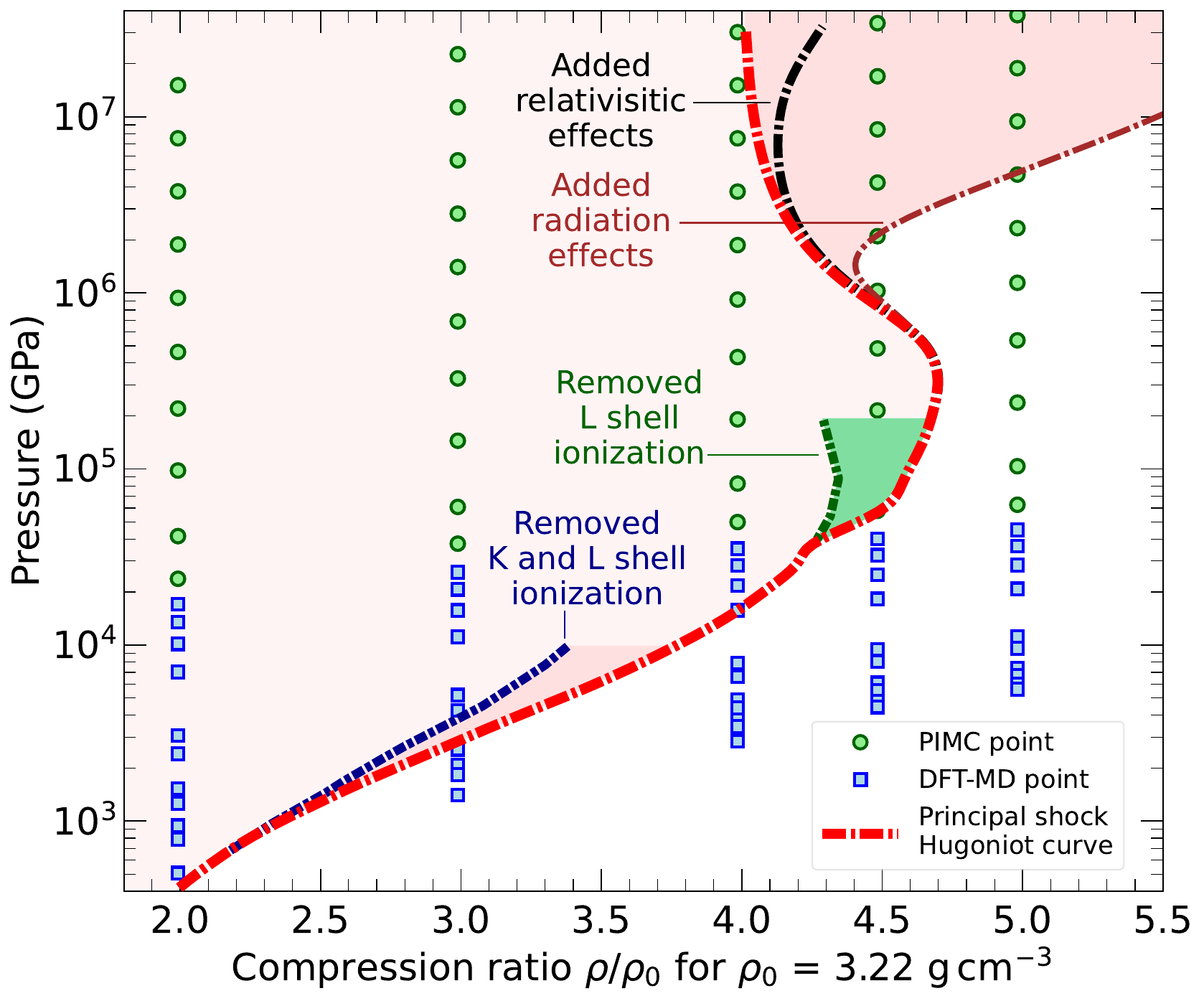}
    \caption{Left: schematic PIMC configuration in the $x$-$\tau$ plane; the red and green paths show electrons and protons. 
    Center: PIMC results for the electronic ITCF $F_{ee}(\mathbf{q},\tau)$ of warm dense beryllium 
    in the $q$-$\tau$-plane (colored surface), compared to an XRTS measurement at the NIF~\cite{Doeppner_nature_2023} (dashed blue) 
    \cite{dornheim2024unraveling}.
    Right: Shock Hugoniot curve of MgSiO$_3$ that was computed by combining PIMC and DFT-MD results into a single, consistent EOS stable. Ionization, radiation and relativistic effects were analyzed in Ref.~\cite{gonzalez2020path}.
    }
    \label{fig:pimc_fig1}
\end{figure*}

We consider the canonical partition function, for the hamiltonian (\ref{eq:general-hamiltonian}), 
in the coordinate representation for $N$ identical fermions in a volume $\Omega$ at an inverse temperature of $\beta=1/k_\textnormal{B}T$ that plays the role of {\em imaginary time},
\begin{eqnarray}\label{eq:Z_PIMC}
    Z_{N,\Omega,T} = \frac{1}{N!} \sum_{\sigma\in S_N} \textnormal{sgn}(\sigma) \int_\Omega \textnormal{d}\mathbf{R}\ \bra{\mathbf{R}} e^{-\beta \hat{H}} \ket{\hat{\pi}_\sigma \mathbf{R}}\ ,
\end{eqnarray}
where $\mathbf{R}=(\mathbf{r}_1,\dots,\mathbf{r}_N)^T$ contains all $N$ coordinates.
The antisymmetry of the thermal density matrix under the exchange of particle coordinates gives rise to the sum over all possible permutations $\sigma$ of the respective permutation group $S_N$, which are realized by the corresponding exchange operator $\hat{\pi}_\sigma$. The sign function in front of the integral is positive (negative) for an even (odd) number of pair exchanges. The generalization of Eq.~(\ref{eq:Z_PIMC}) to multiple species (e.g., spin-up and spin-down electrons) is straightforward.

The immediate problem with Eq.(\ref{eq:Z_PIMC}) is that the thermal density operator does not factorize into kinetic energy and potential energy contributions, $\hat{\rho}=e^{-\beta \hat{H}}\ne e^{-\beta \hat{K}}e^{-\beta \hat{V}}$, because $\hat K$, $\hat V$ do not commute. Therefore, the matrix elements of $\hat\rho$ cannot be directly evaluated exactly. However, a number of approximations exist that become increasingly accurate at high temperature, which leads to the founding concept of path integral calculations, 
$\hat{\rho}=\prod_{\alpha=0}^{P-1}e^{-\epsilon\hat{H}}$, with $\epsilon=\beta/P$ a step in imaginary time. Even the \emph{primitive factorization}, $e^{-\epsilon\hat{H}}\approx e^{-\epsilon\hat{K}}e^{-\epsilon\hat{V}}$, becomes exact in the limit of high temperatures, $P\to\infty$, although more sophisticated factorizations exist~\cite{cep,sakkos_JCP_2009,Chin_JCP_2002,Zillich_JCP_2010}. This so-called \emph{Trotterization}~\cite{trotter} is straightforward for systems with purely repulsive interactions, such as the UEG, but some extra care is needed for the Coulomb attraction between electrons and nuclei~\cite{Po88,filinov_jpa03,filinov_pre04,kleinert2009path,Militzer_HTDM_2016,Boehme2023}.

Since a number of comprehensive introductions to the PIMC method are available in the literature~\cite{cep,boninsegni1,Dornheim2019c}, we restrict ourselves here to the canonical partition function in Eq.~(\ref{eq:Z_PIMC}), $Z_{N,\Omega,T} = \sumint \textnormal{d}\mathbf{X}\ W(\mathbf{X})$,
where $\sumint \textnormal{d}\mathbf{X}$ combines the integration over coordinates and the sum over permutations. 
The basic idea of PIMC is then to stochastically sample all possible configurations $\mathbf{X}$ according to their statistical weight $W(\mathbf{X})$, which is straightforward to evaluate numerically~\cite{cep,boninsegni1,boninsegni2,Mezzacapo_PRA_2007,Dornheim_PRB_2021_nk} with modern extensions of the celebrated Metropolis algorithm~\cite{metropolis}. In the left panel of Fig.~\ref{fig:pimc_fig1}, a schematic configuration $\mathbf{X}$ of a PIMC simulation of hydrogen is depicted, where particles have been mapped onto a path in imaginary time $\tau$. This is often referred to as the \emph{classical isomorphism}~\cite{Chandler_JCP_1981}, since the original quantum many-body system has been mapped onto an classical ensemble of interacting ring polymers. All paths are closed loops as long as diagonal matrix elements of the density matrix are evaluated in Eq.~(\ref{eq:Z_PIMC}). Permutations combine individual paths into longer loops. With decreasing temperature, these paths become longer and more quantum mechanical in nature. For bosonic systems, this eventually leads to Bose-Einstein condensation~\cite{cep} while for fermonic systems, the cancellation of positive and negative terms in Eq.~(\ref{eq:Z_PIMC}) leads to Pauli exclusion. 

The thermal de Broglie wavelength, $\lambda_T = \sqrt{2\pi\beta\hbar^2/m}$, characterizes the magnitude of the quantum fluctuations around a classical path in Fig.~\ref{fig:pimc_fig1}. So the electron paths are substantially more delocalized than the paths of the much heavier protons. The two rightmost electrons are connected via a permutation that leads to a negative sign of the configuration weight $W(\mathbf{X})$. This is the root cause of the notorious \emph{fermion sign problem} (FSP)~\cite{Loh_PRB_1990,Troyer_PRL_2005,Dornheim_PRE_2019,Lyubartsev_2005}, which manifests itself by an exponential increase in the required computer time with respect to key parameters such as $N$ or $\beta$ due to the near perfect cancellation of an increasing number of positive and negative terms. 

\begin{figure}
    \centering
    \includegraphics[width=1.00\linewidth]{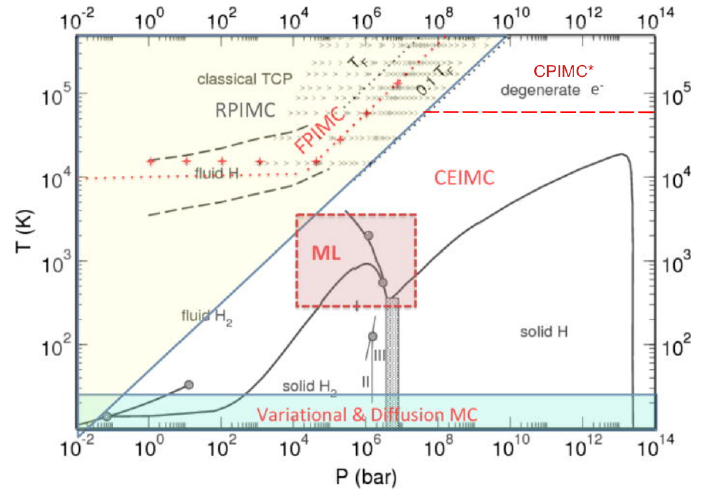}
    \caption{Overview from Ref.~\cite{Bonitz_POP_2024} of different types of quantum Monte Carlo simulation methods for WDM (modified). Density, pressure and phases refer to warm dense hydrogen. 
    }
    \label{fig:pimc}
\end{figure}

\subsection*{State of the art}

Numerous strategies have been introduced to deal with the FSP~\cite{Schoof_PRL_2015,Dornheim_NJP_2015,Hou_PRB_2022,yilmaz_jcp_20,Hirshberg_JCP_2020,Dornheim_Bogoliubov_2020,Blunt_PRB_2014,Malone_PRL_2016,Joonho_JCP_2021}, see also the overview in Fig.~\ref{fig:pimc}. Because of their relevance to WDM, we discuss the following four methods:

(i) In a number of seminal papers~\cite{Ce91,Ceperley_PRL_1992,Ce96}, Ceperley and co-workers introduced the fixed-node approximation that {\em restricts} all paths, $\mathbf{R}(t)$, to remain in the positive domain that is spanned by a trial density matrix, $\rho_T[\mathbf{R}(t),\mathbf{R}(t=0)]>0$. Like DFT, the RPIMC method is formally exact. Ceperley showed that the FSP can be completely avoided if the exact nodal structure could be computed in polynomial time. For interacting systems, one is required to work with approximate forms of $\rho_T$ like free-particle nodes. This {\em restricted} PIMC (RPIMC) method has first been applied to hydrogen~\cite{Pierleoni1994,Ma96,Mi99,MC00,Militzer2001} before Militzer and coworkers simulated hot, dense helium~\cite{Mi06,Mi09} and then extended it step by step to heavier elements \cite{Driver2015,Driver2015b,zhang2016path,zhang2017equation,Driver2018,gonzalez2020equation} up to silicon, which required the incorporation of bound states into the nodal structure~\cite{militzer2015development}. The PIMC predictions were combined with DFT-MD results in order to construct coherent EOS tables over a wide range of pressure-temperature conditions. Compounds like CH$_x$~\cite{zhang2018path}, BC~\cite{zhang2020benchmarking}, BN~\cite{zhang2019equation}, MgO \cite{soubiran2019magnesium}, and MgSiO$_3$ were directly simulated before results were compiled into the first-principles EOS (FPEOS) database~\cite{FPEOS}. One goal was the prediction of shock experiments. The right panel of Fig.~\ref{fig:pimc_fig1} illustrates how the ionization of different electronic shells increase the shock compression. With the FPEOS database, shock states of mixtures can be predicted with the additive volume approximation that has been shown to work well for T$>$20~eV~\cite{militzer2020nonideal}. 

(ii) A number of authors have suggested the use of anti-symmetric imaginary-time propagators (determinants)~\cite{Dornheim_PhysReports_2018,Lyubartsev_2005,Dornheim_NJP_2015,filinov_ppcf_01,Takahashi_Imada_PIMC_1984,Chin_PRE_2015,Dornheim_CPP_2019,Filinov_CPP_2021}. These automatically take into account some cancellation between positive and negative terms, which alleviates the FSP but does not fully remove it. Dornheim and co-workers~\cite{Dornheim_NJP_2015,Dornheim_JCP_2015} suggested to combine determinants with a fourth-order $\hat{\rho}$ factorization  that allows for sufficient accuracy with a small number of time slices. In this formulation the FSP is substantially less severe for small $P$, thus the resulting permutation blocking PIMC (PB-PIMC) approach has extended PIMC towards higher densities and lower temperatures~\cite{Dornheim_PhysReports_2018,Dornheim_NJP_2015}. In combination with the configuration PIMC (CPIMC) approach, PB-PIMC simulations have been pivotal for the construction of accurate parametrizations of the UEG exchange--correlation free energy~\cite{Groth_PRL_2016,Karasiev_PRB_2019}, see Sec.~\ref{section_15}. A.~Filinov and Bonitz~\cite{Filinov_CPP_2021, Filinov_PRE_2023} have used an approach similar to PB-PIMC in the grand ensemble to simulate the UEG and warm dense hydrogen~\cite{Bonitz_POP_2024}.

(iii) Bonitz and collaborators have suggested an alternative paradigm by formulating PIMC within the second quantization. The configuration PIMC (CPIMC) method~\cite{Schoof_PRL_2015,yilmaz_jcp_20,schoof_cpp11,Schoof_CPP_2015,Groth_PRB_2016,groth_jcp} is highly efficient at weak coupling, even at strong quantum degeneracy, but becomes less efficient with increasing coupling and localization, making it complementary to direct PIMC, PB-PIMC and RPIMC. The combination of CPIMC ($r_s$$\lesssim$1) and PB-PIMC ($r_s$$\gtrsim$1) has allowed for the construction of a UEG parametrization that covers the entire relevant WDM parameter space without using fixed nodes.
In addition, Schoof \emph{et al.}~\cite{Schoof_PRL_2015} have used CPIMC to probe the accuracy of the RPIMC data by Brown \emph{et al.}~\cite{Brown_PRL_2013}. Improvements in the exchange-correlation energy have been made at high density and low temperatures that have since been confirmed by independent studies~\cite{Malone_PRL_2016,Joonho_JCP_2021}.

(iv) Xiong and Xiong~\cite{Xiong_JCP_2022,Xiong_PRE_2023} 
used path integral MD to compute fermionic properties by simulating fictitious identical particles. This was adapted to PIMC simulations of WDM~\cite{dornheim2024unraveling,Dornheim_JCP_xi_2023,Dornheim_JCP_2024,Dornheim_MRE_2024,dornheim2025fermionicfreeenergiestextitab,Dornheim_JPCL_2024,dornheim2025taylorseriesperspectiveab}, where it allows for the simulation of up to $N=1000$ electrons at weak to moderate degeneracy.
In contrast to RPIMC or PB-PIMC, this $\xi$-extrapolation technique retains full access to all imaginary-time correlation functions (ITCF)~\cite{Dornheim_JCP_ITCF_2021}. The ITCF leads one to linear and non-linear density response functions, see Sec.\ref{section_05}, which has recently allowed for the first exact results for the density response in warm dense hydrogen~\cite{Dornheim_MRE_2024}. In addition, XRTS experiments with WDM~\cite{Glenzer_revmodphys_2009}, see Sec.\ref{section_13}, give access to the ITCF, which allows for a direct comparison between measurement and exact theory that is shown in the central panel of Fig.~\ref{fig:pimc_fig1} for warm dense beryllium~\cite{Doeppner_nature_2023,dornheim2024unraveling}.

\subsection*{Challenges \& outlook}

From a theoretical perspective, the FSP will remain the greatest challenge in PIMC simulations of WDM in the foreseeable future, which renders the development of improved nodal surfaces for RPIMC important~\cite{militzer2015development}. One venue are variational techniques that minimize the free energy~\cite{militzer_pre_00,JML-1-38}. Improved nodes would allow one to perform simulations of heavier elements and for lower temperatures by, e.g., alleviating the \emph{reference point freezing} that has been seen for high electronic degeneracies. Currently, the RPIMC method has difficulties under conditions where K and L shells are occupied. A practical pseudopotential approach that has made DFT so successful has yet to be developed for RPIMC. 

Other potential strategies include the development of alternative quantum mechanical representations that lead to a complementary manifestation of the FSP (such as CPIMC) or improved blocking strategies (like PB-PIMC). There are two types of goals. On the one hand, approximate but practical computational methods are needed that can characterize real materials over the broad range of WDM conditions. On the other hand, asymptotically exact PIMC simulation methods are also needed that are applicable to more idealized situations and can serve as benchmarks.

A second PIMC frontier concerns the computation of dynamic properties, cf. Sec.~\ref{section_05}. The analytic continuation of ITCF results is possible in principle but it constitutes a difficult, ill-posed problem~\cite{Jarrel_PhysReports_1996}. Recent PIMC results for the dynamic structure factor $S_{ee}(\mathbf{q},\omega)$ of the warm dense UEG~\cite{Dornheim_PRL_2018,Groth_PRB_2019} are promising, but they relied on a number of exact constraints that are only known for the UEG. The development of new, system-agnostic analytic continuation tools~\cite{chuna2025,Otsuki_PRE_2017,Bertaina_AdvPhysX_2017,Fei_PRL_2021} will give access to dynamic properties of light elements~\cite{dornheim2024unraveling,Dornheim_MRE_2024}.

Thirdly, we mention PIMC calculations of the equation of state properties of WDM that go beyond the existing calculations of the internal energy and the pressure. This includes alternative observables such as the free energy~\cite{Dornheim_JCP_2015,dornheim2024directfreeenergycalculation,dornheim2024eta} and the chemical potential~\cite{dornheim2024chemicalpotentialwarmdense}. This frontier might also lead to better finite-size corrections~\cite{Holzmann_PRB_2016} that so far have been mostly focused on the UEG~\cite{Chiesa_PRL_2006,Drummond_PRB_2008,Dornheim_PRL_2016,Dornheim_JCP_2021}.
From a more practical perspective, for warm dense hydrogen~\cite{rygg_prl_23} as well as for methane~\cite{Kritcher2020}, shock wave measurements have been compared directly with RPIMC predictions~\cite{Hu_PRB_2011,zhang2017first} and were found to be in good agreement, with remaining deviations on the 1 or 2 $\sigma$ level that need to be resolved. As experiments probe more extreme conditions, measurements for other materials will be compared in the near future. 

To summarize, the unique role of PIMC for WDM comes from its quasi-exact nature, without the need for any external input. PIMC is particularly well suited to rigorously benchmark less accurate but more efficient methods~\cite{Bonitz_POP_2024}. Fermionic PIMC can be used to benchmark the nodes of RPIMC to extend the range of validity of the latter. In the future, fermionic PIMC and RPIMC might be able to test and improve other approaches and, e.g. test exchange-correlation functionals in DFT [Sec.~\ref{section_03}],
benchmark self-energies in Green functions [Sec.~\ref{section_06}], or verify closure relations in hydrodynamics [Sec.~\ref{section_07}].

Moreover, PIMC allows for the construction of accurate data tables of density response functions~\cite{Dornheim2023_density_response} and various other properties~\cite{FPEOS} to be used as input by other approaches and models, and for the determination of atomic properties such as ionization potential depression~\cite{Bonitz_POP_2024,bonitz_cpp_2025,bellenbaum2025estimatingionizationstatescontinuum}. Finally, we mention recent efforts to use PIMC simulations for the model-free interpretation of XRTS experiments~\cite{dornheim2024modelfreerayleighweightxray,dornheim2024unraveling}. 

\newpage 
\clearpage

\section{Density Functional Theory}\label{section_03}
\author{Mohan Chen$^{1}$, Aurora Pribram-Jones$^{2}$, Fran\c{c}ois Soubiran$^{3}$, Samuel Trickey$^{4}$}
\address{
$^1$HEDPS, CAPT, College of Engineering and School of Physics, Peking University, Beijing 100871, People’s Republic of China \\
$^2$University of California Merced, 95343 California, United States \\
$^3$Universit\'e Paris-Saclay, CEA, Laboratoire Mati\`ere sous conditions extr\^emes, 91680 Bruy\`eres-le-Ch\^atel, France \\
$^4$University of Florida, Gainesville, 32611 Florida, United States
}

\subsection*{Introduction}

As presented in the Preamble, the computational exploration of WDM requires solving a quantum $N$-body
problem. The general problem is too complicated to tackle even with the most advanced supercomputers. A first helpful simplification is to assume that the nuclei behave as classical particles. That is the Born-Oppenheimer approximation (see, e.g. Refs.\cite{oppenheimer_zur_1927,zwanzig_nonequilibrium_2001}), which is applicable in a large variety of WDM cases. Even then, the problem remains formidable: computing accurate polyelectronic wavefunctions is difficult and highly demanding of computer resources. In that regard, Hohenberg \& Kohn~\cite{Hohenberg_PR_1964} (HK) offered a paradigm shift by demonstrating that the ground-state variational minimization problem is equivalent to minimizing a functional of the electronic density $n(\mathbf{r})$: density functional theory (DFT) was born.

Kohn \& Sham~\cite{Kohn_PR_1965} soon offered an effective way to minimize the HK density functional. They observed that for any electronic density 
arising from a polyelectronic Hamiltonian associated with a given external potential $v$, there exists an equivalent independent 
electron system with the same electronic density $n(\mathbf{r})$ and ground state in an effective Kohn-Sham (KS) potential:
\begin{equation}
    v_\textrm{eff}(\textbf{r})= v(\textbf{r})+ \int\frac{n(\textbf{r}')}{\left|\textbf{r}-\textbf{r}'\right|} \textrm{d}\textbf{r}'+v_\textrm{xc}(\textbf{r}),
    \label{eq:dft:veff}
\end{equation}
which includes electronic density effects plus an exchange-correlation (XC) potential $v_{xc}$. The difficult interacting quantum $N$-body problem thus becomes a tractable non-interacting fermion one. However, most of the crucial quantum effects are pushed into the XC functional. Its exact form is unknown, and there is no simple way to construct accurate approximations. The great strength of DFT is thus its relative simplicity with the KS scheme, but its dependence on the quality of the XC functional is its great weakness. 

DFT was proven for the ground state. Mermin offered a finite-temperature (FT) extension~\cite{Mermin_PR_1965}, in which the ground-state energy functional becomes the grand potential, inviting the use of DFT in the WDM regime, where temperature is a key parameter. However, this requires FT XC functionals, an area of ongoing research that benefits from tools such as the FT adiabatic connection formula~\cite{Pittalis_PRL_2011,Harding_JCP_2022}, in which the XC free energy is written in terms of the potential contribution alone, and scaling relationships between $n(\textbf{r})$, $T$, and interaction strength~\cite{Pittalis_PRL_2011,Dufty_PRB_2011,Dufty_Mol.Phys_2016}.

Because the KS scheme is computationally costly, especially at high temperatures, an orbital-free (OF) version of KS has had significant success in the hot dense matter regime. OF-DFT amounts to taking the KS mapping at its plainest. No electronic orbitals are computed, but only the electronic density. Naturally, the price paid is an approximation of the KS kinetic energy functional. Most of modern ground-state OF-DFT has been reviewed in Refs.~\cite{Wang_2000,Mi_Chem.Rev_2023}. Free-energy OF-DFT requires both non-interacting functionals and XC functionals.  Typically, these have
been built as extensions of ground-state functionals~\cite{Karasiev_electronic_structure_2025}. A key part of those generalizations concerns the satisfaction of constraints from FT gradient expansions, as well as provable properties of the respective functionals.

\subsection*{State of the art}

\subsubsection*{Molecular Dynamics}

DFT driven molecular dynamics, or {\it ab initio} MD (AIMD, also called Born-Oppenheimer MD and, misleadingly, quantum MD), has yielded numerous breakthroughs in our understanding of matter under extreme conditions.
For instance, equations of state (EOS) have been developed for several materials~\cite{FPEOS,Hu_PRB_2011,Hu_PRE_2014} and even phase diagrams have been constructed via
thermodynamic integration~\cite{Morales_PNAS_2009,Soubiran_PRL_2020,Chen_PRM_2023}. AIMD, whether with conventional KS or OF-DFT~\cite{Fiedler_PRR_2022,Lambert_2013}, has also led to significant advances in the calculation of ionic transport properties (diffusivity, viscosity,...)~\cite{Stanek_PoP_2024,Melton_PoP_2024,Blanchet_PoP_2024,Soyuer_MNRAS_2020}, which are highly important in hydrodynamic simulations. The application of OF MD to asymmetric mixtures~\cite{PhysRevE.93.063208,PhysRevE.95.063202,PhysRevE.100.033213} has shown its efficacy in that area. For the lightest elements, efforts have been made to include quantum nuclear effects through a path-integral molecular dynamics formalism~\cite{Geneste_PRL_2012,KangLuoRungeTrickey2020}.

\subsubsection*{Linear response theory}

In the KS context, it is possible to use the Kubo-Greenwood formalism to extract Onsager coefficients at finite frequency~\cite{Holst2011,dufty2017}. Note that use of mean-field orbitals such as KS in the exact Green-Kubo expression of
course omits any continuum, scattering state contributions (see Section 4). This method has been applied to numerous systems with great success, offering a deeper understanding of planetary interiors~\cite{french_ApJS_2012, french_PRE_2022, Witte_PRL_2017, Soubiran2018,Stixrude2020,21MRE-Liu,preising_ApJS_2023,24B-Liu} and determination of optical properties and reflectivities, which are an important diagnostic in laser shock experiments~\cite{soubiran2013, qi2015, Guarguaglini2021}. The extension to core electrons has enabled computation of X-ray absorption near-edges structure (XANES) spectra ~\cite{Jourdain2020} and has provided very robust estimates of the thermal conductivity~\cite{Stanek_PoP_2024}.

\subsubsection*{Exchange-Correlation Functional}

\,\,The simplest XC free-energy choice is a ground-state approximate functional evaluated with proper Fermi-Dirac occupations $f_i(\mu, T)$. This ``GSA'' is common in the literature.  Experience suggests that it works reasonably well up to $T_{\rm e} \approx 1$eV.  However, the GSA is an uncontrolled approximation atop whatever approximations were built into the functional. For the ``lower-rung''~\cite{PerdewSchmidtJacob} functionals, the local density approximation (LDA, dependent on $n({\mathbf r})$ alone) and generalized gradient approximations (GGA, dependent also on
$s:=|\nabla n|/2(3\pi^2)^{1/3}n^{4/3}$), the GSA $T$-dependence is unambiguous, depending on occupations only in the density. For meta-GGA functionals that depend on the KS kinetic energy density, the $T$-dependence is complicated by the occupations found in the kinetic energy density.

Beyond the GSA, the development of XC free-energy functionals has tracked the ground-state LDA, GGA, meta-GGA hierarchy~\cite{PerdewSchmidtJacob} with a considerable reliance on gradient expansion results in order to identify the critical temperature dependencies~\cite{Karasiev_electronic_structure_2025}. In short, the LDA rung has been developed in Ref.~\cite{KSDT2014PRL} based on the parametrization of quantum Monte Carlo data for the uniform electron gas (details in Sec.~\ref{section_15}, with detailed refinements using the same analytical framework but based on improved QMC data given in Ref.~\cite{Groth_PRL_2016}) and then in Ref.~\cite{KDT16}. The second rung of the ladder is a fully thermal GGA. The KDT16 GGA~\cite{KDT16} uses a generalization of the $s$ variable to a $T$-dependent reduced density gradient consistent with the $T>0$ gradient expansion for exchange, while two approximate GGA prescriptions seek to correct the zero-$T$ PBE XC~\cite{Perdew_PRL_1996} form with thermal XC effects at the LDA level, additively~\cite{Sjostrom_2014} or multiplicatively~\cite{kozlowski2023generalized}. As seen in the ground state, FT meta-GGAs come in two forms, orbital-dependent and orbital-independent. The more common orbital-independent forms typically depend on FT counterparts to the density, the dimensionless reduced density gradient $s$ and reduced density Laplacian $q:=\nabla^2n/4(3\pi^2)^{2/3}n^{5/3}$.  Very recently, a full-fledged free energy generalization of the orbital-dependent SCAN~\cite{Sun_prl_2015} meta-GGA functional has been given~\cite{HillekeKarasievTrickey2024}, as well as an orbital-independent approximate meta-GGA~\cite{KarasievMihalovHu2022} that uses an additive thermal correction with a perturbation-like approach.

\subsubsection*{Spectral separation}

In its basic form, the computational cost of KS-DFT scales as $N_B^3$ where $N_B$ is the number of non-trivially occupied states. Since $N_B$ grows
with $T$, for the high $T$ and pressures associated with WDM, this scaling can be prohibitive, especially for use of KS-DFT to drive AIMD (a main motivation of efforts at ML-based accelerations~\cite{ellis_accelerating_2021,fiedler_npj_23}).  
An important physical observation, however, is that the higher-energy occupied KS states are almost indistinguishable from plane waves and have very small occupation numbers $f_i \ll 1$. That motivates spectral separation methods~\cite{Zhang_POP_2016,Blanchet_pop_2020,White_prl_2020,22B-Hollebon,22B-Liu,SadighAbergPask2023,23E-Bethkenhagen}.

In the extended first-principles molecular dynamics (ext-FPMD) method~\cite{Zhang_POP_2016,Blanchet_pop_2020}, the KS orbitals above a
user-selected threshold energy are replaced with plane waves, using a constant potential shift to make the two spectra compatible. ext-FPMD has been used for systems ranging from condensed matter to WDM, hot dense plasmas~\cite{16B-GaoChang}, and for EOS calculations, e.g. deuterium~\cite{Zhang_POP_2016},~lithium deuteride~\cite{Zhang_POP_2016}, helium~\cite{Zhang_POP_2016}, aluminum~\cite{Zhang_POP_2016,Blanchet_cpc_2022}, beryllium~\cite{20PP-Yuzhi}, boron~\cite{22CPP-Blanchet}, poly-$\alpha$-methylstyrene~\cite{21B-Liu}, silicon~\cite{23PRB-Si}, and iron~\cite{Blanchet2025}. A somewhat more refined version of the same concept was introduced by Hollebon and Sjostrom~\cite{22B-Hollebon} who used a Thomas-Fermi density of states (with an effective local potential from their non-local pseudopotential) to determine the above-threshold contribution to the density. For warm dense Fe and LiD EOS, the scheme was more efficient (in that it converged faster to the ordinary $T>0$ KS results) than ext-FPMD. Recently, the physical reasoning underlying these spectral separation schemes has been put on a sound formal basis by Sadigh, $\mathring{\text{A}}$berg, and Pask~\cite{SadighAbergPask2023} by using spectral partitioning analysis, inviting computational exploitation.

A conceptually different route that exploits somewhat the same physics is stochastic DFT (sDFT). It samples the KS single-particle
spectrum via stochastic orbital selection and uses the Chebyshev trace method to achieve linear scaling with system size and number of occupied
bands~\cite{13L-Baer}. Recent advancements particularly relevant to WDM include FT sDFT~\cite{Cytter_Rabani_prb2018,hadad2024stochastic} and mixed stochastic -
deterministic DFT (mDFT) methods. The mDFT method integrates KS orbitals with stochastic orbitals, enhancing computational efficiency \cite{White_prl_2020, 22B-Liu,Sharma_PRE_2023,24MRE-ChenTao,25MRE-ChenTao}. Similar in spirit, the spectral quadrature (SQ) method~\cite{23E-Bethkenhagen} is a density-matrix based $\mathcal{O}(N)$ scheme for solving the KS equations especially suited to high-$T$ calculations.

\subsubsection*{Orbital-Free DFT}

As mentioned already, OF-DFT is the most direct realization of the KS mapping. The difference from conventional KS is that
the OF-DFT version of the universal functional requires \textit{both} an accurate approximation of the non-interacting kinetic energy $T_\textrm{s}[n]$ \textit{and} the usual approximate XC energy $E_\textrm{xc}[n]$ as explicit density functionals. For finite temperatures, these become free energy functionals. Older OF-DFT calculations had known flaws associated with Thomas-Fermi-von Weizsäcker (TFW)-type kinetic energy density functionals (KEDFs), particularly in strongly correlated and inhomogeneous systems~\cite{Wang_2000,Mi_Chem.Rev_2023}. Recently, advances in the development of accurate kinetic energy functionals  have been made~\cite{LKT2020,mi2023orbital,sun2023truncated,Karasiev_electronic_structure_2025}, including exploitation of machine learning (ML) techniques to construct more accurate  $T_\textrm{s}[n]$ approximations~\cite{seino2019semi,imoto2021order,ryczko2022toward,zhang2024overcoming,sun2024machine,sun2024multi}.

\subsubsection*{MLIPs}

Traditional MD simulations differ from AIMD in the use of potential energy surfaces (``force fields''). The cost-benefit trade-off is with the fidelity of the potential energy surface (PES). The application of ML interatomic potentials (MLIPs) to obtain high-quality PES has led to remarkable improvements in both MD accuracy and efficiency
~\cite{18L-deepmd, 18CPC-deepmd, 20CPC-Denghui, 2020GB-Jia,20JPCM-Qianrui,24MRE-ChenTao}. 
The MLIP models gave MD trajectories that well reproduced first-principles results for radial distribution functions, static structure factors, and dynamic structure factors. Similar MLIPs have been used to study the ion dynamics of shock-compressed copper~\cite{22B-Schorner}, the ion-ion structure factor of warm dense Al~\cite{22B-Schorner-Al}, laser-driven microscopic dynamics from solid to liquid~\cite{zeng2023full}, diffusion and viscosity of warm dense CH~\cite{Kumar_PoP_2024b}, shock Hugoniot of various elements~\cite{Kumar_PoP_2024a}, thermal conductivity of MgO at extreme conditions~\cite{qiu2025coupled}, and ultrafast heterogeneous melting of metals \cite{zeng2025ultrafastv}. While some of the ML potentials were trained to tackle a narrow range of temperatures and pressures, methods proposed to overcome the shortcoming include a $T$-dependent version of DPMD in which $T$ is treated as one of the neural network input parameters~\cite{20PP-Yuzhi} and the spectral neighbor analysis potential (SNAP). Both have increased applicability range, with the latter used to design a ML interatomic potential for Al from room temperature to about $60,000\,$K~\cite{kumar_transferable_2023}. Complementing these predictions of the total energy, a ML scheme for predicting the density of states (DOS) with a given set of local environments of atoms was proposed~\cite{22B-Zeng}.
MLIPs trained on AIMD results for modest-sized systems can fail to reproduce the AIMD
results for larger systems, see Ref.~\cite{karasiev2021} and Section 16 for further discussion.  
Discussion of ML for electronic structure more generally is in Section 24.

\subsection*{Challenges \& outlook}

\subsubsection*{T-Dependent XC}

Development of the thermal DFT formalism continues on many fronts. The generalized thermal adiabatic connection approach~\cite{harding_exchangecorrelation_2024} provides means to extract approximate XC entropy, an exact correction to capture explicit temperature dependence, as well as an LDA-like approximate correction~\cite{aguilar-solis_finite-temperature_2025}. Moving toward new finite-$T$ XC approximations, FT generalized KS presents an overarching formalism for recasting the free energy as a partially interacting functional of a temperature-dependent statistical operator. This provides firm footing for existing hybrid approximations and also drives adiabatic connection--driven hybrid development, such as the two-legged approximation~\cite{harding_finite-temperature_2025}. Further work spans the field, pursuing ML avenues~\cite{callow_physics-enhanced_2023,fiedler_machine_2023}, harnessing density-to-potential inversions~\cite{martinetto_inverting_2024}, and including ``higher-rung" effects from the condensed matter and quantum chemistry communities~\cite{crisostomo_exchange-correlation_2024,van_benschoten_removing_2024,bao_exploring_2025}. On the OF XC front, finite-$T$ extensions of zero-$T$ advances~\cite{francisco_reworking_2023,giarrusso_modeling_2025}, via the Pauli potential~\cite{karasiev_framework_2024}, and via semiclassical approaches~\cite{redd_investigations_2024,okun_orbital-free_2024,daas_large-z_2023} may soon appear on the horizon.

\subsubsection*{Pre-trained Model}

The recent development of atomic-scale pre-trained models, such as MACE~\cite{24arxiv-Batatia}, DPA2~\cite{24npj-Zhang}, M3GNet~\cite{24npj-Qi}, ALIGNN~\cite{2021-choudhary}, and CHGNeT~\cite{2023-Deng} represents a promising frontier in computational materials science, offering a transformative alternative to traditional DFT and MD methods for studying high-temperature and high-pressure systems. These models, trained on vast datasets of atomic configurations and their corresponding energies and forces, can capture complex interatomic interactions with remarkable accuracy and efficiency. However, challenges remain, such as ensuring transferability across diverse chemical environments and addressing the scarcity of high-quality training data. Despite these hurdles, the integration of advanced ML techniques offers the possibility that atomic-scale pre-trained models could eventually surpass DFT and MD in simulating materials at extreme conditions, enabling the exploration of previously inaccessible regimes in condensed matter physics and warm dense matter.

\subsubsection*{Heterogeneous Parallel Computing}

\,\,Its advent has significantly enhanced the performance of both DFT and MD simulations. Recent recognition of these advances 
includes the 2020 Gordon Bell (GB) Prize for the development of Deep Potential Molecular Dynamics~\cite{2020GB-Jia}, which 
simulated 100 million atoms with {\it ab initio} precision.
The 2023 GB Prize was awarded for a framework combining quantum many-body methods with DFT to 
to enable simulations of quasicrystals and 
metallic alloys with a performance of 659.7 PFLOPS. 
~\cite{23GB-Das}. In 2024, the GB Prize recognized a breakthrough in biomolecular-scale AIMD using MP2 potentials~\cite{24GB-Stocks}, with a 
performance of simulating over one million electrons and scaling to 1 EFLOP/s. 
These achievements underscore the transformative impact of heterogeneous computing on computational science, paving the way for even more ambitious simulations of WDM.

\newpage 
\clearpage

\section{Time-Dependent Density Functional Theory}\label{section_04}
\author{Andrew Baczewski$^{1}$, Alina Kononov$^{1}$, Zhandos Moldabekov$^{2,3}$, Alexander J. White$^{4}$}
\address{
$^1$Sandia National Laboratories, Albuquerque, 87185 New Mexico, United States \\
$^2$Center for Advanced Systems Understanding (CASUS), D-02826 G\"orlitz, Germany \\
$^3$Helmholtz-Zentrum Dresden-Rossendorf (HZDR), D-01328 Dresden, Germany \\
$^4$Los Alamos National Laboratory, Los Alamos, 87545 New Mexico, United States
}

\subsection*{Introduction}

\begin{figure}[b]
    \includegraphics[width=\columnwidth,trim={0 0.5cm 0 0.5cm},clip]{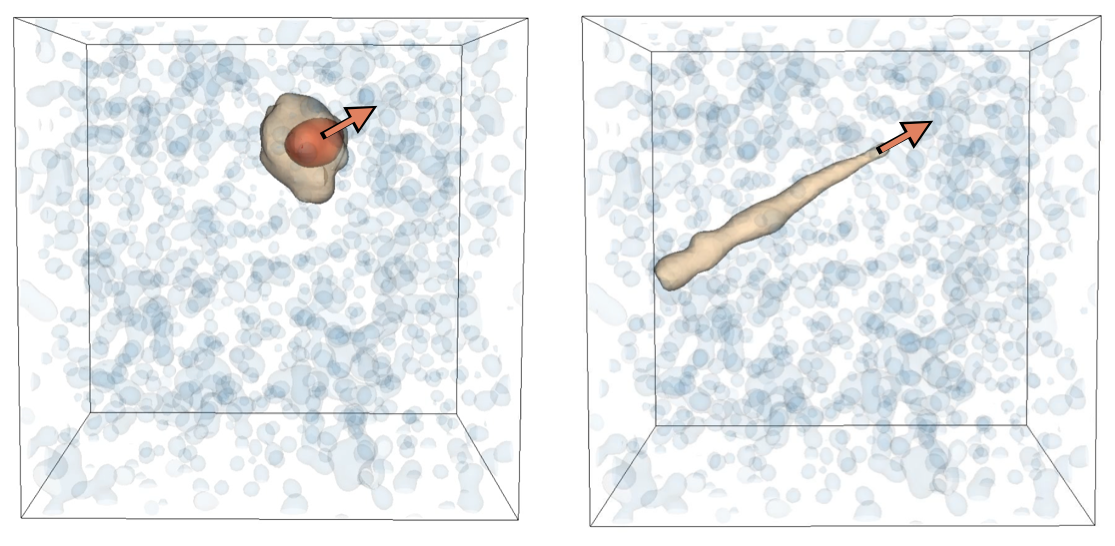}
    \caption{Electronic density response to a 300 keV (left) and 30 MeV (right) proton traversing warm dense deuterium in a TD-DFT stopping power simulation. Arrows indicate the proton's direction of motion, and red and yellow isosurfaces bound regions where the density exceeds its initial unperturbed values (blue). TD-DFT captures both localized and extended electronic excitations created by ions in both velocity regimes.
    \label{fig:tddft}
    }
\end{figure}

Time-dependent density functional theory (TD-DFT) enjoys computational efficiency similar to that of conventional DFT for modeling extended quantum many-body systems (see Sec.~\ref{section_03}) while enabling many calculations beyond DFT's formal scope. In particular, TD-DFT rigorously accesses not only response functions such as the optical conductivity~\cite{bertsch2000real} and dynamic structure factor~\cite{maddocks1994ab} but also quantities related to non-equilibrium electron-ion dynamics such as stopping powers~\cite{correa2012nonadiabatic} (see Fig.~\ref{fig:tddft}). Crucially, these quantities often directly relate to experimental signatures like reflectivity, x-ray Thomson scattering cross sections, and beam energy-loss respectively. Although TD-DFT is formally exact, practical calculations require approximations that often suffice to compare favorably with WDM experiments~\cite{Bonitz_POP_2024, Moldabekov_JCTC_2023}. In addition to improving these approximations, extending feasibility to new regimes and observables remain active research areas.

Like conventional DFT, TD-DFT formally relies on a correspondence between the exact time-dependent electronic density of a quantum many-body system and the time-dependent one-body potential that generates it from a given initial wavefunction~\cite{Runge_Gross_1984,van1999mapping}. 
This mapping facilitates a dramatic computational simplification: rather than storing and operating on the entire time-dependent many-body wavefunction, the same information can be accessed (in principle) from one of its local observables.
Operationally, equations of motion for the density itself can be constructed and solved instead of the many-body Schr\"{o}dinger equation.
Common formulations introduce a fictitious auxiliary system of non-interacting KS orbitals as a proxy for constructing the density.

While conventional DFT relies on similar constructions, the central theoretical advantage that TD-DFT provides is information about experimentally-relevant observables beyond the ground or equilibrium state. Conventional DFT is often used outside of its formal domain of validity, \emph{e.g.}, in computing observables from matrix elements in the basis of KS orbitals. In contrast, TD-DFT provides formal scaffolding for rigorously defining observables away from equilibrium and/or including changes to observables due to electron density reconfiguration in response to a perturbation, \textit{i.e.}, the full linear response and nonlinear response (see Sec.~\ref{section_05}). TD-DFT also naturally captures collective excitations, namely plasmons, which would otherwise need to be manually incorporated into conventional DFT through an auxiliary model (\emph{e.g.}, the Mermin dielectric function~\cite{plagemann2012dynamic,Schoerner_PRE_2023}).

In this section, we begin by reviewing the various formulations of TD-DFT and then describe its state-of-the-art applications in WDM science, focusing on calculations that relate to experimental observables. We then conclude by reviewing the theory's open problems most relevant to high-energy-density physics and prospects for interfacing with lower levels of theory, higher levels of theory, and experiments.

\subsection*{State of the art}

\subsubsection*{Real-time dynamics}
The most general, real-time (RT) formulation of TD-DFT can model the non-equilibrium dynamics of a quantum many-body system subject to arbitrary perturbations. The associated equations of motion are the TD-KS equations,
\begin{equation}\label{eq:KS_tddft}
\imath\hbar\frac{\partial}{\partial t}\phi_{k}({\bf r},t)=\left(-\frac{\hbar^2}{2m}\nabla^2+v_{\mathrm{eff}}\left[n\right]({\bf r},t)\right)\phi_{k}({\bf r},t).
\end{equation}
The one-body potential $v_{\mathrm{eff}}$
contains the total external potential (due to ions and any imposed perturbations) and density-dependent Hartree and XC terms (see Eq.~\eqref{eq:dft:veff}). The unknown KS orbitals ($\phi_k$) are typically initialized through a conventional finite-temperature DFT calculation, in which the electronic density is $n({\bf r},0)=\sum_{k} f_k(\mu,T_\mathrm{e}) |\phi_{k}({\bf r},0)|^2$, where $f_k$ obeys the Fermi-Dirac distribution for chemical potential $\mu$ and electronic temperature $T_\mathrm{e}$. The occupations $f_k$ are fixed throughout a time evolution that is driven by 
a time-dependent external potential that might model an optical or x-ray field, a fast ion, or some other perturbation.

\subsubsection*{Stochastic variants}
At high temperatures or for many electrons, the high-energy tail of the Fermi distribution demands many KS orbitals.
Similar to DFT, stochastic projection can approximately reduce the rank of the KS density matrix. The stochastic orbitals can be propagated and treated as if they were \textit{nonorthogonal} KS orbitals, with orbital weights defined by their normalization factors~\cite{Gao_jcp_15}. This approach reduces the number of propagated orbitals at the expense of stochastic variation in the resulting observables, which can be difficult to tolerate for long-time dynamics. The shape of the Fermi-Dirac function lends itself to a mixed stochastic-deterministic treatment, which explicitly retains high-occupancy KS orbitals while stochastically approximating the rest~\cite{White_prl_2020,White_JPCM_2022}.  

\subsubsection*{Orbital-free methods}
Although OF formulations offer another efficient alternative, extending equilibrium OF-DFT to the time domain is non-trivial~\cite{Michele_PRB_2021,Jiang_prb_2022}. The adiabatic approximation, while relatively successful for KS-TD-DFT XC functionals and common for OF-DFT kinetic energy functionals~\cite{Karasiev_electronic_structure_2025, Mi_Chem.Rev_2023}, leads to singularities in the non-interacting electronic response within OF-TD-DFT~\cite{Neuhauser_jcp_11}. This problem necessitates at least a semi-local in time formulation of OF-TD-DFT, which introduces a current-dependent potential~\cite{White_PhRvB_2018,Jiang_prb_2022} that breaks time-reversal symmetry but approximates relaxation mechanisms. 

\subsubsection*{Linear-response (LR) TD-DFT}
In the LR regime, LR-TD-DFT avoids the real-time propagation of orbitals by solving for dynamic response properties within the frequency domain. This approach can outperform RT-TD-DFT for small systems or relatively small perturbation frequencies. The conventional approach to LR-TD-DFT solves a Dyson equation for the density response function and requires many low-occupancy bands~\cite{LRT_GPAW2}, thus involving costly inversions of large matrices. Alternatively, the Liouville–Lanczos approach to LR-TD-DFT~\cite{Rocca_JCP_2008} overcomes this limitation through a recursive algorithm~\cite{moldabekov2025applyingliouvillelanczosmethodtimedependent}.

\subsubsection*{Observables}
The principle behind \emph{any} TD-DFT calculation is that various observables --- from the dynamic structure factor (DSF) to the force on a fast ion --- are formally a functional of the time-evolved electronic density or current density. One of the original motivations for applying TD-DFT in WDM science was to calculate the DSF~\cite{baczewski2016x}, relevant to XRTS measurements, without defining observables in terms of matrix elements in the KS basis~\cite{plagemann2012dynamic}. While KS matrix elements seem to deliver a good approximation to the average momentum of orbitals within $k_B T$ of the chemical potential (and thus quantities like DC conductivities), they must be augmented by a model dielectric function to capture the plasmonic excitations probed by XRTS measurements in the collective regime~\cite{plagemann2012dynamic,Schoerner_PRE_2023}. Generally, TD-DFT provides satisfactory accuracy for plasmonic and bound-free features in the XRTS spectrum of WDM~\cite{baczewski2016x, Mo_prl_2018, Mo_prb_2020}.

TD-DFT can also access optical properties like the dynamic conductivity. 
RT formulations do this by applying Ohm's law 
to the current response to a uniform external electric field \cite{kononov2025real}. 
For cases obeying a Drude-like frequency dependence, TD-DFT conductivity predictions tend to agree with the more common DFT-based Kubo-Greenwood (DFT-KG) formalism~\cite{andrade_negative_2018, ramakrishna2024electrical}. However, TD-DFT is expected to exhibit more favorable computational properties than DFT-KG, and its formal grounding may modify predictions for more complex systems, \textit{e.g.}, those featuring a partially filled $d$-band~\cite{baczewski2021predictions, Moldabekov_Omega_2024}.

In addition to LR functions, RT-TD-DFT can compute nonlinear properties like electronic stopping power~\cite{correa_calculating_2018}: the energy transfer rate between an energetic charged particle and electrons in a medium (see Fig.~\ref{fig:tddft}). Electronic excitations beyond the Born-Oppenheimer approximation of DFT-MD dominate stopping powers for projectile velocities comparable to typical electron velocities. Obtaining experimentally-relevant results requires careful consideration of the projectile's trajectory, special attention to finite-size errors, and/or averaging results over multiple calculations~\cite{White_JPCM_2022, Kononov2023, kononov_2024}. Although TD-DFT stopping power predictions offer the best agreement with experimental data among a variety of models~\cite{Malko_NatCom_2022, Ding_PRL_2018}, differing trends under increasing temperature remain unreconciled. In conditions accessed by WDM experiments to date, OF-TD-DFT and KS-TD-DFT agree closely~\cite{Malko_NatCom_2022, Ding_PRL_2018},
though OF-TD-DFT accuracy degrades for slower ions~\cite{White_JPCM_2022}. Simulations of high-energy \textit{electron} irradiation further access the non-equilibrium electron mean-free path, a critical quantity for direct-drive ICF~\cite{Nichols_pre_2023, Nichols_pop_2024}.

\subsection*{Challenges \& outlook}

\subsubsection*{Computational cost}
Despite TD-DFT's successes, several challenges remain. First, the large computational expense of TD-DFT calculations often limits feasible regimes and system sizes, precludes higher-accuracy approximations, and hinders more widespread application for WDM research. Most TD-DFT simulations of WDM use local or semi-local adiabatic approximations and the plane-wave pseudopotential approach with frozen core electrons for high-Z materials, placing high-energy properties like the DSF K-edge and x-ray opacity out of reach. Improved GPU and hybrid GPU-CPU utilization~\cite{andrade2021inq} could significantly improve TD-DFT simulation throughput. More efficient implementations of TD-DFT in adaptive-mesh finite-element or full-potential linearized augmented plane-wave codes could provide an alternative to pseudopotentials. Stochastic methods may play a role, particularly for extending feasibility to high temperatures. Each of these strategies for improving efficiency could expand TD-DFT usage for WDM simulations.

\subsubsection*{XC functionals}
A related challenge concerns the central approximation of TD-DFT: the XC functional. Hybrid functionals, which include some fraction of exact exchange, generally deliver higher accuracy in conventional DFT, but their widespread use within TD-DFT requires faster exact-exchange algorithms~\cite{Leveillee_jctc_2025}. Beyond incorporating XC advances from conventional DFT, non-adiabatic or memory effects are difficult to systematically incorporate because they introduce a dependence 
on the density's time history. Lack of memory effects in practical TD-DFT calculations may explain the failure to capture physical thermalization behavior~\cite{kononov2022electron}. More broadly, potential inversion analyses show that steps are a generic feature of non-locality that will be difficult to integrate into TD-DFT~\cite{elliott2012universal}. Non-locality can be incorporated through local approximations in TD-current-DFT~\cite{nazarov2007including}, though applications in WDM have yet to be demonstrated. Other promising approaches for further improving TD-DFT's foundations and practical accuracy include using features from higher levels of theory that more naturally capture non-locality and formalizing thermal TD-DFT~\cite{pribram}.

\subsubsection*{Extending applicability}
Beyond challenges in improving accuracy and efficiency, open problems also remain in extending the applicability of TD-DFT to new observables. For example, TD-DFT thermal conductivity calculations have not been realized because of the difficulty in explicitly representing a thermal gradient. 
Similarly, the influence of external magnetic fields on WDM transport properties --- relevant to magnetic inertial fusion concepts --- remains unexplored within TD-DFT. However, a recently developed TD-DFT framework for nonadiabatic Born effective charges could open new avenues for investigating electronic optical response and extracting average ionizations in mixtures~\cite{PhysRevLett.134.095102}. A similar extension connecting electron-ion couplings to a dynamical property could also be illuminating. Finally, TD-DFT could enable opacity calculations for many-atom systems, but the lack of thermal static correlation (due to the Mermin-KS initialization) and dynamic correlation (due to the adiabatic XC approximation) need to be addressed. 

\subsubsection*{Interfacing with other techniques}
Despite remaining challenges, TD-DFT nonetheless captures degeneracy, coupling, and density effects that often challenge more efficient plasma models.
Thus, given the continued scarcity of experimental data, TD-DFT has emerged as a useful benchmark for guiding improvements to other computational methods~\cite{Stanek_PoP_2024}. For example, dynamic response functions computed with TD-DFT have recently constrained electron-ion collision frequencies entering the Mermin dielectric model for free-electron response~\cite{Schoerner_PRE_2023, Hentschel_pop_2023, Hentschel_PhysPlasmas_2025}.
Further opportunities to augment lower-accuracy models could involve characterizing nonlinear behavior \cite{kononov2025nonlinear} that may help explain discrepancies reported in alpha-particle stopping powers~\cite{Stanek_PoP_2024}.

However, a central tension of TD-DFT is the difficulty of improving XC approximations in the absence of high-accuracy reference data, while WDM pushes the method into regimes where it is often the most accurate feasible method. Recent quantum Monte Carlo (QMC) studies begin to test the limits of TD-DFT for certain observables in warm dense UEG and low-Z systems~\cite{moldabekov2025applyingliouvillelanczosmethodtimedependent, Moldabekov_PRR_2023}. As computing hardware and algorithms continue to improve, other observables and higher-Z systems may become tractable for QMC (see Sec.~\ref{section_02}) or non-equilibrium Green's functions (see Sec.~\ref{section_06}), exposing the weaknesses and highlighting the successes of TD-DFT. Eventually, practically exact quantum simulation algorithms on quantum computers~\cite{proctor2025benchmarking} might also benchmark TD-DFT~\cite{rubin2024quantum}. Recent work showed that some quantum simulation algorithms are asymptotically superior to \emph{any} TD-DFT-like mean-field approach~\cite{babbush2023quantum}, with encouraging prospects for non-equilibrium simulations in the WDM regime. Higher-accuracy methods might someday serve as a new gold standard for WDM modeling.

In addition to interfacing with other levels of theory, exciting opportunities have emerged to connect TD-DFT with experimental efforts.
Continued progress in refining XRTS resolution~\cite{gawne2024ultrahigh} not only enhances prospects for detailed validation of TD-DFT predictions, but also opens new opportunities to exploit TD-DFT's strengths for improved interpretation of measured spectra. Similarly, advances in lower-velocity short-pulse proton sources~\cite{Malko_NatCom_2022} along with simultaneous sample characterization may enable stopping power experiments to distinguish among different theoretical models (see Sec.~\ref{section_11}). Recently developed THz techniques for accessing electrical conductivity~\cite{Ofori-Okai_POP_2024} further extend possibilities for directly interfacing TD-DFT with experiment (see Sec.~\ref{section_14}). The confluence of computational, theoretical, methodological, and empirical opportunities promises further exciting TD-DFT-enabled advances in WDM in the near future.

\newpage 
\clearpage

\section{Density Response Theory}\label{section_05}
\author{Panagiotis Tolias$^1$, Jan Vorberger$^2$}
\address{
$^1$Space and Plasma Physics, Royal Institute of Technology (KTH), SE-10044 Stockholm, Sweden \\
$^2$Helmholtz-Zentrum Dresden-Rossendorf (HZDR), D-01328 Dresden, Germany 
}

\subsection*{Introduction}

Response theory can be formulated for any pair of Hermitian operators $\hat{A},\,\hat{B}$, but we shall focus on the density-density version with $\hat{A}=\hat{\rho}(\boldsymbol{r}),\,\hat{B}=\hat{\rho}(\boldsymbol{r}^{\prime})$~\cite{quantum_theory}. The response to weak time-dependent perturbations can be calculated via response functions which describe the $\delta{n}(\boldsymbol{r},t)$ change in the 
expectation value of $\hat{\rho}(\boldsymbol{r})$ due to an applied potential $V_{\mathrm{ext}}(\boldsymbol{r}^{\prime},t^{\prime})$ that couples to $\hat{\rho}(\boldsymbol{r}^{\prime})$. Assuming that $\delta{n}(\boldsymbol{r},t)$ can be expanded in a power series of the perturbation strength, one obtains the formal definitions of the nonlinear response functions (linear, quadratic, cubic, ...)~\cite{vorberger2024greensfunctionperspectivenonlinear}. For infinitesimal perturbations, the response is described solely by the linear term. The Kubo formula, Lehmann representation and Kramers-Kronig relations are the pillars of linear response theory (LRT), which lead to profound exact results such as the detailed balance relation, fluctuation-dissipation theorem, frequency moment sum rules and stiffness theorem~\cite{quantum_theory,IchimaruBook,PinesBook}.

Density responses are rich in physics and comprise the basis for WDM studies of structure, transport and dynamics~\cite{Bonitz_POP_2024} with an array of advanced formalisms dedicated to their evaluation~\cite{Bonitz_POP_2020,Dornheim2023_density_response}. In particular, the quest for an accurate description of equilibrium and non-equilibrium WDM properties is closely connected with density response functions, since they constitute key ingredients in the determination of x-ray scattering spectra~\cite{falk_wdm,Glenzer_revmodphys_2009}, stopping powers~\cite{Malko_NatCom_2022}, relaxation rates~\cite{Vorberger_PRE_2010}, equations of state~\cite{kremp2005quantum}, and conductivities or opacities~\cite{Bergermann_PRB_2023,reinholz_PRE_2015,PRE_Hu_2014}.

\subsection*{State of the art}

\subsubsection*{Linear response}

The traditional recipe of incorporating correlations and quantum effects in the density response functions reached its climax in the combination of a Mermin-type response with local field corrections (LFCs) so that electron-ion collisions and electron XC effects are included~\cite{Fortmann_PRE_2010}. The problem of the determination of the LFC of the UEG at arbitrary temperatures and densities has been nearly solved by parameterizing static LFCs extracted from PIMC simulations~\cite{Dornheim_JCP_2019,Dornheim_PRL_2020_ESA,Hamann_PRB_2020,Dornheim_PRB_2021_ESA,Dornheim_EPL_2024}. 
The restriction to the static $\omega=0$ limit constitutes an approximation  that only introduces a minor error in many situations of interest. The necessary collision frequencies are nowadays obtained using the Kubo-Greenwood (KG) formula featuring Kohn-Sham DFT orbitals~\cite{plagemann2012dynamic}. It has become possible to check the validity of the UEG approximation, inherent in many response function calculations, against full PIMC simulations without nodal restrictions for light elements such as hydrogen and beryllium~\cite{Dornheim_MRE_2024,Boehme2022,Boehme2023}. Thus, truly first principle calculations of structural properties of WDM have become possible~\cite{Dornheim_JCP_2024}. The extracted material specific static electron-electron, electron-ion, and ion-ion LFCs constitute invaluable benchmark tests for TD-DFT and provide new insights in WDM microphysics.

In the spirit of the direct perturbation method of PIMC simulations~\cite{bowen2}, standard DFT calculations featuring an external static field enable the extraction of the static density responses~\cite{Moldabekov_PPNP_2024,Moldabekov_JCP_2021,Moldabekov_PRB_2022,Moldabekov_prb_2023, Moldabekov_Electronic_Structure_2025, Moldabekov_JCP_averaging_2023}, leading to an improved capability to benchmark XC functionals for DFT~\cite{Moldabekov_JPCL_2023, Moldabekov_jcp_2023_hybrid} and kernels for TD-DFT~\cite{Moldabekov_JCTC_2023} against PIMC. Nowadays, the optical limit of linear response functions is routinely accessed using the KG approach~\cite{Bergermann_PRB_2023,PRE_Hu_2014}, whose combination with the perfect screening sum rule allows the determination of the charge state from DFT simulations~\cite{Bethkenhagen_2020}. Recent efforts to extract it with the PIMC method should be noted~\cite{dornheim2024unraveling}. This quantity is irrelevant for DFT or QMC simulations and cannot be rigorously defined in WDM conditions, but it is intuitive and useful in simplified approaches. In density functional perturbation theory, phonon spectra and electron-phonon interactions can be calculated at the linear response level. They are exceedingly useful for the description of energy and temperature relaxation in WDM after laser irradiation (see Sec.~\ref{section_23})~\cite{Waldecker_PRX_2016,Akhmetov_2023,Zhang_2024}. Moreover, the generalized friction force is a promising concept for the description of relaxation~\cite{Simoni_2019}. In addition, non-equilibrium states identified by deviations of the Wigner distribution from the Fermi or Boltzmann forms might be subject to additional external drivers in the linear regime~\cite{Vorberger_PRE_2018}. Finally, inhomogeneities in the pristine material or laser irradiation often necessitate the consideration of non-uniform densities in the linear response~\cite{Kozlowski_2016,Beuermann_PRE_2019}. 

The self-consistent dielectric formalism combines exact LRT results with an approximate LFC expression as a functional of the static structure factor (SSF)~\cite{SingwiTosi_Review,Ichimaru_PhysReports_1987}. At finite temperatures, the formalism is more conveniently formulated in the imaginary Matsubara frequency domain (structural properties) and the imaginary-time domain (dynamic properties)~\cite{Tolias_JCP_Matsubara_2024,stls}. For the warm dense UEG, the semi-empirical effective static approximation (ESA) yields near-exact results for many properties~\cite{Dornheim_PRL_2020_ESA,Dornheim_PRB_2021_ESA}. Progress in the strongly coupled UEG is highlighted by the semi-classical hypernetted chain (HNC)~\cite{Tanaka_JCP_2016,Dornheim_PRB_2020} and integral equation theory (IET)~\cite{Tolias_JCP_2021,Lucco_Castello_EuroPhysLett_2022} schemes and their quantum counterparts (qHNC, qIET) that feature a dynamic LFC~\cite{Tolias_JCP_2023}. The IET predicts near-exact interaction energies~\cite{Tolias_JCP_2021} and the qIET very accurate SSFs away from Wigner crystallization~\cite{Tolias_JCP_2023}.

The frequency moment sum rules constitute the building blocks of the method of moments for dynamic responses. The truncated Hamburger problem refers to the reconstruction of any spectral function, e.g. the loss function $\mathcal{L}(\boldsymbol{q},\omega)$, from a finite set of its frequency moments based on the Nevanlinna theorem~\cite{tkachenko_book}. The determination of the unknown parameter function requires external input and additional assumptions. In classical dense plasmas, accurate five moment versions have been formulated based on a static approximation combined with either external knowledge of the static limit of $S(\boldsymbol{q},\omega)$~\cite{Arkhipov_2010} or the empirical observation that $S(\boldsymbol{q},\omega)$ exhibits broad extrema at $\omega=0$~\cite{Arkhipov_2017}. For the warm dense UEG, an accurate nine moment version has been recently devised based on the estimation of $\mathcal{L}$ frequency moments up to the eighth order from a two-parameter Shannon information entropy maximization procedure~\cite{Filinov_PRB_2023}. Finally, it is noted that high-order positive DSF frequency moments have been extracted from PIMC simulations exploiting their connection with the high-order ITCF $\tau$-derivatives at $\tau=0$~\cite{Dornheim_moments_2023}.

\subsubsection*{Nonlinear response}

Given the connection between nonlinear density responses and multi-point correlation functions or structure factors~\cite{vorberger2024greensfunctionperspectivenonlinear}, high-order correlations are explicit in nonlinear density response formalisms instead of being subsumed in LFCs. Thus, nonlinear response functions contain a wealth of non-trivial information about the system and are responsible for numerous effects~\cite{Sitenko}. Nonlinear versions of the Kubo formula~\cite{kubo1957statistical,Peterson_1967,Watanabe_2020,Bradlyn_2024}, fluctuation-dissipation theorem~\cite{Golden_1972,Golden_1982,Kalman_1987,Golden_2018}, sum rules~\cite{Golden_1975,Golden_1985,Rommel_1996} and stiffness theorem~\cite{Moroni_1992,Moroni_1995,Dornheim_2023stiff} have a long history and are still being developed. Nonlinear contributions to the stopping power, see the Barkas $Z^3$ effect~\cite{Barkas_1963,Ashley_1972}, have served as a main motivation for the study of nonlinear effects~\cite{Hu_1988,Pitarke_1995,Bergara_1997}. Ideal quadratic and cubic response functions have been utilized in the study of nonlinear phenomena in graphene and the ground state UEG~\cite{Hendry_2010,Mikhailov_2012,Mikhailov_2014}. Based on these early results, it has been shown recently that the non-interacting limit of arbitrary order diagonal harmonic density responses can be generally expressed as the weighted sum of linear ideal density responses evaluated at all multiple harmonics~\cite{Tolias_2023}. 

\begin{figure}
    \centering
    \includegraphics[width=\linewidth]{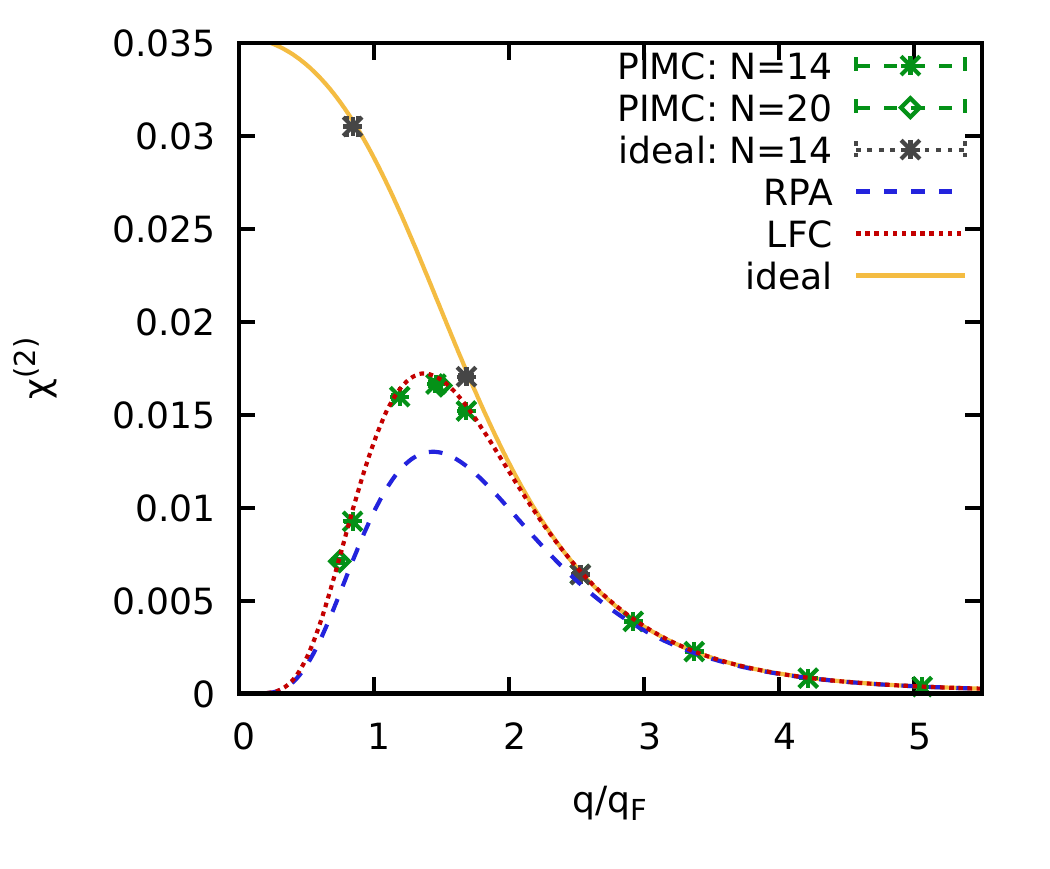}
    \caption{The quadratic static response function at the second harmonic $\chi^{(2,2)}(q,0)$ for the warm dense UEG at $r_s=2$ and $\Theta=1$. The green stars (green diamonds) show PIMC results for $N = 14\,(N = 20)$ electrons, and the grey stars show PIMC data for ideal fermions. The dashed blue and dotted-red lines show theoretical results within the RPA and adopting the static LFC extracted from PIMC~\cite{Dornheim_JCP_2019}. The solid-yellow line shows the ideal response function $\chi^{(2,2)}_0(q)$. (Reproduced from Ref.~\cite{Dornheim_PhysRevRes_2021}).}
    \label{fig:resp_1}
\end{figure}
Nonlinear response functions of the warm dense UEG have been extracted from PIMC simulations with the direct perturbation method leading to  unambiguous results for static perturbations~\cite{Dornheim_PRL_2020}. Static nonlinear response functions have been independently obtained from PIMC simulations with the ITCF equilibrium method, here based on the nonlinear imaginary-time FDT~\cite{Dornheim_JCP_ITCF_2021}, with the advantage that the complete wavenumber dependence is obtained from a single simulation~\cite{Groth_PRB_2019}. Combined with a solid understanding of the equations of motion of the nonlinear response functions~\cite{vorberger2024greensfunctionperspectivenonlinear}, these PIMC studies enabled a characterization of the static nonlinear responses in terms of LFCs, see Fig.~\ref{fig:resp_1}~\cite{vorberger2024greensfunctionperspectivenonlinear,Dornheim_PhysRevRes_2021,Dornheim_JPSJ_2021}. Mode coupling and the excitation of higher harmonics have been demonstrated theoretically in the UEG~\cite{Dornheim_CPP_2022}. Spin resolved nonlinear responses and their cross correlation have been studied~\cite{Dornheim_CPP_2021}. DFT calculations of the UEG nonlinear response have also been successful~\cite{Moldabekov_JCTC_2022,moldabekov2025dftresponse}. Nonlinear effects have even emerged in PIMC studies of warm dense hydrogen, but a deep analysis is pending~\cite{Dornheim_MRE_2024,Boehme2022}.

\subsection*{Challenges \& outlook}

\subsubsection*{Linear response}

\begin{figure}
    \centering
    \includegraphics[width=\linewidth,clip=true,trim={0 2cm 0 3cm}]{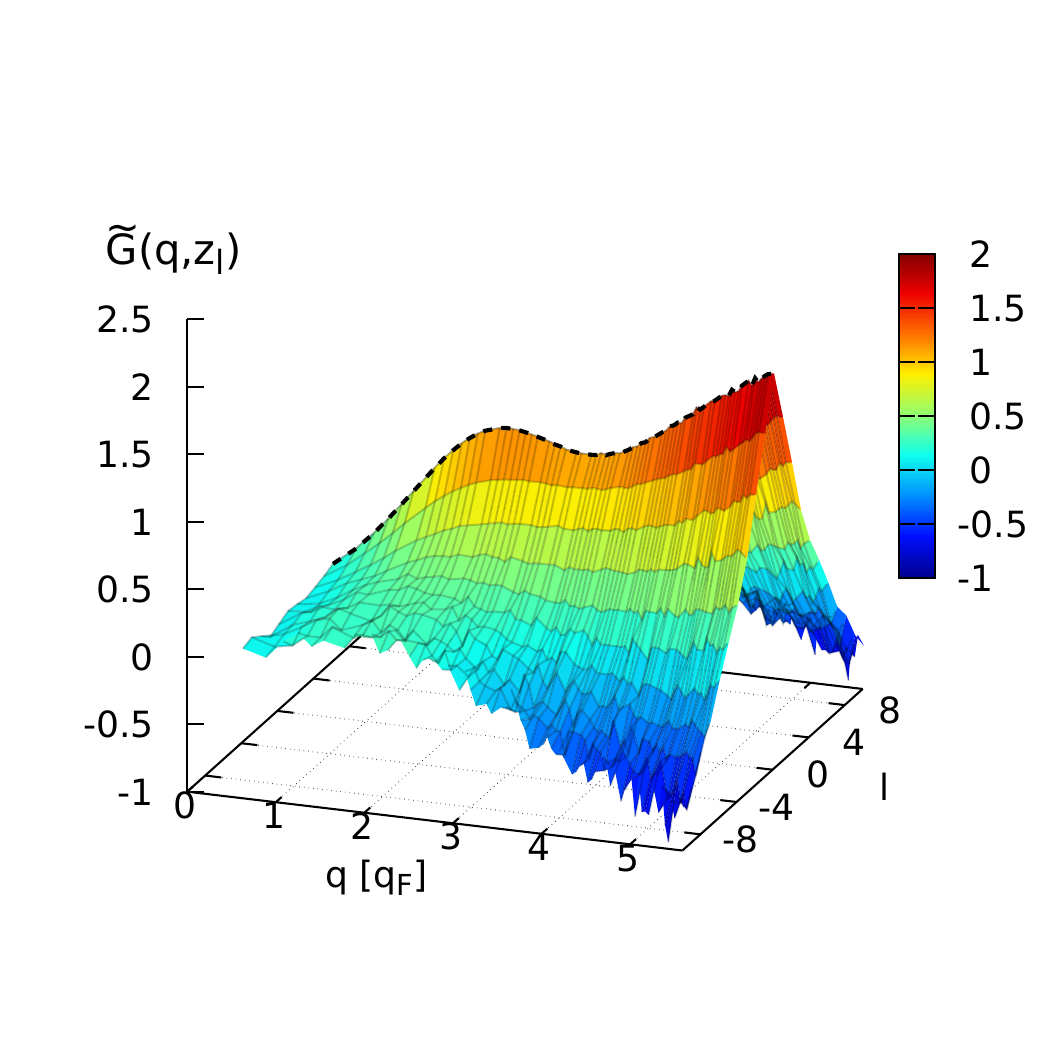}
    \caption{Ab initio PIMC results for the dynamic Matsubara local field correction of the UEG at $r_s=20$ and $\Theta=1$, with $l$ the Matsubara frequency order. (Reproduced from Ref.~\cite{Dornheim_PhysRevB_2024}.)}
    \label{fig:resp_2}
\end{figure}
The missing step towards the exact description of the linear density response of the warm dense UEG is the PIMC extraction of the dynamic LFC, which entails the numerical inversion of a two-sided Laplace transform (ITCF-DSF connection)~\cite{Dornheim_PRL_2018,Groth_PRB_2019}. This analytic continuation is an ill-posed problem with respect to PIMC error bars~\cite{Jarrel_PhysReports_1996,Hui_PhysReports_2023} and novel techniques are required for reliable robust inversion. For instance, the recently derived Fourier-Matsubara series expansion of the ITCF can be inverted to directly yield the Matsubara density response and Matsubara LFC, see also Fig.~\ref{fig:resp_2}~\cite{Tolias_JCP_Matsubara_2024,Dornheim_PhysRevB_2024}. The latter exhibits a smooth behavior in reciprocal space for any state point and could be parameterized~\cite{Dornheim_EPL_2024,Dornheim_PhysRevB_2024,dornheim2024shortwavelengthlimitdynamic}. Moreover, the Nevanlinna analytic continuation method can be used to convert Matsubara LFCs to dynamic LFCs as exemplified in the field of finite temperature Green's functions~\cite{Fei_PRL_2021,Khodachenko_2024}. The availability of a dynamic LFC parametrization would make temperature dependent and fully dynamic kernels available for linear response TD-DFT applications, while the DSF knowledge from PIMC could be used to benchmark XC functionals and kernels used in DFT. Further, since PIMC simulations without nodal restrictions are now possible for low-Z elements~\cite{Dornheim_JCP_2024}, their material and stoichiometry specific linear response functions and LFCs could be obtained leading to new insights as well as to DFT-benchmarking possibilities.

The optical limit of response functions allows to obtain absorption coefficients, opacities, and thermal and electric conductivities. For the critical evaluation of the approximations inherent in the widely used KG formula (lack of electron-electron collisions, use of KS orbitals, see Secs.~\ref{section_03} \& \ref{section_04}) and the quality of existing databases, efforts need to be directed into extending PIMC results to the long-wavelength limit. This can be done by employing larger super-cells and using the $\xi-$PIMC variant~\cite{Dornheim_JPCL_2024} or by utilizing PIMC approved XC functionals and kernels in LR-TDDFT. Another very fruitful research avenue is offered by the imaginary time DFT~\cite{Flamant_2019,McFarland_2021}. Since ITCFs constitute a natural PIMC observable, it seems prudent to insert ITCF information directly into DFT (if possible), without the detour of frequency space, thus circumventing the analytic continuation problem.

\subsubsection*{Non-linear response}

A necessary step concerns the calculation of the non-interacting limit of the off-diagonal density responses of arbitrary harmonic and nonlinearity order. Perturbative approaches lead to apparent poles in all ideal off-diagonal responses (e.g. the cubic response at the first harmonic)~\cite{vorberger2024greensfunctionperspectivenonlinear,Tolias_2023}. This suggests that the series expansion with respect to the perturbation strength becomes divergent. On the other hand, PIMC and DFT simulations demonstrate the finiteness of such responses and quantify their influence in the overall density perturbation~\cite{Dornheim_PRL_2020,moldabekov2025dftresponse}. It is currently unclear whether these singularities can be removed upon a reorganization of contributing terms or whether the iterative solution of the nonlinear equation is inappropriate. Another necessary step concerns the calculation of the non-interacting limit of the multi-point correlation functions in real and reciprocal space~\cite{Tolias_ctpp_2025}, an essential element towards the understanding of the effect of mean field terms and local field factors in the build up of high order correlations. Such general expressions are currently only available at the ground state~\cite{Tanaka_JPSJ_2011}.

The availability of a temperature-density dependent static or dynamic LFC for the warm dense UEG from PIMC simulations can be exploited to compute the static or dynamic quadratic and cubic density response or to determine the quadratic and cubic LFC after successive functional differentiations with respect to the density~\cite{Dar_2023}. Nonlinear responses and LFCs constitute a higher level of approximation and are worth benchmarking against QMC simulations and other theoretical models. In addition, nonlinear response theory allows advanced stopping power calculations with the nonlinear density fluctuation method~\cite{Hu_1988,Pitarke_1995,Bergara_1997}, which is suitable for modern high flux particle beams and should be superior to traditional approaches (linear response or binary approximation) being capable of accounting for strong beam plasma coupling and the crossover from the collisional to the collective regime. Unfortunately, experimental tests of the predictions of nonlinear response theory are lacking. Modern optical lasers as drivers and x-ray free electron lasers as diagnostic tools are more than capable of exciting and probing nonlinear WDM behavior, but preliminary estimates show that modern XRTS signals are unlikely to carry non-linear signatures. Therefore, alternative ways to directly detect nonlinearities are desirable.

Since all existing nonlinear response studies are devoted to the warm dense UEG, considerable effort should be directed towards expanding the scope to the strongly coupled UEG and to real WDM materials. Fortunately, the PIMC method is already capable of such studies for warm dense hydrogen and other low-Z materials are within reach. Finally, given the central role of the imaginary-time domain in probing the nonlinear response, nonlinear extensions of Matsubara frequency domain results should prove beneficial.

\newpage 
\clearpage

\section{Green's functions}\label{section_06}
\author{Michael Bonitz$^{1,2}$, Joshua Kas$^3$, Gerd R\"opke$^{4}$}
\address{
$^1$Institut für Theoretische Physik und Astrophysik, Christian Albrechts Universität zu Kiel, 24089 Kiel, Germany \\
$^2$Kiel Nano and Interface Science (KiNSIS)\\
$^3$University of Washington, Seattle, WA 98195, United States \\
$^4$University of Rostock, 18051 Rostock, Germany
}

\subsection*{Introduction}

Green's functions (GF) methods constitute the most general and powerful approach to quantum many-body systems 
(\ref{eq:general-hamiltonian}) 
that provide direct access to correlations, quantum and spin effects, screening and bound state formation, transport coefficients and optical properties. They can be equally applied to the ground state of many-body systems, to thermodynamic equilibrium, as well as to situations far from equilibrium, including excitation and relaxation behavior. In the context of dense plasmas and WDM, GF methods were  actively used and covered in detail by a number of text books ~\cite{kremp2005quantum,red-book,green-book,bonitz_qkt}. GF methods were originally developed for ground-state problems and later extended to finite temperatures in equilibrium (imaginary time or Matsubara GF, $G^M$, MGF)~\cite{matsubara55,abrikosov-etal.62}. The main advantages, as compared to other approaches, are the fulfillment of important conservation laws, the direct access to the spectral function, $A$, and density of states (DOS), as well as the existence of a systematic approximation scheme that is based on Feynman diagrams~\cite{kadanoff-baym,stefanucci2013nonequilibrium}. The extension of GF methods to arbitrary nonequilibrium situations was achieved by Keldysh~\cite{keldysh_contour,bonitz_pss_19_keldysh} who introduced real-time GFs, the equations of motion of which are known as Keldysh-Kadanoff-Baym equations (KBE)~\cite{kadanoff-baym}. While the GF methods are based on a systematic quantum statistical approach, the perturbation expansion only provides exact results (benchmarks) for some limiting cases. 
In connection with the partial summation of Feynman diagrams, new concepts (quasiparticle and self-energy, $\Sigma$, screening, local field corrections, Pauli blocking, cluster expansions etc.) are introduced. Compared to numerical simulations, such as PIMC [Sec.~\ref{section_02}] or DFT-MD [Sec.~\ref{section_03}], for solving the GF or the associated correlation functions, the analytical solutions based on diagram techniques are approximations that describe physical effects more intuitively and are very efficient in special limiting cases e.g.,  at low or high densities.

The GF approach is formulated in the second quantization based on expectation values of creation and annihilation operators, 
$a_i^\dagger, a_i$, acting on a complete set of single-particle orbitals, $|i\rangle$, where the time dependence arises from the Heisenberg picture. The single-particle GFs (correlation functions) are defined as expectation values with the equilibrium $N$-particle density operator
\begin{eqnarray}\nonumber
    G^<_{ij}(t,t') =\pm\frac{1}{\mathrm{i}\hbar}\left\langle \hat{a}_{j}^\dagger(t')\,\hat{a}_i(t)\right\rangle,
    G^>_{ij}(t,t') 
    =\frac{1}{\mathrm{i}\hbar}\left\langle \hat{a}_i(t)\,\hat{a}_{j}^\dagger(t')\right\rangle\,.
\nonumber
\end{eqnarray}
The Wigner function of a uniform plasma (using momentum states, $|i\rangle \to |\textbf{k}\rangle$), is given by $f_{\mathbf{k}}(t)=\pm i\hbar G^<_{\mathbf{k}}(t,t)$,  which evolves due to external fields and collisions, according to quantum kinetic equations~\cite{bonitz_qkt} that are direct generalizations of Boltzmann's equation (BE) and systematically include quantum and correlation effects. Furthermore, the GF approach gives direct access to spectral properties, such as the spectral function, via the difference $A_{\textbf{k}}(t,t') = i\hbar \left[G_{\textbf{k}}^>(t,t')-G_{\textbf{k}}^<(t,t')\right]$, and the density of states. In analogy to the one-particle GF, one defines the two-particle Green functions $G_2$, that describe, among others, two-particle bound states or particle-hole fluctuations, or, more generally, the $N$-particle GF.

\subsection*{State of the art}

We start with applications to charged particle systems in thermodynamic equilibrium. The stationarity is exploited via center of mass ($\hat T$) and relative ($\tau$) time variables, $G(t,t')\to G(\hat T,\tau)\equiv G(\tau) \to G^M(\omega)$, where the $\hat T$-dependence drops out and the $\tau$-dependence is the origin of the spectral properties (the $\omega$-dependence follows from Fourier transformation).

\subsubsection*{Thermodynamic properties}

\,\,Recently, the cumulant expansion (CEx) for the retarded one-electron GF was extended to finite temperatures~\cite{Kas_PRL_2017,Kas_PRB_2019}. Here, an exponential ansatz for the time-dependent GF, $G(\tau)=G^{0}_k(\tau){\rm exp}[C_k(\tau)]$ leads to an expression for the cumulant function $C(\tau)$ in terms of $\Sigma^{\rm GW}$ -- the self-energy within the $GW$-approximation, where $\Sigma^{\rm GW}$ introduces the temperature dependence. At low $T$, CEx accurately describes complex spectral features, such as multiple plasmon excitations, which are lacking the GW-based Dyson equation expression for the Green's function. Thermodynamic properties can be obtained from $G(\tau)$ through the Galitskii-Migdal-Koltun sum rule~\cite{stefanucci2013nonequilibrium,galitski-migdal_jetp_58} for the energy per particle $\epsilon=E/N$ at temperature $T$, 
$
E(T)=\sum_k\int d\omega\ \frac{1}{2}\left(\omega +\epsilon_{k}^0\right)A_k(\omega)f(\omega),$ where $A_k(\omega)=-(1/\pi)G_k(\omega)
$ 
is the one-electron spectral function, and $f(\omega)$ the Fermi function. The chemical potential $\mu$ is defined self-consistently by enforcing the particle number, $N(\mu,T)=\sum_k\int d\omega A_k(\omega)f(\omega)$. All other thermodynamic properties can be found either from the energy per particle, or the exchange-correlation potential, which is equivalent to the chemical potential for the UEG, which we consider in the following. For example, the free energy per particle is given by $f(n,T)=(T/T')f(n,T')-T\int_{T'}^Td\tau[\epsilon(n,\tau)/\tau^2],$ where $T'$ is large enough that the asymptotic behavior applies. In the UEG, an alternative expression is, $f(n,T)=3r_{s}^3\int_{r_s}^\infty d\bar r_s\,\mu[n(\bar r_s),T]/\bar r_s^4$. Other properties investigated include the pressure, entropy, specific heat, and isothermal compressibility. The cumulant GF approach was applied to the UEG over a wide range of densities and temperatures showing good agreement with RPIMC~\cite{Brown_PRB_2013} (see Sec.~\ref{section_02}) and fitted model data up to $r_s\sim 10$,  over a wide range of temperatures $0<\Theta<10$. Further, the real-space multiple scattering GF approach was recently extended to high electron temperatures $T_e$~\cite{TanHTGF,TanHTSE,TanCOHSEX}, which allows for calculations of X-ray spectra of solids. $T_e$ is decoupled from the lattice temperature, enabling simulation of systems out of equilibrium, such as those formed with X-ray FELs. Calculations of X-ray absorption spectroscopy (XAS) show that at low to intermediate $T_e$ ($\Theta \le 1$), the dominant effects can be interpreted through a frozen DOS approach, where the change in spectrum can be attributed to changes in occupation, and include a shift of the edge accompanied by an increase (decrease) in pre-edge (post-edge) weight. At higher $T_e$, changes in the DOS, in $\Sigma$, as well as the core-level binding energy, must be included in order to explain the transition from reverse saturated to saturated-absorption, as shown for XFEL-created warm-dense Cu~\cite{Mercadier2024}.

\subsubsection*{Transport and optical properties}

Non-equilibrium processes are effectively described by Zubarev's non-equilibrium statistical operator (NSO) approach~\cite{zubarev-book1,zubarev-book2} which is based on the construction of a relevant statistical operator $\rho_{\rm rel}(t)$ that can also include initial correlations. Since $\rho_{\rm rel}(t)$ is a generalized Gibbs distribution, the methods of the MGF can be applied to evaluate the averages. While thermodynamic equilibrium is a trivial case where the equations of state are derived using standard techniques~\cite{green-book}, hydrodynamic equations and transport properties, as well as kinetic equations can also be obtained via a systematic approach. As a special example, thermoelectric transport coefficients are expressed in terms of equilibrium correlation functions (generalized linear response theory, gLRT)~\cite{redmer_phys-rep_97}, as they are known from the fluctuation-dissipation theorem and the Kubo theory~\cite{kubo1957statistical}. Introducing moments of the distribution function, ${\bf P}_n=\sum_k \hbar {\bf k} (\hbar^2 k^2/2 mk_BT)^n f_k$, and the corresponding generalized forces $\dot {\bf P}_n$, the Onsager coefficients, in particular, the electrical conductivity $\sigma$, thermoelectric power $\alpha$, and thermal conductivity $\lambda$, are given by ratios of two matrices. The matrix elements are correlation functions which can be calculated using MGF techniques~\cite{roepke_PRA_1988}. The relation to kinetic equations was shown in~\cite{reinholz_PRE_2012}. For the electrical conductivity $\sigma(n,T)$, a virial expansion was derived, for arbitrary materials with charge number $Z$, including hydrogen~\cite{roepke_tmf_18}. In particular, the contribution of electron-electron collisions was investigated, and the correct Spitzer limit was obtained. This is in contrast to DFT-MD simulations that do not correctly reproduce collision effects~\cite{Bonitz_POP_2024,french_PRE_2022,reinholz_PRE_2015,roepke_PRE_2021,ropke_PoP_2024}, see Secs.~2-4 . The response to a dynamic, wave-number dependent external field was considered, which leads to the optical conductivity, the dynamic dielectric function and the dynamic structure factor ~\cite{Reinholz2005}. Conservation laws were considered and the Mermin formula for the dielectric response was determined~\cite{Selchow_PRE_2001}. The inclusion of bound states in the dielectric function was shown~\cite{ropke_PSS_1979}, and optical properties, in particular bremsstrahlung~\cite{Fortmann_HEDP_2006} and optical line profiles~\cite{Guenter_PRA_1991}, were calculated.

\subsubsection*{Nonequilibrium dynamics}

An alternative nonequilibrium approach concerns the direct solution of the KBE equations of motion of the single-particle GF in the two-time plane~\cite{bonitz-etal.96jpcm}. This requires the computation of initial correlation self-energies~\cite{semkat_99_pre,semkat_00_jmp} and allows to compute the electron dynamics on (sub-)femtosecond scales excited by laser fields~\cite{binder-etal.97prb,kwong-etal.98pss}. Moreover, the calculation of the real-time dynamics allows one to compute the density response (dynamic structure factor, plasmon dispersion) including correlation effects in a sum-rule preserving fashion~\cite{kwong_prl_00} where simple self-energies were shown to reproduce complex kernels of the Bethe-Salpeter equation (BSE) for the two-particle Green functions~\cite{bonitz_99_cpp2}. Nonequilibrium GF simulations were successfully applied to quantum plasmas in strong laser fields, and yielded important results, including inverse bremsstrahlung heating, the generation of higher harmonics of the laser field~\cite{Kremp_1999,haberland_01_pre}, and the modification of the plasmon spectrum by the laser field~\cite{bonitz_99_cpp}.
The real-time approach also allows one to compute the stopping power beyond linear response using a NEGF-Ehrenfest approach~\cite{balzer_prb16} and to investigate correlation effects at low projectile energies, such as the generation of doublons in 2D quantum materials~\cite{balzer_prl_18, borkowski_pss_22,bonitz_pss_18}, as well as the neutralization of highly charged ions upon impact~\cite{balzer_cpp_21,niggas_prl_22}. Note that full solutions of the KBE are very costly scaling cubically with the time step number $N_t$. Here, recently breakthroughs were achieved by applying the generalized Kadanoff-Baym ansatz (GKBA)~\cite{lipavsky_generalized_1986,balzer-book,hermanns_prb14}. Transforming the time-diagonal KBE for $G^<(t,t)$ into the G1--G2 scheme -- a time-local system for the one- and two-particle GFs~\cite{schluenzen_prl_20,joost_prb_20}, reduces the CPU-time scaling to linear order in $N_t$. Another advantage of the G1--G2 scheme is that high-level self-energies, such as T-matrix, GW and dynamically screened ladder approximation (DSL) have become feasible for full nonequilibrium simulations~\cite{joost_prb_22,bonitz_pssb23}.

\subsection*{Challenges \& outlook}

\subsubsection*{Thermodynamic properties}

GF methods  require less computational resources than comparable approaches. Non-perturbative approaches such as the cumulant expansion for the one-electron GF provide a consistent framework capable of predicting equilibrium properties and EOS, as well as excited state properties and spectra. The current cumulant expansion approach has limited accuracy for large correlations, $r_s>10$~\cite{Dornheim_PhysReports_2018} and also lacks thermodynamic consistency, in the sense that different paths to the same thermodynamic quantity do not provide the same results. One possible way forward is to include local fields in the screened Coulomb interaction~\cite{Dornheim_PRL_2018,quantum_theory}. The calculation of X-ray spectra of out-of equilibrium states produced at XFELs is a relatively new endeavor, since experimental results have only recently become available. This provides an essential benchmark of current and newly developed theories and computational approaches. In the case of XAS of XFEL-created WDM, qualitative agreement is found with a combination of BE results and high temperature real-space GF techniques, where the results of the BE solution~\cite{JurecXMDYN_XATOM} are utilized to define an effective $T_e$ for XAS calculation, neglecting any effect of lattice excitation relying on the relatively large timescale (10s of ps versus 10s of fs). However, there is evidence that, at the high intensities reached at XFELs, this approximation may break down, even with short pulse width, and certainly this will be important at longer delay times. Therefore, techniques to accurately describe the system's evolution, including all pertinent degrees of freedom, are important. Another point of importance, in systems at high $T$, and with varying density, is an accurate description of the core-level binding energies, including the effects of screening and continuum lowering~\cite{bonitz_cpp_2025,SonCL}. In XFEL-created WDM, as well as at high pressure such as that found at Earth's core, these shifts can be quite large, and an accurate accounting is essential for the use of X-ray spectra as an EOS probe.

\subsubsection*{Transport and optical properties}

GF methods are complementary to numerical evaluation of correlation functions by PIMC and DFT-MD, see Secs.~\ref{section_02},\ref{section_03}. 
At low densities or high temperatures simulations become very resource consuming and can be complemented by analytical results which are needed, e.g., for thermal conductivity, and for studying the influence of electron-electron collisions~\cite{french_PRE_2022,roepke_PRE_2021,ropke_PoP_2024}. 
Moreover, interpolations and  analytical parametrizations are required as input, e.g., for hydrodynamic simulations, thereby preserving conservation laws~\cite{kadanoff-baym}. GF methods also allow one to treat dynamic properties in a way that is compatible with equilibrium properties. Examples are the dielectric function, dynamic structure factor and dynamic local-field corrections. Optical properties that are related to dynamic conductivity are of major relevance in WDM. GF results for radiation transport and opacity are crucial to improve existing approaches such as OPAL and to identify different physical processes~\cite{roepke_pre_19}. An interesting challenge is the incorporation of relativistic plasmas and finally a QED-based treatment of hot and dense plasmas. Further, GF approaches should be extended to plasmas of different materials and mixtures. 
Here, DFT-MD simulations provide very flexible tools to calculate transport and optical properties, whereas GF methods are of importance to analyze these results and serve as benchmarks, in limiting cases. An important problem is the ionization potential depression (IPD), which is observed in various experiments with compressed matter and still poses a challenge to theory~\cite{Lin_PRE_2017}. Here, recently, novel PIMC results were reported~\cite{Bonitz_POP_2024, bonitz_cpp_2025}, see Sec.~\ref{section_02}. An important input from GF theory is the density dependence of the bound states from solutions of the BSE for the electron-ion two-particle GF~\cite{roepke_pre_19,seidel_pre_95}. 

\subsubsection*{Nonequilibrium dynamics}

The G1--G2 scheme holds great promises for applications to WDM. Until now, it has been successfully applied to lattice models in condensed matter physics and for cold atoms, allowing one to compute the ultra-fast response of correlated systems to femtosecond lasers or to the impact of highly charged ions~\cite{niggas_prl_22}. Applications to WDM are hampered by the large dimension of a momentum basis leading to excessive storage requirements for the two-particle Green function. First applications to quasi-1D geometries are very encouraging~\cite{makait_cpp_23}.
We also mention numerical instabilities due to aliasing effects for which recently a solution could be presented~\cite{makait_cpp_25}.
The storage problem can be mitigated by applying embedding schemes~\cite{balzer_cpp_21,bonitz_pssb23,balzer_prb_23}, as well as with a novel quantum fluctuations approach (QFA)~\cite{schroedter_cmp_22,schroedter_23} that constitutes a systematic generalization of the classical  theory of Klimontovich~\cite{schroedter_cpp_24} and allows one to eliminate two-particle Green functions in favor of correlation functions of single-particle fluctuations. A comparison of the approximations to standard approximations within the BSE was performed in Ref.~\cite{schroedter_pssb23}. The QFA  also gives access to nonequilibrium two-time quantities, such as density correlation functions and the dynamic structure factor~\cite{schroedter_23}. Finally, a particular challenge remains to establish a connection between the short-time regime, described by the G1--G2-scheme, with the long-time limit described by kinetic equations of the Boltzmann or Balescu-Lenard-type, as well as with the intermediate relaxation phase, where electrons and ions (lattice) are described by different temperatures.
\newpage 
\clearpage

\section{Radiation-hydrodynamics and Molecular Dynamics}\label{section_07}
\author{Frank Graziani$^{1}$, Gianluca Gregori$^{2}$, Stephanie Hansen$^{3}$, Tlekkabul Ramazanov$^{4}$, Aidan P. Thompson$^{3}$}
\address{
$^1$Lawrence Livermore National Laboratory, Livermore, CA 94550, United States \\
$^2$Department of Physics, University of Oxford, Parks Road, Oxford OX1 3PU, United Kingdom \\
$^3$Sandia National Laboratories, Albuquerque, 87185 New Mexico, United States \\
$^4$Al-Farabi Kazakh National University, Almaty, Kazakhstan 
}

\subsection*{Introduction}

 Radiation-hydrodynamics (RH) ~\cite{pomraning1973,mihalas1985, Zeldovich1966,  Castor_2004}, is the description of matter when hydrodynamic and radiative processes are coupled and both contribute to material motion. Because RH captures physical phenomena at the continuum scale, it has become the main computational tool for the description of astrophysical, plasma, and fluid phenomena ~\cite{saupe2023, Clark_PoP_2019, Mezzacappa2020}. For example, today, the computer codes that simulate high energy density phenomena in nature (e.g supernova explosions ~\cite{Mezzacappa2020}) or inertial confinement fusion (ICF) ~\cite{Clark_PoP_2019, Marinak_PoP_2024}, are millions of lines long, massively parallel, 1D-2D-3D, multi-physics (e.g. hydrodynamics, radiation transport, magnetic fields, thermonuclear burn) codes. The hydrodynamic equations are of the Braginskii or Navier-Stokes form (with or without magnetic fields) ~\cite{Boyd_2003}, ~\cite{lamb1945, zhou2024, zhou2025instabilities} and radiation transport ~\cite{pomraning1973, mihalas1985, graziani2005} can take on many options, the most commonly used is gray or multi-group diffusion. But both deterministic and Monte Carlo transport, albeit more expensive computationally, is also used. The set of equations require material inputs from LTE (Local Thermodynamic Equilibrium) and nLTE (non-Local Thermodynamic Equilibrium) opacity, equation of state (EOS), electron-ion coupling, thermal and electrical conductivity, magnetic resistivity, viscosity, thermoelectric transport, fusion reactivity models or codes~\cite{Salzmann1998, Atzeni_2009}. These multi-physics components are often reduced (simplified) versions of 
 a more complex (sub-grid) physics that cannot be solved together within the main RH code.

 Recent advances in high performance computing have opened another avenue to a deeper understanding of WDM and at the same time provided insight into the accuracy of kinetic theory and hydrodynamics ~\cite{alder1959}. Classical molecular dynamics (CMD) is a discrete particle simulation method developed in the 1950’s by Alder and Wainwright ~\cite{alder1959, hockney1981, birdsall1991, griebel2007}. The CMD method is simply the numerical integration of equations of motion of a set of particles that are interacting via some potential energy function. The method has been used to study both WDM and hot dense matter relevant to fusion applications.  ~\cite{glosli2008,dimonte2008,graziani2012, Sjostrom2014, baalrud2014, shaffer2017}. MD simulations are computationally intensive and restricted to much smaller length and time scales than RH. However, the results from MD simulations are ``exact" in the sense that given a potential energy function and the equations of motion, the evolution of the many-body system does not depend on any closure relations or any other physical assumptions. 
 However, with the development of fast and efficient multi-processor solvers as well as the availability of large supercomputers, simulations in excess of $10^8$ particles are now becoming a possibility \cite{Komatsu2014}. These simulations have enough particles to be able to directly bridge from the micro to the meso-scale, exploring the full turbulent cascade from the forcing scale down and below Kolmogorov's dissipation scale.
  
 Quantum molecular dynamics ~\cite{car1985,zhang1999,griebel2007} combines CMD for the classical objects (ions) with quantum mechanics for the electrons, see Sec.~\ref{section_03}. Because of the large mass ratio between ions and electrons the Born-Oppenheimer approximation ~\cite{griebel2007} is invoked, the electrons are treated with DFT, and the total energy of the system provides the potential energy function for the classical dynamics of the ions. Specifically, QMD calculates interatomic forces quantum mechanically, usually within the framework of density functional theory. QMD has become a standard simulation method for calculating equations of state and transport coefficients of warm dense matter ~\cite{clerouin2006,Bonev2004, sjostrom2018, Bonitz_POP_2020}. 

 Quantum hydrodynamics (QHD) is an alternative approach to describe warm dense matter systems using the field variables density, momentum and energy of classical hydrodynamics. The efforts to describe quantum systems within a hydrodynamic approach was initiated in its full form by Madelung and Bohm and applied by Manfredi and Haas for many-particle systems and critically assessed by Bonitz, Moldabekov, and Ramazanov ~\cite{haas2011, bonitz_pop_19, manfredi_prb_21}. These efforts were driven by a desire to self-consistently take into account quantum non-locality, spin statistics and correlation (non-ideality) effects from the beginning when developing a hydrodynamic description.  
 
\subsection*{State of the art}

\subsubsection*{Radiation hydrodynamics}

Radiation-hydrodynamics codes are critical for the design, optimization, and diagnosis of WDM and HED experiments. Even relatively simple RH codes can provide initial estimates of energy evolution at the target scale: tracking heating, compression, and diffusive and radiative losses. However, in recent years, there has been increased focus on improving RH models to account for high spatial resolution and 3D effects (e.g. Refs.~\cite{Clark_PoP_2019, liberatore2023}), extended magneto-hydrodynamics (e.g. Ref.~\cite{Seyler_2011}), and inline nLTE for radiation transport~\cite{Scott_2010}. 

Machine learning (ML) or cognitive simulation models for HED experiments have become the research tool to help researchers explore a vast design space. The HED community has developed sophisticated codes and clever handoffs to span the required length and time scales and has begun to employ these methods to address the computational challenges associated with large-scale 3-D RH simulations~\cite{Humbird_2021}. ML methods, in many cases, require running an ensemble of simulations. This means that the RH code has to be fast, robust, and accurate. In addition, sensitivity studies ~\cite{saltelli2000}, which also require running an ensemble of simulations, are a valuable tool to understand what physics models (e.g. opacity, viscosity, thermal conductivity, equations of state, etc.) matter most in a RH simulation and where experimental and computational efforts should be focused.

In alternative approaches, instead, data-driven ML techniques are used to improve our modeling capability by learning directly from data provided either by microscopic-scale numerical simulations or even experiments. This approach has been applied to thermal transport \cite{Miniati:2022,Luo2025}. In particular, the model of Luo et al. \cite{Luo2025} is able to learn the spatio-temporal heat flux kernels directly from particle-in-cell data, capturing dynamic transport behaviors beyond the reach of classical formulations. 
The challenge is now in the development of integrated algorithms capable of coupling these ML kernels into RH codes over a range of different micro-physics, yet preserving the numerical accuracy given the intrinsic noise presented in any neural network scheme \cite{Miniati:2022}.

\subsubsection*{Molecular dynamics}

Historically, quantum and classical molecular dynamics have played very distinct roles. QMD has provided quantitatively accurate predictions for thermodynamic and transport properties of warm dense matter at specified local thermodynamic equilibrium conditions of temperature, density, and elemental composition. In contrast, CMD has provided qualitative representations of non-equilibrium behaviors that emerge at larger length scales, including mass, momentum, and energy transport, as well as time-dependent transient and spatially inhomogeneous processes, including hydrodynamic instabilities, solid deformations, and phase change kinetics ~\cite{dimonte2008,graziani2012, Daligault_PhPl_2018}. However, while CMD interatomic potentials accurately captured properties of specific materials at conditions close to the ground state, it was difficult or impossible to retain correct physics and chemistry behaviors in simulations approaching WDM conditions ~\cite{jones2007analysis}. Over the last decade CMD has undergone a dramatic change. Machine learning interatomic potentials (MLIPs;  see also Sections \ref{section_03}, \ref{section_22} and \ref{section_24})) have emerged that are trained to match sets of small scale QMD simulations at WDM conditions.  Much of the progress in this area has been driven by the ability to distribute the relatively high computational load of MLIPs over large numbers of processors via the widely used LAMMPS code~\cite{thompson2022}. This has allowed large-scale CMD simulations to retain much of the quantitative accuracy of QMD, enabling the use of CMD to make quantitative predictions for specific materials at specific conditions, including mapping out phase diagrams for pure and multi-component solid, liquid, and vapor systems (via rigorous free energy sampling methods), studying large-scale dynamic processes (phase change, mass transport, heat transport), as well as the exploration of entirely new emergent phenomena.  Good examples of this are recent CMD simulations of pure diamond~\cite{Nguyen-Cong2024} and carbon/hydrocarbon mixtures~\cite{Cheng23} at planetary core pressures and temperatures up to and beyond the melting point of diamond. An important caveat to this is that for sufficiently high temperatures ($T > 0.5$~eV), many properties measured in QMD can be recovered from CMD simulations using simple radial pair potentials fit to QMD forces, obviating the need for MLIPs~\cite{Stanek2021}.

Somewhere in between CMD and QMD is the technique called wave-packet molecular dynamics (WPMD). This was originally proposed by Heller \cite{heller1975time,feldmeier2000molecular} after the observation that a Gaussian wave function remains a Gaussian in a quadratic potential, and its time evolution follows classical equations. In more recent approaches WPMD corresponds to a family of models in which the electron dynamics are computed explicitly, by restricting their wave function to a parameterised functional form, while simulating hundreds to thousands of particles over ionic time scales. 
In the formulation of Svensson et al. \cite{svensson2023development} the WDM regime is modeled by taking wave packets that can be elongated in arbitrary directions. Because of the ability to treat a large number of particles, while retaining quantum dynamics, this approach is well-suited for the 
calculation of dynamic properties as well as transport quantities in WDM \cite{svensson2024,svensson2025}.

\subsubsection*{Quantum hydrodynamics}

 QHD can be formulated in terms of an average density $n(r,t)$, velocity $\boldsymbol{v}(r,t)$ and the generalized force $-\nabla\mu(r,t)$ were derived in Refs.~\cite{Banerjee2000b, Michta2015} for the case of fully degenerate electrons (zero temperature limit) on the basis of a semiclassical approach:

 \begin{equation}
    \frac{\partial}{\partial t} n\left( \mathbf{r}, t\right)+{\nabla} \left[n\left( \mathbf{r}, t\right)v\left( \mathbf{r}, t\right)\right]=0,
\label{QHD1}
\end{equation}
\begin{equation}
    m_e \frac{\partial}{\partial t} \mathbf{v}\left(\mathbf {r}, t\right)+m_e \mathbf{v}\left(\mathbf {r}, t\right)\mathbf{\nabla} \mathbf{v}\left(\mathbf {r}, t\right)=-\mathbf{\nabla} \mu \left( \mathbf{r}, t\right). 
 \label{QHD2}
\end{equation}
To extend equations (\ref{QHD1}),(\ref{QHD2}) to finite temperatures, the grand canonical ensemble should be used where the free energy functional $F[n]$ consists of the ideal (non-interacting) and the exchange correlation parts. In Refs.~\cite{bonitz_pop_19,zhandos_pop18} , the potentials associated with Fermi pressure and the Bohm potential were derived using the inverse polarization function of electrons at any degeneracy.  This analysis pointed out the limitations of the previously used models.

\subsection*{Challenges \& outlook}

\subsubsection*{Radiation hydrodynamics}

There are several critical challenges faced by RH codes to reach the precision needed for truly predictive simulations that could inform, e.g., integrated designs for inertial fusion energy and EOS experiments in the WDM regime ~\cite{Marinak_PoP_2024, Hurricane16, Hurricane_RMP_2023}. First, there is the massively multiscale nature of the simulations, which ideally would capture phenomena at the larger than meter driver length scale through the centimeter-scale target (gross material motion and instability development), through the micron-scale meso-physics (turbulence and mix), and finally through the atomic-scale response (EOS and transport coefficients). A second challenge is the internal consistency of the atomic-scale physics: at present, most RH codes rely on calibrated sub-scale models for mesoscale physical effects and on data tables for constitutive material properties ~\cite{sterne, starrett2016equation, hansen2023}. However, there is often no enforced consistency among the datasets for EOS, transport, and opacity or between those fundamental constitutive properties and mesoscale models. Further, most existing tabulated data is computed in local thermodynamic equilibrium (LTE), which assumes steady-state and thermal distributions among ion, electron, and radiation distributions at a single temperature. However, the optical, x-ray, and pulsed-power drivers used to create WDM in the laboratory often impart non-thermal sources of energy to only one of the constituents of a plasma (ions, electrons, collective modes, or radiation), and most HED plasmas have a fundamental imbalance between radiation and material temperature that can lead to profound deviations from LTE. The dynamics of equilibration among plasma constituents are also not generally resolved in RH models. Finally, there is a need for synthetic diagnostics from integrated RH codes that can be directly compared to measured data ~\cite{moore2024}.

\subsubsection*{Molecular dynamics simulation}

The emergence of quantum-accurate MLIPs driving large-scale CMD simulations has effectively solved the problem of accurate sampling of thermodynamic and transport properties in condensed matter at mild WDM conditions. Nevertheless, this does not address the effects of electronic temperature, electron excitation, electron-ion coupling, electronic spin, and ionization, all of which play a large role in WDM conditions. All of these phenomena require that the MLIPs and the codes that run them be extended beyond the Born-Oppenheimer approximation that is implied under conventional CMD. While some of these effects have been treated qualitatively in the past (spin dynamics~\cite{Tranchida2018}, semi-classical electron particles~\cite{morozov2009localization, su2007excited}, two-temperature models~\cite{Duffy2007}, linearized Bohmian trajectories~\cite{larder2019fast}), new approaches start to emerge. For example, the inclusion of a Bohmian trajectory scheme within smoothed particle hydrodynamics \cite{Campbell2025} provides a new avenue for a simulation scheme that retains the speed of CMD but allows for rich inclusion of several quantum-mechanical effects, including the ability to model the bound state wavefunctions correctly. 

In Bohmian mechanics and in WPMD, one of the main challenges is that at sufficiently high temperatures, the wave function expands indefinitely, requiring regularization, and the effect of this regularization on the overall dynamics must be assessed. In addition, a problem of all MD schemes is that ionization is not self-consistently accounted for within the model, even if solutions to this problem have recently been presented \cite{Plummer_PRE_2025}.

More systematic approaches are starting to emerge. This will result in large-scale CMD simulations of matter at WDM conditions that are consistent with QMD simulations, providing access to phenomena such as propagation of energy in electronic modes at speeds greater than mechanical shockwave propagation. Kinetic Theory Molecular Dynamics (KTMD)~\cite{Graziani2014}, offers a hope of physics informed interatomic potentials for MD by using a quantum kinetic equation to solve for the local electron density which is then convolved with the bare interatomic potentials to yield a quantum informed interatomic potential suitable for non-equilibrium phenomena. 
 
\subsubsection*{Quantum hydrodynamics}

Currently, QHD is not a widely used method for performing simulations of WDM. Unfortunately, QHD as formulated for a single particle has been applied to many-body quantum systems, populating the literature with incorrect results ~\cite{bonitz_pop_19}.  QHD can be properly formulated for many body systems, such as plasmas, so that the exchange potential, degeneracy pressure, and mean-field Hartree-Coulomb terms are present ~\cite{Michta2015, zhandos_pop18}. Properly formulated, QHD will play an important role in addressing macroscopic problems such as turbulence and mix, plasma oscillations, collective modes, shocks and charged particle stopping in WDM ~\cite{michta2020, graziani_cpp_21}.

\newpage 
\clearpage

\section{Average Atom and Chemical Models}\label{section_08}
\author{Dirk Gericke$^{1}$, G\'erard Massacrier$^{2}$, Ronald Redmer$^{3}$, C. E. Starrett$^{4}$}
\address{
$^1$CFSA, Department of Physics, University of Warwick, Coventry CV4 7AL, United Kingdom \\
$^2$Univ Lyon, Univ Lyon1, Ens de Lyon, CNRS, Centre de Recherche Astrophysique de Lyon UMR5574, F-69230, Saint-Genis-Laval, France \\
$^3$University of Rostock, Institute of Physics, D-18051 Rostock, Germany\\
$^4$Los Alamos National Laboratory, Los Alamos, 87545 New Mexico, United States
}

\subsection*{Introduction}

Average atom (AA) and chemical models (CMs) exploit physical approximations in order to calculate the properties of warm dense matter in a computationally efficient way.  With a history spanning more than a century, there have been many variations of these approaches. Drawing a clear separation between these two classes of models can only be indicative. 

As a rule of thumb, the basic building block of the AA models is a single nucleus immersed in the plasma with its accompanying electronic cloud. The bound and free components of the (usually) spherically symmetric electronic density are obtained from quasi-classical considerations or from the solutions of a one-electron hamiltonian in some effective potential. Besides the nucleus Coulomb field, the latter includes the Hartree energy of the electronic cloud and some exchange-correlation correction.  It must be obtained  self-consistently with the electronic distribution. The earliest AA model is known as the Thomas-Fermi-Dirac (TFD) model~\cite{thomas1927, Fermi1928, feynman1949equations}. The nucleus of charge $Z$ is placed at the center of a sphere with a radius that is dictated by the material density. It is made charge neutral by $N=Z$ electrons which are distributed according to Fermi-Dirac statistics.  The density of these electrons is calculated in a semi-classical approximation.  This TFD model can still be useful today~\cite{wjphysics}, often giving reasonable estimates of the equation of state (EOS), the electronic density, the average ionization and other properties. Another celebrated AA model has been built by Stewart \& Pyatt (SP) along the same ideas~\cite{Stewart_ApJ_1966}. Their work provides an expression for the ionization potential depression (IPD) that has been largely used in simulations of experiments. It has been called into question in the light of recent experimental results~\cite{Ciricosta_PhRvL_2012, Hoarty_PhRvL_2013}, some of which have rejuvenated a modified version of the Ecker \& Kr{\"o}ll model~\cite{ecker-kroell_63}.

In CMs, atoms, molecules, ions with different charge degrees, and possibly excitation states, as well as electrons are considered as separate species from the outset. Accordingly, dissociation, ionisation, and excitation are seen as (chemical) reactions with $\rm{ion}^{j+} \rightleftharpoons \rm{ion}^{(j+1)+} + \rm{e}^-$ being a typical example. The equilibrium requires equal chemical potentials on both sides which, in the easiest case, leads to the Saha equation determining the densities of ion species~\cite{kremp2005quantum}
\begin{equation}
\frac{n_j}{n_{j+1}} = \frac{g_j}{g_{j+1}} \, \frac{n_e \Lambda_e^3}{2} \,
                                    \exp\left( - E_j^{\rm \,eff} / k_B T \right) \,.
\end{equation}
Here, $g_j$ are the statistical weights of the states $j$ and $j+1$, $\Lambda_e$ is the thermal wave length of electrons, and $E_j^{\rm \,eff}$ is the effective ionisation energy. In WDM, $g_j$ and $E_j^{\rm \,eff}$ are modified by the interactions with other particles. Determination of the nonideality corrections is a major task for WDM. Examples for effective ionisation energies are the approaches of Debye~\cite{DH_1923}, Stewart \& Pyatt~\cite{Stewart_ApJ_1966}, Ecker \& Kr{\"o}ll~\cite{ecker-kroell_63}. Similarly, rate equations can be set up if the charge state distribution is not yet in equilibrium~\cite{kremp2005quantum}. The central quantity to determine the state of WDM is the Helmholtz free energy $F$. Within CMs, it is constructed by adding correlation parts to the ideal contributions for the mixture of all species. These contributions are mainly of two types. The various interactions between species are taken into account via effective potentials often with hard cores to render the finite sizes of the atomic states. The internal partition functions of ions and atoms are modified to mimic energy levels shifts and pressure ionization. The minimization of the free energy yields thermodynamic quantities as with AA models, but also the populations of the different species.

\subsection*{State of the art}

\subsubsection*{Average Atom models}

Progress in AA models can be viewed as improving or as removing approximations in the TFD model, which is actually an application of orbital-free, finite-temperature DFT~\cite{Mermin_PR_1965,blenski2007variational}.  Therefore, improvement can be made by using Mermin-Kohn-Sham DFT, this was first achieved by Liberman~\cite{Liberman1979} in 1979.  Research on this model continues to the present day, both improving the numerical techniques~\cite{wilson2006, starrett2019wide, starrett2023average} and fundamental derivation (finally solved in Ref.~\cite{blenski2007variational}).  This model and similar models are used to this day to calculate EOS 
\cite{starrett2019wide, ma2024equation, swift2019atom, ovechkin2021equation, faussurier2019pressure, starrett2016equation, hou2015equations, massacrier2011equation, piron, Faussurier2010, sterne, pain2007shell}, conductivity~\cite{shaffer2020model, sterne, wetta2023average, perrot1987electrical} and optical properties~\cite{Johnson2006Optical, hansen2023, shaffer2017free, piron2013variational}.  Other improvements have focused on the treatment of band structure~\cite{Hou_PhPl_2006, MassacrierBoehmeVorbergerSoubiranMilitzer2021, rozsnyai1972relativistic, Callow_PRR_2022, Starrett_PRE_2020},  including the effects of ionic structure through the pair correlation function $g(r)$~\cite{Faussurier2010, starrett2012fully, chihara1991unified, perrot1990dense}, coupling to mixing rules to allow treatment of mixtures~\cite{starrett2012average}, and including explicit excited states~\cite{SonCL, hansen2023, perrot1995equation, faussurier2018density, starrett2024dense}.

One application of AA models is the calculation of the EOS over a very wide range of temperatures and densities. This application is challenging from a numerical and an algorithmic point of view.  Much effort has been expended in making codes as robust as possible for the range of conditions and materials necessary. In this vein, the \texttt{Purgatorio} code was developed at LLNL~\cite{wilson2006}.  The authors developed several innovations. This includes an adaptive mesh refinement technique for evaluating energy integrals quite difficult to evaluate accurately due to the challenge in finding all weakly bound states and the appearance of narrow resonance structures in the continuum.  An alternative solution to this numerical problem was presented in Ref.~\cite{starrett2019wide}, where it was pointed out that instead of using the orbital approach, one can directly evaluate the one-electron, time-independent, Green's function.  Since the Green's function is analytic in the complex plane, the energy integral can be transformed into a contour integral.  It turns out that this guarantees a smooth integrand that is easily evaluated accurately, meaning that resonances do not need to be resolved and weakly bound states are guaranteed to be included.  Further, and as envisioned by More 40 years ago~\cite{more1985pressure}, Siegert states can be used to evaluate the AA model~\cite{starrett2023average} with the benefit that the continuum states are replaced by a sum over discrete, complex energy, states.

A major advance in the theoretical underpinning of AA models was achieved by the presentation of a first variational derivation~\cite{blenski2007variational}.  This formally correct way to derive AA models removed the ambiguity with regards to EOS calculations, present since Liberman's original model~\cite{Liberman1979}.  It also demonstrated that 
the Liberman model is functionally variationally consistent provided the Friedel oscillations in the electron density are damped outside the ion sphere~\cite{piron2013variational}.

In the early days of AA models, Rozsnyai~\cite{rozsnyai1972relativistic} presented a model that used alternative boundary conditions to estimate the band width which is zero in Liberman's model. Recent work has explored this idea further~\cite{Hou_PhPl_2006, MassacrierBoehmeVorbergerSoubiranMilitzer2021, Callow_PRR_2022}.  Comparisons of a new model, \texttt{AvIon}, with DFT-MD results showed good agreement on band structure width and valence-conduction gaps.  

AA models can also be extended to predict accurate ion-ion pair correlation functions.  Early work on this was led by Chihara~\cite{chihara1991unified} and separately by Perrot and Dharma-wardana~\cite{perrot1990dense, dharma1983effective} which was generalized recently~\cite{starrett2014hedp}.  The basic idea is that the screening cloud of electrons around an ion can be calculated in an AA calculation.  The electron density of this screening cloud can then be used as a closure relation for the quantum Ornstein-Zernike equations, which are then solved to give the ionic structure.  This model was later validated by experiment~\cite{starrett2015models}.  The validity of the model is limited to plasmas that behave like simple metals (i.e., are hot enough).  For plasmas that have significant deviations from a nearly free electron density of states, the results are inaccurate due to an assumption inherent in the model.  For example, if transient bonding is significant, the predictions of the model are relatively poor~\cite{starrett2014integral, Dharma-wardana_PRE_2025}. Interestingly, this work predicts an ion-ion pair interaction potential that can be used in classical MD simulations.  This Pseudoatom molecular dynamics (PAMD) model has been used to calculate EOS data and ion transport coefficients, with very promising results (notwithstanding the limitation mentioned above)~\cite{ma2024equation, starrett2015pseudo, Heinonen_ApJ_2020, falkov2022pseudoatom}.  An alternative scheme, with similar aims is presented in Ref.~\cite{hou2021ionic}, and a recent variational model has also been presented~\cite{blenski2023variational}.

AA models have also been used to calculate both electrical and thermal conductivity.  Recent advances include coupling to advances in kinetic theory and the Kubo-Greenwood formula~\cite{shaffer2020model, Johnson2006Optical}, new applications and models~\cite{ovechkin2016transport, faussurier2019electrical}, as well as the effect of the definition of average ionization~\cite{wetta2023average, Burrill2016}.  This last topic, the definition of average ionization, has been addressed many times over the decades, and originates from the fact that there is no unique definition of this quantity.  Many reasonable definitions have been proposed which differ~\cite{starrett2019wide, sterne, wetta2023average, callow2023improved, petrov2021collisional, faussurier2021carbon}.  We note that in Liberman's model~\cite{Liberman1979}, and most other AA models (eg.,~\cite{blenski2007variational, wilson2006, starrett2019wide}), the EOS and electronic structure do not depend on the definition of the average ionization, it is an output of these models (for a given definition).

Recent efforts have focused on the calculation of X-ray Thomson scattering spectra using AA models~\cite{baczewski2016x, johnson2016average, Souza2014}.  In particular the work of Ref.\cite{baczewski2016x} shows generally good agreement, with only minor differences, compared to much more computationally expensive MD-DFT results. 
Other recent efforts have also been directed at making the AA less averaged, in the sense of going beyond Fermi-Dirac occupations for the electron states.  These recent models~\cite{SonCL, hansen2023, piron2013variational, starrett2024dense, faussurier2018density} have begun to resolve occupations into individual integer-electron configurations.  This is in part motivated by the failure of DFT based models to predict realistic absorption coefficients in hot plasmas~\cite{gill2021time, karasiev2022first}. 
This is also mandatory to deliver an IPD that goes beyond the mean value of the AA model, and that takes into account the different ionic charge stages in the plasma.


\subsubsection*{Chemical models} 

In the 1970's and 80's, Ebeling and co-workers developed efficient Pad{\'e} formulae for the correlation contribution of charged particles to the free energy  which interpolate between the well-known Debye-H\"uckel (low-density) and Thomas-Fermi (high-density) limits~\cite{green-book}. Considering also the interactions between the neutral particles based on, e.g., Fluid Variational Theory (FVT)~\cite{ross_jcp_83, juranek_jcp_02} and the polarization interactions between charged and neutral particles~\cite{redmer_phys-rep_97}, the dissociation and ionization equilibrium can be solved and the EOS is determined for dense plasmas or WDM. In particular, such CMs predict an abrupt non-metal-to-metal transition with a thermodynamic instability below a second critical point which is known as Plasma Phase Transition (PPT). The PPT was extensively discussed for hydrogen~\cite{ebeling-richert_85, saumon_aps_95, reinholz_PRE_95, norman_ufn_21} and helium~\cite{Foerster1992, Preising_CoPP_2021}. For a recent review of the hypothetical PPT and the actual liquid-liquid phase transition (LLPT) derived from high-pressure experiments and \textit{ab initio} simulations, see Ref.~\cite{Bonitz_POP_2024}.

Plasma environment effects were introduced in a more refined way in the internal partition functions of ionic species. As part of the Opacity Project (OP)~\cite{Seaton_JPhB_1987, Seaton_MNRAS_1994}, aimed at solving the so-called opacity problem in stellar and solar physics, the Mihalas-Hummer-D\"appen EOS used an occupation probability formalism to populate internal atomic states~\cite{hummer-mihalas_88, mihalas2_aj_88, daeppen_aj_88, Mihalas_ApJ_1990}. This formalism has recently been extended to include plasma electric microfields in a superconfiguration approach~\cite{Loboda_CoPP_2009, Ovechkin_PRE_2023}.  Atomic structure codes for the isolated ion have also been adapted to include the screening of the free electrons cloud~\cite{Massacrier_JQSRT_1994, Belkhiri_HEDP_2013, Gu_PRA_2020, Zeng_AA_2020}. Contrary to AA models, this screening was however not computed self-consistently. This was done for separate ion stages in Refs.\cite{massacrier2011equation,Potekhin_PRE_2005}, but at the expense of modifying energy levels only perturbatively. Some recent papers~\cite{Zeng_PRE_2025, Gao_PRE_2025} go one step further in coupling the atomic structure calculations with self-consistent screening potentials.

Much progress has also been made in CMs concerning the modeling of interspecies interactions. Recent applications include partially ionized hydrogen~\cite{French2017} and water plasma~\cite{Schoettler2013}  for conditions as found in planetary interiors. As part of a global effort at LANL to produce opacity tables with the \texttt{ATOMIC} code, the \texttt{ChemEOS} model was improved~\cite{Kilcrease_HEDP_2015} for a better description of the Coulomb interaction among particles, extending previous work~\cite{green-book, Chabrier_PRE_1998} for totally ionized plasmas to partially ionized ones. A new framework based on the thermodynamic integration of the excess free energy obtained through MD simulations of interacting species was recently applied in Ref.~\cite{Plummer_PRE_2025}. A systematic approach to include any number of components interacting through arbitrary potentials was developed in Ref.~\cite{Davletov_NJPh_2023} giving hope to a fruitful exploration of potential parametrizations that can, in turn, be applied in diagnostics and reduced models for transport and relaxation in WDM.

\subsection*{Challenges \& outlook}

In the last few years, AA and CM have mutually fertilized each other. They have been driven by the need to explain the results of increasingly complex experiments~\cite{Ciricosta_PhRvL_2012, Hoarty_PhRvL_2013, glenzer_PRL_2007, Bailey_Natur_2015}. They now share some common ingredients to describe the thermodynamics and the optical properties of mixtures of ions, the structure of which is strongly modified in the warm dense matter regime. Though far more complex than their historical premises, they are still quite efficient when compared to DFT-MD simulations, while being able to address the richness of atomic structure.

Current efforts to extend AA models into configurationally resolved models have opened up a new frontier in research, in particular to determine the 
EOS, IPD, optical properties and conductivities~\cite{SonCL, hansen2023, piron2013variational, starrett2024dense, faussurier2018density}.  Exploitation of AA models for electronic and ionic transport coefficient tables and fits is ongoing and  will continue to improve our understanding of inertial confinement fusion~\cite{haines2024charged} and astrophysical objects like white dwarfs~\cite{saumon2021current, piron2018average}.  

Alternative boundary conditions is also a promising area for improvement, with the potential to make the models more physically realistic~\cite{MassacrierBoehmeVorbergerSoubiranMilitzer2021}.  In this context, the connection of AA models to the Korringa-Kohn-Rostoker method~\cite{Starrett_PRE_2020} has been demonstrated and investigated~\cite{ottoway2021effect}, and makes clear the connection to multi-center DFT calculations.

In the same vein, a paradigm shift is taking place concerning electronic states. Instead of the usual distinction between bound and free electrons, the meaningful characteristics of wave functions start to emerge as being localized versus delocalized~\cite{Gao_PRE_2025, Gawne_PRE_2023, Perez-Callejo_PRE_2024}, closer to a condensed matter view. This refers back to the choice of boundary conditions in the quantum solutions of AA models or modified atomic structure codes. State-of-the-art DFT-MD simulations have shown that \textit{bound states} in dense plasmas or WDM are rather short-lived correlations than stable species as assumed in CMs~\cite{Collins1995, Murayama2025}. This has profound consequences on the very meaning of related quantities such as the average ionization degree or the ionization potential depression, their use, e.g., in  the interpretation of X-ray Thomson scattering experiments based on the Chihara decomposition~\cite{Doeppner_nature_2023, Witte_PRL_2017, Schoerner_PRE_2023}, or for transport coefficients~\cite{grabowski2020}.

All these recent and somewhat patchy developments still have to be integrated in a consistent derivation, as was done for the single AA in~\cite{blenski2007variational}. This would give a coherent view of how a mixture of electrons and nuclei still build ions under WDM conditions, which effective potentials should be used for their interactions in the free energy construction, and how they respond to external probes as, e.g., in XFEL experiments.

\newpage 
\clearpage

\section{Diamond Anvil Cells}\label{section_09}
\author{Zuzana Kon\^{o}pkov\'{a}$^1$, Alexander Goncharov$^2$}
\address{
$^1$European XFEL GmbH, DE-22869 Schenefeld, Germany \\
$^2$Carnegie Institution of Washington, Washington, DC, 20015, United States}

\subsection*{Introduction}
Warm Dense Matter (WDM) represents an intermediate regime between condensed matter and plasma physics, characterized by extreme conditions of temperature and pressure. One of the methods to study high-pressure states relevant to planetary interiors is static compression using diamond anvil cells (DAC). DACs allow for controlled compression of materials to pressures comparable to those found in Earth's outer core. However, the mechanical limitations of diamond anvils, particularly their confinement failure at very high pressures~\cite{OBannon2018}, restrict the P-T (pressure-temperature) conditions that can be explored, allowing only the lower boundary of WDM states to be studied. Despite these limitations, DACs provide a unique advantage for studying exotic states of matter, such as superionic phases, on extended time scales. These states, which exhibit ionic conductivity and structural transformations, can persist for long enough for detailed investigation. To reach high-temperature states within DACs, laser heating is commonly employed, using focused pulsed or continuous near-infrared (NIR) lasers~\cite{Goncharov2009,Anzellini2020,Konopkova2021,Konopkova2016}. The technique is also widely used for the processing of materials and the synthesis of new materials~\cite{Alabdulkarim2022}. Over time, DACs have been coupled with X-ray probes at synchrotron sources and Free Electron Lasers~\cite{Cerantola2022} and various spectroscopy techniques~\cite{Jiang2020}. Extreme P-T states relevant to the WDM regime have also been achieved in shock compression experiments, where samples were precompressed in a DAC~\cite{Jeanloz2007,Brygoo2015}. This approach enables access to a wider range of states beyond the principal Hugoniot.  

\subsection*{State of the art}

Diamond anvil cells coupled with optical spectroscopy techniques have allowed the investigation of electronic structures in various systems, including simple molecular materials, noble gasses, and compounds of the lower mantle ~\cite{Lobanov2021,Jiang2018,Jiang2020,Prakapenka2021,McWilliams2015}. These experiments, which employ pulsed laser heating in combination with fast optical probes utilizing broadband pulsed lasers and gated data acquisition, have demonstrated the metallization of simple molecular compounds under conditions approaching those of WDM. 

\subsubsection*{Noble gases} 

\begin{figure}
\includegraphics[width=\linewidth]{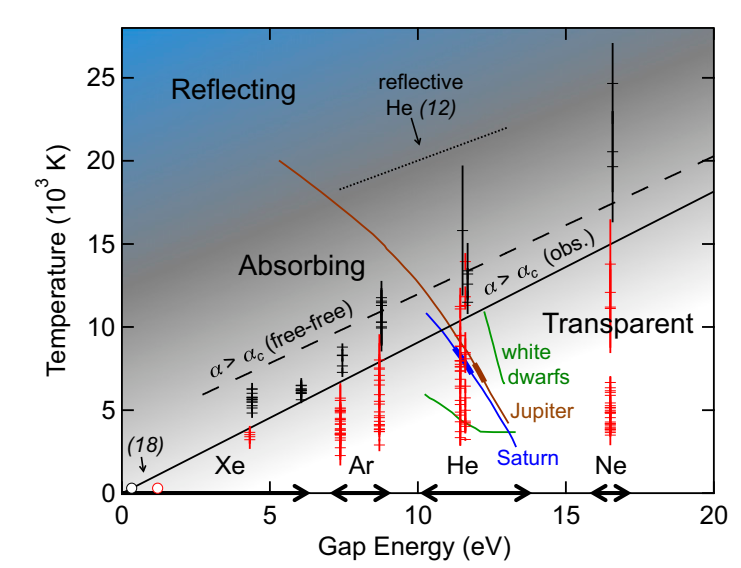}
\caption{Optical properties of noble gases at 1.15-2.76 eV as a function of band gap, from Ref.~\cite{McWilliams2015}.}
\label{noble}
\end{figure}

In noble gases (He, Ne, Ar, Xe), the insulator-to-metal transition has been observed at temperatures of 4,000–15,000 K and pressures of 15–52 GPa (Fig.\ref{noble})~\cite{McWilliams2015}. The onset temperature of high absorption ($T_C$) was found to exhibit a systematic increase with the band gap ($E_g$), following $k_{\mathrm{B}}T_C \approx 0.078 E_g$. The resulting high-temperature states exhibit low electron mobility, resembling those found in amorphous semiconductors or poor metals. The electronic changes in the warm dense noble gases were found to play an important role in the miscibility question - for example, the onset of the conducting states corresponds with the conditions where helium should be soluble in metallic hydrogen. These findings provide new insights into the physical processes that occur in white dwarf stars \cite{Bergeron2022} and giant planets such as Jupiter and Saturn - phase separated He should tend to dissolve just above Saturn's core, however, Ne may remain exsolved in the deep interior of Saturn, while in Jupiter's core, both He and Ne should undergo insulator-to-conductor transformation and hence become more soluble \cite{McWilliams2015}.

\subsubsection*{Hydrogen}

\begin{figure}
    \centering
    \includegraphics[width=\linewidth]{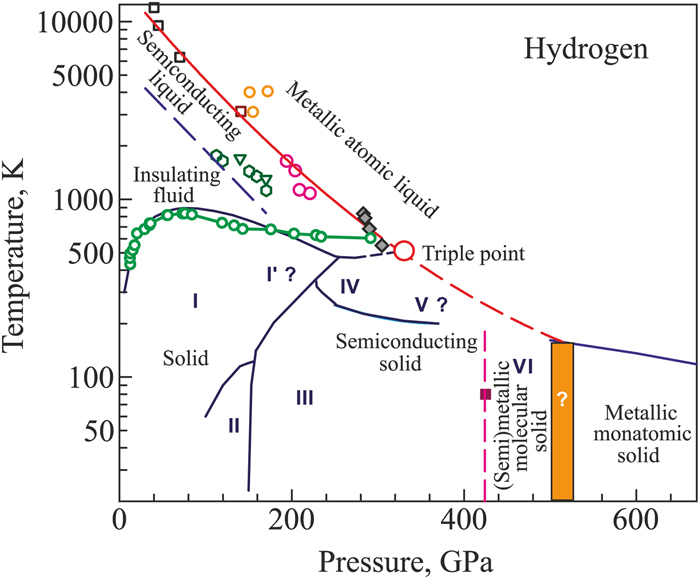}
    \caption{Phase diagram of hydrogen at high pressure and various temperatures, from \cite{Goncharov2020}. The phase lines are from static and dynamic experiments of Refs. \cite{Goncharov2011,Howie2015,eremets_nat-phys_19,loubeyre_nat_20,Zha2017}. A hypothetic transition to an atomic metallic phase is shown by an orange box. At higher pressure theory predicts an atomic metallic phase with a declining melt line shown by a solid blue line \cite{Chen2013}. At high temperature, the experiments and theory show two almost parallel boundaries corresponding to a transition into a semiconducting state (dashed blue line) and insulator-metal transition (solid red line).}
    \label{fig:hydrogen}
\end{figure}

At conditions approaching the WDM regime, hydrogen exists in a dense fluid or plasma state. As hydrogen transitions into the fluid phase, it loses its molecular character, undergoes chemical bonding changes, and becomes conductive. Theoretical calculations predict that the corresponding phase boundary terminates at a critical point below which, at lower pressures and higher temperatures, hydrogen undergoes gradual ionization. The transition of insulator-to-metal molecular dissociation has been studied in a wide T-P range (Fig. \ref{fig:hydrogen}) using dynamic compression and DAC techniques \cite{Loubeyre2012,Celliers_Science_2018,Knudson_Science_2015,weir_prl_96,fortov_prl_07,Zaghoo2016,Zaghoo2017,Jiang2020,McWilliams2016}. At low temperatures (80 K), synchrotron infrared absorption spectroscopy suggests hydrogen transition to a metallic state above 420 GPa \cite{loubeyre_nat_20}. However, at high temperatures, optical absorption and reflectivity measurements in DACs have observed metal-like reflectance at P$>$150 GPa and T$>$3000 K. Extrapolations suggest that the insulator-to-metal phase boundary intersects the melting line at a triple point, beyond which solid hydrogen (phase IV or V) is expected to transition into a metallic liquid. However, this regime remains largely unexplored. DAC experiments on hydrogen are particularly challenging because of its high mobility under extreme pressures and temperatures. As a result, many of the reported experimental findings are unique and often significantly contradict each other.

\subsubsection*{Water}

\begin{figure}
    \centering
    \includegraphics[width=\linewidth]{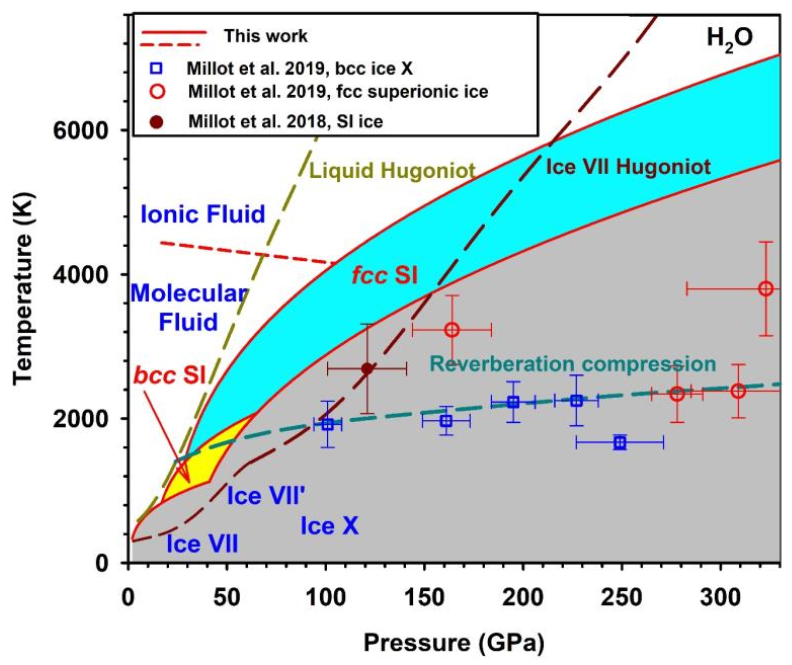}

    \caption{Phase diagrams of water in dynamic and static DAC experiments from \cite{Millot2018} and \cite{Prakapenka2021}, respectively.}
    \label{fig:ice}
\end{figure}

DACs also contributed to the study of superionic (SI) water phases under high-pressure and high-temperature conditions. In SI states, protons exhibit high mobility within a rigid oxygen lattice, giving rise to extraordinary ionic conductivity. These states can also undergo structural phase transitions. The superionic and conducting states of liquid water have been observed using dynamic and static compression techniques \cite{Millot2018, Prakapenka2021} (Fig. \ref{fig:ice}). Discrepancies remain in the P-T conditions, with static compression experiments indicating significantly higher temperatures for the stability of the face-centered cubic (fcc) SI phase. Based on these findings, fcc SI ice is expected to exist in water-rich giant planets such as Neptune and Uranus.

A recent breakthrough in static compression techniques concerns their integration with X-ray free electron lasers (XFELs), where heating is achieved through the absorption of intense femtosecond high-photon-energy X-ray pulses \cite{Zastrau_JSyncRad_2021,Liermann2021,Ball2023}. This method allows for precise control over heating rates and enables studies at higher temperatures where conventional laser heating is problematic. Using an array of densely spaced pulses, rapid sequential heating is achieved before significant cooling occurs between pulses, effectively maintaining high temperatures over extended timescales.

A pioneering XFEL experiment \cite{Husband2021} demonstrated an innovative method for heating water samples at high pressures through indirect X-ray absorption by a surrounding gold absorber. In this process, core-electron excitation in gold generates an initial high electron temperature on a femtosecond timescale. Over hundreds of picoseconds, electron-ion thermal equilibrium is established, leading to hydrodynamic expansion of the gold sample and subsequent heat diffusion into the adjacent water sample. This mechanism enables the water sample to reach several thousand kelvins, sustaining these temperatures for hundreds of nanoseconds before cooling. If successive X-ray pulses arrive before complete cooling, the high-temperature state can be maintained over tens of microseconds, forming a cycle of heating and cooling that facilitates the study of thermodynamic pathways. Such experimental setups, when combined with high-repetition-rate detectors like the AGIPD at EuXFEL \cite{Allahgholi2019}, enable structural analysis at each temperature step with unprecedented temporal resolution. This has allowed researchers to probe the phase transition kinetics of SI water at high pressures, revealing insights into polymorphic transitions. Notably, SI water exists in two polymorphic forms: one with a body-centered cubic (bcc) oxygen sublattice and another with a face-centered cubic (fcc) arrangement, with the mobile protons resulting in high conductivity (Fig. \ref{fig:ice}). 
However, in contrast to conventional laser-heating experiments, which observe the formation of the SI-fcc phase at pressures as low as 30 GPa, XFEL-driven experiments have primarily observed SI-bcc up to 50 GPa \cite{Husband2024}. This discrepancy highlights the role of distinct thermodynamic pathways in phase formation—specifically, crystallization from melt during cooling versus phase formation during heating. Furthermore, the vastly different experimental timescales in XFEL experiments, which operate orders of magnitude faster than traditional methods, may favor nucleation of phases with lower solid-fluid interfacial free energy, differing from those predicted by equilibrium phase diagrams based on long-duration heating experiments. These findings underscore the need to reconsider conventional phase stability diagrams when studying matter under extreme conditions using advanced XFEL techniques.

\subsubsection*{Iron}

The melting and phase diagram of iron under conditions relevant to Earth's inner core has been extensively studied to interpret seismic observations that reveal exotic core properties, such as strong elastic anisotropy, low viscosity, high sound attenuation, and a low shear modulus. Experimental research consistently identifies the hexagonal close-packed (hcp) phase as the last stable solid form of iron before melting at pressures exceeding 200 GPa. In particular, along the iron Hugoniot, the melting line is crossed at approximately 250 GPa, yet most dynamic compression experiments indicate the persistence of the hcp phase under these PT conditions \cite{Balugani2024,Turneaure2020,Kraus2022}. Theoretical predictions suggest that other phases with similar Gibbs free energies may become stabilized under these conditions, particularly if impurities are incorporated. Recent MD studies propose that the bcc structure, stable at core pressures below melting temperatures, exhibits rapid atomic diffusion that mechanically stabilizes the phase \cite{Belonoshko2017,Li2024,Ghosh2023}. Such diffusion, occurring via collective atomic movements along specific crystallographic planes, has been suggested as a mechanism explaining the Earth inner core low viscosity and high elastic anisotropy. However, despite these predictions, the experimental detection of such a state has remained elusive. Recent high-pressure experiments aimed to address this gap by probing statically compressed hcp-Fe using DACs at pressures exceeding 200 GPa. These samples were heated using pulse trains from the EuXFEL with progressively increasing pulse energies. 
However, further investigations are needed to clarify the mechanisms driving phase transformations induced by ultrafast X-ray heating. The unique conditions created by XFEL pulses—particularly the high electron temperatures resulting from intense X-ray-driven electronic excitation—can affect phase stability in ways not observed with other compression techniques \cite{Azadi2024}. This highlights the need to explore the interplay between electronic excitation, atomic diffusion, and phase transitions in iron under extreme conditions.


\subsection*{Challenges \& outlook}

Recent experiments have demonstrated that a variety of DAC techniques can be employed to investigate materials under conditions relevant to the WDM regime. However, this remains a challenging domain due to the difficulty of reaching and maintaining the extreme pressure-temperature (P-T) conditions of WDM, as well as performing reliable in situ measurements. Achieving and sustaining such states often requires sophisticated experimental setups and refined technical solutions.

Most DAC-based approaches operate in very short time domains—typically in the nanosecond to microsecond range. These rapid protocols help minimize unwanted chemical reactions and enable data acquisition before the sample loses confinement. However, operating on such short timescales also means that materials may not reach thermodynamic equilibrium, making it necessary to account for nonequilibrium states and the kinetics of phase transformations. In this context, close collaboration with theory is essential for interpreting experimental results.

DAC techniques continue to evolve, enabling access to higher pressures \cite{Dewaele2018,Jenei2018}. A significant recent development is the integration of DACs with ultrabright radiation sources such as X-ray free-electron lasers (XFELs), offering new opportunities to probe material properties in the WDM regime. In particular, the use of pump-probe beamlines with variable time delays at XFEL facilities allows for time-resolved studies of transient states and dynamic processes in WDM. Another promising direction involves combining ultrabright XFEL or synchrotron sources with pulsed laser heating—both to generate extreme P-T conditions and to perform spectroscopic measurements, thereby enhancing our ability to study the structure and dynamics of WDM.

n\newpage 
\clearpage

\section{Laser Creation of WDM}\label{section_10}
\author{David Riley$^1$, Brendan Kettle$^2$}
\address{
$^1$Queen's University of Belfast, Belfast BT7 1NN, United Kingdom\\
$^2$The John Adams Institute for Accelerator Science, Blackett Laboratory, Imperial College London, London, SW7 2AZ, United Kingdom\\
}

\subsection*{Introduction}

The extreme conditions of warm dense matter (WDM) generally mean that we cannot create samples in a quasi-steady state and we need to rely on target inertia to maintain solid or near-solid density for long enough to make measurements.  This means delivering the energy needed (typically $>10^{11}$ Jm$^{-3}$) in timescales faster than the hydrodynamic expansion time. For this reason, high power lasers, which can deliver up to kJ of energy in ns or sub-ns timescales, have become a widely used methods of creating WDM in the laboratory. 

\begin{figure}[tbh]
    \centering
\includegraphics[width=0.9\linewidth]{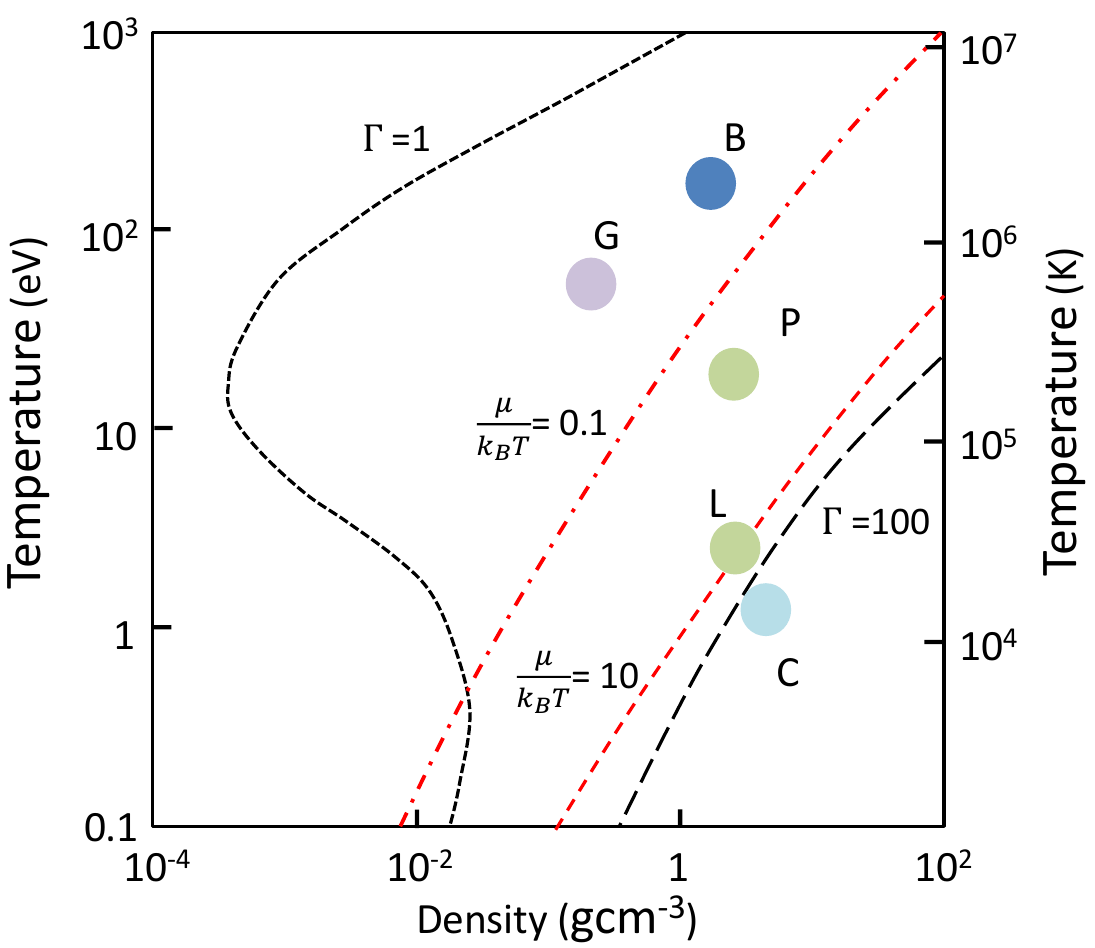}
\caption{Schematic showing where some experiments referenced lie in the WDM regime. G: Glenzer et al \cite{glenzer2003} using radiative heating. P: Patel et al \cite{patel2003} and L: L\'evy et al \cite{levy2009}, both using proton heating. B: Booth et al \cite{booth2015} with fast electrons. C: Coleman et al \cite{coleman2022} using shock compression. The black curves show the contours of values of ion-ion coupling parameter that bracket the WDM regime, calculated using a Thomas-Fermi model for Z=12. The red lines show contours of degeneracy parameter that roughly bracket the WDM regime.}
\label{fig:wdmregime}
\end{figure}

In Fig. \ref{fig:wdmregime}, we see typical conditions created by laser-driven WDM experiments discussed in this article, mapped onto temperature-density space. We see that laser-driven experiments can produce samples ranging from moderately to highly coupled systems with ion-ion coupling parameters, ($\Gamma=Ze^{2}/ak_{B}T_{i}=1-100$, where $a$ is the ion-sphere radius). We can also see electron degeneracy parameters, ($\mu/k_{B}T$) ranging from  0.1 to 10, where $\mu$ is the electron chemical potential. For the purposes of this article, we classify experiments into three broad types.

Firstly, it has long been demonstrated that ns duration lasers can be used to drive shock waves with pressures in excess of 100 GPa, either by direct laser ablation \cite{veeser1978,trainor1979} or indirectly, using laser generated x-rays \cite{lower1994,zhao_2013,zhao_2017}. Such pressures are capable of compressing solid samples up to a few times their ambient density with accompanying shock heating to temperatures of the order of $\sim$1 eV. These conditions cover the high density and low temperature region of the WDM regime as outlined in figure \ref{fig:wdmregime}. The speed of shock at the $>$ 100 GPa regime is typically of order $10^{4}$ ms$^{-1}$ and with ns duration pulses we can typically compress targets of 10s $\mu$m thickness.

Secondly, high-power ns laser pulses can be used to create intense sources of x-ray photons \cite{kania1992}, in the sub-keV to a few keV range, that can heat a separate sample via photo-absorption. This approach has been used to create large relatively uniformly heated WDM samples with temperatures from $\sim$1 eV to 10's eV at densities close to solid \cite{glenzer2003,kettle2016,Kettle_2015}. 
    
Thirdly, for shorter pulse experiments with ps and fs lasers, beams of energetic electrons and protons can be generated that can be used to heat solid targets. In the former case, the "fast" electron propagation into the target causes a return current of lower energy electrons that resistively heat the solid \cite{bell1997} and temperatures in excess of  $\sim$100 eV can be created by such means \cite{booth2015}. With appropriate target foils, intense laser pulses can create energetic beams of protons \cite{clark2000,Snavely2000_a} by mechanisms such as target-normal sheath acceleration (TNSA) with proton kinetic energies usually in excess of 10 MeV, and recent results showing energies up to 150 MeV \cite{ziegler2024}. The spread of proton energies generated can lead to quite uniform heating over 10s of $\mu$m in a target. A key feature of the generation of proton beams is that, in contrast to fast electron heating, these can be used to heat a solid target separated from the proton generation foil. Early experiments along these lines \cite{patel2003} showed that by using a curved acceleration target, proton beams could be focused to enhance the localised heating, with up to $\sim$20eV achieved. An important consideration in experiment design is that the spread of energies leads to a temporal dispersion from sub-ps initial pulses to 10s ps at the separate target.
    
Finally, we might consider experiments in which, far from wanting to study an equilibrated sample, we wish to explore time-dependent effects such as electron-ion equilibration e.g. \cite{Mahieu2018}. Since the advent of sub-100 fs laser pulses it has been possible to consider the generation of a uniform slab of WDM created by direct laser heating of electrons e.g \cite{Ng2005}. For thin foils heated to energy densities that are not too high, we might have hydrodynamic expansion times that are shorter than the typically ps or sub-ps timescales for microscopic processes such as electron-ion equilibration.

\subsection*{State of the art}

\subsubsection*{Laser driven shocks}

Since the initial experiments, e.g. \cite{veeser1978,trainor1979} generation of strong shocks in solids has benefited greatly from the development of phase plate technology \cite{kato1984} that allows a smoother distribution of intensity in the focal plane. Developments since then, for example, continuous phase plates,  e.g. \cite{marozas2016} have allowed millimeter sized focal spots with noncircular, flat-topped profiles to be deployed  e.g. \cite{riley2023}. This provides more lateral uniformity in the shock pressure and, with spots up to mm in scale, allows a more one-dimensional experiment that is more readily compared to simulations. An alternative to the use of phase plates in generating a smooth pressure profile is to use the intense x-ray fluxes that can be generated from high Z foils or hohlraum targets to drive a planar shock. This technique relies on using the short absorption length of sub-keV photons in an absorption layer on the surface of the sample to be shocked. The subsequent heating and pressure rise drives a shock into the sample. This technique generally requires higher laser energies due, in part, to the loss of efficiency in converting laser energy to x-ray energy. Nevertheless, pressures of up to 2 TPa have been generated \cite{lower1994}.

        The development of pulse shaping technology with sub-100ps resolution \cite{dinicola2024}, has been important in providing for temporal control of the pressure history. Using such techniques we can, for a given peak pressure, vary the compression from a single shock, on the Hugoniot, to  stepped shocks and quasi-isentropic ramp compressions at high density and lower temperature, thus exercising a degree of control over the WDM region sampled. This is significant as the conditions in planetary interiors are generally off the Hugoniot, which for a given  compression will have too high a temperature to match the desired conditions.
   
   A significant advance in the last two decades has been the marriage of high-power laser systems in the $\sim$10-100 J energy regime to x-ray free-electron sources \cite{nagler2015matter,Balugani2024}. The latter provides a tuneable, ultra-short and highly collimated probe with which to make full and rich use of x-ray diffraction and x-ray Thomson scattering techniques. Such facilities are equipped with high-quality VISAR and streaked optical pyrometry (SOP) systems that allow estimates of shock pressures, speeds and temperatures to complement the measurements of microscopic structure, such as melting and phase changes \cite{White2020} coming from the x-diffraction measurements and measurements of temperature, electron density and ionisation from the x-ray Thomson scattering data \cite{Glenzer_revmodphys_2009}. These facilities, with a repetition rate of order 10 shots or more per hour, for the optical laser, have been used to make detailed measurements of shocked phases of solids at pressures relevant to, or bordering on the WDM regime e.g. \cite{briggs2017}.
    As with free-electron lasers, a great deal of progress has been made in recent years in combining synchrotron sources with high-power laser systems \cite{Turneaure2020}. Despite the broader spectral range compared to an x-ray FEL source and the longer pulse durations, (which are generally $\sim$100ps as opposed to $\sim$0.1 ps in XFELS), Coleman et al \cite{coleman2022} have, for example, studied shock compressed Ag to 330 GPa, making high quality x-ray diffraction measurements to extract density measurements.

\subsubsection*{WDM creation with short-pulse lasers:}

As noted above, high intensity ps lasers have been used to create WDM samples via two basic mechanisms that involve particle heating. The first is the generation of a supra-thermal electron population at the surface of a solid target. This happens via mechanisms such as resonance absorption and \textbf{J}$\times$\textbf{B} heating. The effective temperature of these electrons can range from 10s keV to several MeV. As the electrons stream into the solid target, the strong return current causes resistive heating that can easily reach the WDM regime of 1-100 eV and beyond. A drawback of this technique is that there can be strong gradients in the conditions if the target is thick and the WDM sample is closely adjacent to a hot strongly emitting laser-created plasma which may interfere with diagnostics. A key diagnostic of this kind of experiment is observation of the spectrum of K-$\alpha$ emission from the heated material \cite{hansen2005}. This is generated by inner-shell ionisation caused by the fast electron beam. Thus the heating mechanism and diagnostic are intimately tied and time dependent effects essential to consider in the analysis.
    By contrast, proton heating experiments have relied on generating a loosely collimated source of MeV protons from one foil by mechanisms such as TNSA, which are then used to irradiate and heat a separate sample foil. Complex experimental arrangements have been implemented to allow diagnostic probing of the dense matter state created beyond measurement of optical emission from the sample surface. For example, L\'evy et al \cite{levy2009}, implemented XANES measurements alongside optical emission measurements and surface expansion measurement by frequency domain interferometry. They were able to probe the evolution of the absorption spectrum and, by fitting to models, track the evolution of the sample temperature.

For ultra-short pulse direct heating of a thin-foil experiments have mostly concentrated on using optical diagnostics to explore the electron-ion equilibration times, connecting reflectivity and conductivity. For example, Chen et al \cite{Chen2013} used such a technique to show electron-ion equilibration times in Au that were much longer than predicted by theoretical calculations.

\subsubsection*{Radiatively heated WDM samples:}

An experimental advantage to using x-ray heating to create WDM samples is that the mean free path of X-rays in the keV regime, can be 100s of microns for lower Z elements. This allows for mm sized samples with low heating gradients \cite{glenzer2003}. As a consequence, hydrodynamic expansion times can be more than a few nanoseconds, allowing for quasi-steady state conditions to be probed. For example, for the conditions generated by Glenzer et al \cite{glenzer2003}, the sound speed would be expected to be $\sim4\times10^{4}$ ms$^{-1}$ and with a sample size of 0.6 mm we expect a decompression time, estimated as size divided by sound speed, of about 15 ns which allows time for equilibration, for example, of the electron and ion temperatures. A drawback of creating such large samples is that, even with good conversion of laser-energy to x-ray energy, large laser energies up to the kJ regime are needed to reach the required energy densities. A further challenge is that for moderate to high Z samples, higher energy photons for both heating and probing may be needed. Development of the latter has been carried out \cite{barrios2013} and upgraded XFEL facilities can operate at $>$10keV.

\subsection*{Challenges \& outlook}

A major challenge for laser-created WDM remains the desire to create samples with temperature and density conditions and materials that are appropriate to planetary interiors. An important aspect of this challenge includes the need to have samples that are quasi-steady-state. For laser generation of WDM, we often heat either the ions or electrons and rely on electron-ion energy exchange to come closer to thermal equilibrium. For ns pulses this would normally be justified. On the other hand, with ultra-fast optical pulses in the femtosecond regime, we can deliberately heat on timescales much shorter than e-i equilibration and with appropriate diagnostics explore important processes such as thermal and electrical conductivity that depend on the rates of microscopic processes. A discussion of time-dependent WDM has been presented, for example, in \cite{white_dynamic_2023}.

Despite challenges, the outlook is positive for the use of high-power lasers for WDM creation. A key reason for optimism is the already successful and ongoing integration of lasers with x-ray FELs and synchrotron facilities. For example, a key feature that has made x-ray FELs successful, in a WDM diagnostic role, is that the highly collimated, quasi-monochromatic, tuneable source is almost ideal for techniques such as x-ray Thomson scattering  which can provide detail including ionisation, electron and ion-temperature \cite{Glenzer_revmodphys_2009}.  X-ray diffraction techniques, likewise, can be implemented with the ability to probe for clear indications of melting and phase changes in multi-megabar shock experiments. The ultra -short ($\sim$0.1 ps) duration of the XFEL means it can be coupled to ultra-short laser-driven samples to probe the microscopic structural and temperature evolution of the sample after heating on a similar timescale and provide data complementary to the optical reflectivity data used previously. For synchrotron sources, the  naturally broad and smooth spectrum is well suited to techniques such as EXAFS and XANES.

The use of laser-plasma x-ray sources for volumetric heating requires larger lasers of $\sim$kJ and above, especially if we desire the creation of larger samples with longer decompression timescales, where we can approach the quasi-steady state conditions. There are, worldwide, several such facilities, e.g. NIF \cite{Moses_NIF}, Omega, Omega-EP, Orion, Gekko, LMF \cite{Rozanov_2016} and the Shenguang series including Shenguan II and III. New proposed facilities such as NSF-Opal \cite{bromage_2025},will further extend the possibilities for WDM research. These large facilities generally have a low repetition rate. This is a challenge for carrying out systematic experiments. However, technologies such as the DiPole diode pumped laser systems \cite{Mason2018}, deployed at XFEL in Hamburg \cite{Gorman_JAP_2024}, allow 10 Hz repetition rates for pulses over 100 J and may in future be extended to kJ systems. Such inter-laboratory collaboration is an important feature for development of future facilities as is a multi-diagnostic approach. The latter is key as WDM diagnostics are often quite theory dependent for their interpretation and, as noted above, the use of multiple diagnostics, such as XRTS and XANES, discussed in other chapters of this roadmap, alongside diagnostics such as VISAR and SOP can create a more complete, better constrained picture of a WDM sample. In parallel, collaboration between modelers in molecular dynamics and radiation hydrodynamics can be important in assembling a complete picture of an experiment from macroscopic to microscopic structure and dynamics.
 
There are also specific, challenging, experiments that can developed, such as recent experiments on Si shocked up to melting that have explored the ion temperature via ultra-high resolution ($\sim$0.1eV) x-ray Thomson scattering techniques e.g. \cite{McBride_RevSciInstrum_2018}. The extension of such measurements to ion-acoustic waves in WDM will be a challenge as background noise may be higher and much of the probe energy is lost in achieving the necessary meV bandwidth. Such experiments are at the frontier of WDM exploration with laser created samples.
\newpage 
\clearpage

\section{Particle beams for WDM generation and stopping power studies}\label{section_11}
\author{Witold Cayzac$^{1}$, Sophia Malko$^2$, Paul Neumayer$^3$}
\address{
$^1$CEA, DAM, DIF, F-91297 Arpajon, France \\
$^2$Princeton Plasma Physics Laboratory, Princeton, New Jersey 08540, United States\\
$^3$GSI Helmholtzzentrum f\"ur Schwerionenforschung GmbH, 64291 Darmstadt, Germany
}

\subsection*{Introduction}

When a charged particle (projectile) at high kinetic energy penetrates matter (sample), it transfers energy to the sample, thereby slowing down. The energy transfer is dominated by Coulomb collisions with the electrons in the sample and scales with the  projectile's charge $\propto Z_p^2$.  
The increasing interaction time as the projectile slows down leads to a gradual increase of the longitudinal energy transfer, finally resulting in a rapid rise at the so-called Bragg-peak just before the projectile comes to rest. Note that this peaked energy deposition is the basis for tumor treatment with charged particles, e.g. \cite{Schardt_RMP_2010}. 

The range is directly related to the initial projectile energy $E_0$, and scales approximately $\propto E_{0}^{2}$. As an example, the range of a 250 keV electron is 300 $\mu$m, while that of a 10 MeV proton is 630 $\mu$m (in aluminum at solid density). The energy lost by the projectile is transferred to secondary electrons, which in turn distribute their energy in a collision cascade to other electrons in the sample. Rapid equilibration (femtosecond time scale) within the electronic subsystem followed by electron-ion equilibration (typically within picoseconds) leads to heating of the sample material along the projectile's path. Applied to the problem of generating WDM states in the laboratory, heating by charged particles thus presents a means to homogeneous, volumetric heating of a sample. Provided the duration of the projectile pulse is short compared to the hydrodynamic timescale, the heating can be considered isochoric, thus resulting in a sample with known density, well suited for measuring WDM properties. 


The ion stopping power is a fundamental transport property that reflects the excitation modes of the sample. It plays a key role in ICF, where the $\alpha$-particle stopping both in the hot-spot and the surrounding ultra-dense, partly degenerate fusion fuel crucially determines the propagation of the burn wave. Contrary to ion stopping in ideal highly ionized plasmas where numerous theoretical models are available \cite{Cayzac_PRE_2015,Cayzac_NatCom_2017}, stopping in WDM involves additional effects like partial plasma ionization and dense plasma effects, such as electron coupling and electron degeneracy. While most models agree for projectile velocities $v_p$ much larger than the thermal electron velocity $v_{th}$ ($v_p \gg v_{th}$), discrepancies increase for lower velocity ratios and are largest around the Bragg peak where $v_p/v_{th} \approx$ 1 and the stopping power reaches its maximum \cite{Gericke_2002,Ding_PRL_2018,White_PhRvB_2018}. In addition to these theoretical challenges, very few experimental data are available to date for benchmarking the models. Experimental measurements are necessary in the whole range of $v_p/v_{th}$ values, and are most difficult and most important in the Bragg peak region. Indeed, the latter is most sensitive to the plasma temperature and ionization, which requires a precise target characterization. Experiments are based on pump-probe setups and necessitate a well-known mono-energetic ion source, a well-characterized, uniform WDM sample that is quiescent during the ion beam duration, and a high-resolution detector.

\subsection*{State of the art}

Powerful bunches of energetic charged particles are typically produced by large-scale particle accelerators. For example, at the GSI Helmholtzzentrum für Schwerionenforschung (Darmstadt/Germany), the heavy-ion synchrotron \textit{SIS18} can deliver higly-charged uranium ions ($U^{73+}$) at several hundred MeV/u with up to $5\times10^9$ particles per bunch, with pulse durations down to 250 ns. The ion beam can be focused down to spot sizes of approx. 0.5 mm, reaching temperatures of up to 0.35 eV in a high-Z target \cite{Hesselbach2025}. The combination with GSI's high-energy laser PHELIX~\cite{Major_HPLSE_2024}, delivering nanosecond laser pulses up to $200\,$J at $527\,$nm to generate intense laser-driven x-ray sources, allows for x-ray probing of such heavy-ion heated matter, e.g. by x-ray diffraction or x-ray Thomson Scattering \cite{Hesselbach2025, Luetgert2024}.

Over the last decade, numerous studies have focused on the generation and optimization of particle beams, produced by ultra-intense laser pulses, garnering particular interest as compact, high-brightness particle sources. Laser systems, with energies ranging from a few J to kJ and pulse durations as short as $25$ fs and up to a few ps, can be focused to “relativistic” intensities of $10^{18} - 10^{21} \, \text{W/cm}^2$, where the corresponding ponderomotive potential reaches MeV and above.  In the interaction with overdense targets, e.g. samples at solid density, efficient collisionless absorption can reach values well above 50\% \cite{Ping_PRL_2008}, resulting in a significant fraction of laser energy converted to relativistic electrons \cite{Wilks1992, Kemp_PRL_2012}. These electrons are highly divergent, with a half-angle of approximately $30^\circ$, which scales with the laser intensity \cite{Beg1997,Key1998}. Transport of such intense relativistic currents can no longer be calculated by particle tracking Monte-Carlo codes, as they neglect the influence of the strong self-generated electric and magnetic fields \cite{Davies_PRE_2002}. Correctly modeling particle transport and target heating then requires hybrid models that includes field generation \cite{Honrubia_LPB_2006, Robinson_PoP_2007}. A large body of work has been performed to understand the transport of intense currents of relativistic electrons in the context of Fast-Ignition (e.g. \cite{Strozzi_PoP_2012,Green_NatPhys_2007}), and to improve beam collimation and reduce divergence, by application of external \cite{BaillyGrandvaux2018} and self-generated fields \cite{Scott2012, Malko2019, Kar_PRL_2009}. At MeV energies, the range of such electrons is of order millimeter, and consequently, the energy is spread over a large volume. However, when using mass-limited targets, electron refluxing within the target, due to reflection from the surface sheath fields, effectively confines the hot electrons to the target volume (e.g. \cite{Myatt_PoP_2007}). Thus, a large fraction of laser energy can be deposited within picoseconds in a small target volume, leading to isochoric heating to temperatures above 100 eV \cite{Nilson_PRL_2010,Neumayer_PoP_2012}.

When using mass-limited targets, in particular thin foils, the relativistic electron population results in a charge separation and the formation of strong electrostatic sheath fields at the target boundaries. This field accelerates protons and heavier ions from the target surface via \textit{Target Normal Sheath Acceleration} (TNSA). The resulting proton beams typically exhibit a half-angle divergence of about $20^\circ$ and a broadband energy spectrum extending up to several tens of MeV, depending on the laser and target parameters \cite{Macchi2013,Snavely2000_a}, with record energies now reaching above 100 MeV \cite{ziegler2024}. Beyond TNSA, alternative ion acceleration mechanisms such as \textit{Radiation Pressure Acceleration} (RPA) \cite{Gonzalez2022}, \textit{Breakout Afterburner} (BOA), and \textit{collisionless shock acceleration} \cite{Fiuza2012,Fiuza2013,Liu2016} have been discovered in recent years. Advancements in targetry \cite{Prencipe2017}, including liquid and near-critical-density gas jets \cite{Bohorquez2024}, have played a key role in these developments. 

One of the key applications of interest has been the generation of matter at high energy density conditions. Extensive experimental and computational work has investigated schemes to focus laser-accelerated protons in order to increase the specific energy deposition. This was first demonstrated at the Janus laser facility at LLNL \cite{patel2003} using hemispherically shaped foils (also called \textit{“hemis”}). As in the TNSA-mechanism the acceleration is along the normal to the target surface, the protons were accelerated radially inward toward a focal point near the geometrical center of the hemisphere. Using streaked optical pyrometry and XUV emission to measure the thermal emission of a sample foil heated by the accelerated protons, significantly increased heating was demonstrated when using hemi-targets.
In addition, the proton focusing dynamics were inferred using the mesh radiography technique, which provided the position of the virtual focus and the size of the proton focal spot \cite{Bartal2012}. A few experiments have explored proton focusing using various structured targets — such as hemispheres and half-hemispheres, extending the results obtained at Janus with a 10 J laser to hundreds of joules at other laser facilities \cite{Offermann2011,Bartal2012,Foord2012,Kar2016}. 
Proton focusing with hemispherical targets is particularly attractive for proton fast ignition in inertial confinement fusion, where a proton beam is used to deposit energy within a 10 $\mu$m spot and heat the DT fuel. In such a scenario, the hemisphere would be attached inside a cone to protect it during compression. The addition of the cone was found to alter the focusing dynamics due to self-generated fields at the cone tip \cite{McGuffey2020}. 

Aiming at the benchmarking of ion stopping power modeling in dense plasmas, three main kinds of experiments have been performed over the last 10 years. First, nearly mono-energetic protons created from $\mathrm{D^3He}$ fusion reactions have probed X-ray heated light-element probes in metal-coated tubes. Initial measurements were done for Beryllium samples \cite{Zylstra_PRL_2015} and lacked a reliable temperature characterization. Follow-up experiments using an upgraded platform with X-Ray Thomson scattering were performed with Boron samples which were diagnosed to feature a $\approx$ 10\,eV temperature \cite{Lahmann_PPCF_2023}. These experiments probe high-velocity stopping ($v_p/v_{th} >$ 10), where thermal and degeneracy effects are not significant. Therefore, classical models like Zimmerman's one \cite{Zimmerman_LLNL_90} including partial ionization are sufficient to reproduce the data. In this beam-plasma regime, ion stopping can be used as a diagnostic for the ionization potential and the WDM electronic structure models. TD-DFT calculations \cite{Ding_PRL_2018, White_PhRvB_2018} are in good agreement with the data, and they predict a reduction of the stopping power compared to classical models at lower $v_p/v_{th}$ ratios. 
Second, an experiment using two short-pulse lasers has probed the stopping power at significantly lower velocity ratios $v_p/v_{th}$ down to $\approx$ 3, i.e. in the intermediate velocity regime approaching the Bragg peak \cite{Malko_NatCom_2022}. The short pulses generated both the proton probe beam (by TNSA) and the warm-dense carbon target to be probed. A specifically developed magnetic device was used to select a monoenergetic proton beam of around 500\,keV energy from a broadband TNSA spectrum as the projectiles \cite{Apinaniz_NatSR_2021}. The warm dense carbon sample was generated by irradiating a 1\,$\mu$m thick carbon foil by a 0.5\,J 200\,fs laser beam. The laser focal spot was much larger than the proton beam size for ensuring a one-dimensional plasma expansion. The plasma temperature was measured using two separate spectrometers and reached $\approx$ 5--10\,eV, benchmarking the hydrodynamic simulations. The measured energy loss appears smaller than any prediction, about 40\,\% below the classical models calculation, and still nearly 20\,\% below the TD-DFT calculations \cite{White_JPCM_2022} which are closest to the data. 
A third approach consists in measuring reaction-in-flight (RIF) neutron spectra generated in implosion experiments and comparing them with predictions assuming various stopping power models for the ions created in secondary fusion reactions and transported in dense and degenerate matter conditions \cite{Hayes_PoP_2015,Hayes_NatPhys_2020}. In contrast to the first two mentioned approaches, this latter one is an indirect measurement and, although it can test various models in degenerate conditions, it does not allow a precise model benchmarking.

\subsection*{Challenges \& outlook}
After recent upgrades and optimization, the \textit{SIS18} at GSI has set a new record in particle numbers with up to $4\times10^{10}$ uranium ions per bunch. This was achieved using ions at a lower charge state ($U^{28+}$), due to the higher space charge limit. Focusing of such high-rigidity beams will, however, require high-field super-conducting focusing magnets.
High beam-current accelerators can in principle be scaled to Mega-Joule pulse energies, at pulse durations down to 10 ns, parameters required for heavy-ion driven inertial fusion \cite{Hofmann_MRE_2018}. While such an ignition-scale accelerator facility is not being considered at the moment, a new generation of high-intensity heavy-ion accelerators is currently under construction in Germany and China. The HIAF facility \cite{Yang_NIMB_2013} is located near Huizhou, and is currently under commissioning. The international FAIR facility \cite{Spiller_IPAC_2014} is built alongside GSI, using the existing SIS18 as injector, and will go into operation within the next years. These facilities are expected to provide ion bunches of up to $5 \times 10^{11}$ high-Z ions (e.g. uranium), in pulses below 100 ns duration. Focused to sub-mm spot sizes, this will allow for heating to electron-volt temperatures, i.e. well into the WDM regime. Consequently, both facilities plan to entertain an extensive HED experimental program\cite{Cheng_MRE_2018,Schoenberg_PoP_2020}, see Sec.\ref{section_26} for more details.

While significant progress has been made in understanding proton focusing from hemispherical targets for heating samples, key questions remain unresolved in both simulations and modeling—such as identifying the optimal target-laser parameters for focusing and achieving high conversion efficiency. In addition, the connection between the virtual focus measured by mesh radiography and the physical focal spot has not yet been established. Therefore, new methods for measuring the effectiveness of proton focusing are required to further enhance the heating capabilities of proton beams using this technique.

While WDM samples are readily produced by intense-laser generated particle bunches, use of such samples to measure WDM-properties still remains a significant challenge, due to the short timescales, and rather small sample volumes, but also due to strong background produced in relativistic laser-matter interaction. Notable exceptions are for example measurements of electron-ion-relaxation in graphite by means of x-ray diffraction, heated both by laser-accelerated protons \cite{White_SciRep_2012} and hot electrons \cite{White_PRL_2014}, or measurements of the EOS along isentropes, using XUV-radiography of expanding wire-targets, heated by TNSA-protons \cite{Hoarty_HEDP_2012}.
Here, the advent of x-ray free electron lasers, with dedicated end-stations dedicated to HED experiments, featuring high-intensity laser systems, could prove to be a game changer. The unprecedented spatial and temporal resolution achieved by FELs will allow to perform precision measurements even on the micron-scale samples produced \cite{laso2024cylindrical}.


Concerning ion stopping-power studies, experiments using proton beams generated by fusion reactions and WDM samples generated isochorically by X-ray heating can be extended to lower $v_p/v_{th}$ ratios for probing interaction regimes where discrepancies between models are more important. One way is to increase the WDM temperature and ionization by optimizing the heating drive and by implementing lighter Lithium targets if feasible. Also, lower-velocity projectiles like 3\,MeV protons from DD reactions might be considered, but the consequently reduced target thickness needs to be managed in view of the quicker target expansion. 
Platforms using two short laser pulses are promising for measuring the stopping power in the Bragg-peak region in WDM, all the more as they allow a high-repetition rate for accumulating data. In a first step, data at 500\,keV proton energy should be refined by improving the statistics and the precision of the measurements. Then, by varying the selected proton energy and the plasma conditions, a wide range of $v_p/v_{th}$ values can be investigated, including the Bragg peak region, for constituting an experimental database in the long run. Yet, lowering significantly the proton energy requires an optimization of the energy selector and of the proton transport to reduce the proton beam bandwidth and time spread, and to mitigate the angular straggling and maximize the proton collection. Higher-Z materials than carbon with a more complex ionization distribution can also be envisioned.  
Studies of ion transport in moderately-coupled and degenerate conditions in the frame of direct-drive implosions are also in progress. Measurements of the energy-loss of knock-on deuterons from one-dimensional DT gas-filled CD-capsule implosions have been performed at the OMEGA laser facility and will be compared to various stopping models with simulations using a Monte-Carlo particle tracing code \cite{Kunimune_APS_2023}.

\newpage 
\clearpage
\section{Mechanical Shocks}\label{section_12}
\author{Gilbert Collins$^{1,2,3}$, Marcus D. Knudson$^4$}
\address{
$^1$Department of Physics and Astronomy, University of Rochester, Rochester, New York 14627, USA\\
$^2$Laboratory for Laser Energetics, University of Rochester, Rochester, New York 14623, USA\\
$^3$Department of Mechanical Engineering, University of Rochester, Rochester, New York 14627, USA
$^4$Pulsed Power Sciences Center, Sandia National Laboratories, Albuquerque, New Mexico 87185, USA
}

\subsection*{Introduction}

\begin{figure*}[t]
\centering
\includegraphics[width=0.85\linewidth]{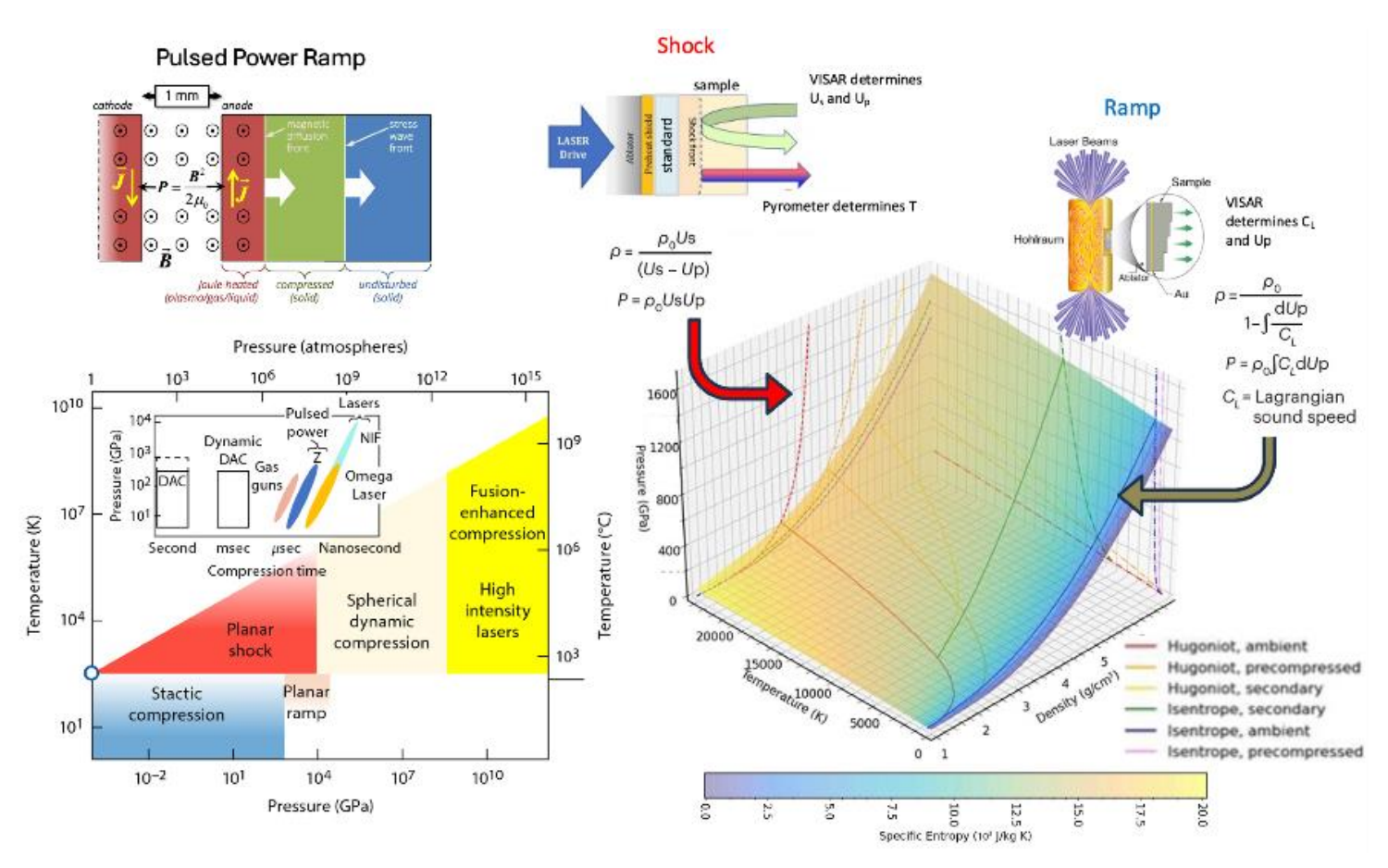}
\caption[width=0.98\linewidth]{Experimental techniques for accessing warm dense matter. The insets show target sketches for (top left) tuned current profiles to produce ramped $JXB$ compression wave through a sample, (top center) direct laser shock experiment in a transparent sample, and (top right) ramped laser intensity producing x-rays in a hohlraum to drive a ramped compression wave through a stepped target. Lower left inset shows general timescales for the different drivers~\cite{Duffy2019,NAS}.}
\label{fig:eosstates}
\end{figure*}

Developing a laboratory understanding of warm dense matter, i.e., in the deep interiors of planets, is the result of more than a century of work. From Bridgman’s early static experiments~\cite{Slater1965}, to Birch’s gun technology developed during the Manhattan Project~\cite{Birch1979}, to Van Valkenburg’s diamond anvil cell, (and many others around the world)~\cite{Fortov2016, Stishov} to today’s high energy density (HED) facilities, scientists have developed numerous ways to recreate and characterize matter at the same conditions produced by the crushing force of gravity in compact astrophysical objects. In the past few decades, significant advancements have been made in utilizing HED facilities such as Omega~\cite{boehly1997initial}, NIF~\cite{ICF_PRL_2024}, Z~\cite{Sinars}, and combined HED and x-ray sources~\cite{DCS,Gorman_JAP_2024,nagler2015matter}, to generate mechanical stress waves to explore materials in the warm dense matter regime. Different compression techniques access different pressure and temperature regimes, and each have intrinsic timescales for the measurement (Fig.\ref{fig:eosstates}). 


The general principle for these platforms is the same; available energy is used to deliver a shaped impulse that creates a tailored mechanical stress wave that drives a material into the WDM regime (Fig.\ref{fig:eosstates}). For gas guns, an impactor is accelerated by pressurized gas and impacts a sample, resulting in a strong mechanical stress wave in the sample. For large laser facilities, an intense laser pulse (between $10^{13} - 10^{15}$ W/cm$^2$) with a tuned temporal history is absorbed by an ablator material, resulting in a rapidly heated and expanding plasma. Momentum conservation results in a mechanical wave that propagates in the sample~\cite{fernandez_prl_19}. In other experiments, laser energy is converted to X-rays using a Hohlraum, resulting in similar ablation and stress waves~\cite{Marshall2019}. For pulsed power, large, shaped current pulses (10s of MA) create large magnetic fields (100s of T), and the resulting Lorentz force generates intense magnetic pressure (100s of GPa)~\cite{Davis2023, Brown2023}. For high conductivity metals (i.e., aluminum and copper) the resulting mechanical wave outruns the magnetic diffusion front, thereby enabling dynamic compression experiments. In other experiments, the Lorentz force is used to magnetically accelerate aluminum or copper flyer plates to ultra-high velocities (up to 40+ km/s) for more traditional plate impact experiments~\cite{Lemke2011}.

A steady shock experiment measures a single point on the Hugoniot – the locus of states accessed by a single shock. The Hugoniot relations~\cite{Zel’dovich2002}, determined from conservation of mass, momentum, and energy, are a set of 3 equations with 5 unknowns requiring measurement of 2 variables; often the shock velocity ($U_s$) and particle velocity ($u_p$) are measured.  Because it is often easier to measure $U_s$, other types of experiments measure $U_s$ of the sample in question and that of a material whose equation of state (EOS) is well known (referred to as an EOS standard). Impedance matching is then used to infer the sample $u_p$. Calibrating EOS standards at several TPa is an ongoing effort~\cite{Brygoo2015, Celliers2005, Knudson_2009, Desjarlais2017, Marshall2019}.

Most EOS experiments are performed in planar geometry so that waves are steady and diagnostic access is straightforward.  While a steady shock is important for measuring the pressure, density, and internal energy, a decaying shock~\cite{Duvall1978} is often used to sample a continuous sequence of Hugoniot states and is useful for determining phase transitions~\cite{Bradley2000, Celliers2000, Celliers2000a, Eggert2010, Hicks2006, McWilliams2012}. For even higher pressures, in excess of 100 Mbar, convergent shock waves are used to amplify the pressure; a spherically converging shock exhibits a dramatic amplification in pressure and density. The material ahead of the shock, is undisturbed (with the exception of preheating effects~\cite{Nilsen2020}) and subsequent shocked layers are on the principal Hugoniot, albeit at higher pressures as the shock converges~\cite{Döppner2018, Kritcher2016, Kritcher2020, Swift2012}. 

The high-pressure Hugoniot is not directly relevant to planets because a shock creates significant dissipative heating, producing temperatures well above most planetary temperature-pressure profiles~\cite{Wallace1981, Wallace1981a, Wallace1982}; these are often closer to a material's principal isentrope. Ramp compression is a technique that approaches isentropic compression (dissipative processes in materials result in moderate heating). Here the applied pressure continuously increases such that at every point the speed of the compression wave does not exceed the material's sound speed~\cite{Bradley2009, Fratanduono2021, Hall2001, Rothman2005, Smith2007a, Smith2014, Davis2023, Brown2023}. Because material response is nonlinear, ramped compression waves steepen with propagation. Monitoring evolution of the compression wave as a function of material thickness allows one to infer the sound speed as a function of pressure, and stress as a function of density.

Intermediate to shock and ramp compression includes multiple shock compression~\cite{Guarguaglini2019, Guarguaglini2021, Hansen2021, Holmes1995}, shock compression of precompressed materials~\cite{Brygoo2021, Celliers2010a, Eggert2008, Loubeyre2012, Loubeyre2004, Tabak2024}, shock compression of different polymorphs~\cite{Fratanduono2018, Millot2015, Hicks2006}), and shock-ramp compression~\cite{Smith2023, Seagle2020}. All of these techniques are attempts to control entropy production (and therefore temperature rise), where in shock-ramp experiments, the strength of the initial shock sets the entropy (i.e., temperature increase), while the subsequent ramp increases the pressure and density along a quasi-isentrope. This technique is the most flexible and allows access to the largest region of phase space, constrained only by the available energy at the HED facility.

\subsection*{State of the art}

Here we provide two examples of the state of the art in using mechanical waves to probe warm dense matter. The first illustrates high-precision Hugoniot and decaying shock experiments to probe the melt behavior of carbon. The second illustrates cryogenic shock, precompressed shock, multiple shock, and shock-ramp experiments to probe the metallization of hydrogen over a broad range of temperature and pressure.

\begin{figure}[tbh]
\centering
\includegraphics[width=0.98\linewidth]{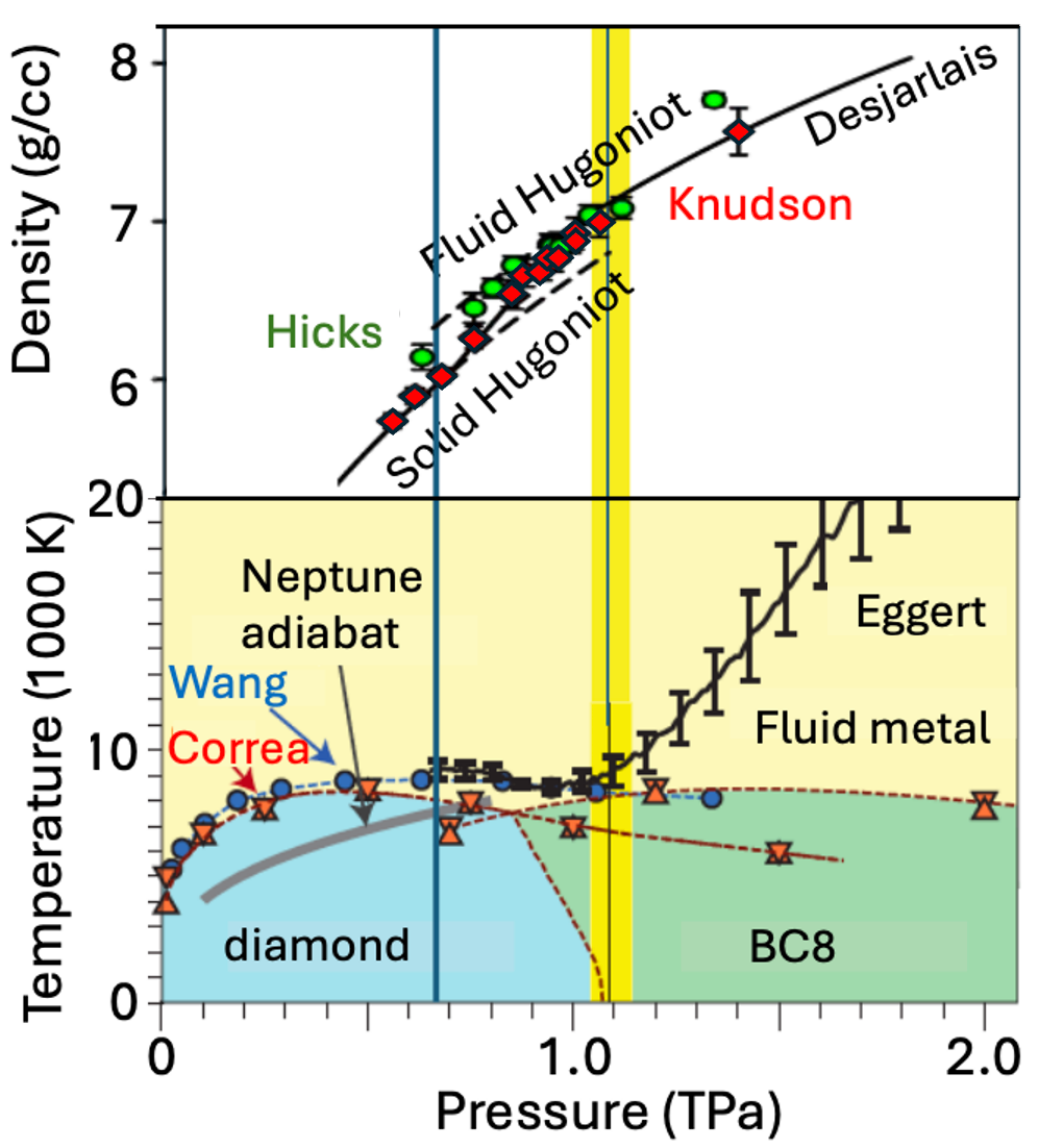}
\caption{Diamond Hugoniot data from Z~\cite{Knudson2008} and Omega~\cite{Hicks2008} (top) and shock front temperature data~\cite{Eggert2010} (bottom). Shock compressed diamond reflectivity increases above a few percent at $\sim$600 GPa to $\sim30\%$ at P $>$ 1100 GPa. The mechanical measurements from Z, the plateau in temperature measurements, and evolution of the reflectivity provide a self consistent picture of carbon melt along the Hugoniot.}
\label{fig:diamondshock}
\end{figure}

The melt behavior of carbon is important to inertial confinement fusion~\cite{Mackinnon2014} and planetary physics~\cite{Ross1981}. Further, ab-initio molecular dynamics (AIMD) simulations~\cite{Correa2006, Knudson2008} suggest the presence of a diamond-BC8-liquid triple point (Fig.\ref{fig:diamondshock}) along the diamond Hugoniot. This interesting behavior has been explored at large HED facilities using ultra-high velocity flyer plate experiments~\cite{Knudson2008} and decaying shock experiments~\cite{Eggert2010}.

High-precision Hugoniot experiments exhibit subtle slope changes in the pressure-density plane (Fig.\ref{fig:diamondshock}); densification at the onset of melt (anomalous melt of diamond) and a decrease in slope at the triple point (normal melt of BC8). These subtle slope changes are in good agreement with AIMD simulations. Decaying shock experiments~\cite{Eggert2010, Bradley2004}, show a dramatic slope change (plateau) in temperature as the Hugoniot crosses the melt curve due to the latent heat of fusion. There is a subtle temperature depression in the vicinity of the predicted triple point. The reflectivity also increases continuously as a function of melt fraction~\cite{Bradley2004}.  While shock measurements provide strong evidence for the existence of a diamond-BC8-liquid triple point on the melt boundary of carbon, ramp compressed diamond diffraction data~\cite{Lazicki2021} show the diamond phase persists to at least 2 TPa. Similarly, ramp compressed EOS data~\cite{Smith2014} show no volume discontinuity to 5 TPa, consistent with a stable HCP phase. This stability of the HCP phase in the solid contradicts the triple point determination from the shock wave data, an important question yet to be resolved.

As a second example we turn to the metallization of hydrogen. The earliest metallization experiments were performed on light gas guns using reverberating shocks~\cite{weir_prl_96,Knudson2018}. These experiments drove cryogenic liquid samples to 100s of GPa and 2-4 kK. DC conductivity measurements exhibit monotonic increase with pressure and temperature that ultimately reach a plateau consistent with minimum metallic conductivity at $\sim$140 GPa and $\sim$3 kK.

Early experiments on HED facilities focused on the Hugoniot of cryogenic liquid deuterium~\cite{Celliers2000a}. However, shock compression of the fluid results in rapid temperature increase and a maximum density compression of slightly over 4-fold~\cite{knudson_prl_01, Knudson_2004, dasilva_prl_97, collins_science_98}. Later experiments utilized shock compression of precrompressed samples. These experiments increase the achievable compression by an additional factor of 2~\cite{Loubeyre2012, Brygoo2015}. Both sets of Hugoniot experiments probed metallization at relatively low density, where the transition is predominately temperature driven.

\begin{figure}[tbh]
\centering
\includegraphics[width=0.98\linewidth]{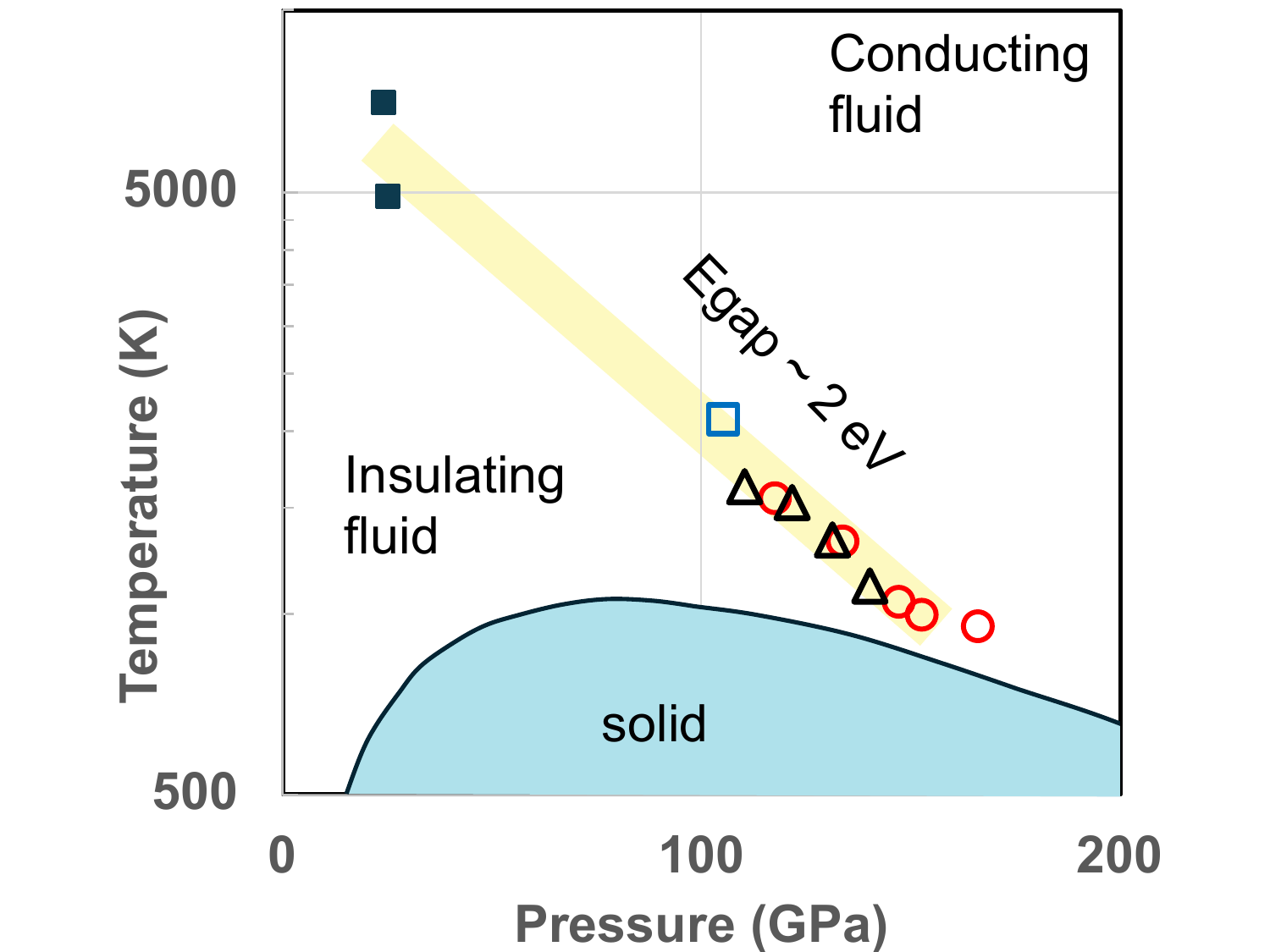}
\caption{Contour of constant bandgap in deuterium. Black open triangles~\cite{Knudson_Science_2015} and red open circles~\cite{Celliers_Science_2018} show transition from transparent to opaque under shock-ramp compression at photon energies of 2.1 and 1.9 eV, respectively.  Black closed squares~\cite{Celliers2000a,Loubeyre2012} show where shock compressed reflectivity is near background values(few percent) at 2.1 eV photon energy.  Open blue square shows where band gap is $\sim$2.0 eV for shock reverberation experiments~\cite{weir_prl_96}. Below these data, deuterium is a molecular insulator. Above these data, deuterium transforms through a semiconducting phase to an atomic, metallic fluid.}
\label{fig:hmetal}
\end{figure}

Recent shock-ramp experiments explored the metallization transition at lower temperatures and higher densities~\cite{Celliers_Science_2018,Knudson_Science_2015}. In these experiments the bulk of the entropy (temperature) increase occurred during the initial shock, with subsequent compression driving the sample to high pressure-density near isentropically. These experiments probed metallization at relatively low temperature, where the transition is predominately density driven. Fig.\ref{fig:hmetal} shows existing dynamic compression data at conditions where the electronic energy gap is near 2 eV (where the fluid transitions from transparent to opaque). Combined, these data provide important benchmarks for metallization of fluid hydrogen.

\subsection*{Challenges \& outlook}

A new generation of extreme matter experiments are underway~\cite{NAS}. Improvements in entropy control~\cite{Celliers_Science_2018, Knudson_Science_2015, Schwemmlein2024} are enabling access to potential new quantum states of matter such as superconducting superfluid hydrogen~\cite{Babaev2004}. Recent experiments reveal chemical bonding can engage core as well as valence electrons, and simple metals at ambient conditions can transform into topological insulators at TPa pressures~\cite{Pickard2011, Polsin:2022a}.  Such behavior may be quite general at extreme pressures, leading to question if the Thomas Fermi limit is ever realized, and if so, at what conditions.

Ignition brings new possibilities, extending pressures into the 100s if not 1000s of billions of atmospheres~\cite{Abu-Shawareb2022}. Moreover, advanced data analysis techniques~\cite{Ruby2021} and characterization capabilities, including x-ray lasers, synchrotrons, inverse Compton~\cite{Rinderknecht2022}, and potentially laser undulators~\cite{Fuchs2009}, combined with pressure drivers are changing the landscape of measurements including, elastic~\cite{briggs2017} and inelastic scattering~\cite{Fletcher_NP_2015}, and imaging~\cite{Schropp2015} measurements. Engaging phase enhancements will enable accurate atomic to macroscale understanding of compact astrophysical objects throughout the universe.

\newpage 
\clearpage

\section{%
X-ray Scattering Diagnostics%
}
\label{section_13}
\author{%
Dave Chapman$^1$,
Luke Fletcher$^2$,
Thomas Kluge$^3$
}
\address{%
$^1$First Light Fusion Ltd., Oxford OX5 1QU, United Kingdom\\%
$^2$SLAC National Accelerator Laboratory, Stanford University, Menlo Park, CA, United States\\%
$^3$Helmholtz-Zentrum Dresden-Rossendorf (HZDR), D-01328 Dresden, Germany%
}

\subsection*{Introduction}

The investigation of WDM requires precise experimental techniques to characterize its microscopic structure and atomic-scale properties, necessitating diagnostic tools capable of elucidating its many-body interactions and collective behaviors. X-ray diagnostics, particularly scattering and spectroscopic methods \cite{sheffield2010plasma}, have emerged as primary investigative tools, producing quantitative measurements of these phenomena \cite{Glenzer_revmodphys_2009,Ciricosta_PhRvL_2012,kraus_xrts,Doeppner_nature_2023}. The non-perturbative nature of high-energy radiation is critical for maintaining the sample state during measurements, which is essential for model validation. Both spectrally resolved and angularly resolved X-ray scattering techniques provide complementary capabilities. These methods have been systematically applied to WDM research over decades, producing reproducible results that have advanced our understanding of this enigmatic and distinct thermodynamic regime \cite{glenzer_PRL_2007,Kraus2016b,Descamps2020,Doeppner_nature_2023}.

\subsection*{X-ray Thomson Scattering}

From its origins in the early 2000s using high-power laser-produced plasmas \cite{glenzer2003}, X-ray Thomson scattering (XRTS) has flourished into one of the most versatile and adaptive front-line WDM diagnostics. It has not only 
provided insights into phenomena that are directly relevant for planetary and astrophysics \cite{Kritcher_POP_2009, Kraus_PhysRevLett_2013,Doeppner_nature_2023} and high-energy-density matter \cite{Fletcher_PhysRevLett_2014, Chapman_NatCommun_2015, Fletcher_Frontiers_2022}, and has played a crucial role in generating data key to the recent successes of the NIF \cite{Kraus2016b, Döppner2018}. The development and rapid advancement of high-brilliance, narrow-bandwidth, and high-repetition-rate sources at X-ray free-electron laser (XFEL) facilities \cite{Glenzer_JPhysB_2016, Zastrau_JSyncRad_2021, Descamps_JSyncRad_2022, Gawne_JAP_2024} has further accelerated groundbreaking designs of novel experiments with unprecedented signal-to-noise levels and reproducibility \cite{Fletcher_NP_2015,gawne2024ultrahigh}.

The core diagnostic capability afforded by XRTS is the direct measurement of dynamic correlations between electrons. Depending on the scale length probed by the radiation, the information obtained provides insight into single-particle properties, yielding measurements of the electron distribution function in the momentum/energy space or their various 
or collective excitations: plasmons \cite{glenzer_PRL_2007, Doeppner_HEDP_2009, Neumayer_PhysRevLett_2010} and phonons \cite{Gregori_PhysPlasmas_2009, Murillo_PhysRevE_2010, McBride_RevSciInstrum_2018}. Electronic transitions \cite{Johnson_PhysRevE_2012, Mattern_PhysPlasmas_2013}, collisional physics \cite{Doeppner_JPhysConfSeries_2010, Sperling_PhysRevLett_2015, Faussurier_PhysPlasmas_2016, Hentschel_PhysPlasmas_2025}, band structure effects \cite{Gamboa_PhysPlasmas_2015}, species miscibility in multicomponent systems \cite{Wuensch_2011, frydrych2020demonstration}, and exotic nonequilibrium behavior \cite{Chapman_PhysRevLett_2011, Chapman_HEDP_2012, Vorberger_PRE_2018, Vorberger_PhysLettA_2024} can also be inferred. 

Under reasonable and commonly fulfilled experimental constraints the power spectrum of scattered radiation from a volume $\mathcal{V}$ is given by \cite{Crowley_NewJPhys_2013, Crowley_HEDP_2014b, Dornheim2023_density_response}
\begin{equation}
    \label{eq:intensity_definition}
    \frac{\partial^{2} P_{\text{s}}}{\partial \Omega\partial \omega_{\text{s}}}
    \approx \,
    I_{0} r_{e}^{2} G(\theta,\phi)
    n_{e} \mathcal{V} \left(1+\frac{\omega}{\omega_{0}}\right)^{2} S_{ee}(\mathbf{k},\omega)
    \,,
\end{equation}
in which $r_{e}\approx2.82\,\mathrm{fm}$ is the classical electron radius, $G(\theta,\phi)$ is a term related to the scattering geometry and properties of the probe beam, having intensity $I_{0}$, frequency $\omega_{0}$ and wave vector $\mathbf{k}_{0}$. The term $S_{ee}(\mathbf{k}, \omega)$ is the dynamic structure factor (DSF), which encodes all dynamic correlations between electrons \cite{kremp2005quantum}
\begin{eqnarray}
    \label{eq:dsf_definition}
    S_{ee}(\mathbf{k}, \omega)
    & =
    \frac{i\hbar}{2\pi n_{e}}
    \int_{-\infty}^{\infty}
    \mathrm{d}\tau \,
    e^{i\omega\tau}
    \left\langle\delta\varrho_{e}(\mathbf{k},\tau)
    \delta\varrho_{e}(\mathbf{k},0)\right\rangle\, , \quad\quad
\end{eqnarray}
with $\mathbf{k}$ and $\omega$ denoting the wave vector and frequency of density fluctuations, respectively. Corrections due to gradients in inhomogeneous samples have been considered before \cite{Belyi_PhysRevLett_2002, Kozlowski_2016,  Belyi_SciRep_2018}, although these can mostly be neglected for WDM and simplified treatments \cite{Fortmann_HEDP_2009, Chapman_PhysPlasmas_2014} are reasonable.

The DSF is typically decomposed into three terms following a semi-classical approach \cite{Chihara_2000}, later generalized to multicomponent plasmas~\cite{Wuensch_PhysRevE_2008}. These correspond to scattering from: (1) the quasielastic ion feature; (2) inelastic 
transitions between continuum states; (3) inelastic transitions between continuum and core states. 
Furthermore, the impact of free-bound transitions have also recently been considered \cite{Bohme_PhysPlasmas_2025}, which modify the blue wing of the Rayleigh peak for partially ionized systems.

The spectral shape of each feature is sensitive to plasma conditions to different degrees, with specific sensitivities dictated by the Salpeter parameter $\alpha = 1/k \lambda_{\text{scr}}$, in which $\lambda_{\text{scr}}$ is the screening length and $k =|\mathbf{k}|$. Thus, with suitably accurate theoretical descriptions of $S_{ee}(\mathbf{k},\omega)$, forward modeling of the scattered power spectrum combined with rigorous statistical techniques \cite{Poole_PhysRevRes_2024, Suazo_Betancourt_RevSciInstrum_2024, Hentschel_PhysPlasmas_2025,Kasim_POP_2019} provides a powerful diagnostic of conditions of dense plasmas. This is the core idea behind forward modeling tools \cite{Gregori_PRE_2003, Chapman_thesis_2015, Chapman_AWE_PPN_2017} and more sophisticated \emph{ab initio} approaches that post-process time-dependent DFT \cite{baczewski2016x,Mo_prl_2018,Schoerner_PRE_2023,gawne2025stronggeometrydependencexray} and PIMC simulations \cite{dornheim2024unraveling}.

Whilst forward modeling has been used to great effect in the past, it is often dependent on model assumptions such as Chihara's semi-classical ansatz. Recently, however, a novel, fully model-free approach was developed \cite{Dornheim_NatCommun_2022} based on the insight that the two-sided Laplace transform of the DSF 
\begin{equation}
    \label{eq:dsf_laplace_transform}
    F_{ee}(\mathbf{k},\tau)
    =
    \int_{-\infty}^{\infty} 
    \mathrm{d}\omega\,
    e^{-\omega\tau}
    S_{ee}(\mathbf{k},\omega)
    \,,
\end{equation}
where $\tau=i\hbar\beta$ is the imaginary time, has a well-defined relationship to the temperature in equilibrium via the properties of imaginary time correlation functions (ITCF) \cite{Dornheim_NatCommun_2022, Dornheim_MRE_2023,Dornheim_PTR_2023}. Further application of this formalism to the spectral moments of the DSF~\cite{Dornheim_moments_2023} can be used to infer the absolute intensity of the signal via the $f$-sum rule \cite{Dornheim_SciRep_2024}, allowing calibration of the static electronic structure factor $S_{ee}(k) = \int_{-\infty}^{\infty}\mathrm{d}\omega\,S_{ee}(\mathbf{k},\omega)$. 

The ITCF approach is limited by the requirement that the plasma be in equilibrium, such that a well-defined temperature is conceptually meaningful. This can be checked by examining the asymmetry of the ITCF in the $\tau$ domain \cite{Vorberger_PhysLettA_2024}; in the case of nonequilibrium one must still appeal to forward modeling. Another current restriction of the technique is that the approximate relationship $k = 2\omega_{0}/c \, \sin(\theta/2)$ must be fulfilled, such that the term $(1+\omega/\omega_0)^2 \approx 1$ in Eq.\,(\ref{eq:intensity_definition}). This condition is strongly fulfilled when the dynamic range of the spectrum is small compared to the energy of the probing radiation \cite{Dornheim2023_density_response}, as is the case for high-energy laser-driven line emission sources and modern XFEL facilitites. Leveraging ab initio simulations, forward modeling tools and the ITCF approach together, XRTS provides an as-yet-unparalleled level of insight into the conditions of dense, strongly interacting matter.

\subsection*{X-ray Diffraction}

X-ray diffraction (XRD) is a powerful technique for probing the atomic and molecular structure of materials. Diffraction arises from the interaction of X-rays with atomic electron clouds. The scattered intensity is governed by the atomic form factor, $f(k)$, which describes how an individual atom scatters X-rays as a function of the scattering wave number $k$. This atomic form factor is the Fourier transform of the electronic density around a nucleus \cite{James_1962}.
%
The total scattered intensity is given by the structure factor  
\begin{equation}
    I(\mathbf{k}) = \sum_{i,j} f_i(\mathbf{k})f_j^*(\mathbf{k})e^{i \mathbf{k} \cdot \mathbf{r_{ij}}}
\end{equation}
accounting for scattering contributions from all atoms in the system. In a periodic crystal, the total scattered intensity results from the sum of atomic form factors across all unit cells. Constructive interference at specific scattering angles leads to diffraction peaks, described by Bragg’s law 
$n\lambda = 2d\sin\theta$
where $d$ is the interplanar spacing, $\lambda$ is the X-ray wavelength, $\theta$ is the scattering angle, and $n$ is an integer \cite{Bragg_1913}.  

However, WDM lacks long-range crystalline order due to thermal motion and partial ionization, disrupting periodicity. As a result, diffraction patterns are dominated by broader features indicative of short-range order rather than sharp Bragg peaks. In such disordered systems, the scattered intensity is described by the static structure factor, $S(k)$, which characterizes atomic correlations in liquids or amorphous solids \cite{Zernike_1927}
\begin{equation}
\label{eqn:Sq}
    S(k) = 1 + n \int \mathrm{d}\mathbf{r}\, e^{-i\mathbf{k}\cdot\mathbf{r}} \, g(r)
\end{equation}
where $g(r)$ is the radial pair distribution function and $n$ is the atomic number density.   

In ionized materials the elastic scattering intensity $I(k)$ is modified due to the presence of free electrons. In this regime, the total scattering amplitude includes contributions from both the atomic form factor $f(k)$, representing bound electrons, and the screening form factor $q(k)$, describing the cloud of free electrons screening the effective ionic charge. 
The resulting scattered intensity is given by \cite{Chihara_2000}

\begin{equation}
    I(k) 
    = | f(k) + q(k) |^2 S_{ii}(k)
    \,,
\end{equation}
where $S_{ii}(k)$ is the ion-ion static structure factor, defined analogously to Eq.\,\eqref{eq:dsf_definition}. The screening term $q(k)$ modifies the scattering cross-section, leading to deviations from neutral atom scattering models. In highly ionized WDM, Coulomb interactions and collective effects influence $S_{ii}(k)$, which in turn affects the measured diffraction pattern. These modifications provide insight into charge screening and structural properties under strongly coupled conditions.

Measuring the full elastic scattering response $I(k)$ in WDM provides insights into local order, phase transitions, melting, and crystallization. However, ionization and strong Coulomb interactions modify $I(k)$, distinguishing WDM from classical liquids. Peaks in $S_{ii}(k)$ correspond to characteristic interatomic distances, revealing structural details under extreme conditions \cite{Fletcher_NP_2015}. XRD is crucial for determining the density and temperature of laser-shocked and heated materials, mapping phase boundaries, and identifying lattice structures at high pressures \cite{briggs2017, Gorman_JAP_2024, McBride_2019}.  

Capturing $I(k)$ in WDM is challenging due to extreme temperatures, high densities, and the transient nature of these states. Advances in X-ray sources, as previously described,
have enabled high-precision measurements. XFELs provide the ultra-bright short pulses required to capture snapshots of the liquid structure before atomic motion disrupts coherence, facilitating the study of transient structural correlations in ionized liquids. XFELs generate high-quality diffraction profiles in a single femtosecond pulse, aligning with the timescale of laser-driven dynamic compression and ultra-fast heating \cite{Kraus2017, Tateno2010, Gleason2015, Briggs2019, Coleman2019, Albertazzi2017, Yang2024, Morard2024}. This allows XRD data to be collected from extreme pressure-temperature states, significantly advancing our understanding of WDM with unprecedented temporal resolution\cite{Pascarelli2023}.

XFEL experiments combining XRD with diamond anvil cells (DACs) have recently been used as a novel approach to dynamically study warm dense materials \cite{Pace2020, Meza2020, Liermann2021, Husband2021, Frost2024}. XRD provides real-time information on lattice parameters, crystal symmetries, and phase boundaries in response to applied pressure, while simultaneously utilizing short-lived pulsed X-ray sources that can dynamically heat materials to extreme conditions \cite{Ball2023}. This method yields precise structural information before the sample can react with the DAC target environment—a common limitation in extended heating experiments using synchrotron sources or optical excitation mechanisms \cite{Prakapenka2004, Morard2018, Pascarelli2023}. The DAC/XRD platform enables direct observations of temperature-induced chemical reactions, phase transition stability, and material synthesis under extreme pressure states before hydrodynamic expansion occurs. When integrated with complementary techniques such as X-ray Emission Spectroscopy (XES), these experiments enhance our understanding of the interplay between structural modifications and electronic properties at atomic scales \cite{Kaa2022}.

In a crystal lattice, atoms vibrate around their equilibrium positions due to thermal energy and are intrinsically linked to the material phonon frequency by the Debye temperature. Recent ultra-fast XRD techniques have been employed to investigate variations in lattice dynamics via Debye-Waller theory as a function of temperature and excitation conditions \cite{Decamps2024, Heimann2023}. When metals like gold are subjected to intense optical laser irradiation over fs-timescales, electrons are heated to several electronvolts while ions remain relatively cold near solid ambient densities. This creates a stiffening of the interatomic potential due to strengthened metallic bonding, causing an increase in phonon frequencies across the Brillouin zone \cite{Ernstorfer2009, Recoules2006, Smirnov2020}. The observations of phonon hardening in gold can potentially extend to other face-centered cubic (fcc) metals such as aluminum, copper, and platinum, as similar mechanisms govern the material response to extreme conditions \cite{Recoules2006, Smirnov2020}. Such research opens avenues for designing advanced materials with tailored properties for high-temperature applications as well as advancing our understanding of thermal-acoustic transport properties in warm dense materials.

\subsection*{Small-angle X-ray Scattering}

Small-angle X-ray scattering (SAXS) and the related method of grazing-incidence SAXS (GSAXS) have emerged as a powerful special case of XRD for probing the nanoscale structural dynamics of WDM \cite{Kluge2014,PhysRevResearch.4.033038}. 
While SAXS is typically measured in transmission, GSAXS looks at the reflection of a surface where the incoming angle is below the critical angle. 
In that case, the signal is sensitive to only a few skin-depths. 

With keV photon energies at X-ray free-electron lasers, SAXS is sensitive to the mesoscale with $\sim1-10\;\mathrm{nm}$ resolution.
SAXS is sensitive to density fluctuations and nanoscale inhomogeneities within a sample, enabling the characterization of features such as roughness, voids, clusters, compressions, and phase transitions. 
In WDM research, SAXS has been instrumental in visualizing the ultrafast evolution of electron distributions at interfaces during intense laser-solid interactions, e.g., due to (non-thermal) melt, ablation and compression. 

Sharp interfaces give rise to scattering streaks (or isolated peaks for periodic structures) that have been used to probe phase transitions and shock dynamics inside solid density targets subjected to high-energy or high-intensity laser irradiation \cite{Kluge2017,2023_Kluge_et_al}. 
These studies highlighted the ability of SAXS to resolve shock front propagation and associated instabilities with high precision. 
The sharpness of the interface is connected to the signal reduction at larger scattering angles through a Debye-Waller-like signal suppression, which allows to quantify interface scale-length changes from melt, ablation or compression \cite{Gorkhover2016,Kluge2018}. 

Capturing laser-generated instability structures, e.g. from Weibel or Rayleigh-Taylor-like instabilities in the bulk or at the surfaces is challenging due to their transient nature and low signal strength. Efforts have to be taken to suppress bremsstrahlung and other background especially in high-intensity laser experiments when surpassing $10^{19}\;\mathrm{W/cm^2}$. 
First indications for laser generated instabilities during relativistic laser irradiation show unprecedented possibilities to characterize them in-situ during their growth, by far exceeding the resolution of direct imaging methods \cite{Ordyna2024}. 

SAXS has also proven effective in studying phase separation and material de-mixing under extreme conditions. 
For example, He et al. combined SAXS with X-ray diffraction to observe diamond formation in shock-compressed C–H–O samples, a process relevant to planetary interior modeling \cite{doi:10.1126/sciadv.abo0617}. 
The SAXS data provided detailed insights into the kinetics of diamond nucleation and growth, highlighting the technique's power for investigating complex processes in WDM.

Another significant branch of SAXS is resonant scattering, where the form factor $f(k)$ becomes complex when the X-ray energy matches that of atomic bound-bound or bound-free transitions \cite{Kluge2016}. 
This generates asymmetries in the scattering pattern $I(k) \neq I(-k)$ that can be used to reconstruct the resonant strength and therefore the plasma ionization and temperature. This approach allows simultaneous measurements of temperature, ionization states, and nanometer-scale expansion dynamics, contributing to a more comprehensive understanding of laser-induced plasma behavior \cite{Gaus2021}.

\newpage 
\clearpage

\section{%
X-ray Absorption Spectroscopy and Imaging Techniques
}\label{section_13b}
\author{%
Brendan Kettle$^1$, 
Ulf Zastrau$^2$
}
\address{%
$^1$The John Adams Institute for Accelerator Science, Blackett Laboratory, Imperial College London, London, SW7 2AZ, United Kingdom\\%
$^2$European XFEL GmbH, Holzkoppel 4, 22869 Schenefeld, Germany%
}

\subsection*{Introduction}

In addition to X-ray scattering diagnostics, techniques that are based on the absorption of X-rays can be powerful tools for the investigation of warm dense matter in the laboratory. 
These include spectroscopic methods that interrogate the samples on an atomic scale, as well as microscopic imaging of shock fronts.
Benefits include the ability to probe aperiodic materials, as well as provide information on both the electronic and atomic structures simultaneously, which makes it ideal for investigating the dynamics of non-equilibrated states. 

\subsection*{X-ray Absorption Spectroscopy}

\begin{figure}
    \centering
    \includegraphics[width=0.7\linewidth]{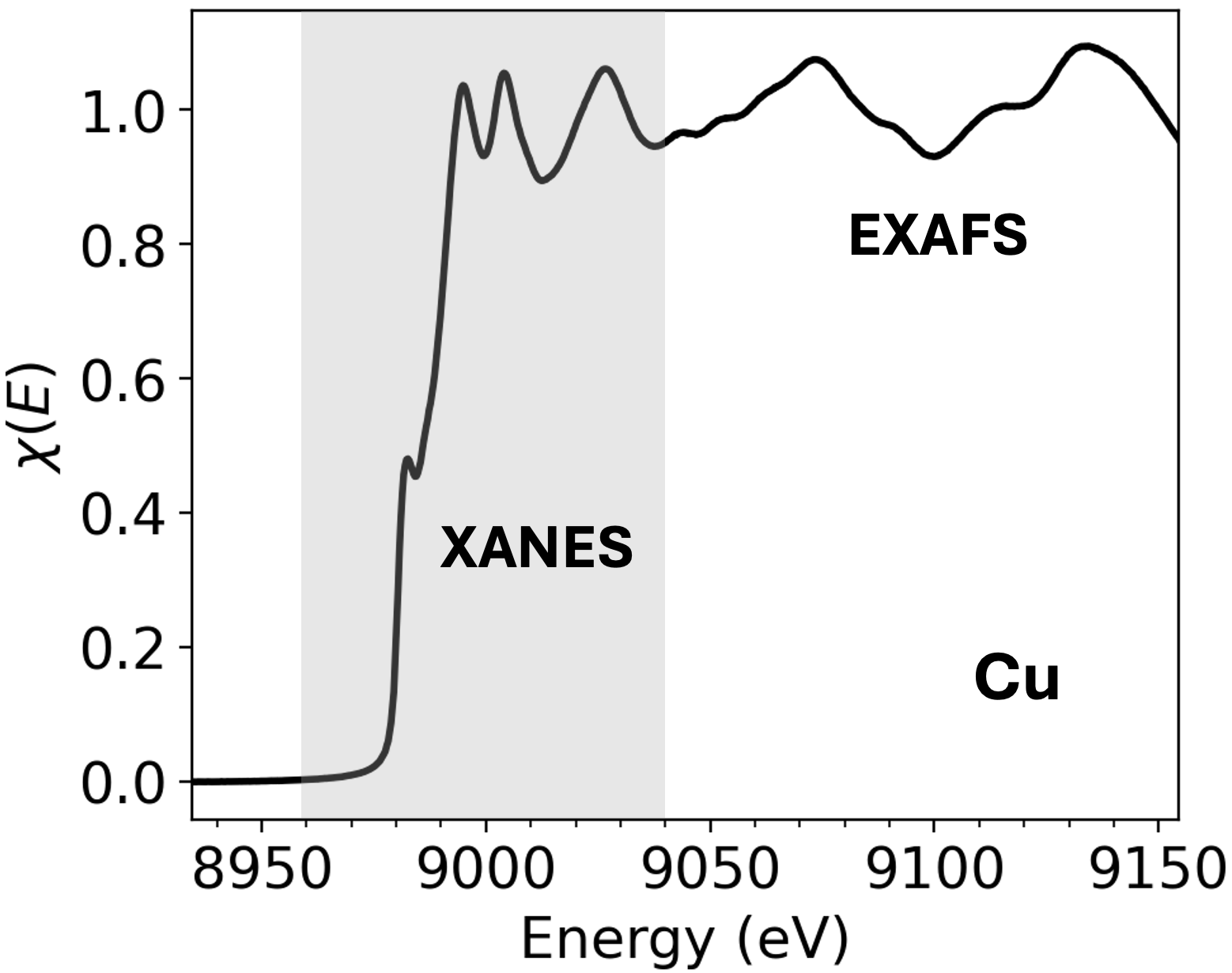}
    \caption{A typical X-ray absorption spectrum of the K-edge of a Copper sample. The XANES profile near the edge details the electronic state (highlighted in grey), while the EXAFS profile after the edge details the local atomic structure. }
    \label{fig:XAS_example}
\end{figure}

The absorption of X-rays within a sample provides a direct measurement of the current state of its electronic structure.
By analyzing the spectral contents of a given X-ray probe, one can observe line transitions, atomic edge shifts, ionization effects and more.
Predominant techniques involve X-ray Absorption Near Edge Structure spectroscopy (XANES) and Extended X-ray Absorption Fine Structure spectroscopy (EXAFS)~\cite{KoningsbergerPrins1988, RehrAlbers2000}.
In these techniques, the absorption, scattering and interference of ejected photoelectrons from neighboring atoms manifest as modulations in the absorption profile near resonant edges. 
See Figure~\ref{fig:XAS_example} for an example. 
$\chi (E) = (\mu (E)) - \mu_0(E)) / \Delta\mu_0 $ represents the normalized absorption profile, where $\mu(E)$ is the absorption coefficient as a function of energy E, $\mu_0(E)$ is the smoothly varying atomic-like background absorption (often fitted with a spline function), and $\Delta\mu_0$ is the jump in absorption at the edge.
These modulations are unique signatures for each material and are directly linked to not only the local electronic structure but also the atomic structure and vibrational properties of the sample.
The electronic structure information is contained near the edge itself (XANES), where the absorption and scattering of transitions within a single atom dominate.
In effect, the near-edge profile can be used to infer the electronic density of states.
For probing warm dense states, one can rely on the fact that the vacancy factor is obtained from Fermi-Dirac statistics and is intrinsically dependent upon the electron temperature.
Broadening of the absorption edge can be used to deduce the electron temperature for high-density low-temperature WDM~\cite{DorchiesRecoules2016}.
Resonant transitions can also appear as pre-edge features, where visibility can be used to understand the density of states further. 
Shifts in the edge energy are used to diagnostic charge-state and oxidation effects, through screening of the core binding energies.
The local ionic structure information is contained in the oscillatory features after the edge (EXAFS), which are created from the multiple scattering events between local atomic sites.
Changes in the atomic locations alter this profile, and for example, the ion temperature can be deduced from the magnitude of the oscillations.
Ab Initio calculations are able to calculate spectra with a high degree of precision~\cite{Jourdain2020}.
Since these calculations relate to the electronic and atomic properties of a sample, other macroscopic attributes can be derived, such as the equation of state, electron transport, optical and magnetic properties etc.

For undertaking X-ray absorption spectroscopy measurements, the broadband X-rays provided by Synchrotron facilities offer the ideal spectral smoothness, and indeed measurements of samples under ambient conditions has proved extremely successful over the last few decades across a vast range of sciences, including biology, chemistry, material science, environmental science and more.
However, when it comes to studying warm dense matter, implementing the required level of sample pumping within a synchrotron end-station can prove difficult, as the co-location of a high-energy driver has conventionally not been available.
However, in recent years, high-energy laser systems have become available at a few end-stations, enabling shock driven measurements to be undertaken~\cite{Hernandez2024, Torchio2016}.
For example, using a 100 ps probe from a synchrotron, new constraints have been put on the melting temperature and phase stability of shocked Iron, up to 270 GPa~\cite{Balugani2024}, ruling out a predicted transition to a high-temperature bcc phase.
The evolution of the electronic structure of warm dense copper has also been studied, using a 150 fs pump laser at the ALS synchrotron ~\cite{Cho2011}. 


Suitable laser-plasma driven backlighters, can offer an alternative to synchrotron sources. 
The major benefit being that they are often co-located with other high-energy lasers for pumping samples to dense energetic conditions.
Solid target backlighters have been successfully demonstrated using mid-Z elements, that exhibit a host of M-L transitions. 
For example, a Bismuth backlighter (${\approx2.8}$ keV) has been used to produce a 200 ps duration probe for a shocked chlorinated plastic target, raised to few times solid density and ${\approx10}$ eV temperature.
An observed shift in the K-edge position allowed testing of ionisation potential depression models~\cite{riley2023}.
In another study, EXAFS has been performed on the OMEGA laser system at LLE to investigate Iron at pressures up to 560 GPa~\cite{Ping2013,Ping2013a}.
However, these investigations require a large high-energy laser system to produce the backlighter, and are generally limited in probe duration to ${>100}$ ps, similar to that of the synchrotron facilities.
There have been some successful investigations with few picosecond resolution, using less energetic laser-driven plasma backlighters, where averaging over many shots allowed the investigation of phase changes of femtosecond heated warm dense matter~\cite{Leguay2013,Dorchies2011}.

Recently, the broadband X-rays generated from laser-plasma wakefield accelerators have offered new opportunities in undertaking X-ray absorption spectroscopy measurements on a rapid timescale.
A high-intensity laser pulse (usually sub-picosecond) is focused through a gas medium to form a plasma channel.
As the laser pulse travels, electrons are expelled by the Ponderomotive force, creating an ion cavity in its wake. 
The extremely strong electric field inside can subsequently accelerate electrons to GeV energies in just a few centimeters.
While in the back of the ion cavity the electrons perform betatron oscillations around the laser axis, producing high-energy X-rays~\cite{Kneip2010, Corde2013}.
These X-rays have a smooth, synchrotron-like emission spectrum, and pulse duration similar to that of the electron bunch; down to 10's of femtoseconds.
The source size is on the order of microns and the emission is directed in a tight cone along the propagation axis, with a divergence of ${<20}$ mrad FWHM.
These unique properties make it ideal to perform X-ray absorption spectroscopy measurements on an ultrashort timescale, and it is paving the way for a host of novel investigations of warm dense matter, in particular for non-equilibrated samples, where femtosecond probe times are required.

For example, the non-equilibrium dynamics of a copper foil brought from solid to WDM by a femtosecond laser pulse have been studied~\cite{Mahieu2018}.
Looking at the Copper L-edge (930 eV) a sub-100 fs rise-time of the electron temperature was measured using the magnitude of a pre-edge peak feature.
The electrons were then seen to dissipate their energy to the ions over the next ${\approx5}$ ps.
In another study of the Copper L-edge, ultrafast electron energy transport is investigated in laser-heated warm dense copper~\cite{Grolleau2021}.
By changing the thickness of the foil, the ballistic transport of the hot electrons was compared to thermal conduction across the bulk sample.
While these ultrafast examples required many shots to be integrated, single-shot XANES and EXAFS has recently been demonstrated using high-flux laser-plasma accelerator sources~\cite{Kettle2019, Kettle2024}. 
This capability will enable low shot rate experiments with samples driven to high pressures and temperatures.
This will become particularly advantageous for planned future facilities, where the usefulness of a tandem ultrashort pulse laser, linked to a high energy system has become clear. 
For example, the scheduled Vulcan 20-20 facility in the UK will have a 1 PW ultrashort pulse laser capable of providing a high-brightness betatron backlighter, coupled to more than 10 kJ of drive energy in nanosecond beamlines, as well as another 20 PW ultra-high intensity beamline.
This will enable novel ultrafast X-ray absorption spectroscopy investigations of matter driven to extremely high densities and pressures, and also non-equilibrium states.

While the narrow bandwidth and stochastic nature of the emission spectrum of XFELs is not ideal, they are still extremely bright, femtosecond duration probes that can be used for X-ray absorption spectroscopy.
If a suitable reference beam is split, the spectrum normalisation can be accounted for and absorption measurements undertaken, albeit over 10's to 100's shots to build a profile of suitable accuracy.~\cite{Katayama2013}. 
For example, XANES of laser shock compressed iron up to 420 GPa and 10,800 K was performed at the Linac Coherent Light Source (LCLS). A modification of the pre-edge feature suggests that iron is molten at pressures and temperatures higher than 260 GPa and 5680 K along the principal Fe Hugoniot.
In another study, the XFEL acted as both pump and probe, and again the Copper L-edge was exploited by XANES spectroscopy to investigate the behaviour of both free and bound electrons during the formation of WDM~\cite{Mercadier2024}.
It exhibited non-linear effects transitioning from reverse saturable absorption, to saturable absorption at higher intensities.

\subsection*{X-ray Imaging}

X-ray imaging, sometimes referred to as x-ray shadowgraphy or backlighter, is a technique as old as the discovery of x-rays. Shock wave experiments require precise design of the target and specific tailoring of the spatial and temporal laser profiles to reach the highest pressures, and being able to observe the shock wave is an important diagnostic tool. Before the availability of the first hard x-ray free electron lasers and instruments such as MEC at the LCLS, e.g. pre-2012, laser-driven plasma sources of K$\alpha$ or the longer-lived He$\alpha$ were employed to create transmission-type images of warm dense matter. In particular, line densities and density variations in shock wave and implosion experiments were investigated as early as 1990~\cite{hammel1994x}. While these backlighters are kind-of monochromatic tens-of-micron diameter point sources with $> 20~$ps duration, their extended bandwidth, lack of coherence and directionality soon became a limiting factor for the retrieval of absolute density data. 

With the advent of hard x-ray free-electron lasers, a tunable x-ray source with laser-like properties such as monochromaticity (few tens of eV down to sub-eV bandwidths), $< 50~$fs pulse uration (e.g. shorter than a phonon period) and high degree of coherence and directionality was at hand to design a new class of precision imaging.

A new field of science was explored by measuring quantitatively the local bulk properties and dynamics of matter under extreme conditions, in this case by using the short XFEL pulse to image an elastic compression wave in diamond~\cite{Schropp2015}. The experiment was conducted at the MEC instrument at the Linac Coherent Light Source (LCLS)~\cite{nagler2015matter, nagler2016phase}. The elastic wave was initiated by an intense optical laser pulse and was imaged at different delay times after the optical pump pulse using magnified x-ray phase-contrast imaging. The temporal evolution of the shock wave can be monitored, yielding detailed information on shock dynamics, such as the shock velocity, the shock front width and the local compression of the material. The method provides a quantitative perspective on the state of matter in extreme conditions. 

The experimental geometry uses point-projection: A sub-micrometer x-ray focus was created by means of beryllium compound refractive lenses, and characterized in phase and amplitude by ptychography~\cite{Schropp2015}. Later experiments of this type were able to demonstrate simultaneous phase-contrast imaging (PCI) and X-ray diffraction from shock compressed matter~\cite{seiboth2018simultaneous}. By utilizing the chromaticity from compound refractive X-ray lenses, the researchers focussed the 24.6~keV 3rd order undulator harmonic of the LCLS to a spot size of 5~$\mu$m on target to perform X-ray diffraction. Simultaneous PCI from the 8.2~keV fundamental X-ray beam was used to visualize and measure the transient properties of the shock wave over a 500~$\mu$m field of view. Furthermore, this experiment showed the ability to extend the reciprocal space measurements by $5\,$\AA$^{-1}$, relative to the fundamental X-ray energy, by utilizing X-ray diffraction from the 3rd harmonic of the LCLS.

Progress was made by modifying the imaging setup, moving away from the point-projection scheme, and using a beryllium compound lens stack downstream of the sample, very much similar to an objective in an optical microscope, to obtain a $\sim 200$nm resolved phase and amplitude contrast image on a thin scintillator detector~\cite{laso2024cylindrical}. This experiment was carried out at the HED-HIBEF instrument of the European XFEL~\cite{Zastrau_JSyncRad_2021}. With 35~fs temporal resolution, the researchers show that by the irradiation of a thin wire with single-beam Joule-class short-pulse laser, a converging cylindrical shock is generated compressing the wire material to conditions relevant to the above applications.  The data collected for Cu wires is in agreement with hydrodynamic simulations of an ablative shock launched by highly impulsive and transient resistive heating of the wire surface. The subsequent cylindrical shockwave travels toward the wire axis and is predicted to reach a compression factor of 9 and pressures above 800~Mbar. Simulations for astrophysical relevant materials underline the potential of this compression technique as a new tool for high energy density studies at high repetition rates.

However, in these experiment to contribution from refractive index of the material to phase and absorption were still intertwined and had to be forward-modeled using initial assumptions. To overcome this limitation, Talbot interferometry has recently been introduced at XFELs. The method uses a two dimensional phase diffraction grating placed in the beam, producing at a specific distance behind the grating a self-image of the grating. The setup is essentially the same as the direct imaging setup, with the addition of a Talbot grating~\cite{galtier2025x}, and the reconstructed image yields independent data for absorption contrast and phase. 

Finally, the photon energy of the XFEL can be tuned to a resonance transition in warm dense matter. E.g., the K-L transition of 21$^+$ ions in a Cu plasma was a transition energy of around 8.3~keV. This line will therefore show an increased absorption when interacting with Cu~21$^+$ ions as compared to any other state of Cu. First, still unpublished, experiments exploit this concept to track the spatio-temporal evolution of specific ionic states in dense plasmas with sub-micron accuracy.

In summary, it is currently possible to measure---independently but simultaneously---the phase and amplitude contrast images of matter in extreme conditions with 200~nm spatial and 35~fs temporal resolution, with the option to tune the absorption contrast to a certain ion state.


\newpage 
\clearpage

\section{Optical, VISAR, THz Spectroscopy, absorption}\label{section_14}
\author{Peter Celliers$^{1}$, Benjamin Ofori-Okai$^{2}$, Tommaso Vinci$^{3}$}
\address{
$^1$Lawrence Livermore National Laboratory, Livermore, CA 94550, United States \\
$^2$SLAC National Accelerator Laboratory, Stanford University, Menlo Park, CA, United States \\
$^3$LULI, Ecole Polytechnique, Sorbonne Université, F-91120 Palaiseau Cedex, France
}

\subsection*{Introduction}

In addition to the X-ray diagnostics discussed earlier (see Secs.\ref{section_13},\,\ref{section_13b}), the exploitation of other regions of the electromagnetic spectrum has led to an improved understanding of the WDM regime. The optical (visible) region ($\lambda\approx400-700$ nm) as well as sections of the near-infrared region (NIR, $\lambda\approx 800-1000$ nm) have been used to generate and diagnose WDM.  More recently, terahertz (THz) radiation has been explored as a probe of WDM with the goal of measuring the electrical conductivity.

\subsubsection*{Optical and VISAR:} 
Optical recordings of strong shocks in laser-driven targets were among the first examples of experimental studies of WDM plasmas. For instance, an early experiment employed optical shadowgraphy to diagnose shock propagation \cite{VanKessel1974}, while later experiments recorded self-emission signals from the rear surface of laser-driven samples \cite{Mclean1980, Obenschain1983}.

The self-emission signals in some cases indicated preheat in laser-fusion targets, but also shock transits. Measurements of shock transits were used to infer the shock parameters\cite{veeser1978, trainor1979, Cottet1984} and to test 
the consistency of equation of state models \cite{Koenig1995}. One early study recorded the thermal spectrum onto a streak camera slit in order to diagnose the shock temperature of the sample \cite{Ng1985}. The latter method is challenging for opaque metallic samples because high time resolution is required to capture the signal as the shock emerges from the free surface and before the cooling plasma in the release wave screens the emission from the shock state.  A related pioneering development, active probing using a co-timed pulsed laser to probe shock-released plasma profiles, provided some of the first experimental datasets to test theoretical models of electrical conductivity in WDM plasmas \cite{Parfeniuk1986, Ng1986, Celliers1993}. Another test of transport models came from the interpretation of emission signals from strongly shocked Si which provided evidence of a finite electron-ion equilibration rate in WDM shocks \cite{Celliers1992, Ng1995, Loewer1998}. These investigations spurred the development of new theoretical approaches\cite{Dharmawardana1998,Dharmawardana2001,Gericke_pre_2002}.

Parallel developments in ultra-short laser pulse technology enabled the creation of WDM states through direct irradiation of samples with sub-picosecond heating pulses \cite{Milchberg1988}. This technique provided one of the earliest optical tests of models of dense plasma conductivity, such as those developed by Lee and More \cite{Lee1984} and by Rinker \cite{Rinker1985A, Rinker1985B}. Detailed analysis showed that hydrodynamic expansion can affect the interpretation of the results even for sub-picosecond irradiation \cite{Fedosejevs1990, Ng1994}. A closely related experimental development was the sub-picosecond irradiation of free standing samples of sub-wavelength thickness to create a near-uniform plasma
layer that could be interrogated by optical transmission and reflectivity measurements with different polarizations, wavelengths and oblique angles of incidence \cite{Widmann2004}. This type of experiment, articulated by Ng {\it et al.} as the ``idealized slab plasma,''\cite{Ng2005} provides a unique platform for a variety of investigations of non-equilibrium phenomena in WDM studies \cite{Ao2006, Ping2006, Ping_PRL_2008}.

A notable development in WDM studies with ultra-short laser pulses is the use of chirped pulse probes.  One configuration, similar to a displacement interferometer, produces the interference pattern in the spectral domain and uses
spectrally-resolved detection;\cite{Geindre2001} this method is commonly known as frequency domain interferometry.  A related interferometric method, analogous to a velocity interferometer, was used to demonstrate ultrafast observations of shock states in precompressed samples \cite{Armstrong2010}.

\subsubsection*{THz Spectroscopy:}

There has been long-standing interest in high energy density sciences to successfully predict the zero-frequency (DC, or electrical) conductivity of WDM, as it is often an input for modeling processes including predicting the growth of Rayleigh-Taylor instabilities in inertial confinement fusion (ICF) implosions \cite{Lindl_PoP_1994} or for modeling magnetic fields produced by planetary dynamos \cite{french_ApJS_2012, Nettelmann_ApJ_2012}. While ideal-plasma-based models exist for calculating the conductivity, in the WDM regime the fact that the thermal energy is comparable to the electron energy at the Fermi level combined with the strong Coulomb coupling results in significant deviation from the ideal plasma behavior. Recently, terahertz time-domain spectroscopy (THz-TDS), a widely used technique in other disciplines \cite{Schmuttenmaer_ChemRev_2004, Neu_JAP_2018} has been adapted for probing WDM states with the goal of extracting the electrical conductivity. The electric field of light with frequency $\nu$ = 1 THz (1 THz = 10$^{12}$ Hz = 1 ps$^{-1}$) oscillates slowly compared to electron scattering time, $\tau=1/\nu_{e}$ (i.e. $\omega\tau\ll1$). In the context of the Drude model, where the frequency dependent conductivity is given by
\begin{equation}
    \tilde{\sigma}(\omega)=\frac{n_{e}q_{e}^{2}}{m_{e}\nu_{e}(1-i\omega\tau)},
\end{equation}
with $n_{e}$ the free electron density, $q_{e}$ the charge on an electron, and $m_{e}$ the free electron mass, when the previous condition is met, one finds
\begin{equation}
    \omega\tau\ll1 \Rightarrow\tilde{\sigma}(\omega)\approx\frac{n_{e}q_{e}^{2}}{m_{e}\nu_{e}}=\sigma_{0}
\end{equation}
with $\sigma_{0}$ the DC conductivity. While this argument is presented in the context of the Drude model, the implications on the use of THz measurements to infer the electrical conductivity hold broadly. This, combined with the fact that THz pulses can have ps durations, means that THz measurements directly capture information about DC-like material properties using an ultrafast probe sufficient for probing highly transient states WDM states that are produced in a laboratory.

\subsection*{State of the art}

\subsubsection*{Optical and VISAR}

The possibility of combining active probing with Doppler interferometry led to the development and widespread use of
VISAR as a diagnostic for WDM plasmas \cite{Barker1972, Celliers1998, knudson_prl_01, Knudson2003, Knudson_2004, Celliers2004a, Knudson2008, Celliers2023}. This diagnostic is particularly fruitful for studying transparent dielectrics of scientific and technological importance \cite{Celliers2000, Hicks2003}, because it can be combined with pyrometry to diagnose shock
state (velocity), temperature and electrical conductivity (optical reflectance) simultaneously and with high precision \cite{Koenig2003, Celliers2004b, Hicks2006}. Among the most impressive examples of this capability were the
measurements of the melt curves of diamond, \cite{Bradley2004,Eggert2010} several silica polymorphs \cite{Millot2015}, signatures of the superionic transition in water \cite{Millot2018} and the observation of the metallisation of deuterium at the liquid-liquid metal-insulator transition \cite{Knudson_Science_2015, Celliers_Science_2018}.

VISAR diagnostics are now used routinely at high energy density facilities. At the National Ignition Facility, the first application of VISAR was to provide data needed for high precision shock tuning of the initial shock sequence launched
during an ICF implosion \cite{Munro2001,Boehly2009,Robey2012}. Another key application of VISAR is to determine absolute equation of state of materials along a quasi-isentropic compression path \cite{Reisman2001, Davis2005, Smith2006, Smith2014, Kraus2016a, Smith2018, Fratanduono2021, Smith2023}. Both line-imaging and fiber-coupled point VISAR have become
standardized tools for diagnosing compression paths (shocks and ramps) at high energy density science facilities world-wide. In many applications, the VISAR diagnostic often complements a microscopic probe, such as an x-ray source, to provide a measurement of the pressure while the x-ray probe provides information on the microscopic structure of the sample. At laser facilities, the x-ray source can be produced from a laser-produced plasma \cite{Rygg2012, Coppari2013, Lazicki2015},
while at the light-source facilities the WDM state is produced by a pulsed laser source synchronized with both the VISAR diagnostic and the x-ray probe \cite{Kraus2016b,Gleason2015,Kraus2018}. Many other examples are given
in a recent review \cite{Celliers2023}.

The imaging capabilities of VISAR and displacement interferometers have been extended beyond one-dimension (using streak camera recording) to two-dimensions using framing camera sensors \cite{Celliers2010, Oh2021}. Initial application
of two-dimensional VISAR was motivated by a need to understand inhomogeneities in ICF ablators \cite{Ali2018a, Ali2018b}. A similar imaging approach using a displacement interferometer configuration was used to study heterogeneous response in shocked solids \cite{Greenfield2007}. Two-dimensional VISAR has seen applications in understanding heterogeneous flow
in shock compressed samples \cite{Smith2013}, in understanding drive uniformity in ICF ablators \cite{Oh2021} and in laser-driven samples fielded at light-source facilities \cite{Gorman2022}.

\subsubsection*{THz spectroscopy}

To date, two approaches to using THz measurements for studies of WDM have been demonstrated: (1) A transmission mode geometry to study thin films heated with a short-pulse laser \cite{Ofori-Okai_RSI_2018, Chen_NatComms_2021}; (2) A reflection mode geometry to study compressed matter \cite{Ofori-Okai_POP_2024}. In both cases, a pulse of THz radiation that has interacted with a solid driven to WDM conditions is measured as the time-domain electric field, $E(t)$ using a nonlinear optical technique known as electro-optic sampling \cite{Nahata_APL_1996, Planken_JOSAB_2001}. WDM studies have only recently become possible because of the maturation of single-shot measurements of the THz time-domain waveform with sufficient a signal-to-noise ratio\cite{Ofori-Okai_POP_2024, Ofori-Okai_RSI_2018, Teo_RSI_2015}. 

For studies with THz probing WDM in the transmission geometry, samples have been heated using femtoscond pulses NIR pulses from a table-top laser system, or using XUV radiation produced by an FEL. For near-single-cycle THz pulses, the time-domain field is converted into the frequency domain through the Fourier Transform, resulting a complex-valued spectrum, $\tilde{E}_{s}(\omega)$. By capturing the spectrum of the incident pulse, measured without the sample, $\tilde{E}_{0}(\omega)$ a complex-valued transmission coefficient $\tilde{t}(\omega)$ can be calculated according to $\tilde{t}(\omega)=\tilde{E}_{s}(\omega)/\tilde{E}_{0}(\omega)$. For a sufficiently thin, free-standing film, the conductivity can be related to the transmission according to
\begin{equation}
    \tilde{\sigma}(\omega)=\frac{2}{Z_{0}d}\left(\frac{1}{\tilde{t}(\omega)}-1 \right),
\end{equation}
where $Z_{0}= 377$ $\Omega$ is the impedance of free space and $d$ is the film thickness \cite{Tinkham_PR_1956}. When probing using multicycle THz pulses, each individual cycle of the THz field can be used as an essentially independent probe pulse. As such, multicycle THz pulses can be used to study the full time evolution of the sample as it transitions into the WDM regime following ultrafast heating \cite{Chen_NatComms_2021}.

Measurements in a reflection geometry have been employed to study samples undergoing dynamic compression \cite{Ofori-Okai_POP_2024}. Here, targets can be designed to provide \textit{in-situ} referencing of the THz pulse by adding an additional window material (e.g. lithium fluoride) to produce an additional reflection at an interface that remains unperturbed. Using this to normalize measurements of undriven and driven samples, changes in THz reflectivity, $\tilde{r}$ can be measured. This reflectivity can be related to the sample refractive index, which can in turn be connected to the conductivity.

\subsection*{Challenges \& outlook}

\subsubsection*{Optical and VISAR}

\textbf{High repetition rate}. Future applications of optical diagnostics will increasingly take place at facilities that operate at high repetition rate, for example, 10 or 20 experiments per hour or more. This is much higher than the repetition rate at a high energy laser facility (one per hour). Current analysis methods are performed with bespoke software tools that are not easily adapted to the high bandwidth environment of high repetition rate experimentation.
The future challenge will lie in automation of data collection, data analysis and control of the experimental configuration (parameter scans) to maximize the new information that can be revealed during a finite experimental run.  An example of such a strategy in the field of Ptychography was pioneered by Cherukara {\it et al.} \cite{Cherukara2020}.

\textbf{Diagnostic improvement.} Current implementations of the two-dimensional VISAR are restricted to collecting a single frame of data.  A capability to collect multiple frames spanning a relevant time scale would provide a wealth of information on target dynamics, for example, in the study of viscous response of rippled shocks \cite{Miller1991} and rippled interfaces \cite{Opie2017} at WDM conditions. Similarly the deployment of phase extraction and machine learning methods to infer the wavefront phase in addition to the velocimetry phase will add to the information content in two-dimensional VISAR datasets.  

\textbf{Bandwidth and complementary measurement modes.} Current implementations of VISAR and pyrometry are limited to narrowband measurements.  Both VISAR \cite{Erskine1995} and pyrometry \cite{Bailey2008, Gregor2016} can be extended to the broadband domain, possibly with spectral resolution, to reveal new information on electronic structure and optical properties at WDM conditions.  While such capabilities have been demonstrated they are not yet in routine use as production diagnostics.  Finally, the combination of a VISAR probe with other optical measurement modes such as Raman or
optical polarization sensitivity may be used to reveal new information on chemical structure and magnetic response at WDM conditions.

\subsubsection*{THz Spectroscopy}

As THz measurements have only recently been robustly demonstrated in WDM, there are several exciting opportunities for this technique. Table-top laser systems can be used to perform THz measurements on WDM provided that THz pulses with sufficient field strength can be generated. The workhorse approach is to use non-linear mixing of the frequency components of a femtosecond laser pulse in a non-linear medium \cite{Auston1988, Hu1990, rice_terahertz_1994, yeh_generation_2007}. These approaches are limited in that the non-linear medium has a damage threshold and so as laser energy is increased the THz output energy cannot grow with it. Irradiation of metal foils with intense short pulse lasers have also been shown to produce THz radiation \cite{Gopal2012, Gopal2013, Gopal2013b}, and these approaches could also be investigated as a way to leverage existing infrastructure. Additionally, the use of self-replenishing targets has been considered as well \cite{jin_observation_2017}, and this could alleviate the challenges of material damage thresholds.

Going beyond table-top experiments, as a probe it can be incorporated into other existing experimental facilities, thereby taking advantage of the capabilities of different laser drivers to access different WDM states. Beyond ultrafast heating with femtosecond lasers, the use of X-ray sources to isochorically heat targets will extend the range of samples that may be studied with this technique. Additionally, the use of XFEL techniques to constrain the structure, density, and temperature by XRD and XRTS will permit benchmark quality conductivity data of a well-characterized WDM state. Even synchronous THz and optical probing of WDM provides direct access to both the DC and AC conductivities, which can be used to unambiguously determine the electron density and scattering rate.

In comparison with studies using optical probing, there are a number of outstanding measurements required to even analyze THz data on WDM. For example, details of the refractive index of materials under WDM conditions is essential for determining window layers for dynamic compression studies. Details of the eV-range of material bandgaps have been measured using optical probes, and such data remains absent in the THz regime.

\newpage 
\clearpage
\section{Uniform Electron Gas}\label{section_15}
\author{$\footnotemark$\footnotetext{This section is dedicated to Travis Sjostrom.}Tobias Dornheim$^{1,2}$, Valentin V. Karasiev$^3$, Shigenori Tanaka$^4$, Samuel Trickey$^5$}
\address{
$^1$Center for Advanced Systems Understanding (CASUS), D-02826 G\"orlitz, Germany \\
$^2$Helmholtz-Zentrum Dresden-Rossendorf (HZDR), D-01328 Dresden, Germany \\
$^3$Laboratory for Laser Energetics, University of Rochester, 250 East River Road, Rochester, New York 14623-1299, United States \\
$^4$Kobe University, Kobe 657-8501, Japan \\
$^5Dept. of Physics, $University of Florida, Gainesville, 32611 Florida, United States
}

\subsection*{Introduction}

The uniform electron gas (UEG) constitutes the archetypal quantum mechanical many-electron system. Its importance is illustrated by Giuliani and Vignale's tome, where the UEG occupies 60 pages of introduction alone~\cite{quantum_theory}. Major insight into physics as diverse as the Wigner crystallization (strong coupling)~\cite{Clark_PRL_2009,Drummond_PRB_2004,Azadi_PRB_2022}, the Fermi liquid theory, and the integer and fractional quantum Hall effect (among many others) has originated from UEG studies.

Detailed knowledge of UEG properties has been and continues to be particularly important for both WDM theory and computation. This dates back to Feynman, Metropolis, Teller's finite-$T$ Thomas-Fermi theory~\cite{feynman1949equations}. In the last 15+ years, carefully computed, quantitatively accurate UEG properties have been critically important for the construction of free-energy exchange-correlation (XC) functionals for finite-temperature DFT~\cite{KSDT2014PRL,Groth_PRL_2016} (see Sec.\ref{section_02}) and XC-kernels for linear-response time-dependent DFT~\cite{Ramakrishna_PRB_2021} (see Secs.\ref{section_03},\ref{section_04}).  X-ray Thomson scattering interpretation has exploited UEG properties~\cite{Fortmann_PRE_2010,Dornheim_PRL_2020_ESA}. Ionization potential depression estimates have used UEG calibration for a machine-learned local field correction (LFC)~\cite{Zan_PRE_2021}. That LFC also has been used to calculate improved (over Singwi-Tosi-Land-Sj\"olander LFC) low-$Z$ stopping power~\cite{Moldabekov_PRE_2020,Faussurier_POP_2025}. Earlier work examined effective ion-ion potentials~\cite{Moldabekov_PRE_2018,Moldabekov_PRE_2019} in the context of comparisons of STLS with UEG data. Staying with structure factors, in addition to the intriguing physical phenomena mentioned above, the UEG has others that are not so well-known. An example is a \emph{roton-type feature} in its dynamic structure factor~\cite{Dornheim_ComPhys_2022,Dornheim_PRL_2018,koskelo2023shortrange,Takada_PRB_2016}.

Given its combination of elegant simplicity but rich physical behavior, the UEG has been studied with a plethora of methods. Much of the basic literature can be found in Ref.~\cite{quantum_theory}. Here we mention the main methodological themes of recent years, namely Quantum Monte Carlo simulations (QMC)~\cite{dornheim_pop17}, quantum-to-classical mappings~\cite{Perrot_PRB_2000,Dutta_EPL_2013,Liu_JCP_2014}, dielectric theories \cite{Tanaka_JCP_2016,Tolias_JCP_2021,Tolias_JCP_2023,Sjostrom_PRB_2013,Tanaka_CPP_2017,Tolias_PRB_2024,tolias2024dynamicpropertieswarmdense}, and Green's functions~\cite{Kas_PRL_2017,Kas_PRB_2019,kwong_prl_00}.

\begin{figure}
    \centering
    \includegraphics[width=\linewidth]{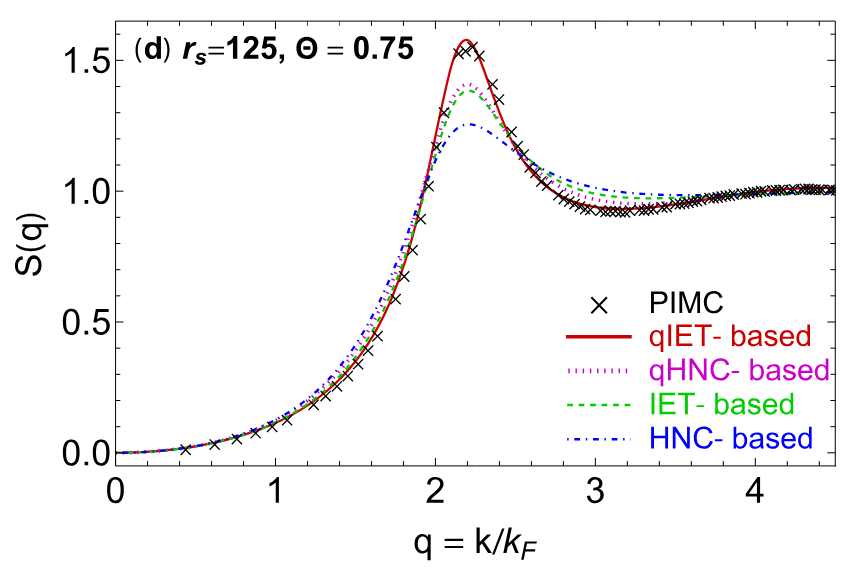}
    \caption{Static structure factor $S(\mathbf{q})$ of the strongly coupled electron liquid at $r_s=125$ and $\Theta=0.75$. Black crosses: quasi-exact PIMC reference data; solid red: quantum IET~\cite{Tolias_JCP_2023}; dotted magenta: quantum HNC; dashed green: classical IET~\cite{Tolias_JCP_2021}; dash-dotted blue: classical HNC~\cite{Tanaka_JCP_2016}. Taken from Tolias \emph{et al.}~\cite{Tolias_JCP_2023} with the permission of the authors.
    }
    \label{fig:UEG_SSF}
\end{figure}

\subsection*{State of the art}

The self-consistent dielectric formalism~\cite{SingwiTosi_Review,Ichimaru_PhysReports_1987} has long provided a dependable, theoretical scheme to calculate the correlational and thermodynamic functions of UEG, where the electron correlation effects beyond the random-phase approximation (RPA) are described in terms of the LFC which is related to the wavenumber-dependent structure factor. For the UEG, the integral-equation strategies were primarily employed on the bases of the approximations due to STLS~\cite{Ichimaru_PhysReports_1987,stls,Tanaka_CPP_2017}, Vashishta-Singwi (VS)~\cite{Sjostrom_PRB_2013,Tolias_PRB_2024}, modified convolution~\cite{TI1989}, and hypernetted chain (HNC)~\cite{Tanaka_JCP_2016,Dornheim_PRB_2020}. More recently, novel schemes to consider strong correlation effects have been proposed, for instance, going beyond the HNC approximation via bridge functions~\cite{Tolias_JCP_2021,Lucco_Castello_EuroPhysLett_2022} and the dynamic (frequency-dependent) LFC by employing the quantum version of self-consistent truncation for kinetic equations~\cite{Tolias_JCP_2023}. The former scheme, called the integral equation theory (IET), works well in the strong-coupling regime; for example, the static structure factor calculated with the quantum version IET (qIET) accurately reproduces the PIMC result even in the strongly coupled electron liquid regime with $r_s=125$ and $\Theta=T/T_F=0.75$, see Fig.~\ref{fig:UEG_SSF}. 
One of the advantages of the dielectric formalism compared to the PIMC approach is its applicability to the low-temperature (even zero-temperature) limit, thus providing complementary information on the correlational and thermodynamic properties of the UEG. Another advantage of dielectric formalism is the capability to describe the small-wavenumber behaviors of correlation functions if the compressibility sum rule~\cite{Tanaka_JCP_2016,TI1989} is sufficiently satisfied. On the other hand, most of the self-consistent dielectric theories fail to retain the positivity of the radial distribution function at and around the origin~\cite{stls,Tanaka_JCP_2016}, which can be remedied by an improved, effective static approximation (ESA)~\cite{Dornheim_PRL_2020_ESA,Dornheim_PRB_2021_ESA,Tolias_PRB_2024} (see below) formulated consistently with the PIMC results.

\begin{figure}
    \centering
    \includegraphics[width=\linewidth]{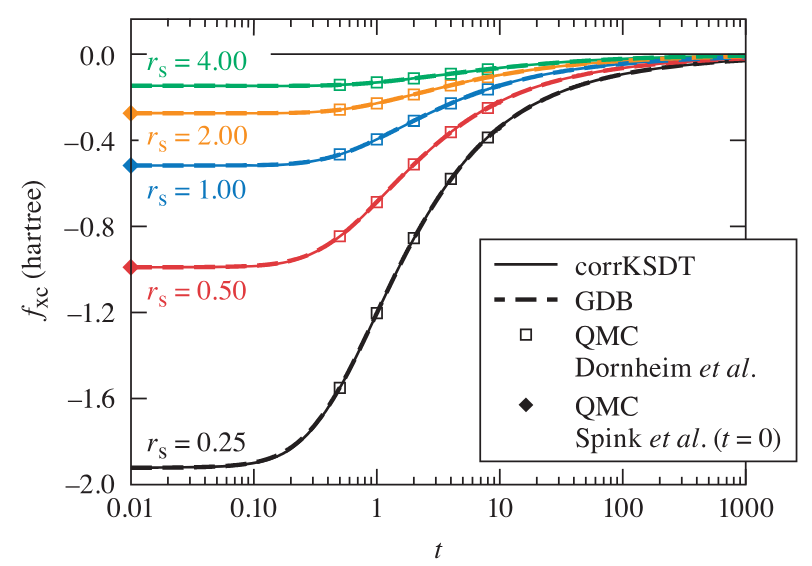}
    \caption{Temperature dependence of the XC-free energy of the UEG (with $t=\Theta=k_\textnormal{B}T/E_\textnormal{F}$) for different densities. Solid and dashed lines: corrKSDT~\cite{KDT16} and GDSMFB~\cite{Groth_PRL_2016} parametrizations; squares: finite-$T$ QMC results from Refs.~\cite{Groth_PRL_2016,Dornheim_PRL_2016}; diamonds: ground-state QMC data by Spink \emph{et al.}~\cite{Spink_PRB_2013}. Taken from Karasiev \emph{et al.}~\cite{Karasiev_PRB_2019} with the permission of the authors. 
    }
    \label{fig:UEG_status}
\end{figure}

The most accurate parametrizations of the XC-free energy per particle of the UEG $f_\textnormal{xc}(r_s,\Theta;\zeta)$, with
$\zeta=(N^\uparrow-N^\downarrow)/N$ being the spin-polarization, are based on PIMC calculations. Karasiev \emph{et
al.}~\cite{KSDT2014PRL} (KSDT) explored different routes towards $f_\textnormal{xc}$ from PIMC data and developed an analytical fit form that continues in use.  They parametrized the restricted PIMC results of Brown \emph{et al.}~\cite{Brown_PRL_2013}, which turned out to have systematic finite-size correction errors.  There also were possible nodal limitations~\cite{Schoof_PRL_2015,Malone_PRL_2016,Joonho_JCP_2021}. Subsequently, Dornheim \emph{et al.}~\cite{Dornheim_PRL_2016} presented alternative CPIMC and PB-PIMC (see Sec.\ref{section_02}) results using a substantially improved extrapolation to the thermodynamic limit. Those data constitute the basis of the parametrization (of the same analytical form) by Groth \emph{et al.}~\cite{Groth_PRL_2016} (GDSMFB) and also inform the corrected KSDT (corrKSDT) version by Karasiev \emph{et al.}~\cite{KDT16}, which are virtually indistinguishable, see Fig.~\ref{fig:UEG_status}. These representations of $f_\textnormal{xc}(r_s,\Theta;\zeta)$ directly allow for thermal DFT simulations on the level of the local density approximation (LDA)~\cite{Moldabekov_JCTC_2024,karasiev_importance_2016,kushal}, constitute the basis for more advanced functionals on higher rungs of Jacob's ladder of functional approximations (see Sec.\ref{section_03})~\cite{KDT16,kozlowski2023generalized,KarasievMihalovHu2022}, and constitute key input for a large number of other applications~\cite{23PRB-Si,Dornheim_PhysRevB_2024,dornheim2024shortwavelengthlimitdynamic,wetta2023average,Roepke_PRE_2024,Callow_PRR_2022,Wetta2025}.
Recently, Dornheim \emph{et al.} have presented alternative routes to the free
energy~\cite{dornheim2025fermionicfreeenergiestextitab,Dornheim_free_energy,dornheim2024eta}
and the canonical chemical potential $\mu$~\cite{dornheim2024chemicalpotentialwarmdense} from PIMC simulations. These independent cross-checks have verified the GDSMFB result, further substantiating the high quality of these state-of-the-art parametrizations.

\begin{figure}
    \centering
    \includegraphics[width=0.8\linewidth]{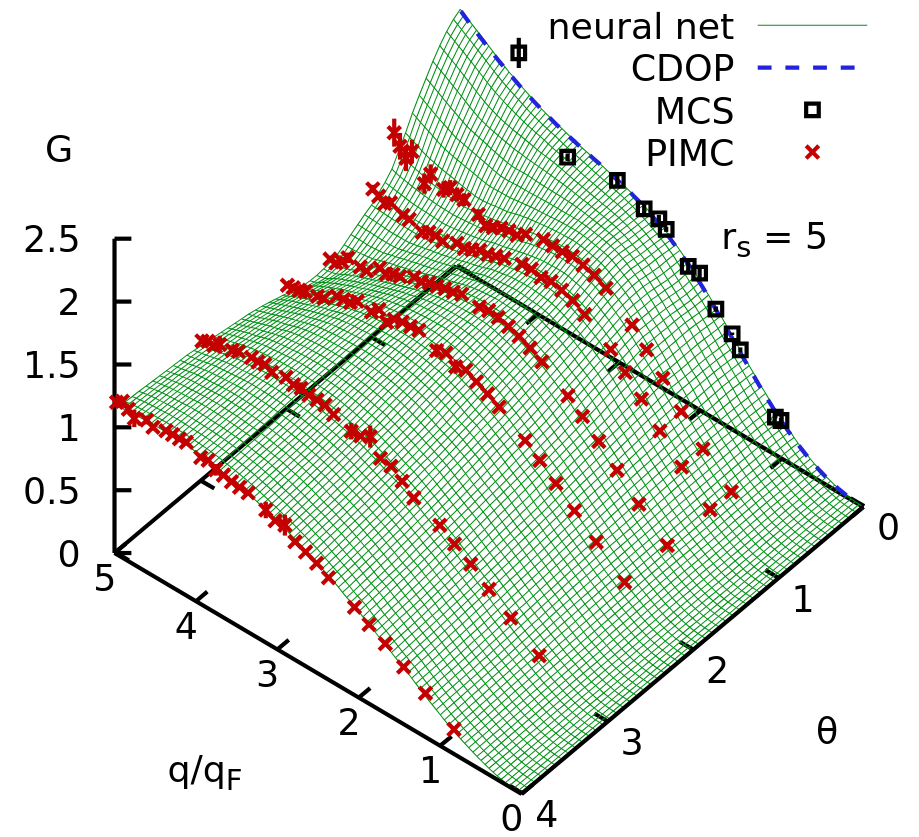}
    \caption{Static local field correction $G(q;r_s,\Theta)$ in the $q$-$\Theta$-plane for $r_s=5$. Red crosses: finite-$T$ PIMC results~\cite{Dornheim_JCP_2019}; black squares: $T=0$ QMC results by Moroni \emph{et al.}~\cite{Moroni_1995}; dashed blue: ground-state parametrization by Corradini \emph{et al.}~\cite{Corradini_PRB_1998}; green surface: neural network representation of $G(q;r_s,\Theta)$.
    Taken from Dornheim \emph{et al.}~\cite{Dornheim_JCP_2019} with the permission of the authors.
    }
    \label{fig:UEG_LFC}
\end{figure}

An additional class of interesting properties of the warm dense UEG is given by its linear response to an external perturbation, which is also at the heart of the dielectric formalism introduced above. After initial proof-of-principle studies using the PB-PIMC~\cite{Dornheim_PRE_2017} and CPIMC methods~\cite{groth_jcp}, first comprehensive results have been presented by Dornheim \emph{et al.}~\cite{Dornheim_JCP_2019} based on the imaginary-time version of the fluctuation--dissipation theorem~\cite{Dornheim_MRE_2023}. These PIMC data have been combined with available results for the ground-state limit~\cite{Moroni_1995,Corradini_PRB_1998} to train a neural network representation of the static LFC $G(r_s,\Theta;q)$ that covers the entire relevant WDM parameter space, and which was subsequently validated by independent QMC results from Hou \emph{et al.}~\cite{Hou_PRB_2022}; see Fig.~\ref{fig:UEG_LFC} for a cutout in the $q$-$\Theta$-plane for $r_s=5$. An additional subtlety concerns the difference between the exact static limit of the dynamic LFC $G(\mathbf{q},0)$ and an effectively static theory for the LFC, $G(\mathbf{q})$.
While the \emph{static approximation} $G(\mathbf{q},\omega)\approx G(\mathbf{q},0)$ works well even for estimation of dynamic properties for a given individual $q$~\cite{Dornheim_PRL_2018}, it introduces a systematic bias for large wavenumbers that accumulates for integrated properties such as the interaction energy $w$ or $f_\textnormal{xc}$~\cite{Dornheim_PRL_2020_ESA}. 
An improved \emph{effective static approximation} (ESA) has been introduced in Ref.~\cite{Dornheim_PRL_2020_ESA} that removes these inconsistencies, and which has subsequently been parametrized analytically in Ref.~\cite{Dornheim_PRB_2021_ESA}.

Other exact PIMC results have been presented for a host of observables, including effective forces in the UEG~\cite{Dornheim_JCP_2022}, the momentum distribution function $n(q)$~\cite{Dornheim_PRB_2021_nk,Militzer_PRL_2002,Hunger_PRE_2021,Dornheim_PRE_2021} (which gives insights to an XC-induced lowering of the kinetic energy at finite $T$ for some densities), various linear~\cite{Dornheim_EPL_2024,Dornheim_PRB_2020,Dornheim_PhysRevB_2024,dornheim2024shortwavelengthlimitdynamic,Dornheim_HEDP_2022,Dornheim_PPCF_2020,Dornheim_PRR_2022} and non-linear static density response properties~\cite{Dornheim_PhysRevRes_2021,Dornheim_PRL_2020,Dornheim_CPP_2021,Dornheim_CPP_2022} as well as related higher-order ITCFs~\cite{Dornheim_JCP_ITCF_2021} (see also Sec.\ref{section_05}).

Finally, we mention the dynamic equilibrium properties of the UEG, which can be computed approximately from Green functions theory~\cite{kwong_prl_00} (see Sec.~\ref{section_06}) and dielectric theories~\cite{Holas_PRB_1987,Dabrowski_PRB_1986}, or computed approximation-free via the analytic continuation of PIMC results for the ITCF $F_{ee}(\mathbf{q},\tau)$~\cite{Dornheim_PRL_2018,Groth_PRB_2019,chuna2025,Filinov_PRB_2023,Dornheim_PRE_2020}, cf.~Eq.~(\ref{eq:dsf_laplace_transform}) in Sec.\ref{section_13}. While the numerical inversion of Eq.(\ref{eq:dsf_laplace_transform}) constitutes an exponentially ill-posed problem~\cite{Jarrel_PhysReports_1996}, the reconstruction problem has been made tractable by enforcing a number of exact constraints that are known  for the finite temperature UEG~\cite{Dornheim_PRL_2018}. This has resulted in the first, highly accurate PIMC results for the DSF $S_{ee}(\mathbf{q},\omega)$ over a broad range of densities, revealing i) an XC-induced red-shift with respect to the RPA even in the WDM regime ($r_s=2$), ii) a remarkable accuracy of the \emph{static approximation} $G(\mathbf{q},\omega)\approx G(\mathbf{q},0)$ for $r_s\lesssim5$, and iii) a roton-type feature in the dispersion relation for $r_s\gtrsim10$ at $\Theta=1$. The latter has been subsequently explained in terms of the spontaneous alignment of pairs of electrons~\cite{Dornheim_ComPhys_2022}, and constitutes an active field of research also in the $T=0$ limit~\cite{koskelo2023shortrange,Takada_PRB_2016,Takada_PRL_2002}. In addition, accurate knowledge of $S_{ee}(\mathbf{q},\omega)$ gives one access to the dynamic dielectric response and conductivity, which has been explored by Hamann \emph{et al.}~\cite{Hamann_PRB_2020,Hamann_CPP_2020}.

\subsection*{Challenges \& outlook}

Somewhat obscured by the high quality of the KSDT / GDSMFB parametrizations, there is a problem in them related to the evaluation of thermodynamic derivatives.  One example is the electron specific heat 
$c_{\rm V}(r_s,\Theta)=-(\Theta/T_{\rm F})\partial^2 f(r_s,\Theta)/\partial \Theta^2\vert_{r_s}=(1/T_{\rm F})\partial \varepsilon(r_s,\Theta)/\partial \Theta\vert_{r_s}$ that behaves badly for about $r_s \ge 10$~\cite{Karasiev_PRB_2019}. Here $f(r_s,\Theta)$ and $\varepsilon(r_s,\Theta)$ are 
the total free- and internal energies, given by the sum of the non-interacting free-energy and XC free-energy (for the total free-energy) and by the sum of the non-interacting kinetic and XC internal energies (for the total internal energy). Analysis shows that for $r_s\ge 10$ a near-total cancellation between the X contribution to $c_{\rm V}$, known exactly, and the C term occurs, such that a small relative error in the $\partial \varepsilon_{\rm c}/\partial \Theta$ derivative may yield a large error in $c_{\rm V}$. Furthermore, as $\Theta$ increases, $\varepsilon_{\rm c}$ decreases while $\varepsilon_{\rm x}$ increases, hence they cross and as a consequence $c_{\rm V}$ has a bump.
An associated oddity is that the Fermi-liquid theory effective mass $m^*$ is related as $\Theta\rightarrow 0$ to $c_{\rm V}(r_s,\Theta)$.  Eich et al.~\cite{Eich_PRB_2017} pointed out that the effective mass ratio  $m^*/m$ from KSDT on $0\le r_s \le 1$ is $0.98 \le m^*/m \le 1.2$ whereas other estimates give $0.95 \le m^*/m < 1.0$.  The discrepancy could be resolved with accurate very low $\Theta$ data.  

An additional frontier of the UEG is posed  by the comprehensive description of its dynamic properties. While the results presented in Refs.~\cite{Dornheim_PRL_2018,Groth_PRB_2019} constitute a solid foundation, considerable additional effort will be needed to construct a rigorous parametrization of the dynamic LFC $G(\mathbf{q},\omega;r_s,\Theta)$ to cover the entire WDM regime. Such a result would be of high value as an XC-kernel for TDDFT (see Sec.\ref{section_04}), for the interpretation of XRTS experiments~\cite{Fortmann_PRE_2010,Dornheim_PRL_2020_ESA,Poole_PhysRevRes_2024} (see Sec.\ref{section_13}), and for the construction of advanced, non-local and consistently thermal XC functionals~\cite{pribram,Patrick_JCP_2015} (see Sec.\ref{section_03}). In this context, we mention recent progress in general purpose analytic continuation methods~\cite{chuna2025,Otsuki_PRE_2017,Fei_PRL_2021}, as well as novel QMC methods that give one direct access to the dynamic XC-kernel for selected parameters~\cite{Leblanc_PRL_2022}. A potentially less involved alternative that is sufficient to estimate frequency-integrated properties is given by the dynamic Matsubara density response (see Sec.\ref{section_05}), which can be obtained directly from the PIMC results for the ITCF~\cite{Moldabekov_PPNP_2024,Dornheim_EPL_2024,Tolias_JCP_Matsubara_2024,Dornheim_PhysRevB_2024,dornheim2024shortwavelengthlimitdynamic}. 

Future efforts might also focus on the optical response properties of the warm dense UEG. These are difficult to study with PIMC methods that are limited to $q\geq 2\pi/L$ (with $L$ the box length) hence require large simulations (though this might be feasible using the $\xi$-extrapolation technique in PIMC~\cite{Dornheim_JPCL_2024}). Alternative approaches unafflicted by this limitation have appeared recently~\cite{Hou_PRB_2022}. We also mention the possibility to estimate higher-order imaginary-time correlation functions from PIMC. In addition to their direct connection to the UEG non-linear density response, these are connected to generalized many-body (e.g., three-body, four-body, etc) dynamic structure factors~\cite{Dornheim_JCP_ITCF_2021}, which remain poorly understood for any interacting quantum system.

Finally, we point out the near absence of reliable results for the $2D$ UEG at finite temperatures~\cite{Clark_PRL_2009,Fisher_PRB_1982,Monarkha_LowT_2012,Rani_2023}.
Indeed, while realistic applications of the $2D$ UEG have traditionally been limited to the ground-state limit~\cite{quantum_theory}, advances in semiconductor applications will likely require taking into account the electronic temperature in the future. In addition, we expect the finite-$T$ $2D$ UEG to offer a plethora of rich physics, including questions about Wigner crystallization and possible roton modes in $S(\mathbf{q},\omega)$.

\newpage 
\clearpage

\section{Hydrogen}\label{section_16}
\author{Maximilian P.~B\"ohme$^{1}$, Suxing Hu$^{2,3,4}$, Philip A. Sterne$^{1}$, Jan Vorberger$^5$}
\address{
$^1$Lawrence Livermore National Laboratory, Livermore, CA 94550, United States \\
$^2$Laboratory for Laser Energetics, University of Rochester, Rochester, New York 14623-1299, United States \\
$^3$Department of Physics and Astronomy, University of Rochester, Rochester, New York 14627, United States \\
$^4$Department of Mechanical Engineering, University of Rochester, Rochester, New York 14627, United States \\
$^5$Helmholtz-Zentrum Dresden-Rossendorf (HZDR), D-01328 Dresden, Germany 
}

\subsection*{Introduction}

A century after the initial formulation of quantum mechanics, an exhaustive understanding of the simplest and most abundant element in the universe, hydrogen, has not been achieved. In particular, the properties of hydrogen under warm dense matter conditions are not yet fully understood. For example, we still do not have a definitive answer of whether or not hydrogen is a superconductor at $\rho=10$ g/cm$^3$, $k_BT=1000$ K. Despite its simplicity, hydrogen exhibits a complex phase diagram. Besides its fundamental importance to condensed-matter physics, better understanding could have tremendous implications to planetary science~\cite{Burrow_RMP_2001}, astrophysics~\cite{saumon2021current}, and fusion energy science~\cite{Hu_PRB_2011,Hu_PRL_2010, Mihaylov_prb_2021,kritcher_gain_2025}.    

For planetary science, the hydrogen EOS plays a crucial role in the accurate modeling of planetary interiors of gas giants such as Jupiter and Saturn. The \textit{Juno} and \textit{Cassini} satellite missions gained insight into the gravitational fields of both planets~\cite{Iess2019,Durante2020}. Understanding of the hydrogen phase diagram is imperative to interpret the measurements of both missions. Recent models of Jupiter's interior (see Sec.~\ref{section_17}) introduce a hydrogen / helium phase separation or a diluted core to accommodate the inhomogeneous structure indicated by measurements~\cite{Morales_PNAS_2009,Lorenzen2009,Schoettler2018,Wahl2017a,Militzer_2022}.

Recent technological breakthroughs in the field of inertial confinement fusion (ICF)~\cite{Abu-Shawareb2022,Kritcher_PRE_2024,ICF_PRL_2024}, further motivate the study of highly compressed hydrogen and deuterium-tritium mixtures. ICF especially relies on an accurate understanding of the underlying physical processes as it is increasingly challenging to directly measure all governing parameters. Thus, one has to heavily rely on multi-scale modeling of the physical process, where either average atom (AA) models, density functional theory (DFT), or quantum Monte Carlo (QMC) methods are employed to calculate the EOS and transport properties, which are then used to inform macroscopic methods such as radiation-hydrodynamics, which provides predictions for the neutron flux during the ignition phase.  

A comprehensive review has recently appeared~\cite{Bonitz_POP_2024}. State-of-the-art computational methods for hydrogen studies include DFT-MD~\cite{Collins1995, Holst_PRB_2008, Caillabet_2011, Hu_PoP_2015} with recent advanced temperature dependent exchange-correlation functionals~\cite{ Mihaylov_prb_2021} and QMC such as Path integral Monte Carlo (PIMC)~\cite{Boehme2022, Boehme2023, Dornheim_PRE_2023, Hu_PRB_2011, Hu_PRL_2010, MC00, Filinov_PRE_2023, Dornheim_MRE_2024, Dornheim_JCP_2024, Militzer2001} or coupled electron-ion Monte Carlo (CEIMC)~\cite{Pierleoni2006,Tubman_2015}. Despite significant advances in the development of {\sl first-principles} methods, there are still disagreements between calculations of hydrogen properties and experiments.

\subsection*{State of the art}

Over the past $25$ years, hydrogen/deuterium shock experiments have been performed using static compression, e.g. diamond anvil cells (DACs), and dynamic compression techniques such as laser-driven or pulse-power-driven shocks~\cite{wang_dac_review_2024,Bonitz_POP_2024}. In DACs, currently accessible parameters are generally $\lesssim5000\,$K and $\lesssim5$\,Mbar with the temperatures at the maximum pressures well $<1000$\,K. Under such conditions, diamonds tend to break easily. DAC experiments on hydrogen additionally suffer from the high hydrogen diffusivity and thus loss of containment, see Sec.~\ref{section_09}. XRD diagnostics on hydrogen is very difficult (strong diamond signal), so that most experimental results for the location and crystal structure of hydrogen high-pressure phases have been based on Raman scattering~\cite{wang_dac_review_2024}. 

During dynamic compression, measurements are necessarily performed within a very short time window (ns-$\mu$s), raising questions about equilibration times, overheating, supercooling, and, in general, whether a true equilibrium state has been created~\cite{Fletcher_Frontiers_2022}. Diagnostic tools have been relatively limited with mainly VISAR, Streaked Optical Pyrometry (SOP), shadowgraphy, and recently XRD and XRTS~\cite{falk_wdm}. Even though XRD and XRTS are promising, the very low scattering cross section of x-rays in hydrogen and the reliance (especially of XRTS) on theoretical modeling make it difficult to fully use their capabilities. Basically, all of the published shock-Hugoniot measurements of hydrogen rely on a pressure standard of unknown accuracy and none of the shock experiments have employed a suitable temperature diagnostics.

Regardless of experimental challenges, comparisons with {\sl first-principles} calculations revealed several surprises. A recent experiment has shown that deuterium is ($\sim 5-10\%$) more compressible than any model prediction along its principal shock Hugoniot with pressures above $\sim 5$ Mbar~\cite{fernandez_prl_19}. This softening behavior was also seen in re-shock Hugoniots. There is no good explanation for the observed softening shock behavior of deuterium, even though some speculation was recently made about the existence of electron plasma-wave excitation~\cite{rygg_prl_23}. However, it is still unknown why such plasma waves do not decay given the high collisionality. Another surprising result concerns a lower than predicted sound speed in shocked deuterium~\cite{Fratanduono_PoP_2019}. Again, there is no apparent explanation thus far. Another set of controversies stems from the $P-T$ conditions of the molecular to metallic transition, believed to be a first order phase transition, the so-called liquid-liquid phase transition (LLPT), at temperatures below $\sim 1500$~K. Reconciling different experimental results, from DACs~\cite{Zaghoo_PRL_2016}, laser-ramp compression~\cite{Celliers_Science_2018}, and pulse-power-driven compression~\cite{Knudson_Science_2015} is still ongoing. 

In general, the lack of conclusive diagnostics strongly hampers the experimental progress. A part of the solution are hydrogen jets that continuously deliver cryogenic hydrogen~\cite{Kim_2016}. If combined with a suitable high repetition rate (optical) pump laser and an X-FEL beam for diagnostics, the low signal-to-noise ratio in x-ray scattering in hydrogen can be overcome simply by brute force data collection~\cite{Fletcher_RevSci_2016}. This naturally presupposes the stable operation and timing of the hydrogen jet, pump laser, and X-FEL. Recent advances in model free temperature measurements using XRTS are very promising~\cite{Dornheim_NatCommun_2022,Dornheim_T_follow_up}. However, the density would still need to be determined from theory in any case other than pure isochoric heating~\cite{dornheim2024modelfreerayleighweightxray}.

Several attempts have been made to generate a hydrogen EOS for a wide range of warm dense matter conditions that is extended to lower densities (lower temperatures) to cover the outer envelope of gas planets and also extended to higher densities (higher temperatures) to accommodate inertial fusion experiments. The most widely used EOS is the tabular EOS from the SESAME database~\cite{sesame,kerley2003equations}. The first EOS stemming purely from calculations is probably due to Saumon \& Chabrier~\cite{saumon_aps_95}. More recent tables rely more heavily on DFT-MD~\cite{Wang_2013,Mihaylov_prb_2021} like the H-REOS~\cite{Becker_AJSS_2014} (a combination of fluid variational theory for the lower densities and DFT-MD for high densities) or are based on DFT-MD combined with RPIMC for high temperatures like the FPEOS~\cite{Hu_PRB_2011,Hu_PRL_2010,Hu_PoP_2015,FPEOS}. Over the last decade, several versions of these EOS have been produced, mainly fueled by an increase in computing power that allowed, e.g., larger system sizes or different XC functionals~\cite{Mihaylov_prb_2021}. As none of the employed methods are entirely approximation free (see the XC functional in DFT-MD, the nodal surfaces in RPIMC, finite size effects etc.), the overall quality of all of the available hydrogen EOS is not known. The production of exact benchmark results for the hydrogen EOS is one of the grand challenges the community is facing.

There is much debate on the theoretical side, mainly via DFT-MD simulations, concerning the exact phase diagram location and the nature of the LLPT, which strongly depends on the employed XC-functional. State-of-the-art DFT-MD predicts a first-order phase transition~\cite{Lorenzen2010,Morales2010, karasiev2021,Bergermann_2024}. Machine learning methods, employed to increase the system size that can be simulated, have so far failed to provide further insight due to their inability to reproduce information contained in the training data~\cite{Cheng_Nature_2020,karasiev2021}. 

One of the most desirable goals is the development of an accurate finite temperature XC functional that can be used for warm dense hydrogen. Temperature-dependent XC functionals have been recently developed~\cite{KSDT2014PRL, KDT16, HillekeKarasievTrickey2024, KarasievMihalovHu2022, Groth_PRL_2016, Mihaylov_PRB_2020}. However, even though these advanced XC functionals are based on first principle PIMC simulation results, their quality for warm-dense hydrogen has not been checked universally~\cite{Bonitz_POP_2024}. For low temperatures and high pressures, another challenge is presented by the relatively small mass of the protons. The approximations inherent in DFT-MD, namely the Born-Oppenheimer and a classical description of the nuclei, thus break down. Path-integral molecular-dynamics (PIMD)~\cite{Hirshberg_PNAS_2019} has been developed for studying the quantum behavior of ions and the same applies for the coupled-electron-ion Monte Carlo (CEIMC) method~\cite{Tubman_2015}.  

It is necessary to obtain exact benchmark points for DFT or AA to quantify the systematic uncertainty introduced when performing calculations using XC functionals in the WDM regime of hydrogen. A first such benchmark is the so-called snap-shot PIMC for warm dense hydrogen~\cite{Boehme2022,Boehme2023}. Snap-shot PIMC exactly solves the partition function integral for a Hamiltonian that describes a collection of electrons in the external potential given by the static proton positions as provided by, e.g., DFT-MD simulations. This method is similar to the procedure in DFT. However, unlike DFT, the electronic structure problem in an external potential is solved using established PIMC algorithms and thus includes all exchange and correlation effects. Snap-shot PIMC can access the exact density, static density response function $\chi(\mathbf{q},\omega=0)=\chi(\textbf{q})$, and many more quantities of hydrogen that can also be obtained from DFT~\cite{Boehme2022,Boehme2023}. Moreover, even a direct comparison of free energies has recently become possible~\cite{Dornheim_free_energy}.

A step further concerns the direct PIMC simulation of electrons and protons without any simplifying assumption about statistics or time scale separation~\cite{Dornheim_MRE_2024}. Thus, the exact computation of all $\chi_{ab}(\textbf{q})$ has been achieved for warm dense hydrogen and  delivers critical insight into its XC behaviour through the computation of the static local field corrections $G_{ab}(\textbf{q},\omega=0)$. As these LFCs can be directly translated into the XC kernel formalism $G(\mathbf{q},\omega)=-(q^2/4\pi)K_\textnormal{xc}(\mathbf{q},\omega)$, this allows to check the accuracy of XC-kernels $K_{\mathrm{xc}}$, that are usually employed in the context of time-dependent density functional theory simulations~\cite{ullrich2012time}. A full two-component PIMC simulation of hydrogen naturally also provides the static structure of electrons, protons, and cross-species correlations. This makes for an excellent benchmark for all flavors of different DFT-MD simulations of hydrogen. The newly introduced $\eta$-ensemble method for the calculation of the free energy also paves the way for a thermodynamically consistent EOS of hydrogen from PIMC~\cite{dornheim2024eta,Dornheim_free_energy}.


\subsection*{Challenges and Outlook}

Remaining experimental challenges for hydrogen concern the reliable deployment of a model independent temperature diagnostic. The ITCF XRTS method is very suitable, however, it presupposes a highly accurate characterization of the source and instrument function over several orders of magnitude~\cite{Dornheim_NatCommun_2022,Dornheim_T_follow_up,Gawne_JAP_2024}. It remains to be seen whether the heating of hydrogen can be optimized using a laser (which should also be capable of driving a shock into the system at the cost of highly non-equilibrium electrons) or whether a beam of protons or other heavier ions constitutes a more appropriate choice for (isochoric) heating. On the other hand, the problem of the low x-ray scattering signal of hydrogen can only be solved using data collection over very many shots, which requires a unique stability of target, pump, and probe in space and time.

The grand challenge for finite-temperature PIMC is to overcome the Fermi sign problem when temperature becomes lower than the Fermi temperature \cite{cep}. One recent advance towards lifting the FSP problem has been introduced by Xiong and Xiong~\cite{Xiong_JCP_2022}. The so-called $\xi$-extrapolation scheme generalizes the partition function with respect to the quantum statistics parameter $\xi \in [-1,1]$ such that any observable can be extrapolated from simulations in the sectors with no fermion sign problem  ($\xi \in [0,1]$) to the fermionic sector ($\xi=-1$). This works remarkably well for systems with weak to moderate quantum degeneracy, but breaks down when the impact of the fermionic quantum statistics becomes more substantial. The method works for the UEG but also for two-component systems and gives exact structural properties of warm dense hydrogen and beryllium as it lifts the sign problem with respect to particle numbers for systems that exhibit low degeneracy~\cite{Dornheim_JCP_2024}. The $\xi$ extrapolation scheme has already been applied analyzing X-ray Thomson Scattering experiments undertaken at the NIF~\cite{dornheim2024unraveling,dornheim2024modelfreerayleighweightxray,schwalbe2025density}. 

Even with recent advances in PIMC methods like the $\xi$-extrapolation method, exact PIMC is still limited to elements lighter or equal to beryllium. The number of particles accessible in a simulation is still small when compared to DFT-MD and causes a resolution limit on any structure factor or response quantity such as the static linear response function $\chi(\textbf{q})$, in particular in the $q \rightarrow 0$ limit. Furthermore, the inaccessibility of lower temperatures means that the comparison of PIMC against finite temperature Kohn-Sham DFT-MD becomes challenging.  As shown by Ceperley \cite{Ce91}, the restricted PIMC method is an in principle exact method if the nodal structure of the density matrix is known. It is therefore imperative to further develop methods that provide access to the exact nodal structure of the density matrix for all conditions of warm dense hydrogen~\cite{militzer_pre_00}. 

Another important advance for an improved understanding of warm dense hydrogen would also be the further study of the correspondence of the ITCF to the dynamic structure factor (DSF) and their features. Highly accurate analytical-continuation methods are needed~\cite{chuna2025}, which would enable the extraction of exact frequency resolved LFCs and XC-kernels to inform computational more efficient schemes such as time-dependent density functional theory to enable, e.g., efficient calculations of dynamic structure factors or stopping power. This is connected to the study of collective effects in warm dense hydrogen like plasmons, double plasmons, rotons, and ion-acoustic waves. In particular rotons, as predicted to occur in the electron gas but also in hydrogen, are a very interesting feature that should be accessible with hydrogen jets and XRTS~\cite{hamann_prr_23}.

Both experimentally and theoretically challenging are the determination of the stopping power and the nonlinear density response functions (see Sec.\ref{section_05})~\cite{Dornheim_PRL_2020}. It is expected that LFCs and nonlinear density fluctuations, as can be extracted from PIMC simulations of warm dense hydrogen~\cite{Dornheim_MRE_2024}, will constitute input in TD-DFT calculations of the stopping power. Stopping power experiments in hydrogen can benefit from these predictions but also from XRTS-based temperature and density diagnostics.

Finally, as more and more exact benchmark results for hydrogen will become available from PIMC simulations (see Sec.\ref{section_02}), XC functionals as well as XC kernels can be validated and the most appropriate one can be employed in DFT-MD (see Sec.\ref{section_03}) and TD-DFT simulations (see Sec.\ref{section_04}) as a first step towards larger system sizes and larger time-scales. Further improved opacities and other transport quantities (for instance for ICF applications, see Sec.\ref{section_21}) should then become available. Another layer of multi-scale approaches is given by effective models that are based on the average atom concept (see Sec. \ref{section_08}), which promise fast results at acceptable accuracy levels.

\newpage 
\clearpage

\section{Gas giants -- H-He mixtures}\label{section_17}
\author{Martin Preising$^{1}$, Nadine Nettelmann$^{1}$, Ronald Redmer$^{1}$}
\address{ 
$^1$University of Rostock, Institute of Physics, D-18051 Rostock, Germany 
}

\subsection*{Introduction}

Hydrogen (H) and helium (He) are the most abundant elements in the universe and the basic components of main-sequence stars, brown dwarfs (BDs), and gas giant planets (GPs). The study of H-He mixtures is therefore key for a better understanding of the formation processes, interior structures, thermal evolutions, and magnetic-field configurations of these compact objects. This section focuses on GPs such as Jupiter and Saturn in our solar system but also on extrasolar Jupiter-like planets that were detected in large numbers~\cite{ncomms_2023}. The corresponding interior conditions represent prototypical warm dense matter (WDM) states, i.e., pressures up to a few tens of megabars and temperatures up to several 10~000~K.

GPs are up to about an order of magnitude less massive than BDs, and although they are more enriched in heavy elements and colder than BDs, GPs are larger. This is because the fraction of their mass where the equation of state (EOS) of H-He is temperature-sensitive is higher. This region extends to $\sim$10~Mbars, which is at 0.6~$R_J$ in Jupiter and at 0.24~$R_S$ in Saturn, see Fig.~\ref{fig:tortenJS}. The atmospheres of GPs and BDs smoothly transition into the warm dense deep interior. Atmosphere and interior are generally assumed to be of same composition, except for species that condense out into clouds or rain out due to phase separation. The most important types of clouds for cool gas planets like Jupiter and Saturn are water clouds at a few bars, while rock clouds are assumed at a few bars in the hot atmospheres of close-in exoplanets.
Phase separation requires sufficiently low temperatures and abundant species. In Jupiter and Saturn, H-He demixing at 1--2 Mbar levels is most important; see, e.g., Refs.~\cite{Schoettler2018, Howard2023, Brygoo2021}.

Pioneering contributions to the H-He EOS and its application to the modeling of the interior composition and structure, the thermal evolution, and the magnetic field configuration of Jupiter and Saturn, were given, e.g., by Stevenson \& Salpeter~\cite{Stevenson1977a, Stevenson1977b}, Saumon, Chabrier, \& Van Horn~\cite{saumon_aps_95}, Guillot~\cite{Guillot1999}, Fortney \& Hubbard~\cite{Fortney_2003}, Nettelmann \textit{et al.}~\cite{Nettelmann2008}, and Wahl \textit{et al.}~\cite{Wahl2017a}. Furthermore, accurate observational constraints (gravity data) that are needed in interior models to fit free parameters are now available from the Juno~\cite{Howard2023,Militzer_2022} and Cassini-Huygens missions~\cite{Ingersoll2020}.

\subsection*{State of the art}

\subsubsection*{H-He mixtures}

The last decade saw enormous progress in calculating the EOS and material properties of H-He mixtures beyond chemical models~\cite{saumon_aps_95} using density functional theory coupled with molecular dynamics (DFT-MD) simulations for WDM for conditions as relevant for Jupiter~\cite{french_ApJS_2012}, Saturn~\cite{preising_ApJS_2023}, and BDs~\cite{Becker2018}. 
DFT-MD (see Sec.\ref{section_03}) in combination with DFT-based Kubo-Greenwood (DFT-KG) methods
\cite{Holst2011, kubo1957statistical, greenwood1958boltzmann, Demyanov2022} (see Sec.\ref{section_05}) provides a means to calculate the EOS (EOS), thermodynamic, transport, and optical properties. The starting point is a planetary model that provides the mass density, the pressure, the temperature, and the helium ($Y$) and heavy ($Z$) element mass fractions over the radius coordinate. Calculation of the EOS and related quantities can either happen by applying the linear mixing approximation of given EOSs for hydrogen, helium, and heavy elements~\cite{french_ApJS_2012} or by constructing simulation boxes containing appropriate mixtures of the said elements~\cite{preising_ApJS_2023} to reproduce the model's mass fractions. Due to our limited knowledge on the type of heavy elements present at each radius coordinate, a common choice to account for a given $Z$ is introducing oxygen or silicon into the box.

\begin{figure}
\centering
\includegraphics[width=1.0 \linewidth]{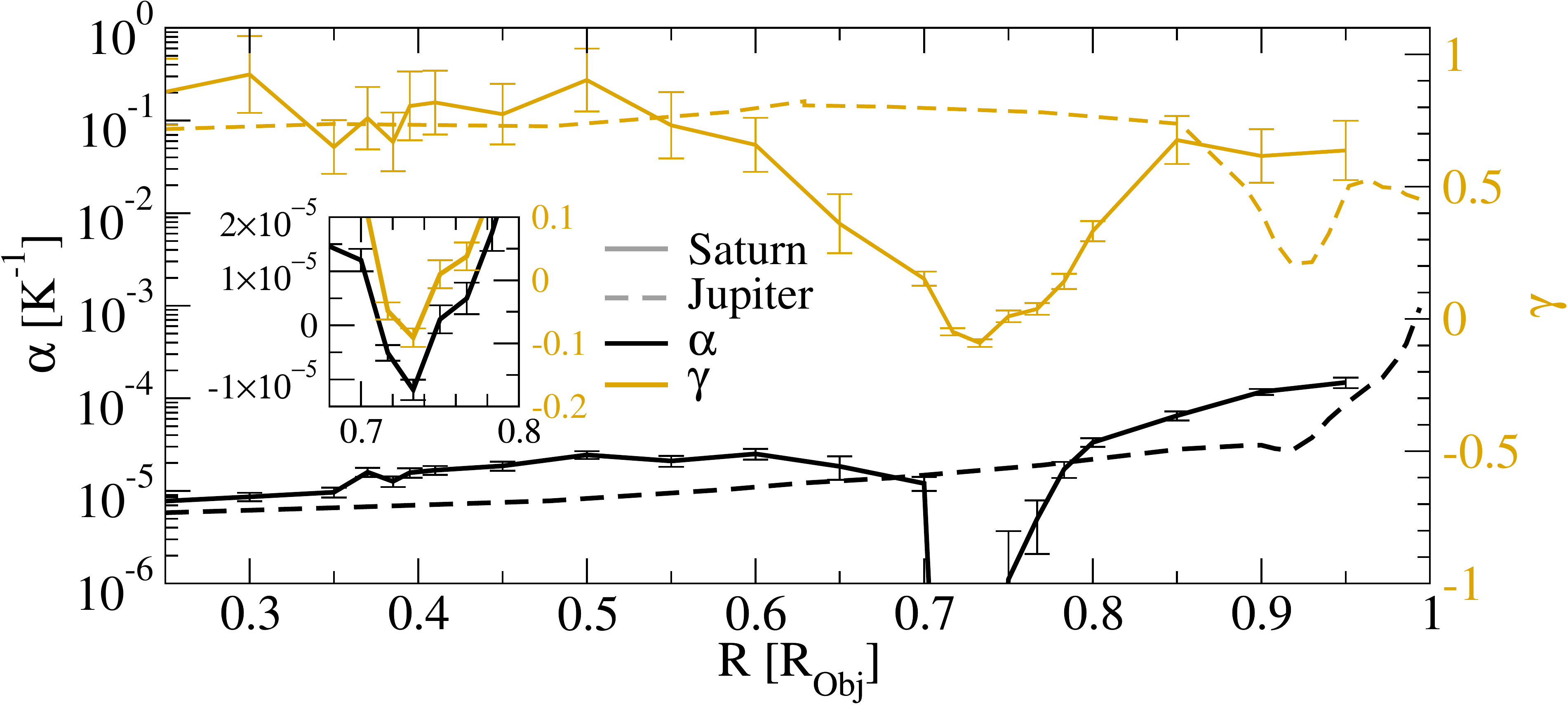}
\caption{\label{fig:alpha_gamma}Thermal expansion coefficient $\alpha$ (black) and Grüneisen parameter $\gamma$ (orange) in Jupiter~\cite{french_ApJS_2012} (dashed) and Saturn~\cite{preising_ApJS_2023} (solid). }
\end{figure}

\begin{figure}
\centering
\includegraphics[width=0.8 \linewidth]{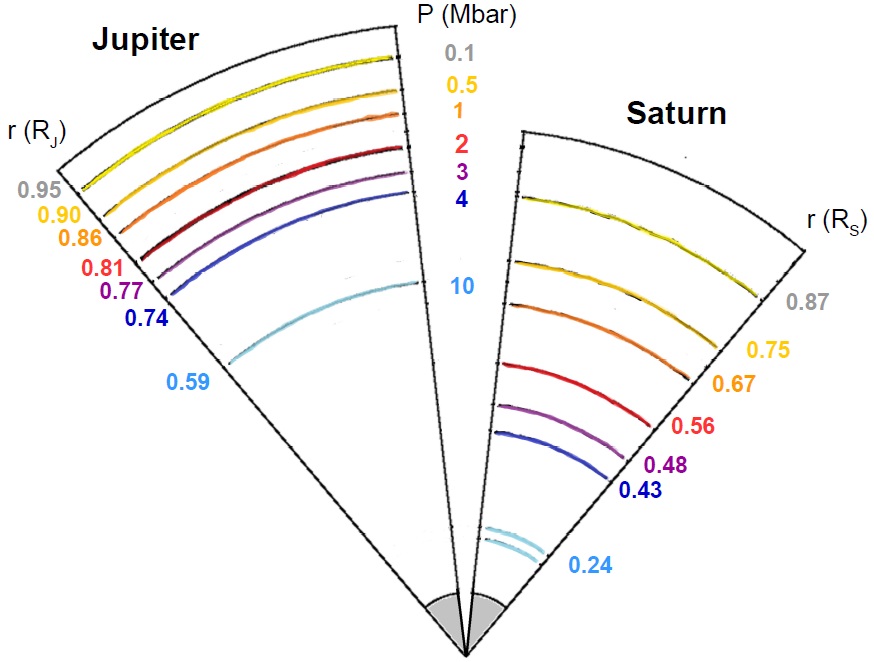}
\caption{\label{fig:tortenJS}Pressure (Mbar)--radius ($R_p$) relations in Jupiter and Saturn. }
\end{figure}

One result of a DFT-MD simulation for a given temperature $T$, mass density $\rho$, and elemental mixture in the simulation box is the pressure $P$ and the internal energy $u$. In order to access the partial derivatives required for some quantities, a total of four additional simulations are mandatory for each chosen sampling point along the model's $P$-$T$ path; two simulations at a higher and lower temperature to evaluate properties along isotherms, and two simulations at higher and lower densities for properties along isochores. 2~-~5\% difference in pressure and temperature with respect to a given sampling point is generally sufficient. The initial values for $\rho$ and $T$, combined with the resulting $P$ and $u$, as well as the derivatives of all the quantities mentioned allow the calculation of the thermal expansion coefficient $\alpha$, the isothermal compressibility $\kappa_T$, the heat capacities at constant volume $c_v$ and constant pressure $c_P$, the sound velocity $c_s$, and the Grüneisen parameter $\gamma$. The warmer the planetary interior, the less pronounced the impact of phase transitions on the slope of thermodynamic functions. The planetary profiles of Jupiter as well as Saturn do not cross the conditions under which the dissociation of hydrogen (see Fig.~\ref{fig:hydrogen}) is a first-order phase transition. However, Saturn's interior is cooler than Jupiter's, resulting in closer proximity to the critical point of hydrogen dissociation, and a more pronounced impact on the slope of the thermodynamic functions. In particular, the thermal expansion coefficient $\alpha$ as well as the Grüneisen parameter $\gamma$ become negative around $\approx$0.71--$0.74~R_S$ in Saturn (see Fig.~\ref{fig:alpha_gamma}), promoting a stably stratified layer due to continuous hydrogen dissociation.

Ionic transport properties result from evaluating the ionic velocity autocorrelation function (self-diffusion coefficients $D_x$ of species $x$), off-diagonal elements of the pressure tensor (shear viscosity $\eta$), and generalized heat currents (ionic thermal conductivity $\lambda_i$)~\cite{French2019}. Sufficient statistical data at each $P$-$T$-point are either the result from one very long MD or multiple shorter MD runs. Depending on the quantity, more than 100,000~MD steps may be required to reach a convergence of around 20\%. The Kubo-Greenwood formalism~\cite{Holst2011, kubo1957statistical, greenwood1958boltzmann, Demyanov2022} evaluates the frequency-dependent Onsager coefficients $L_{mn}(\omega)$ and gives the frequency-dependent dielectric function and derived quantities such as the electronic thermal conductivity $\lambda_e$ and the static electrical (DC) conductivity $\sigma$. This approach requires DFT simulations on individual ionic snapshots, typically 20 to 40, captured from DFT-MD runs in thermodynamic equilibrium.

The ambitious program described above to calculate the EOS and the material properties of H-He mixtures via DFT-MD simulations was executed recently along the $P$-$T$ conditions of the Jupiter~\cite{french_ApJS_2012} and Saturn adiabat~\cite{preising_ApJS_2023}. This data can be used as input for corresponding evolution and dynamo models. Understanding the thermal evolution of planets requires results for the ionic and electronic thermal conductivity, either from DFT-MD or experiments. Understanding the origin and surface morphology of planetary dynamos requires knowledge of the DC conductivity and viscosity, as well as deep knowledge of processes leading to stably stratified layers. The helium rain layer and the hydrogen dissociation region may be stably stratified inside Saturn. Therefore, only the more viscous and less electrically conducting deep helium-layer, the layer of mainly metallic hydrogen between the stably stratified layers, and a shallow region with sufficient DC conductivity above the dissociation region can contribute to the planetary magnetic field.

The recent orbiter missions, Cassini at Saturn (until 2017) and Juno at Jupiter (since 2016), delivered a rich set of observational data that enabled improved interior, thermal evolution, and dynamo models.

\subsubsection*{Saturn and its dynamo}

Cassini constrained the axisymmetry of Saturn's magnetic field to an astonishing $<$0.1 degrees. This is considered evidence for a thick stable layer above the dynamo region, which would axisymmetrize the field that is generated underneath~\cite{Stevenson1980}. The mechanism that is usually considered to lead to stable stratification is helium rain, predicted by H-He phase diagrams published over the past $\sim$15~years; for recent reviews, see Refs.~\cite{Nettelmann24, Howard24}. However, Saturn interior models with helium rain find that helium-droplets would rain all the way to the core~\cite{Howard24, Mankovich20}, leaving no room for an intermediate dynamo region. Models with helium rain also predict a strong helium-depletion in Saturn's atmosphere, although quantitative estimates also depend on the H-He EOS and the water enrichment at depth~\cite{Nettelmann24}. 

Independent evidence for stable stratification in Saturn's deep interior comes from oscillations in Saturn's C ring, also observed by Cassini. In particular, the ring oscillations of azimuthal order $m=-2$ are explained by the $l=2$, $m=-2$ $g$-mode, which would live in a stably stratified dilute core that extends out to 0.7~$R_S$ (1~Mbar) but would quickly decay in a convective region. Due to a frequency overlap with the $l=2$ $f$-mode, which would live in the outer convective region, this $g$-mode can couple to the $f$-mode and split its frequency. The density perturbations cause gravity field perturbations, which, in turn, cause the detected density perturbations in the ring at the split frequencies.
Interestingly, the stably stratified region inferred from ring oscillations nearly coincides with the electrically conducting region where the dynamo can operate. This suggests that Saturn's external magnetic field may be generated and axisymmetrized in a stably stratified extended dilute core, which could be in a state of layered double-diffusive region~\cite{Fortney23} due to gradients in the heavy element abundances rather than in helium concentration. 

A third possibility for a stable layer is a region of negative thermal expansion coefficient $\alpha_T$~\cite{preising_ApJS_2023}, which would actually occur above the dynamo region at about 0.71--0.76~$R_S$ (50--70 GPa).

\subsubsection*{Jupiter, its dynamo and helium rain}

Unifying magnetic field and structure models under one umbrella is also a challenge for Jupiter.
The power spectrum measured by Juno implies a dynamo core radius precisely at 0.81~$R_J$ (2~Mbars)~\cite{Connerney22}. However, the electrical conductivity continues to be sufficient for dynamo generation out to 0.9~$R_J$, beyond which point the steeply decreasing conductivity leads to strong Ohmic dissipation~\cite{Wicht19}. At $\sim$0.97~$R_J$, the equatorial zonal jets decay. There, a stable layer is required to slow the winds and allow breezes at high latitudes~\cite{Wulff22}. Perhaps the deviations from dipolarity of Jupiter's magnetic field, most visible by the presence of the "Great Blue Spot" and its interruption at 0.81~$R_J$ (2~Mbar), are indicative of two dynamos separated or modified by one or more thin stable layers? At Mbar pressures, H-He phase separation would be the obvious candidate. However, under the classical assumption of an adiabatic $P$-$T$ profile, the helium rain region would extend deeper, down to $\sim$4~Mbars/0.74~$R_J$. A narrower helium-region requires super-adiabaticity. But why would this region be super-adiabatic in Jupiter? 

Markham \& Guillot~\cite{MarkhamGuillot24} argue that helium droplet condensation in upward moving parcels leads to a substantial release of latent heat. Warmer temperatures permit a higher equilibrium helium concentration (the helium abundance on the helium-poor side of the phase boundary at given $P$, $T$) than in the surrounding regions. This leads to stable stratification. The higher the latent heat, the steeper the concentration and temperature gradients in the helium rain region, making it thinner. Clearly, this proposal hinges on our poor quantitative knowledge of the latent heat from helium droplet condensation and of heat transport through the helium rain region.

\subsubsection*{Exoplanets and H-He EOS}\,\, Moderately irradiated giant exoplanets appear to be non-inflated and diabatic~\cite{Thorngren16}. However, quantitative estimates of the degree of inflation depend on the H-He EOS used. Denser H-He adiabats, predicted by more recent DFT-MD EOSs, compared to the chemical-picture-based SCvH-EOS from the 1990s~\cite{saumon_aps_95}, decrease the predicted heavy element content and increase the apparent inflation. A statistical study updating the SCvH-EOS based work of Ref.~\cite{Thorngren16} on the radius inflation and inferred metallicity of gaseous exoplanets around Sun-like stars is still missing. However, the predicted heavy-element enrichment of massive hot Jupiters decreases to 1--3$\times$ the stellar value~\cite{Howard25} with the DFT-MD-based CMS19 H-He EOS~\cite{Chabrier2019}, consistent with predictions from planet formation models. The former values based on SCvH-EOS of 5--10$\times$ presented a challenge. This important new trend is caused by the application of improved DFT-MD H-He EOS and will probably alter the irradiation threshold beyond which exoplanets appear inflated.

\subsection*{Challenges \& outlook}

\subsubsection*{Saturn}

Coherent structure and magnetic field models should be developed that aim to explain Saturn's axisymmetrized, dipolar field. H-He phase separation is proposed to cause a stable layer that would do that job, but a successful model is still missing. As Saturn's dynamo may operate in a double diffusive dilute core, new dynamo models are required with high radial resolution. On the observational side, the knowledge of the atmospheric abundances of helium and neon are most important as tracers for helium rain at depth. This requires an entry probe down to a few bars.

\subsubsection*{Jupiter}

Coupled magnetic field and structure models are needed to explain the observed dynamo radius at 0.81~$R_J$ and the high atmospheric metallicity of $3\times$ solar. Both properties relate to the H-He phase diagram and the H-He EOS.

\subsubsection*{H-He mixtures}

It is important to know the H-He phase diagram and the amount of latent heat released upon helium-droplet condensation, the diffusivity of helium-droplets in the H-He environment, and the thermal conductivity in different H-He mixtures. Experiments on H-He demixing may be high-risk but are clearly of great benefit for benchmark predictions~\cite{Schoettler2018, Howard2023, Brygoo2021} and to understand cool gas giants. In addition, accurate experimental constraints at the 1\%-level on the density of H-He mixtures at ~20--50~GPa and a few 1000~K are needed to quantify non-ideal mixing effects between H and helium and to constrain the Jupiter adiabat. Finally, the $P$-$T$ range of extrasolar giant planets and BDs is wide. Experimental benchmark points on the H-He EOS are needed between 1~kbar and 1~Gbar. Uncertainties in the H-He EOS matter if we are to understand the apparent radius anomaly and heating processes in hot Jupiters or the mass-metallicity relation and formation of gas giants.


\newpage 
\clearpage

\section{Ice planets -- H,C,O,N}\label{section_18}
\author{Nadine Nettelmann$^{1}$, Ivan Oleynik$^{2}$, Mandy Bethkenhagen$^{3}$, Dominik Kraus$^{1,4}$}
\address{
$^1$University of Rostock, 18051 Rostock, Germany \\
$^2$University of South Florida, Tampa, Florida 33620, United States\\
$^3$LULI, CNRS, CEA, Ecole Polytechnique—Institut Polytechnique de Paris, 91128 Palaiseau cedex, France\\
$^4$Helmholtz-Zentrum Dresden Rossendorf, 01328 Dresden, Germany
}

\subsection*{Introduction}

Hundreds of exoplanets and, with Uranus and Neptune, two planets in our solar system are known whose mass and radius are consistent with a primary composition of water and other volatiles from the hydrogen-carbon-oxygen-nitrogen (HCON) group of elements. These elements can form ices under ambient conditions as well as in the cold mid-plane of protostellar disks. Indeed, planets are thought to form near the ice-lines of condensible species such as H$_2$O and CO, and therefore, are likely to have accreted a large amount of ices. We refer to such HCON-rich planets as "ice planets" but caution that even for Uranus and Neptune, it is not known whether they really consist primarily of elements from the HCNO-group or alternatively, of H/He-gas mixed into anorganic materials.
This is because the ice-to-rock ratio (I:R) can only be indirectly inferred from observations as rocks condense out and form clouds deep inside at $\sim 2000$~K, or $\sim$10 GPa, see Fig.~\ref{fig:torteIG}. Such depths have not been probed by remote sensing yet. 

Adiabatic models of Uranus and Neptune constrained by gravity are consistent with a high I:R of 1--30 x solar~\cite{Nettelmann13,Malamud24}. However, Kuiper belt objects and comets --a class of objects in the outer solar system that once may have served as the planet building blocks-- have a lower I:R$\sim$1/3~\cite{Malamud24}, suggesting that Uranus and Neptune would rather be rock-rich. On the other hand, analysis of cometary material and interplanetary micrometeorites revealed that a substantial fraction of 1/3--1/2 in those rocks could actually be refractory organics like graphite~\cite{Malamud24}. Overall, this suggests a minimum ratio of ices and organics to an-organic rocks of 1--1.7 in icy planets. As a result, Uranus and Neptune could harbor 20--40\% carbon and water each in their deep interiors, in line with interior models that allow for the presence of rocks or iron in addition to water~\cite{Morf24} and with the observed atmospheric methane abundance of $80\pm 20\times$ solar~\cite{Atreya20}, which is about 22\% in mass. An ice planet of such mass fractions is illustrated in Fig.~\ref{fig:torteIG}. 

What thermodynamic phases HCON-rich material adopts and the miscibility of ices with noble gases (He, Ne, Ar) and rocks under high pressures of 0.1--10 Mbars and high temperatures of about 1,000--10,000~K is a fundamental question for understanding ice planets. For water, simulations have predicted a superionic (SI) phase at Mbar pressres for a long time~\cite{Cavazzoni1999,French2009}. This prediction has finally been confirmed in experiments (Fig.~\ref{fig:ice}). This phase may have important implications on magnetic field  generation~\cite{Redmer11}, the heat flow, and the tidal response of ice planets. However, DFT-MD-simulations also find a melting line depression of the SI phase in the presence of pollutants~\cite{Chau11, Darafeyeu24}, rendering its existence in the ice giants an open question. Moreover, DFT-simulations of various HCON-mixtures predict a plethora of chemical structures and rich phase diagrams~\cite{Naumova2021, Conway}.

Beside peculiar magnetic fields~\cite{Stanley04}, the Voyager~2 flyby also revealed a low heat flux of Uranus while an about an order of magnitude higher heat flux for Neptune. These observations from the 80's are still a mystery~\cite{Helled20UN}, especially as these ice giants are otherwise very similar in mass, radius, atmospheric temperature and composition, suggesting similar interiors. In particular, the gravity data are well explained by a strong increase in the abundance of water at a radius of 0.75--0.9 $R_p$, where pressures are a few to few tens of GPa, see Fig~\ref{fig:torteIG}. If the classical assumption of adiabatic P--T profile is applied, such a structure leads to an overestimation of the luminosity of Uranus~\cite{Helled20UN,Nettelmann25}. With the availability of DFT-MD-based equations of state (EOS), the predicted luminosities of both planets have decreased; model predictions passed the observed value of Neptune in the past decade~\cite{Scheibe19}.

\subsection*{State of the art}

Experimental~\cite{Guarguaglini2019, Guarguaglini2019b} and computational~\cite{Naumova2021, Conway} capabilities have advanced to investigate HCON mixtures under planet interior conditions. So far, the bulk of the work is centered around water and its SI phase~\cite{Millot2018, French2009, Millot2019, Andriambariarijaona2025}. 
Recent experiments at the laser shock-compression LULI2000 facility~\cite{Hernandez2023} suggest that also ammonia adopts an SI phase, and DFT-MD simulations find SI phases in ammonia~\cite{Cavazzoni1999}, water-ammonia mixtures~\cite{Bethkenhagen15, Robinson2020} and in HNO mixtures~\cite{Villa:2023a}.
Other DFT-MD simulations of HCON mixtures find that under deep interior $P$--$T$ conditions, C would phase separate from an O-rich fluid and bind to N and H \cite{Militzer24}. Due to their higher mean molecular weight, these compounds may sink and lead to a stably stratified deep layer underneath a water-rich, perhaps superionic envelope. Another pathway to separate water from carbon is diamond formation. 

\begin{figure}
\includegraphics[width=0.98\linewidth]{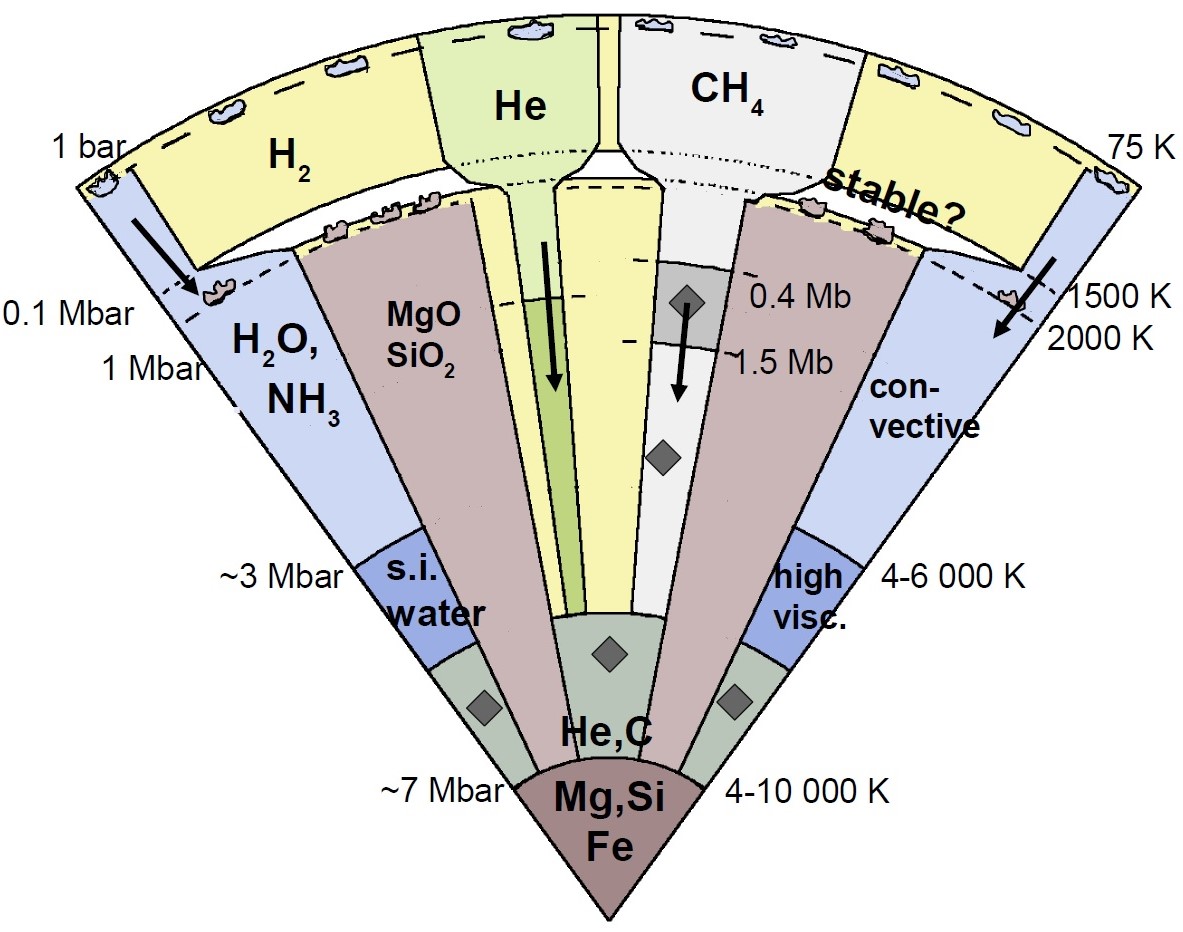}
\caption{\label{fig:torteIG}An Ice Planet Interior composed of 1:1 water (blue) and methane (gray), ice:rock ratio 1.5, otherwise H (yellow) and He (green) with mass fractions indicated by azimuthal angle. Between 0.4 and 1.5 Mbar, carbon might form diamonds that sink down and form an ocean together with rained-out He (gray-green). Water cloud and rock cloud layers are indicated for a Neptune-like ice giant, however, the atmospheric water abundance may be low due to H$_2$/H$_2$O phase separation, leading to a compositional and thermal barrier (white).}
\end{figure}

Diamond has been detected in various experiments ranging from Laser-heated or XFEL-heated DACs~\cite{Benedetti1999, Frost2024} to laser-shock compression~\cite{Kraus2017, doi:10.1126/sciadv.abo0617} and using different C-H or C-H-O mixtures under ice giant interior conditions. The range of pressures of 40 GPa/2500 K and up to 150 GPa where diamonds have been detected is highlighted in Fig.~\ref{fig:torteIG}. Overall, the chemistry is complex and both numerical studies and experiments show that the presence of additional elements, such as N or O, enhance C-H separation and diamond formation. The C-H immiscibility curves across the entire range of C:H ratios are now available thanks to application of machine-learning (ML) \cite{Cheng23}. 

A substantial effort is being directed to investigate carbon at high pressures of several Mbar and thousands of K~\cite{Kraus2016b,Kraus2017,Knudson2008,Smith2014,Brygoo2007,Eggert2010,Benedict2014,Turneaure2017,Lazicki2021,Millot2020} where DFT studies predict the occurrence of a body-centered-cubic BC8 structure between 1.1 and 2.3\,TPa beyond the diamond phase~\cite{Benedict2014,Correa2006,Martinez-Canales2012}. However, ramp compression experiments at NIF using X-ray diffraction find the diamond structure to persist even up to 2 TPa~\cite{Lazicki2021}, and quantum-accurate billion-atom MD simulations~\cite{Nguyen-Cong2024,Young2024} with ML-interatomic potential~\cite{Willman2022} found that diamond remains metastable beyond its thermodynamic stability. 

Several possibilities that relate to the HCON-system are suggested to explain the heat fluxes of ice giants. Owing to the high viscosity in the SI phase of water, convection will be more sluggish and the heat transport be reduced. Primordial heat could be trapped there, while the fluid region above would cool off efficiently. Depending on uncertainties in the heat capacity and adiabatic gradient, such frozen core models can let the planet appear fainter (Uranus) or brighter (Neptune)~\cite{Stixrude21,Stixrude24}. Moreover, tidal dissipation in the superionic core would be enhanced~\cite{Stixrude21}. This accelerates the orbital evolution of the satellites and could explain a few orbital resonance crossings which have previously lacked an explanation. Another possibility to explain the heat flows is by a thermal boundary layer (TBL) at the ice-rich to ice-poor transition, see Fig.~(torteIG). Whether a planet would appear fainter or brighter today depends on the thermal conductivity and heat transport mechanism in the TBL, and its thickness~\cite{Scheibe21}. This is because the TBL first retards the primordial heat inside. Once the TBL has sufficiently built up, the heat will be released and the planet appears brighter than an adiabatic one that has already lost a high portion of primordial heat. A third possibility for Uranus is offered by the latent heat release from methane clouds~\cite{Markham21}. To explain the brightness of Neptune would involve a deep water cloud deck, however, it is currently unknown whether water is sufficiently abundant to form water clouds that inhibit convection~\cite{Nettelmann25}.

The H-rich atmospheres of cold icy planets may actually be water poor. Experiments on 1:1 H$_2$-H$_2$O mixtures revealed that at a few GPa and $\sim$1000~K along the Uranus and Neptune adiabats, H$_2$/H$_2$O phase separation occurs~\cite{bali13}. Application of constructed H$_2$-H$_2$O phase diagrams based on those experimental data and on predictions from DFT-MD simulation at several 10 GPa suggests that the atmospheres of Uranus and Neptune today would contain only about 5--20\% water by mass due to the rain-out, and that the transition to a water-rich deep interior
would be sharp~\cite{CanoAmoros24}, leading to a compositional and perhaps thermal boundary layer.

\subsection*{Challenges \& outlook}

A once-in-a-lifetime highlight would be an Orbiter and Probe mission to Uranus (UOP) as recommended in the 2022 NASA Decadal Survey. The Orbiter would measure the gravitational field including the tidal response~\cite{Parisi24}, the magnetic field, and the global energy balance to high accuracy, while the Probe would measure noble gas and volatile abundances under the methane cloud deck. If equipped with a spectrometer at visible wavelengths, the light from distant stars that is transmitted through the ice particle rings could be measured. From Cassini ring observations at Saturn (see Sec.~\ref{section_17}), it is known that waves in the rings modulate the observable light from the occulted star. The era of analysis of waves caused by planetary oscillations and what wealth of information those would reveal about the interior of Uranus has just begun~\cite{Mank25_Useis}.

\begin{figure}
\centering
\includegraphics[width=0.94\linewidth]{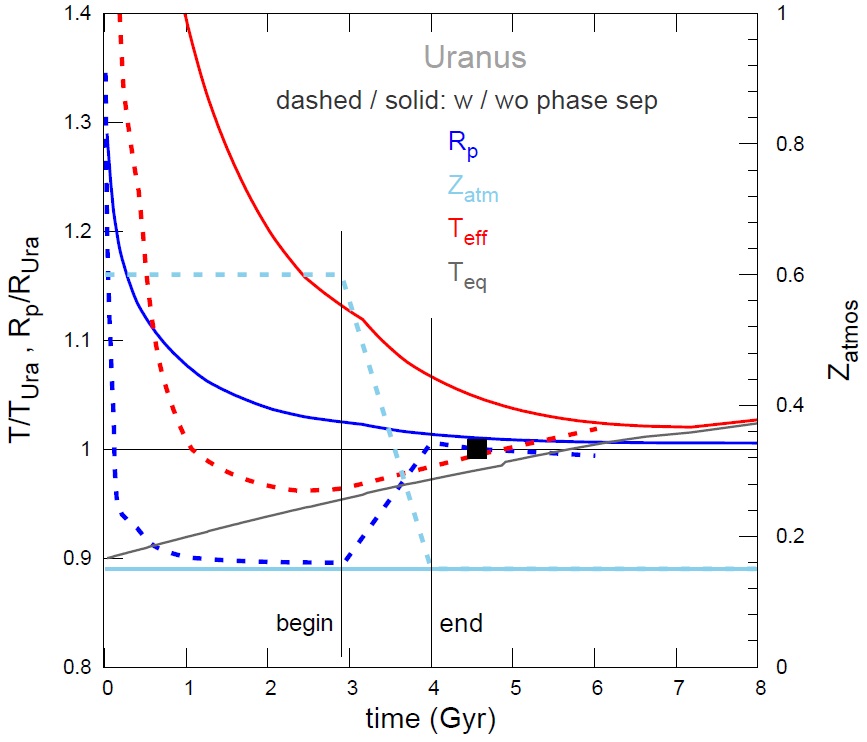}
\caption{\label{figNN:evolU}Outlook on thermal evolution of Uranus with phase separation H$_2$/H$_2$O in the HCON-system (dashed) in comparison to adiabatic evolution without (solid) in terms of $T_{\rm eff}$ (red), which once it reaches the equilibrium temperature $T_{\rm eq}$ (black) follows it onward, radius (blue), and atmospheric ice mass fraction (light blue). Without phase separation, $R_p$ shrinks monotonously while with H$_2$/H$_2$O phase separation, it may rise again  due to the decreasing mass fraction of ices ($Z_{\rm atm}$) in the molecular outer envelope and atmosphere.  
}
\end{figure}

On the computational side, it is important to simulate mixtures of the relevant elements (HCON and He, Ne, Ar). Demixing changes the planetary composition distribution over time (Fig.~\ref{figNN:evolU}). Pollutants influence phase boundaries. The noble gases He, Ne, Ar are observable tracers for processes and the composition deep inside. The recent accomplishments in materials simulations, involving quantum-accurate MD simulations~\cite{Nguyen-Cong2024,Nguyen-Cong2021} employing machine-learning interatomic potentials~\cite{Willman2022,Willman2024}, opens up exceptional opportunities to study the behavior of HCON mixtures in planetary interiors, including the fundamental physics and chemistry of demixing.
 
Laboratory studies of HCON elements remain difficult: static compression experiments, e.g.~with laser-heated diamond anvil cells (see Sec.~\ref{section_09}), cannot guarantee chemical isolation of highly reactive materials~\cite{Frost2023}; dynamic compression experiments, e.g. with laser-driven shock waves, still suffer from limited diagnostic methods for highly transient states (see Sec.~\ref{section_10}), although the experimental capabilities, in particular at XFEL  facilities but also at OMEGA and NIF, have tremendously improved in recent years. This mostly concerns diagnostics such as in situ X-ray diffraction~\cite{Kraus2017,Millot2019,Kraus2023}, small angle X-ray scattering~\cite{Kraus2018,doi:10.1126/sciadv.abo0617}, and spectrally resolved X-ray scattering~\cite{frydrych2020demonstration} (see Sec.~\ref{section_13}). In addition, velocimetry and reflectivity measurements provide valuable macroscopic benchmarks for the EOS of relevant mixtures~\cite{Guarguaglini2021,Millot2018,Hernandez2023,Luetgert2021} (see Sec.\ref{section_14}). A substantial increase of the repetition rate of corresponding drive systems, e.g., the DiPOLE 100X laser at European XFEL~\cite{Gorman_JAP_2024} will enable high precision measurements of phase transitions, chemistry, and electronic properties of relevant materials at a large number of different concentrations. Superposition of numerous experiments will allow us to apply photon hungry inelastic X-ray scattering techniques to precisely measure temperature~\cite{Descamps2020}, conductivity~\cite{ranjan2023toward}, and chemistry~\cite{Voigt_POP_2021}. These experiments will advance our understanding of the interiors and formation of icy planets and the most abundant elements in the universe that form organic molecules --the foundation of life. Lastly, the experimental detection of the BC8 phase of carbon would be a major breakthrough. This phase appears experimentally accessible only through a narrow $P$-$T$ window of metastable liquid carbon~\cite{Kraus2025}. Dedicated multishock-experiments will be conducted in the near future at the NIF~\cite{Osolin2023}.

On the planetary modeling side, it will be important to understand how potentially observable atmospheric abundances, e.g., of He, Ne, and methane can be linked to deep interior abundances in the presence of intermediate stable layers. Structure models need to be updated steadily for the advances in simulated and experimentally probed HCON-systems. Similar to H/He-demixing in the evolution of the gas giants~\cite{Mankovich20,Howard24} (see Sec.\ref{section_17}), demixing within the HCON-system may play a major role in the evolution of ice planets but has not been included yet. Fig.~\ref{figNN:evolU} shows how the loss of water from the outer envelope due to H$_2$/H$_2$O demixing~\cite{CanoAmoros24} may affect the planet radius and luminosity over time, where luminosity is $L=4\pi R_p^2\, F$ and heat fluxes $F$ are expressed in terms of temperatures, $F= \sigma_B T^4$.  If a previously water-rich envelope rains out, the remaining H-He gas expands, and the planet radius may increase over time (blue dashed curve) despite the planetary cooling. Uranus is special as its current heat loss $\sigma_B T_{\rm eff}^4$ almost equals its heat gain by irradiation $\sigma_B T_{\rm eq}^4$. Phase separation and rain-out may provide an explanation (red dashed curve). The influence of the phase diagrams of HCON-systems on the evolution of the ice giants will play a major role if we are to understand the observed radii of individual planets as well as of the population over orbital distance, time, and stellar properties as a whole. Warm Neptune-sized planets are predicted~\cite{Matuszewski23} to be the most common class of planets that the upcoming PLATO mission~\cite{Rauer24PLATO} can detect.

\newpage 
\clearpage

\section{Rocky planets -- MgO-FeO-SiO$_2$-Fe and mixtures}\label{section_19}
\author{Federica Coppari$^{1}$, Alessandra Ravasio$^{2}$, Gerd Steinle-Neumann$^{3}$}
\address{
$^1$Lawrence Livermore National Laboratory, Livermore, CA 94550, United States \\
$^2$Laboratoire LULI, CNRS - Ecole Polytechnique - CEA - Sorbonne Universite,  91128 Palaiseau, France \\
$^3$Bayerisches Geoinstitut, Universit\"at Bayreuth, 95440 Bayreuth, Germany 
}

\subsection*{Introduction}

\begin{figure}
\includegraphics[width=1.05\linewidth]{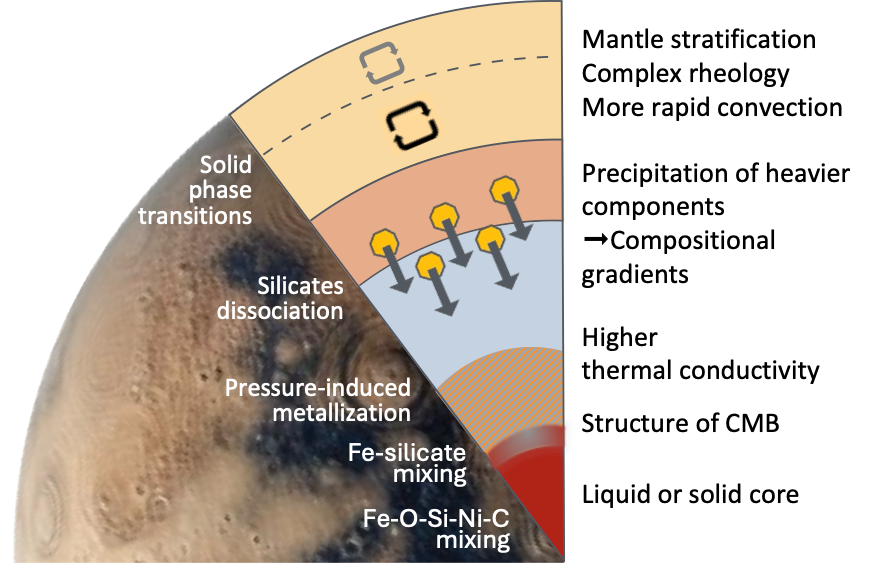}
\caption{\label{fig:rock}Schematic representation of the possible complex interior structure of a super-Earth type exoplanet, and plausible internal processes.}
\end{figure}

Among the terrestrial planets in the solar system (Mercury, Venus, Earth, Mars), Earth is the largest, and the geophysical and geochemical characterization of its interior has educated our first-order understanding on the internal structure of the others. On the basis of mass and moment of inertia as well as cosmochemical considerations, these planets are understood to contain a metallic (iron-based) core and silicate-dominated outer portions (crust and mantle)~\cite{Breuer2023}. Whether the transition from the mantle to the core occurs sharply or as a fuzzy boundary (Fig.~\ref{fig:rock}), depends on many factors, including the miscibility~\cite{Young2024}. For the Earth, the very limited miscibility between metals and silicates~\cite{Arveson2019} is likely to make the exchange of material between the core and mantle minimal. Also, in the Earth, the current mantle and core thermal profiles follow separate adiabats due to gravitational self-compression to first order, with a $T$ jump at the core-mantle boundary (2891 km depth or 136 GPa) from 2500-3000\,K~\cite{Chust2017} to 4000-4500 K~\cite{Wagle2017}.

The study of terrestrial-type planets and their materials has significantly expanded over the past two decades with the discovery of a large number of extrasolar planets that are significantly larger than Earth, but fall on a similar mass-radius trend as the terrestrial planets in our solar system, so-called super-Earths \cite{Lichtenberg2025}. Therefore, their interior $P$ and $T$ are expected to exceed those inside our planet \cite{Boujibar2020}. For example, a super-Earth with five times the mass of the Earth and Earth-like structure can be expected to reach a $P$ of approximately 1 TPa at their core-mantle boundary and 2 TPa at the center. The temperatures are difficult to estimate reliably, but they are significantly higher than in the center of our planet~\cite{Young2024,Boujibar2020}.

Within the Earth, discontinuities in the density and the seismic (acoustic) velocities exist due to phase transitions, both in the mantle and the core. In the mantle, silicates become more densely packed as the pressure increases and undergo phase transitions and decomposition. At $P$ corresponding to 660\,km depth, Mg$_2$SiO$_4$ in the spinel structure (ringwoodite) breaks down into MgSiO$_3$ in the perovskite structure (bridgmanite) and MgO (periclase)~\cite{Ishii2019}. At pressures exceeding those in Earth’s mantle, MgSiO$_3$ in the post-perovskite structure and MgO are predicted to recombine to a crystal with Mg$_2$SiO$_4$ composition and different symmetry from ringwoodite, before eventually breaking down into the oxide components MgO and SiO$_2$~\cite{Umemoto2017}. Similarly, MgO and FeO, two mineral phases forming ferropericlase, a solid solution with a B1 structure at the pressures of the Earth mantle, have been shown to transform from the B1 to B2 phase at conditions of super-Earth mantles~\cite{Coppari2013, Coppari2021}. 

These phenomena might introduce heterogeneity, discontinuities in the density profile, unexpected rheological properties and conductivities that would make large rocky exoplanets very different from Earth. Silicate dissociation~\cite{Umemoto2017,Mazevet2015} and metallization~\cite{Soubiran2018, Stixrude2020,Guarguaglini2021,Millot2015,Mazevet2015, Scipioni2017} might lead to enhancement of electrical conductivity and, for sufficiently high values, the mantles of large rocky exoplanets might contribute to magnetic field generation, something that does not happen in the modern Earth, but may have occurred in early stages of planetary evolution with a magma ocean at the base of the mantle~\cite{Soubiran2018,Stixrude2020,Mazevet2015}.

The main questions associated with the core of a super-Earth are related to the melting temperatures of iron at a given pressure, the incorporation of light elements into liquid iron and how that affects the melting $T$, and the extent to which iron and rocks mix (Fig.~\ref{fig:rock}). These properties determine if the planet has a solid or liquid core and its size, which in turn affects magnetic field generation. Characterizing the planet's dynamic regime, thermal state, evolution and magnetic fields generation requires challenging experimental and computational work on phase equilibria, equations-of-state (EOS), and transport properties of the main constituents at conditions pertinent to rocky planets interiors.

\subsection*{State of the art}

Over the past decade, high-pressure science has seen a remarkable development of experimental and computational capabilities.

Of particular interest to planetary interior conditions, is the development of advanced dynamic compression schemes, including ramp and multiple shock compression~\cite{Swift2005} (see Sec.~\ref{section_12}). They allow reaching off-Hugoniot conditions (i.e., $P$-$T$ conditions different from those obtained by a single shock wave~\cite{Zeldovich1966}), expanding the $P$-$T$ space that can be probed experimentally, approaching the adiabatic profiles of planetary interiors. The combination of nanosecond dynamic compression with ultra-fast x-ray sources at multi-kJ laser facilities~\cite{Rygg2012, Foster2017, Rygg2020, Denoeud2021} represents significant progress for the study of rocky planet interiors (see Sec.~\ref{section_13} \& \ref{section_13b}). Synchrotron and x-ray free electron laser facilities have adopted the dynamic compression scheme which is now available at dedicated beamlines such as DCS at the APS~\cite{Broege2019}, MEC at the LCLS~\cite{nagler2015matter}, HPLF at the ESRF~\cite{Hernandez2024}, and HED-HIBEF at the European XFEL~\cite{Zastrau_JSyncRad_2021}. This effort has enabled measurements such as {\it in-situ} x-ray diffraction with unprecedented resolution, opening the way to the study of crystalline structure, phase transitions, melting and detailed liquid structure characterization~\cite{Morard2024} at conditions exceeding those obtainable in static compression. 

Advances in DAC and laser-heating technologies have also enabled the study of matter under static compression up to $\sim$4000-6000 K at a few Mbar~\cite{Mao2016, Weck2020, Sinmyo2019} (see Sec.~\ref{section_09}). Even though the achievable $P$-$T$ conditions are still generally lower than those reached with dynamic compression, such developments have greatly contributed to closing the gap between the two fields. Compression with DACs has also been combined with x-ray heating from XFELs to reach $T$ exceeding 10000 K for timescales from ps to ns~\cite{Meza2020}.

Lastly, the development of interatomic potentials based on MLIP of KS-DFT results present a significant step forward in computational materials science, including in the WDM regime (see Sec.~\ref{section_03}). DFT-MD simulations are technically -- and to some extent fundamentally -- limited in cell size and run duration, limitations that can be overcome with the use of MLIPs. Simulations with MLIPs therefore provide access to properties that are size-dependent or rate-limited such as bulk or grain boundary diffusion~\cite{Peng2025, Yuan2023}, melting~\cite{Yuan2023,Deng2023} or structure factors near the hydrodynamic limit (wavenumber $q \rightarrow 0$)~\cite{22B-Schorner-Al}.

These advanced tools have enabled the characterization of phase transitions, melting, EOS and transport properties of several planet-forming materials, providing some insights into the interior structure of rocky planets more massive than the Earth. The majority of the studies, however, have focused on single-component or relatively simple systems so far.

\subsubsection*{MgO-FeO}

The B1-B2 phase transition in magnesium and iron oxides has been widely investigated both experimentally and with computations, however, only few studies are available for the (Mg,Fe)O system and over a limited $P$-range~\cite{Du2017, Fu2018, Zhang2018a, Deng2019, DellaPia2022}. MgO has long been predicted to undergo this phase transition, although different computational studies do not always agree on the transition $P$ (see Refs.\cite{Bouchet2019, Soubiran_PRL_2020} and references therein). Experimentally, this transition has been observed at $\sim$600 GPa in dynamic compression of MgO~\cite{Coppari2013, Wicks2024} and at $\sim$230 GPa in static and dynamic compression of FeO~\cite{Coppari2021,Ozawa2011a}. The occurrence of the B1-B2 transition might have consequences for the miscibility of the two oxides, impacting the composition, the density profile as well as the rheological properties of the mantle, with the B2 phase expected to have lower viscosity compared to the B1, by several orders of magnitude~\cite{Coppari2021, Karato2011, Ritterbex2018}.

\subsubsection*{SiO$_2$}

Silica has also been extensively investigated as a prototypical mantle component. Experimental and computational studies have highlighted a remarkable structural complexity at moderate and extreme pressures that is not yet fully understood~\cite{Liang2015, Liu2021}. While it is generally accepted that SiO$_2$, with $\alpha$-quartz stable at ambient conditions, transforms to denser phases at high pressures, there is poor agreement in terms of the phase transition sequence and transition pressure between the static and dynamic compression experiments~\cite{Tracy2020}. Stishovite, one of the high-pressure polymorphs of SiO$_2$ that is recoverable at ambient conditions, exhibits remarkable metastability during shock compression, persisting up to 300 GPa~\cite{Schoelmerich2020}, which lies well into the stability field of the pyrite structure~\cite{Kuwayama2005, Kuwayama2011}, possibly pointing to the importance of kinetic effects in short timescale (ns) experiments. Melting of SiO$_2$, extensively investigated~\cite{Millot2015, GonzalezCataldo2016, Andrault2022}, has also important implications not only for the composition of planets, but also for the transport properties. DFT-MD simulations~\cite{Soubiran2018, Stixrude2020, Scipioni2017} suggest that silica and silicate melts become semi-metallic and they could therefore contribute to the magnetic field generation. Conductivity measurements for SiO$_2$ at Mbar pressures and high temperatures~\cite{Guarguaglini2021,Millot2015} support these predictions, confirming the potential of super-Earth mantles to generate and sustain magnetic fields.

\subsubsection*{Fe}

Despite four decades of intense studies on pure iron, uncertainty remains on crucial information, such as high-pressure melting, phase stability and transport properties, limiting our understanding of the Earth’s core and our ability to model larger rocky planets. Static compression studies have reached a consensus on the melting temperature at core-mantle bpundary pressure, with recent results placing it at 4000-4500 K~\cite{Morard2018}. Nevertheless, uncertainties persist regarding the melting temperature at the pressure of Earth’s inner core boundary (ICB) and beyond, which is critical to obtain an upper bound on the core geotherm and also to describe the interior structure of large terrestrial exoplanets. Most available data derive from dynamic compression with shock waves generated either by plate impacts~\cite{Li2020} or laser drivers~\cite{Balugani2024, Kraus2022, White2020}, and they generally yield melting temperatures above 6000 K at inner-core conditions, in agreement with recent DFT-MD calculations~\cite{Sun2018, GonzalezCataldo2023}, but in contrast with extrapolation from DAC studies~\cite{Sinmyo2019}. A multi-phase EOS model has been developed to reconcile inconsistencies in high-$P$ melting data~\cite{Wu2023}.

Another major question concerns the stable crystal structure of pure iron at inner-core conditions, which has direct implications for the interpretation of seismic data. While  hexagonal close-packed (hcp) iron is found to be the dominant phase in dynamic experiments coupled with in situ x-ray diffraction~\cite{Turneaure2020, Kraus2022, Singh2023} or absorption~\cite{Balugani2024, Ping2013} and the phase coexisting with the liquid in static experiments to $\sim$300 GPa~\cite{Sinmyo2019}, some computational studies claim a re-entrance of the body-centered cubic (bcc) phase at high $P$~\cite{Ghosh2023,Belonoshko2021} or its stabilization through the presence of light elements in the core~\cite{Li2024}. 

At higher $P$, the advent of MJ lasers enabled experiments up to 1.4 TPa~\cite{Smith2018}, the central pressure of a super-Earth with 3.5 to 4 times Earth’s mass~\cite{Smith2018, Boujibar2020}. Furthermore, DFT-MD calculations in the TPa-range suggest that super-Earth cores may crystallize through pathways different than Earth’s core, influencing the thermal evolution and magnetic field generation of these planets~\cite{GonzalezCataldo2023}.

\subsubsection*{Mixtures}

\,\,\,Materials inside planets are not pure elements or endmembers of mineral groups, such as the Fe, MgO, FeO and SiO$_2$ discussed above, or MgSiO$_3$ and Mg$_2$SiO$_4$. Owing to condensation from the solar nebula and planetary formation processes, planetary building blocks are mixed, forming alloys, mineral solid solutions (such as MgO-FeO) as well as aggregates (rocks) whose properties may be dramatically different from endmember minerals. For example, iron in the Earth's core is alloyed with $\sim$5 wt\% Ni \cite{Fischer2025}, with the addition of some other light elements such as C, Si, O, S~\cite{Fischer2025, Hirose2021, Miozzi2022}. Multivariant melting / freezing of the alloy leads to the depression of the liquidus $T$ at which crystallization initiates~\cite{Morard2017, Mori2017, Edmund2022, Hirosaka2022, Dobrosavljevic2023}. Experimentally, this effect has mostly been studied in the DAC over a limited $P$-$T$ range to $\sim$100 GPa and a few thousands K, much lower than the conditions of Earth's core.

While at conditions of the Earth's lower mantle, silicates and iron are immiscible~\cite{Arveson2019}, leading to the sharp boundary between core and mantle, oxides and Fe metals may mix at high $T$, as suggested by DFT simulations~\cite{Wahl2015, Insixiengmay2025}. In super-Earths with larger internal $T$, such a behavior could lead to a gradual transition between its mantle and core (Fig.~\ref{fig:rock}), vastly different from terrestrial planets in our solar system, and similar to the recently developed scenario of a diffuse boundary between the hydrogen envelope and silicate core in mini-Neptunes~\cite{Young2024}.

\subsection*{Challenges \& outlook}

Significant challenges remain in the study of material properties at conditions of planetary interiors.

While conditions of the Earth’s mantle can reliably be achieved and controlled in static and dynamic experiments, conditions in the Earth’s core are far more challenging. The differences in melting $T$ for iron at inner core conditions between static and dynamic experiments remain and they need to be resolved. The measurement of relevant transport properties is also challenging in the DAC and even more so in dynamic compression. Simulations using KS-DFT or MLIPs can aid the experiments, although discrepancies remain, for example, in terms of the thermal conductivity of iron at core conditions~\cite{Konopkova2016, Kleinschmidt2023, Zhang2020, Saha2020, Lobanov2022,PhysRevB.107.115131, Hasegawa2024}.

The novel dynamic techniques have significantly expanded the $P$-$T$ range accessible, reaching conditions of the interior in super-Earths. Pioneering experiments have shown great promise in terms of phase transitions and EOS characterization, however kinetic effects may be a limiting factor in the nanosecond timescale of laser-driven compression or picosecond XFEL techniques and they need to be better understood. MLIP-based simulations on large systems may help in that respect~\cite{Yuan2023, Misawa2020}. 

Studies need to be expanded to multi-component systems, such as MgO-FeO or MgSiO$_3$-based bridgmanite, and include volatile elements to more realistically represent the composition of terrestrial planets. The predicted miscibility closure in MgO-Fe system at high $P$-$T$ conditions~\cite{Insixiengmay2025} awaits experimental confirmation, and requires temperatures that are not yet accessible in static experiments. A possible miscibility between silicates and metals would change our understanding of planetary structure even more dramatically.

\newpage 
\clearpage

\section{Dwarf stars and neutron-star crusts}\label{section_20}
\author{Isabelle Baraffe$^{1,3}$, Simon Blouin$^{2}$, Gilles Chabrier$^{3,1}$, Nicolas Chamel$^{4}$, Andrea Kritcher$^{5}$}
\address{
$^1$University of Exeter, Exeter, EX4 4QL, UK \\
$^2$University of Victoria, Victoria, BC V8W 2Y2, Canada \\
$^3$Ecole Normale Supérieure de Lyon, CRAL, CNRS UMR 5574, 69364, Lyon Cedex 07, France \\
$^4$Université Libre de Bruxelles, 1050 Brussels, Belgium \\
$^5$Lawrence Livermore National Laboratory, Livermore, CA 94550, United States
}

\subsection*{Introduction}

Brown dwarfs (BDs) are objects not massive enough to sustain hydrogen burning in their core and thus to reach thermal equilibrium, defined as $L = L_{\mathrm{nuc}}$ where $L_{\mathrm{nuc}}$ is the nuclear luminosity due to proton fusion. BDs are now routinely discovered with masses down to a few Jupiter masses ($\sim 3$-4 M$_{\mathrm{Jup}}$), notably thanks to the spectral signatures identified with the JWST telescope~\cite{Luhman2024a,Luhman2024b}. Since the present review is devoted to the physics of warm dense matter, we will focus on the interior structure of BDs and thus on the main recent improvements in the calculation of the H/He equation of state (EOS) under typical BD conditions. For abundances characteristic of the solar composition, $Y_\odot=0.27$, $Z_\odot=0.02$, this minimum mass for hydrogen burning is M$_{\mathrm{HBMM}}=0.075$ M$_\odot$ (78.5 M$_{\mathrm{Jup}}$)~\cite{Chabrier2023}. 

White dwarfs (WDs) are stellar remnants that cool down over time, with masses typically around 0.6 solar masses compressed into Earth-sized objects~\cite{saumon2021current}. Their structure consists of three main regions: a degenerate, fully ionized core (most commonly C/O); an envelope where WDM conditions prevail (most commonly H/He, but often with mid-$Z$ impurities and sometimes C/O dominated); and a thin atmosphere where the spectrum forms. Understanding the WDM regime in the envelope is crucial as it mediates both energy transport during cooling (key for using WDs as cosmic chronometers~\cite{Winget_ApJL_1987,Fontaine_PASP_2001}) and the diffusion of elements accreted from disrupted planetary bodies (our only window into the bulk composition of extrasolar rocky objects~\cite{Farihi_NAR_2016,Doyle_Science_2019}). Some WDs also exhibit pulsations~\cite{Corsico_AApR_2019}, causing detectable variations in brightness. By analyzing these fluctuations with theoretical stellar models, we can probe the star's internal structure~\cite{Giammichele_Nature_2018}. This pulsation analysis, which depends sensitively on the physical properties of the WDM envelope, provides insights into prior stages of stellar evolution, and in the special case of the C-atmosphere hot DQ WDs~\cite{Dufour_Nature_2007}, could illuminate the outcome of stellar mergers~\cite{Dunlap_eurowd_2015}.

Neutron stars (NSs) are even more extreme compact objects formed from gravitational core-collapse supernova explosions. They have a maximum mass of about twice that of the Sun but are packed into a radius of about ten kilometers. Surrounded by a thin atmosphere, no more than ten centimeters thick and composed mainly of hydrogen and helium, the surface is probably covered by an ocean of molten iron, a relic of the progenitor star. The region beneath is solid and contains an underground neutron ocean in its innermost part. The core of a NS consists of a liquid mixture of neutrons, protons, leptons, and possibly more exotic particles~\cite{blaschke_assl_2018}. The crust plays a key role in various astrophysical phenomena associated with NSs~\cite{chamel_lrr_2008}. Some NSs, so-called magnetars, are endowed with very strong surface magnetic fields reaching $10^{15}$~G, and possibly even more extreme fields in their interior~\cite{Esposito_2021}.

\subsection*{State of the art}

\subsubsection*{Brown dwarfs}

Central conditions for massive BDs are $T_c \sim 10^5$ K and $\rho_c \sim 10^2$-$10^3$ g cm$^{-3}$. Under these conditions, the average ion electrostatic energy $(Z e^2 )/d$, with $d$ the mean interionic distance, is several times the average kinetic $k_BT$, revealing a strongly coupled ionic plasma with a coupling parameter $\Gamma_i = (Ze)^2/dk_BT > 1$. The temperature is of the order of the electron Fermi temperature $k_BT_F$ and the average inter-electronic distance $d_e$ is of the order of both the Bohr radius, $d_e\simeq a_B$, and the Thomas-Fermi screening length $d_e\simeq \lambda_{T_F}$. We are thus in the presence of a strongly coupled ion plasma immersed in a partially degenerate, strongly correlated, polarisable electron fluid. The envelope temperature is $kT \lesssim 1$ Ryd, so electronic and atomic recombination is expected to take place. The electron average binding energy is of the order of the Fermi energy $Ze^2/a_B \sim E_{\mathrm{F}}$ so that pressure-ionisation is taking place along the internal density profile. In order to address this complex quantum N-body problem the most recent calculations of BD interiors include equations of state for hydrogen/helium mixtures that combine analytical calculations, in the molecular/atomic low density regime and the fully ionised high density regime, with quantum molecular dynamics (QMD) simulations in the in-between pressure ionisation domain \cite{Chabrier2019}. However, as illustrated by the central part of the temperature-density diagram portrayed in Fig.1 of Ref.\cite{Chabrier2023} (labeled QMD), essentially all BD cooling tracks enter the domain where interactions between H and He species can no longer be ignored and thus will affect the BD mechanical and thermal properties, i.e. their structure and evolution. This implies that the calculations for the mixture can no longer rely on the ideal (or ‘additive’) volume law (AVL) approximation, which means that interactions between the hydrogen and helium species cannot be ignored and the entropy of mixing has nonideal parts contributing the free energy of the mixture. Chabrier \& Debras~\cite{ChabrierDebras2021} took into account the impact of these interactions on the thermodynamic properties of the H–He mixture, extending the aforementioned calculations beyond the AVL, by incorporating the QMD calculations performed by Militzer \& Hubbard~\cite{MH13} into the H/He EOS. This state-of-the-art EOS was incorporated in  BD structure and evolution calculations~\cite{Chabrier2023}. As shown by these authors, the H-He interactions have an important impact on the structural and cooling properties of the most massive BDs, improving the agreement between evolutionary models and accurate observations of BD dynamical mass determinations,  helping to resolve part of the previously observed discrepancy between models and observations of binary BDs for which the mass can be determined very accurately. Furthermore, a noticeable consequence of this improvement of the dense H-He EOS is that it yields a larger H-burning minimum mass, now found to be at 0.075 M$_\odot$ for solar composition.

\subsubsection*{White dwarfs}

\begin{figure}
    \centering
    \includegraphics[width=0.8\linewidth]{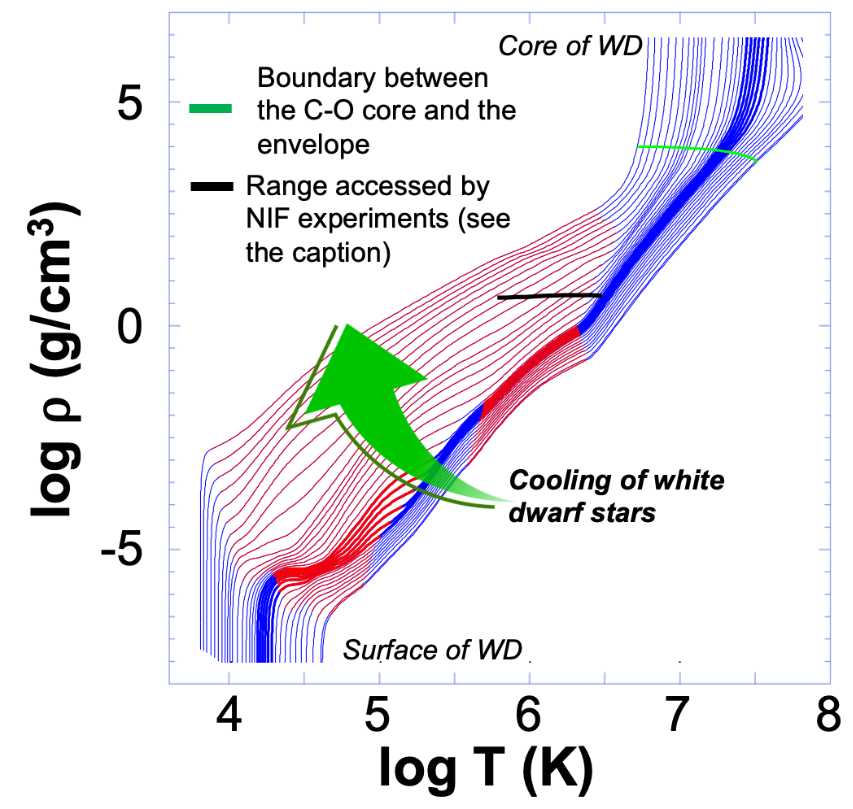}
    \caption{Interior modeling of a WD with a C/O core and pure C envelope \cite{Fontaine_PASP_2001}. The temporal sequence of cooling is represented by multiple lines, with each line corresponding to a snapshot in time. The red regions indicate areas of convective instability, linked to the partial ionization of C. Noted is a region of temperature and density probed by NIF experiments \cite{Kritcher2020}, which reach conditions where the core electrons of C are partially ionized, similar to those at the bottom of the convection zone in hot DQ WDs. This region is mainly responsible for driving pulsations and exhibits the greatest variability in EOS models.}
    \label{fig:WDM_Dwarfs_Fig1}
\end{figure}

The WDM envelope of a WD presents significant modeling challenges. The degree of ionization in this regime is highly uncertain~\cite{saumon2021current}, and the relevant conditions ($\rho \sim 1-10^3\,$g/cm$^3$, $T\sim 10^5 - 10^7\,$K) are extremely challenging to replicate and measure in laboratory settings. Planar shock wave experiments have probed this detailed physics for many decades but were limited to pressures of $\sim 60$ Mbar~\cite{Smith2014}, and as a result, theoretical models in the pressure regime of WD envelopes had not been validated by experimental data. More recently, spherical shock wave experiments at the National Ignition Facility (NIF)~\cite{Kritcher2020} have enabled the creation and study of these conditions, leading to constraining experimental measurements of the EOS relevant for WD envelopes at pressures exceeding 100 million atmospheres. This new capability allows for the first time to test theoretical models in this regime, providing crucial insights into the deep conditions within the convection zone of hot DQ WDs (see Fig~\ref{fig:WDM_Dwarfs_Fig1}).

\subsubsection*{Neutron stars}

\begin{figure}
    \centering
    \includegraphics[width=0.45\linewidth]{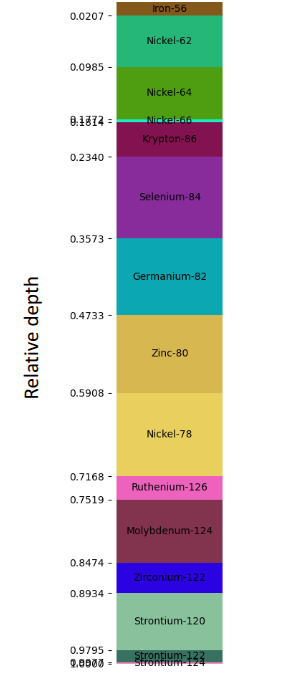}
    \caption{Stratigraphic column of the outer crust of a cold non-accreted NS combining atomic mass measurements and predictions from self-consistent mean-field calculations~\cite{Chamel_2020}. Depths are relative to the neutron-drip transition marking the boundary between the outer and inner crusts.}
    \label{fig:WDM_Dwarfs_Fig2}
\end{figure}

Gravity in a NS is so huge that a few tens of cms below the surface, atoms are crushed and fully ionized in a dense plasma of bare nuclei and free electrons, as in the WD core. The outer crust of a cold isolated NS is expected to be stratified into pure crystalline layers. Progress in experimental nuclear mass measurements over the past decades has allowed one to determine their composition up to a density $6\times 10^{10}~$g~cm$^{-3}$~\cite{Wolf_2013}. Deeper, nuclei become neutron rich up to the so-called neutron-drip point beyond which free neutrons are present at a density above $4\times 10^{11}$~g~cm$^{-3}$, found using nuclear mass models. An example of recent calculations is shown in Fig.\ref{fig:WDM_Dwarfs_Fig2}. The inner crust composition is more uncertain and model-dependent. Some models even predict the existence of a mantle consisting of very exotic configurations collectively termed ``nuclear pasta''~\cite{Ravenhall_1983}. The crust dissolves at a density $\sim 10^{14}$~g~cm$^{-3}$. The presence of strong magnetic fields could substantially alter the internal constitution of the crust and its properties. For magnetic fields $B\gg B_{\rm at}=m_e^2 e^3 c/\hbar^3 \approx 2.35\times 10^9$ G, the atoms present in the NS atmosphere are expected to be very elongated along the field lines and form linear chains. Due to attractive interactions, these chains are predicted to condense at sufficiently low temperatures~\cite{Lai_2001}. Thus, magnetars may have no atmosphere and their surface could be a highly incompressible solid with a density $\rho_s\approx 1.8\times 10^7$ g~cm$^{-3}$ for iron and $B=10^{15}$~G. For magnetic fields exceeding $B_{\rm rel}=m_e^2 c^3/(\hbar e)\approx  4.41\times 10^{13}$~G, nuclei in the crust are expected to become less neutron-rich and the EOS to be stiffer because of Landau-Rabi quantization of electron motion perpendicular to the field~\cite{Mutafchieva_2019,Jiang_2024}. 

\subsection*{Challenges \& outlook}

\subsubsection*{Brown dwarfs}

Although significant improvements have been accomplished recently in the calculation of the H-He EOS under BD interior conditions, some uncertainty remains for this EOS. Indeed, the QMD calculations are still hampered by uncertainties inherent to the exchange–correlation functional used in the calculations. For liquid hydrogen, for instance, pressures obtained with different functionals can differ by as much as $\sim 10$–20\% for a given density in the present domain of interest (see e.g., Refs.\cite{Morales2013H, Mazzola2018}). For the H-He mixture, the knowledge of the EOS would strongly benefit from further first-principle numerical explorations of the thermodynamic properties of the mixture. From the experimental point of view, recent experiments, coupling static compression (Diamond Anvil Cell) to laser shocks, probed the properties of H-He mixtures under the conditions of Jupiter's interior, showing a region of immiscibility along the Hugoniot~\cite{Brygoo2021}. Extending these experiments to higher densities and temperatures would open the route to probing BD interior conditions in laboratories, a fantastic achievement.

\subsubsection*{White dwarfs}

A key challenge concerns the calculation of the diffusion coefficients of trace mid-$Z$ elements (like Si, Ca) through warm dense He or H in the envelope. These coefficients are essential for reconstructing the bulk composition of accreted planetary material~\cite{Koester_AA_2009}. While traditional approaches based on the Boltzmann equation with screened Coulomb potentials work well in weakly coupled plasmas~\cite{Paquette_ApJS_1986,Stanton_PRE_2016}, the WDM conditions in WD envelopes require more sophisticated treatments. Recent work combining effective potential theory with average atom models has revealed significant corrections to transport properties~\cite{Heinonen_ApJ_2020,starrett2014hedp}. A key uncertainty remains the determination of the charge states of these trace elements in the envelope~\cite{saumon2021current}. Also challenging and important is electron thermal conductivity in the envelope, particularly in the regime of partial degeneracy and moderate coupling. Previously, the electron--electron contribution was treated with an interpolation between the degenerate and the classical limits~\cite{Cassisi_ApJ_2007}. The quantum Landau--Fokker--Planck formalism now allows the extension of calculations of the electron thermal conductivity into the partially degenerate regime~\cite{Starrett_HEDP_2017,Daligault_PhPl_2018,shaffer_PRE_2020}. These calculations suggest lower conductivities in this regime, though there is ongoing debate about the magnitude of this correction for $\Theta < 1$~\cite{Blouin_ApJ_2020,Cassisi_AA_2021}. The uncertainty in conductivity affects WD cooling times by up to 1--2\,Gyr~\cite{Salaris_MNRAS_2022,Pathak_arxiv_2024}.

\subsubsection*{Neutron stars}

The internal constitution of all NSs is generally determined under the hypothesis that their dense matter is cold and fully catalyzed, i.e., at the end point of thermonuclear evolution~\cite{Harrison_1965}. Initially very hot with temperatures of order $10^{11}$~K, NSs rapidly cool down and their temperature drops to $10^9$~K within a few days~\cite{Potekhin_2015}. The youngest known NS, lying in the supernova remnant of Cassiopeia A, is about 300 years old and is thus already relatively ``cold''. However, depending on the rates of the various nuclear reactions, the composition of the NS material may thus be frozen at some finite temperature. The full thermodynamical equilibrium of the outermost layers of the NS is unlikely to be maintained after crystallization at temperatures below $10^9$~K, meaning that a more realistic picture of the NS crust is that of a multi-component Coulomb solid~\cite{Fantina_2020}. One of the key challenges is to determine the final abundances of the different nuclear species and the solid structure they form~\cite{Caplan_2020}, which likely vary from one NS to another depending on their history. The formation of a nuclear pasta mantle remains an active field of research. The composition of the crust may be further altered by the fallback of material during the supernova explosion, and more importantly by the accretion of matter from a stellar companion. The final composition is established by various accretion-induced nuclear reactions~\cite{Meisel_2018}. These uncertainties propagate to all crustal properties, including elastic and transport coefficients~\cite{Kozhberov_2023}. In turn, this impacts the cooling of NSs and the evolution of their magnetic field and the formation of ``mountains'' that are sources of continuous gravitational waves~\cite{Gittins_2024}. The composition of the crust is also important for the rapid neutron capture nucleosynthesis process in the ejected material from magnetars and NS mergers, at the origin of elements heavier than iron in the Universe~\cite{Arnould_2007}.

\newpage 
\clearpage

\section{Inertial Confinement Fusion}\label{section_21}
\author{Alexis Casner$^{1}$, Tilo D\"oppner$^{2}$, Omar Hurricane$^{2}$}
\address{
$^{51}$CEA, DAM, DIF, F-91297 Arpajon, France \\
$^2$Lawrence Livermore National Laboratory, Livermore, CA 94550, United States 
}

\subsection*{Introduction}

By far the least difficult to access fusion reaction is deuterium (D) plus tritium (T) fusion,
\begin{equation}
D+T \rightarrow \alpha(3.5~\textrm{MeV}) + n(14~\textrm{MeV}) \label{DT}
\end{equation}
\noindent where the outgoing helium-4 nucleus ($\alpha$-particle) carries away 20\% of the produced energy and the neutron (n) 80\%. A variety of laboratory-based DT fusion schemes have been studied for decades including concepts that do not use large laser drivers such as magnetized liner inertial fusion \cite{slutz2011maglif,gomez2014experimental,gomez2020performance,ruiz2023exploring}. However, the only fusion scheme to demonstrate fusion ignition \cite{Abu-Shawareb2022,Kritcher_PRE_2022,Zylstra_PRE_2022,Hurricane_RMP_2023} and significant energy gain \cite{ICF_PRL_2024,hurricane_prl_24,Pak_PRE_2024,Kritcher_PRE_2024} has been x-ray-driven inertially confinement fusion (ICF). 

\begin{figure}
    \centering
    \includegraphics[width=1.0\linewidth]{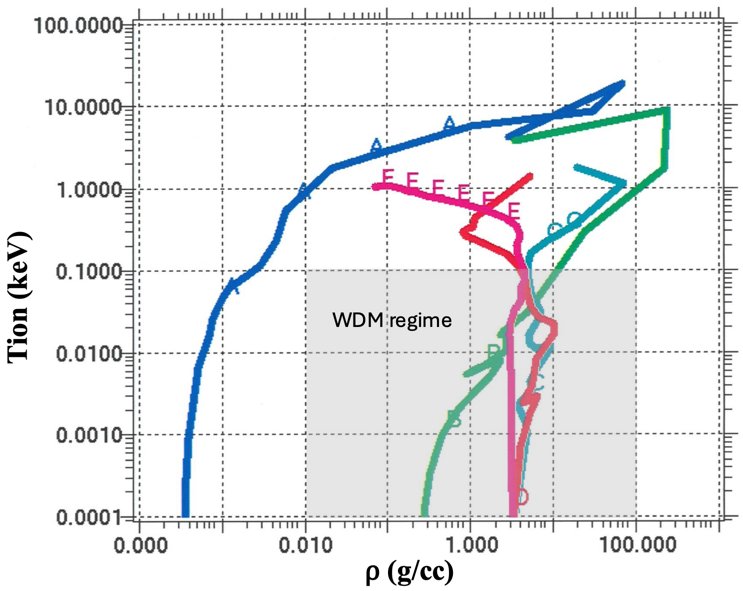}
    \caption{Traces of the average density vs. ion temperature of the DT gas (blue, curve A), DT fuel (green, curve B), inner high-density carbon (HDC) ablator layer (cyan, curve C), tungsten-doped HDC ablator layer (red, curve D), and outer HDC layer (magenta, curve E) over the duration of a characteristic igniting implosion are shown. The WDM regime is shown as gray box. Early in time, the materials in the capsule are heated from a cryogenic state by converging shock-waves and compressive $pdV$ work. The majority ($\sim 95$\%) of the ablator material is vaporized by x-ray-driven ablation causing its average density to drop as it is heated, while inner layers of the ablator and the DT compress and heat at the same time.}
    \label{fig:WDM_ICF_Fig1}
\end{figure}

Typically in ICF, an implosion accelerates a thin layer of cryogenic, solid-density DT fuel inward upon itself \cite{LindlPoP2004,Haan_PoP_2011}. The DT fuel and the capsule material surrounding the DT fuel are in a warm dense matter (WDM) state, as it accelerates to high velocity ($\sim 400$ km/s) while imploding (see Figure \ref{fig:WDM_ICF_Fig1}) and as the volume of the fuel capsule rapidly decreases. The ablation pressures driving the inward acceleration of the implosion typically ramp up from $\sim 1-12$ Mbar (depending upon ablator material) to $\sim 100-150$ Mbar in indirect-drive ICF. Ultimately, the DT fuel runs out of anywhere to go, as it "stagnates" on the high-pressure DT plasma that has developed in the center of the implosion. The deceleration of the imploding shell of DT fuel and remaining ablator material during the stagnation phase converts the kinetic energy of that shell into internal energy of the plasma (i.e. $pdV$ work is done on the plasma), thus generating a rapid increase in the DT plasma pressure. The WDM regime is exited as the ultra-high pressure ($\sim 100$'s Gbar) core of DT is created. The physics role of the implosion is that of a nonlinear pressure (and power) amplifier, and as such the materials involved pass through many states, from initially solid materials at cryogenic temperatures, to fluids, to WDM, then finally to high-energy-density plasmas akin to those at the center of stars.

The DT ``hotspot" at the center of the implosion will generate an explosive thermonuclear feedback-loop where, based upon Eq.(\ref{DT}), $\alpha$ particles generated by DT fusion thermalize with electrons in the DT plasma heating it further to generate even more fusion reactions, if the areal density ($\rho R$) and temperature ($T$) are sufficiently high and if the configuration holds together for sufficiently long time.  If the self-heating feedback loop is robust, "ignition" is achieved, a regime now demonstrated to exist at $\rho R T > 2.2$ g $\cdot$ keV/cm$^2$. In such a case, ignited DT will burn into the surrounding DT fuel, growing the hotspot, and generating more yield -- a process termed ``burn propagation." The degree of fusion fuel burn-up, or whether ignition occurs at all, critically depends upon inertial confinement and on how the implosion evolved during the WDM state earlier in time.

\subsection*{State of the art}

\begin{figure}
        \centering
        \includegraphics[width=1\linewidth]{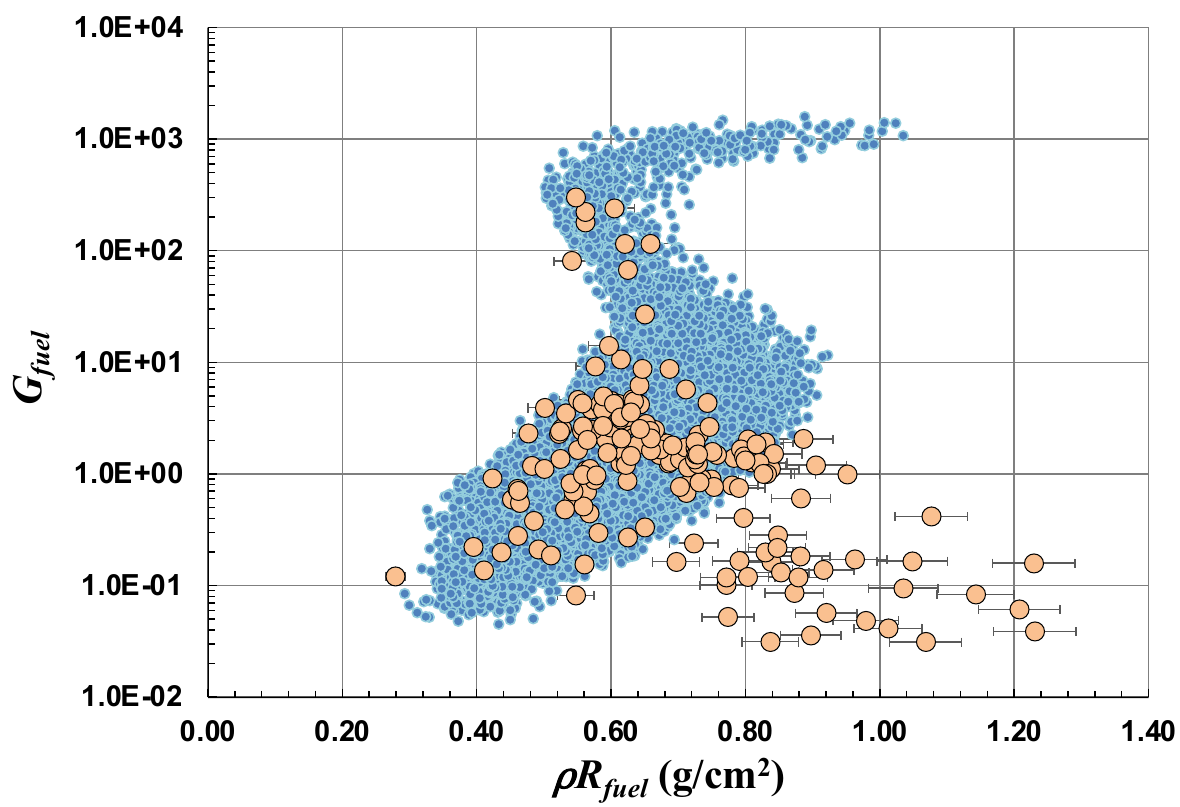}
        \caption{Fusion fuel energy gain vs. fuel $\rho R$ from simulations (blue points) and DT shot data from NIF experiments (yellow points).  While data and simulation are consistent at lower $\rho R$, they are inconsistent at higher $\rho R$. The inconsistency between simulations and data is not completely understood. Figure taken from Ref. \cite{Hurricane_PPCF_2025}}
        \label{gain_vs_rhoR}
\end{figure}

Whereas fusion ignition has been achieved and realized under thermodynamic plasma conditions expected from theory, the amount of laser energy required to obtain ignition on the National Ignition Facility (NIF) \cite{Moses_NIF,DiNicola_NF_2019} using indirect-drive required significantly more laser energy than originally projected before the NIF was built ($\sim 2 \times$). The physics basis for the increase in required laser energy has been established over the past 14 years of study and experiments on the NIF -- the principal factors are the following:
\begin{itemize}
    \item 
    High-compression implosions (with burn-on fuel $\rho R>0.8$ g/cm$^2$) in indirect and direct drive failed to produce significant fusion yield or behavior consistent with simulations (see Fig.\ref{gain_vs_rhoR})  \cite{Goncharov_PoP_2014}, thus forcing a strategy that used modest compression implosions that behaved more predictably. Less compression demands more kinetic energy ($KE$) from the implosion in order to increase $T$ at stagnation to meet the ignition conditions -- the essential scaling being $KE \cdot C_R^2 \sim const.$, where the convergence ratio $C_R$ is the ratio of initial DT fuel center-of-mass (COM) radius divided by the final COM DT fuel radius.  
    \item 
    Residual asymmetry in NIF implosions results in reduced kinetic-to-internal energy conversion at the moment of maximum compression \cite{Hurricane_PoP_2020,Hurricane_PoP_2022}. Thus, some kinetic energy of the implosion that would have been used to compress and heat the DT if the implosion was one-dimensional (1D) is wasted in the form of residual kinetic energy (RKE)~\cite{Kritcher_PoP_2014}.
    \item 
    Less than perfect control of Rayleigh-Taylor (RT) instability \cite{zhou2025instabilities, Casner_2019} in NIF ICF implosions results in mixing of ablator material into the DT fuel and central hotspot \cite{Divol_PoP_2024}. Mixing of high-Z materials into the hotspot enhances Bremsstrahlung radiation losses \cite{Pak_PRL_2020,Bachmann_PRE_2020}, cooling the hotspot, and thus more kinetic energy is needed to compensate to keep the temperature~\cite{hurricane_prl_24}. Mixing into the DT fuel makes the fuel less compressible and more energy is also needed to compensate for that.
\end{itemize}

Ignition and significant fuel energy gain were achieved with modest compression as compared to the original high compression design of the National Ignition Campaign (NIC) \cite{Haan_PoP_2011,LindlPoP2014} which ended in 2012.  The sacrifice made with modest compression was that of energy, where the laser energy required for ignition was significantly more than what was projected for the high compression implosions on the NIF, thus limiting the potential energy gain from fusion. So, it would be beneficial to better understand the origin of the discrepancy shown in Fig.\ref{gain_vs_rhoR}, one hypothesis being that hydrodynamic instability is seeded at the ablator-fuel boundary \cite{Weber_PRE_2023} while in the WDM state. 

Ablator materials and their properties play a critical role in ICF implosions as they are the intermediary between the drive and the fusion fuel. They have to efficiently convert the energy from laser beams or the x-ray drive into a high-pressure shock wave. In addition, the ablator shell usually contains a layer that is doped with a higher-Z element near their inner radius. The thus increased opacity reduces the transmission of the part of the x-ray drive spectrum that could pre-heat the DT fuel and hence reduce its compressibility. The majority of studies to date were done with low-Z ablator materials up to carbon as the ablation efficiency scales inversely with atomic number~\cite{LandenHEDP2024}. The first experiments during the NIC \cite{LindlPoP2014} used plastic ablator shells due to their relative ease of production and versatility to dope with a variety of elements. Beryllium was tested next as it promised to be the most efficient ablator candidate \cite{KlinePoP2016, ZylstraPoP2019},  but these experiments were challenged by microstructure of the Be that potentially gave rise to hydrodynamic mixing. Later, the ICF program at LLNL pivoted to high-density carbon (HDC) shells. Compared to plastic shells, HDC shells were about three times thinner thanks to their high mass density, which allowed shorter laser drives due to reduced shock transition time through the ablator, which in turn help with symmetry control in implosions. However, HDC requires a fairly high first-shock pressure ($\approx$12 Mbar) in order to get above the melting point, which leads to a significant increase in fuel entropy and ultimately limits the compression that can be achieved with HDC-based designs. Boron carbide is currently being developed as a future ablator candidate \cite{ MooreHEDP2016, Bayu_JVST2023} as it provides a very promising compromise between lower melt pressure more similar to plastic, higher mass density, glassy structure (reduced seeds for hydrodynamic instabilities) and good versatility in the choice of dopant materials. Plastic foams are a new class of ablator materials, which show particular promise for future inertial fusion energy (IFE) concepts \cite{callahan2024prospectus} as they offer a path towards holding a liquid DT layer (wetted foams) \cite{Sacks_1987,Olson_2016,OlsonPoP2021,KempPoP2025,nagai2018review}, which allows higher repetition rate experiments compared to the sophisticated DT ice layering scheme currently used at the NIF.

\subsection*{Challenges \& outlook}

Initial attempts to investigate and mitigate instability at the ablator-fuel interface \cite{Hinkel_HEDP_2020} were frustrated by the sensitivity of ICF implosions to shock-timing and the less than precise repeatability in laser pulse delivery -- newer experimental efforts are underway. In addition to the problem of the high compression implosions exhibiting poor fusion performance, another facet of this problem is that the fusion fuel in ICF experiments often compressed less than expected for a given entropy \cite{Landen_PoP_2021}. The hypotheses for this ``compression problem" range from ablator-fuel instability, severe three-dimensional asymmetry (e.g. \cite{Clark_PoP_2019, Hurricane_PoP_2022, Marinak_PoP_2024}), to imperfect understanding of material equations of state (EOS) at WDM conditions, especially for materials that have mesoscale features and microstructure.

The world's current flagship facilities for laser-driven ignition experiments, NIF \cite{Moses_NIF,DiNicola_NF_2019,dinicola2024} and its French counterpart Laser Megajoule \cite{casner2015lmj}, were funded and built for Stockpile Stewardship applications. Additional and larger MJ-scale laser facilities are under construction in Russia \cite{Rozanov_2016} and China \cite{sui2024driver}, which will also enable novel implosion concepts \cite{he2016hybrid,li2025amplifier}. While indirect drive currently is the only feasible approach for achieving ignition, it is not the most efficient one with respect to its energy balance. Indeed, for 2.2 MJ of invested laser energy, upon conversion into x-rays, only a small fraction of it (250 kJ) is absorbed by the fuel shell and drives the capsule implosion. Direct drive ICF (laser beams directly illuminating the imploding capsule) \cite{Craxton_PoP_2015} is expected to be up to 4 times more efficient \cite{betti2016inertial}. Despite this, the 30 kJ OMEGA laser facility \cite{boehly1997initial} remains the only laser world-wide dedicated to developing direct drive ICF.  A revival of the former HiPER \cite{dunne2006high} project was recently proposed at the European level by some researchers \cite{batani2023future}.  Based on hydrodynamic scaling \cite{gopalaswamy2024demonstration}, a burning plasma regime is anticipated for larger scale targets and higher laser energies for conventional hotspot ignition. Meanwhile for the shock ignition approach \cite{batani2014physics, theobald2015spherical}, extreme WDM states were achieved in solid spheres experiments~\cite{Nora2015}. A novel direct drive concept -- the dynamic shell design -- uses a liquid DT sphere encased in a wetted-DT foam \cite{trickey2024physics}. A high-density shell is then formed dynamically in-flight via a series of laser pulses that compress the target, allowing it to rebound, then decelerate the expanding plasma, forming a shock that develops into the shell. This approach will certainly face WDM challenges and remains to be validated at MJ scale \cite{igumenshchev2023proof}. 

Successful future IFE concepts will not only require new laser facilities, but will also crucially depend on an improved understanding of the microphysics in the ablator and the fuel at WDM conditions (see Fig. \ref{fig:WDM_ICF_Fig1}). This provides a unique opportunity for the wider HED science community to make meaningful contributions, through either theory or experiments.   
A multi-facility, multi-platform approach for direct-drive IFE should be developed, in which the microphysics is refined and confidence in target design is advanced by fielding cryogenic implosions on a sub-MJ laser facility. Novel diagnostics \cite{do2021x, valdivia2022current, faenov2018advanced, AllenNC2025} and experimental approaches, using x-ray Free Electron Laser (XFEL) facilities \cite{galtier2025x, rigon2021micron}, are being developed for an improved understanding of this unexplored regime. 

EOS tables for ablator materials and the fuel need to be tested and improved with focus on porous materials including low-density foams \cite{koenig1999equation, paddock2023measuring} and wetted-foam materials. In addition, there are significant uncertainties, e.g., about ionization levels and conductivity in dense plasmas \cite{AllenNC2025,HammelHEDP2010,Ciricosta_PhRvL_2012,Fletcher_PhysRevLett_2014,Kraus_PhysRevE_2016, Doeppner_nature_2023,HallPoP2024}. 
Understanding ionization in dense plasmas is important for correctly modeling the radiation transport through the ablator shell, which affects potential fuel preheat and hence compressibility, and stability against hydrodynamic growth of perturbations. 
Finally, the solid to plasma phase transition is computationally difficult to include in integrated radiation hydrodynamic simulations \cite{duchateau2019modeling,pineau2021improved, kar2020implementing}. Dedicated experiments should be developed \cite{kafka2024imaging, liotard2024solid} to benchmark the solid to plasma transition for ablator materials under consideration for IFE \cite{bourdineaud2025efficient}.

It appears that models (simulation or theory) that improve upon NLTE, EOS, opacity, mixing, heat transport, etc. could all be impactful towards IFE. We currently require gross multipliers in our simulations to match data, which can only be calibrated via iteration between experiment and simulation~\cite{Chen2024, Farmer2025}. With better models, fewer experiments should be required to converge upon a working target design. Generative AI tools have not yet demonstrated their usefulness in ICF as they need to be trained on a relevant dataset. A good 1D physical picture \cite{Hurricane_PoP2020} may be more useful than a model that has no understanding of how important kinetic energy and coast-time are for indirect-drive ICF implosions. On the other hand, any application capable of accelerating hydrocodes would be great, as recently demonstrated for NLTE inline opacity evaluations \cite{Kluth:2020}.  

Although many technical challenges still exist in ICF, especially with attention turning to IFE, which will require higher energy gains \cite{ribeyre2025perspectives}, higher repetition-rates, and more robust targets than is feasible with today's ICF technology, we should not forget how remarkable it is to have an existence-proof of fusion ignition in the laboratory.



\newpage 
\clearpage

\section{Novel Materials}\label{section_22}
\author{Siegfried Glenzer$^{1}$, Dominik Kraus$^{2}$, Eva Zurek$^{3}$}
\address{
$^1$SLAC National Accelerator Laboratory, Stanford University, Menlo Park, CA, United States \\
$^2$University of Rostock, 18051 Rostock, Germany\\
$^3$University at Buffalo, SUNY, United States
}

\subsection*{Introduction}

At pressure and temperature conditions typical of the WDM regime, the behavior of the elements as well as the compounds they form can differ drastically from what is observed on Earth's surface. At 1~atm, the pressure-volume contribution to the free energy is negligible. However, around 100~GPa, the magnitude of the pressure-volume term is comparable to the strength of a chemical bond, and by 10-100~TPa it is comparable to the energies of the core orbitals. In some cases, these core orbitals, typically assumed to be inert, may influence chemical bonding, while in other cases orbitals that are unoccupied at ambient pressures may become valence. As a result, pressure can promote the formation of materials with novel geometries and electronic structures, and unprecedented chemical combinations leading to the emergence of unique properties~\cite{zurek:2019k,Zurek:2022k}. Understanding this complexity is a key goal of the \emph{Center for Matter at Atomic Pressures}, an NSF funded physics frontier center~\cite{cmap}.

For example, hydrogen -- a diatomic molecule with a large electronic gap at ambient conditions -- becomes metallic at the conditions found in Jupiter's interior~\cite{Bonitz_POP_2024}. On the other hand, sodium -- a prototypical example of a nearly free electron metal -- becomes a transparent insulator near 200~GPa at low temperatures~\cite{ma2009transparent}. Ramp-compression experiments up to 480~GPa and thousands of degrees Kelvin have also provided evidence for this phase of sodium~\cite{Polsin:2022a}, where electrons are localized in interstitial regions~\cite{Zurek:2023e}. Similar electron bubbles have been predicted in hot dense iron~\cite{Dai:2012a}. The compounds thought to comprise the interiors of rocky planets feature novel coordination environments~\cite{Zurkowski:2022a}, and at conditions typical of the interiors of ice giant planets, superionic states~\cite{Villa:2023a}, where the hydrogen atoms diffuse through a sub-lattice of the heavy nuclei, might be prevalent. Further, DFT calculations have predicted a number of compounds with unusual stoichiometries that become stable only under pressure~\cite{zurek:2019k,Zurek:2016b,Zhang:2017a}. In addition, new materials might be formed by exotic chemical reactions that only become available at pressures exceeding 100\,GPa and temperatures of several 1,000\,K, e.g., carbon hydrogen phase separation~\cite{Kraus2017}, which may be triggered or accelerated by hydrogen metallization~\cite{Kraus2023}. 

\begin{figure}
\includegraphics[width=0.98\linewidth]{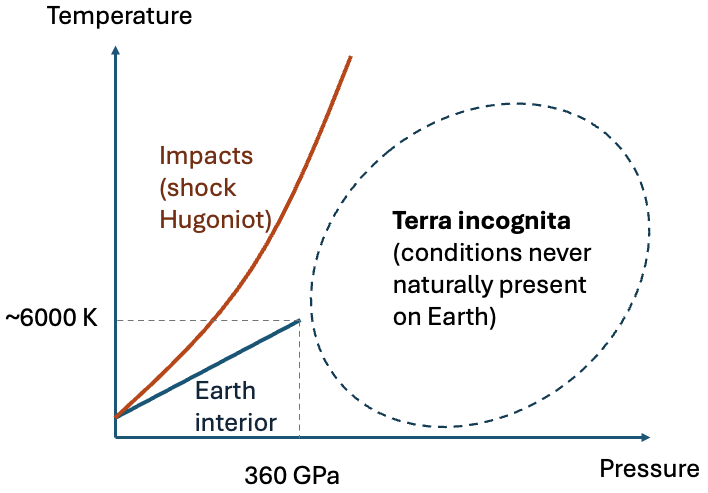}
\caption{\label{fig:terraincognita}Pressure-temperature diagram sketching the terra incognita of conditions that were never naturally present on Earth and where new materials are awaiting discovery. Static pressures and temperatures are limited by those present in the Earth's core. Meteor impacts can achieve higher pressures but at the same time, the temperatures become too high for forming new solid structures at pressures around several 100\,GPa.}
\end{figure}

The parameter space for natural materials synthesis on Earth is limited by the pressure-temperature conditions on the surface and inside our planet. Some well-known materials require extreme environments to be created, but remain metastable at ambient conditions. For instance, natural glass and diamond form during volcano eruptions or meteor impacts, and at high pressures and temperatures well below the planetary surface. Beyond the conditions inside our planet and during meteor impacts, there is a substantial terra incognita in pressure-temperature space starting at $\sim$400 GPa, where so far unknown materials may be formed (see Fig~\ref{fig:terraincognita}). 

While the impact of large meteors or comets can generate higher pressures than present in Earth's interior, the temperatures, which can be estimated by the shock Hugoniot curve of relevant materials, become too high in such events to form new solid structures. Therefore, the discovery of material structures, which might naturally only occur inside planets larger than ours, requires laboratory experiments in combination with theoretical structure prediction techniques. Static compression experiments, e.g., using diamond anvil cells, are usually limited to the stability of diamond around 400\,GPa at 0\,K, with few exceptions~\cite{Dubrovinsky2015}. However, already exceeding 100\,GPa requires substantial effort and the static compression pressure limits decrease rapidly with higher temperatures up to the melting line of diamond. For higher pressure-temperature regions, and thus entering the WDM regime, dynamic experiments are required. One prominent example beyond current static compression limits is the BC-8 structure of carbon (body-centered cubic with 8 atoms per unit cell)~\cite{Yin1984}, which is so far only predicted theoretically. It is assumed to require pressures of $\sim$1\,TPa to form and may exhibit similar hardness as diamond while being substantially tougher due to the absence of cleavage planes. At the same time, calculations suggest that BC-8 carbon is metastable at ambient conditions~\cite{Mailhiot1991}.

\subsection*{State of the art}

\subsubsection*{Theoretical Studies}

DFT-based crystal structure prediction (CSP) techniques, such as random or evolutionary searches, and particle swarm optimization methods, developed $\sim$15-20 years ago, have been extensively applied towards the discovery of novel materials at high pressures~\cite{Zurek:2014i}. So far, most applications have targeted elemental or binary compounds at 0~K, with the effects of temperature typically considered only for the most stable compounds via MD simulations or computation of free energies. Such predictions have been key for inspiring experiments, and in the interpretation of the experimental observables. Recent algorithmic advances have made it possible to use the similarity of an experimental X-ray diffraction (XRD) pattern with a computed one (for an optimally-distorted structure) as an objective in the CSP search, to aid in unravelling the structures of dynamically compressed materials~\cite{Zurek:2024j}. 

\begin{figure}
\includegraphics[width=0.98\linewidth]{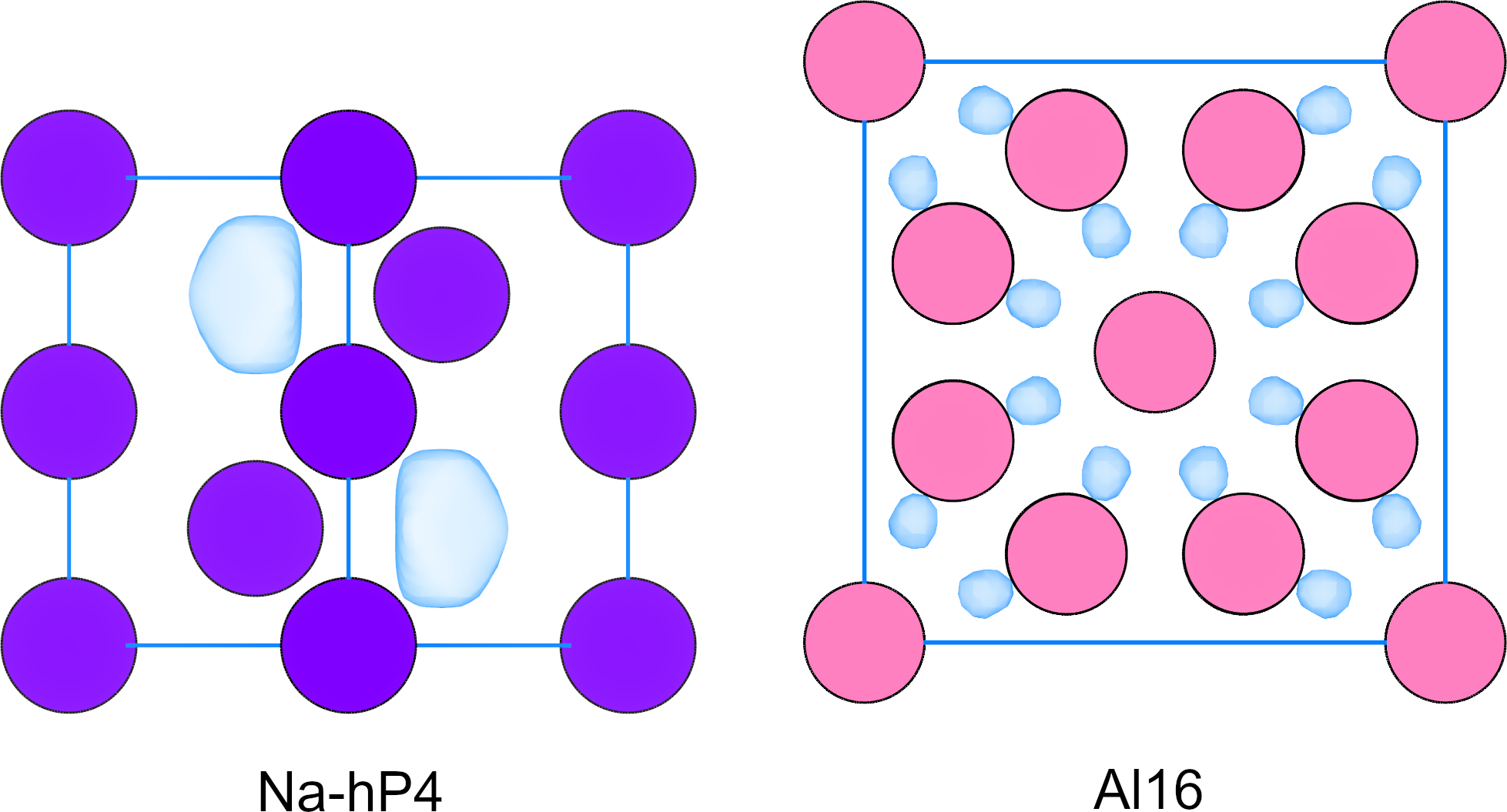}
\caption{\label{fig:electrides} Unit cells of two electrides, Na-$hP4$~\cite{ma2009transparent} (at 200~GPa) and Al16~\cite{Pickard2010} (at 5~TPa), calculated with VASP at the r$^2$SCAN level of theory. The interstitial localized electrons are found in regions where the Electron Localization Function (ELF) is high (isovalue of 0.9 is shown).}
\end{figure}

The Thomas-Fermi-Dirac (TFD) model predicts that at high pressures materials adopt closed-packed metallic lattices. In stark deviation, CSP searches have suggested that there may be a regime where lattices with open voids are preferred. In these high-pressure phases, called electrides, the electrons, trapped in the voids, act as the anion counterparts to the cationic cores~\cite{Zurek:2024p}. Indeed, high pressure phases of Na~\cite{Li2015b,Zurek:2022e}, Mg~\cite{Li2010,Gorman:2022a}, Al~\cite{Pickard2010}, C~\cite{Martinez-Canales2012}, and Ca~\cite{Novoselov2020a,Zurek:2024r} are some of the systems computed to possess electronic distributions consistent with the electride state (see Fig.~\ref{fig:electrides}). Moreover, ramp-compression experiments have provided evidence for the preference of open structures at high pressures and temperatures~\cite{Polsin:2022a,Gorman:2022a}, though it has been difficult to solve their structures, and the electride nature of WDM has so-far only been inferred from DFT calculations.

CSP techniques have also been applied extensively towards the discovery of a variety of novel compounds that may be stable or metastable at planetary pressures including hydrides (e.g.\ NH$_n$ ($n=7,10)$~\cite{Song2019_JPhysChemLett,Zurek:2024d}, FeH$_5$~\cite{Pepin2017,Kvashnin:2018a,Zurek:2018h}), helium-containing species (e.g.\ Na$_2$He~\cite{Dong:2017a}, MgF$_2$He~\cite{Zurek:2017i} and He-NH$_3$ compounds \cite{Liu2020_PhysRevX}), compounds of nickel and iron (e.g.\ XeNi$_3$, XeFe$_3$~\cite{Zhu:2014a,Stavrou:2018a}, and new Fe-Mg-O phases~\cite{Fang:2023a}), as well as unique hydrocarbons (e.g.\ HCNO~\cite{Conway}), to name a few.

\subsubsection*{Experimental Studies}

There have been substantial advances in both the controlled creation of extreme conditions relevant for the synthesis of new materials via WDM conditions and powerful \textit{in situ} diagnostics to discover and characterize the appearance of new structures and their physical properties. In particular, bright X-ray sources are able to penetrate the bulk of transient WDM states in the laboratory~\cite{Pascarelli2023}. Before combining high-energy lasers with accelerator-based X-ray sources, only few pioneering experiments, mostly at large laser systems providing laser-driven X-ray sources such as OMEGA and later also NIF, allowed for both the creation and structural in situ probing of relevant conditions. XRD platforms like PXRDIP~\cite{Rygg2012} and TARDIS~\cite{Rygg2020} were key to identify new material structures in dynamic compression experiments. However, due to the limitations of laser-driven X-ray sources in spectral brightness, collimation and repetition rate, the XRD data is usually not of highest quality in such experiments, which sometimes results in remaining ambiguities such that material structures cannot be uniquely identified. Laser-driven XRD platforms now also exist for dynamic compression experiments at the Z-machine of Sandia National Laboratories~\cite{Ao2023} and isochoric heating with heavy ion beams at GSI/FAIR~\cite{Schoenberg_PoP_2020, Luetgert2024, Hesselbach2025}.

As a flagship case for materials synthesis via extreme conditions beyond static compression, the creation of BC-8 carbon from diamond at the NIF has been attempted for more than a decade~\cite{Smith2014}. However, it appeared that diamond remains metastable also at high-pressure conditions, with experiments reaching 2\,TPa~\cite{Lazicki2021}. Consequently, several more sophisticated pathways to BC-8 carbon have been proposed~\cite{Shi2023,Nguyen-Cong2024}, of which some are currently pursued.

Accelerator-based X-ray sources, X-ray free electron lasers~\cite{Glenzer_JPhysB_2016} and state-of-the-art synchrotron sources~\cite{Wang2019}, in combination with high-energy lasers or gas guns substantially increase the quality of in situ XRD patterns in dynamic compression experiments~\cite{Kraus2025}. This allows to identify very complex crystalline structures, like host guest structures or electrides, liquid structure and chemical reactions of light elements. Examples include the formation of hexagonal diamond from graphite~\cite{Kraus2016b}, defects dynamics like shock-driven twinning in tantalum~\cite{Wehrenberg2017}, and chemical reactions forming nanodiamonds~\cite{Schuster2020}. In addition to high-quality in situ XRD, these sources allow further high-quality in situ X-ray diagnostics, often providing the possibility to use several simultaneously. XANES/EXAFS at synchrotrons can probe the element-specific local electronic and ionic structure in dynamic compression studies~\cite{Sevelin-Radiguet2022}. Phase contrast imaging at synchrotrons and XFELs provides information of material response to WDM conditions on the macro- and mesoscale~\cite{Schropp2015,Olbinado2018,Brennan-Brown2019,Katagiri2023}. Using different harmonics of the XFEL, simultaneous phase contrast imaging (fundamental) and diffraction (3rd harmonic) has been demonstrated~\cite{seiboth2018simultaneous}. Moreover, the extreme spectral brightness of XFEL sources has started to allow spectrally resolved X-ray scattering to characterize dynamic processes and material properties at WDM conditions. Next to the possibility of characterizing temperature and density via X-ray Thomson scattering~\cite{Fletcher_NP_2015,Descamps2020}, this involves electronic properties~\cite{Sperling_PhysRevLett_2015,ranjan2023toward}, chemistry~\cite{frydrych2020demonstration}, and local structure~\cite{Voigt_POP_2021}. With rep-rated driver systems like DiPOLE 100X at European XFEL~\cite{Gorman_JAP_2024}, which allow the accumulation of numerous experiments at the same conditions with high reproducibility, photon-hungry spectroscopy techniques become more accessible. 

The generation of useful materials via extreme WDM conditions, which often can only be achieved transiently in dynamic compression experiments, also requires the successful release and recovery to ambient conditions. Abrupt release from WDM conditions to nearly zero pressure involves complex microphysics like lattice oscillations and plastic work and cannot be modeled by isentropic expansion~\cite{Heighway2019}. To shed more light into this process, in situ measurements now allow for valuable insights into these processes to benchmark atomistic simulations~\cite{Heuser2024}. Finally, dynamic compression experiments at WDM conditions typically accelerate sample materials to velocities of many km/s and are thus comparable to meteor impact velocities. Catching and ensuring survival of exotic material structures formed in WDM environments is therefore challenging~\cite{Schuster2023}. Soft low-density materials like liquids, waxes, gels, and aerogels are standard candidates for catcher materials but also glass or metal plates are used. Overall, the study of material recovery from dynamic experiments reaching pressures exceeding 100\,GPa is still in early stages and will require substantial advances for efficient recovery of potentially interesting materials formed at WDM conditions.

\subsection*{Challenges \& outlook}

\subsubsection*{Theoretical Studies}

Simulations based on quantum-accurate machine learning interatomic potentials (MLIPs; see also Sections \ref{section_03}, \ref{section_07} and \ref{section_24}) will make it possible to carry out CSP studies on large unit cells~\cite{WangMa:2023a}, potentially incorporating the effects of temperature in structural relaxations and free energy calculations. MLIPs will enable long-time MD simulations for the discovery of new materials at WDM conditions and gain atomistic insight into phase transitions, chemistry, and transport properties. Depending upon the application and desired level of accuracy, researchers can train bespoke MLIPs including EDDPs~\cite{Pickard2022}, MTPs~\cite{Shapeev:2016a}, GAP~\cite{Bartok_PRL_2010} and  SNAP~\cite{Rohskopf2023}, or employ universal ones such as MACE~\cite{24arxiv-Batatia}, DPA2~\cite{24npj-Zhang}, M3GNet~\cite{24npj-Qi}, ALIGNN~\cite{2021-choudhary}, CHGNeT~\cite{2023-Deng}, Orb~\cite{rhodes2025orbv3}, and MatterSim~\cite{yang2024mattersim}.

\subsubsection*{Experimental Studies}

Existing X-ray sources at dynamic compression facilities, especially if not XFEL sources, possess numerous limitations, which have hindered progress in understanding the emergence of complex materials properties and quantum phenomena at WDM conditions~\cite{mp3}. The temporal resolution blurs measurements, preventing time-resolved studies for the kinetics of phase transformations. Moreover, the low signal-to-noise ratio and number of discernible diffraction lines can make structural determination difficult. At the same time, the driver systems at XFEL sources are currently rather small-scale, where the available driver energy often limits experiments to lower-pressure conditions. 

Yet, new facilities are being planned that would make it possible to overcome these limitations. One example on the optical laser-only side is the NSF-OPAL~\cite{nsfopal} facility where long-pulse beams will shock and ramp-compress planetary materials to TPa pressures, while femtosecond PW beams will create ultrafast probes for time-resolved X-ray and electron diffraction, radiography/phase contrast imaging, and X-ray emission and absorption spectroscopy. At the same time, substantial upgrades are foreseen at XFEL sources to increase the drive energy from the current max. 100\,J level to multiple kJ (e.g., MEC-U at LCLS or HIBEF 2.0 at European XFEL)~\cite{Kraus2026}. This will allow to access TPa pressures at XFEL facilities to further explore the terra incognita shown in Fig.~\ref{fig:terraincognita}.

\newpage 
\clearpage

\section{Laser matter interaction}\label{section_23}
\author{Dirk O.\ Gericke$^{1}$, Baerbel Rethfeld$^{2}$, \\ Alex P.L.\ Robinson$^{3}$, Thomas G.\ White$^{4}$}
\address{
$^1$ Centre for Fusion, Space and Astrophysics, Department of Physics,
     University of Warwick, Coventry CV4 7AL, United Kingdom\\
$^2$RPTU University Kaiserslautern-Landau, Department of Physics and State Research Center OPTIMAS, Kaiserslautern, Germany \\
$^3$STFC Rutherford-Appleton Laboratory, Didcot OX11 0QX, United Kingdom \\
$^4$Department of Physics, University of Nevada, Reno, Nevada 89557, USA\\
}

\subsection*{Introduction}

Lasers are an indispensable tool in WDM research used for the creation of WDM samples, further heating of matter, direct diagnostics, and the creation of secondary sources such as particle beams and X-rays \cite{Riley_2021}. These diverse applications require very different lasers both in frequency and intensity, and this variability renders different physics aspects important. Due to the multitude of processes with similar energy scales, laser-WDM interactions are not well understood for all parameter sets. 

\begin{figure}[t]
\includegraphics[width=0.98\linewidth]{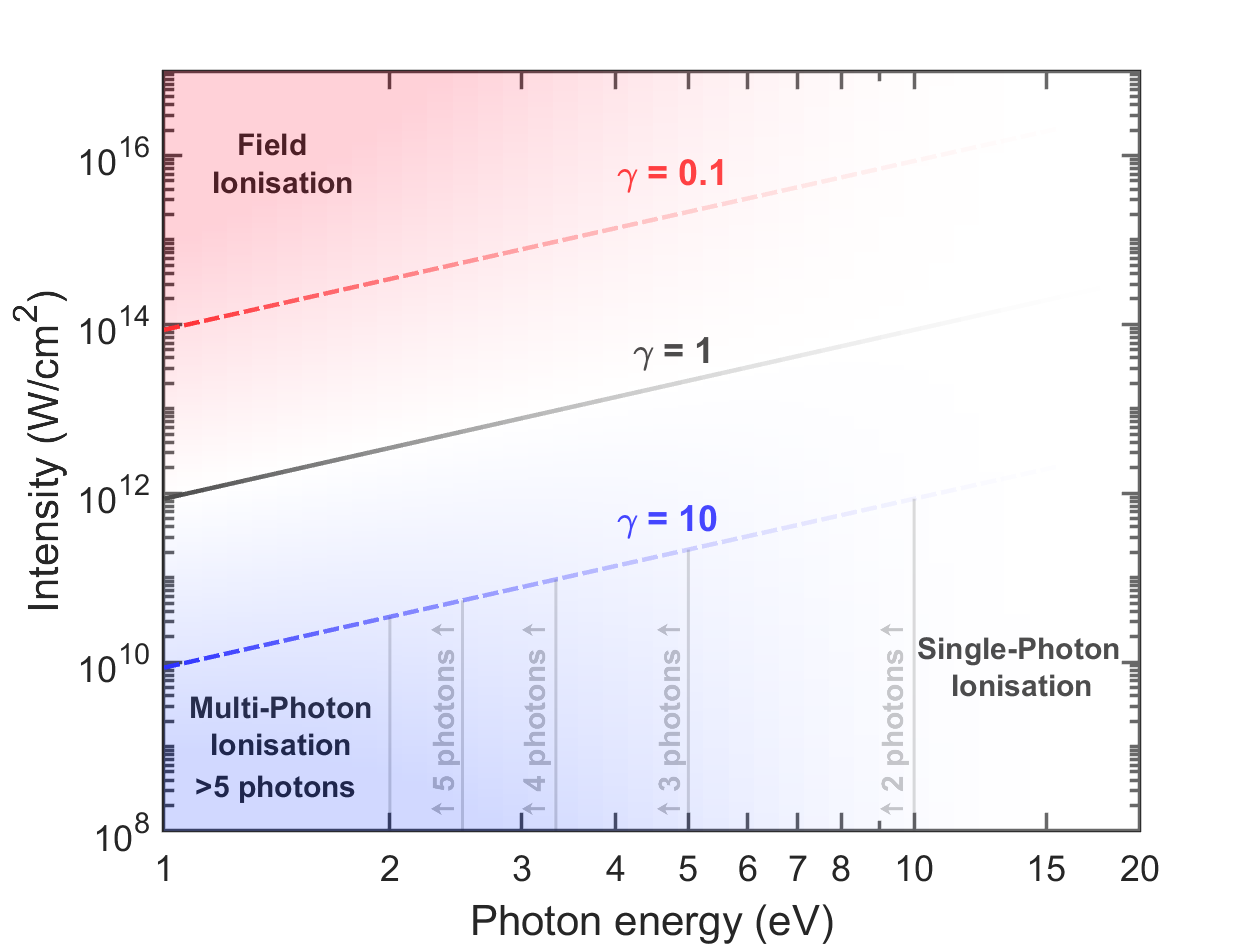}
\caption{Regimes of ionisation for different laser intensities and photon energies. A Keldysh parameter \cite{Keldysh1965} of unity indicates the transition between field ionisation and multi-photon processes. The example given assumes a ionisation potential of 10\,eV which clearly allows for direct ionisation for photon energies higher than 10\,eV making the Keldysh parameter obsolete in this region.
\label{fig:Keldysh:para}}
\end{figure}

The complexity of laser-matter interactions can be slightly reduced by combining laser intensity and frequency scales. For excitations of solids with a band gap or ionisation in plasma-like systems, one may use the ratio between the quiver and field energies $\gamma = \omega_L (2m_e I_0)^{1/2}/(e E_L)$ (Keldysh parameter) in order to quantify the different regimes of ionisation \cite{Keldysh1965}. Here, $\omega_L$ is the laser frequency, $E_L$ is its electric field amplitude and $I_0$ is the band gap/ionisation energy; $\gamma \gg 1$ indicates that multi-photon processes drive the transitions and, thus, quantum effects are important, whereas field ionisation, that allows a classical description of the laser field, dominates for small $\gamma$. Accordingly, the medium must be described: we need a full quantum description for large $\gamma$ while high field strengths often allow a classical, but sometimes relativistic, treatment. The different regions are illustrated in Fig.~\ref{fig:Keldysh:para} for a medium with a band gap or ionisation energy of $I_0=10$\,eV. 

For metals and plasmas, there is no energy barrier to bridge, but the free carriers can interact directly with the laser field. Once again, low intensities may require a description via single-photon absorption. For higher fields, inverse Bremstrahlung or collisional absorption is the dominant process \cite{Seely_1973}. This process considers the disturbance of the electron quiver motion in the oscillating field by collisions with the ions. If the quiver energy is smaller than the thermal or Fermi energy, one reaches the linear regime of Ohmic heating \cite{silin1965}. On the other hand, absorption occurs mainly via collective processes in the relativistic limit.

Laser-matter interactions always drive a cascade of relaxation processes \cite{Rethfeld_2017,white_dynamic_2023}. However, the energy and time scales can vary strongly, requiring fundamentally different theoretical descriptions. In the following, we will briefly discuss (i) small laser intensities that require a quantum mechanical description often based on kinetic theory, (ii) laser intensities generating energy scales similar to the thermal or Fermi energies in WDM and (iii) very high laser intensities that drive the electrons strong enough that correlations and quantum effects quickly become irrelevant. Clearly, the intermediate regime is most closely related to WDM research as the field strength is here on the same scale as thermal, exchange and correlation energies.

\subsection*{State of the art}

\subsubsection*{Low laser intensities}

Laser pulses at low intensities can excite a number of electrons within the conduction band of a metal. Depending on the photon energy, these so-called hot electrons have energies a few eV above the Fermi energy. These electrons can move ballistically through the irradiated sample and lead to homogeneous heating within a thickness of up to about $100\,$ nm, as has been shown for gold \cite{Hohlfeld1997}.

Within a few femtoseconds, electron-electron scat\-ter\-ing establishes a Fermi distribution of high temperature, by far exceeding the lattice (or phonon) temperature, $T_e \!>\! T_i$, \cite{Anisimov1974}. Then electron-phonon coupling drives the temperature equilibration on a picosecond timescale. When the melting temperature is exceeded, a melt front proceeds from the surface into the depth of the material. At considerable superheating, homogeneous melting within the bulk of the material due to statistical fluctuations becomes probable \cite{Rethfeld2002,Siwick2003,white_Nature2025}. The loss of crystalline order often occurs before significant expansion, thus, the target remains at solid density. 

Ultrafast (non-thermal) melting processes are known for covalently bound semiconductors. 
Here, the excitation of electrons into the conduction band strongly affects the interatomic binding, softens the phonon modes, and leads to a rapid loss of crystalline order \cite{Stampfli1994,Sundaram2002}. In metals with highly excited electrons at a few eV temperature, the electronic band structure as well as the stiffness of the solid can change strongly \cite{Recoules2006} and phonon hardening can be observed \cite{Ernstorfer2009,Decamps2024}

From an applied perspective, low-intensity lasers are frequently utilized as diagnostic light sources in pump-probe configurations, where a pump pulse induces moderate excitations in the sample, and a weak laser pulse (probe) examines the transient conditions \cite{Bauer2015,Ndione2022}. When the sample is pumped with a stronger laser pulse causing phase transitions of the sample, it is typically probed with an ultrafast X-ray or electron beam. This methodology has been employed to elucidate the fundamental dynamics of phase transitions, electron-phonon coupling, and other microscopic processes. Well-tailored laser pulses also allow for controlled micro-structuring of surfaces or laser-induced ablation \cite{Sugioka2017,Rethfeld_2017} 

\subsubsection*{Intermediate laser intensities}

When a visible laser pulse with an intensity in the range of approximately $10^{12}-10^{15}$~W/cm$^2$ irradiates a solid material with a large bandgap, the initially transparent material can become highly absorbing as multiphoton- or tunneling processes transfer electrons across the bandgap. The Keldysh parameter introduced above is close to unity and both of these asymptotic processes cannot be  distinguished clearly, compare Fig.~\ref{fig:Keldysh:para}. Independent of the details of the ionisation mechanism, a large density of conduction electrons can be reached by an avalanche of impact ionization following the initial ionisation of seed electrons \cite{Rethfeld2004,Balling2013}. The temporal shape of the laser pulse is crucial for the impact on the material. Depending on their temporal orientation, Airy pulses have been shown to cause or avoid damage to a dielectric solid \cite{Englert2007}.

Increasing the electron density above a critical density leads to dielectric breakdown, which can rapidly transfer the solid into a plasma state. Peaks of moderate intensity in the pedestal of a high-intensity beam can therefore induce phase transitions before the main pulse hits the sample, fundamentally changing the interaction with the laser field. Further heating can cause ablation at the material surface, which, in turn, can drive shocks that compress and heat samples to WDM conditions, often in a more uniform or controllable way than higher-intensity interactions. 

The ablated plasma can also be harnessed as a secondary source of moderate-energy ion or electron beams and robust X-ray emission. In this intensity regime, long-pulse lasers can sustain hot plasmas over nanosecond durations, often producing helium-like line emission in higher-Z materials. By contrast, short-pulse lasers, even at mid-range intensities, tend to generate strong $K_{\alpha}$ lines and a broad bremsstrahlung background through hot electron collisions with nuclei.

Further heating of fluid and plasma samples with lasers at these intermediate intensities is best described by collisional absorption. Early work based on a classical description \cite{DawsonOberman_1962,Decker_1994} showed a break-down for high densities. Thus, the main challenge of any model is to simultaneously include quantum effects and strong electron-ion interactions \cite{Kremp_1999}. Quantum effects can be included within dielectric theory \cite{Grinenko_2009} but good quantitative agreement with simulation data \cite{Pfalzner_1998,Hilse_2005} can only be reached with the inclusion of strong collisions by a hybrid-scheme \cite{Grinenko_PRL_2009} that combines polarization and friction data \cite{Gericke_1999b}. 

\subsubsection*{High laser intensities}

As laser intensities become progressively higher and approach the relativistic threshold ($I\lambda^2 > $ 10$^{18}\,{\rm Wcm}^{-2}\mu{\rm m}^2$), energy absorption is dominated by the large (classical) fields, see Fig.~\ref{fig:Keldysh:para}, and the generation of supra-thermal electrons becomes a central feature of the interaction.  The generation of a population of highly energetic electrons is important since the range of these electrons will be relatively large even in cold dense matter.  This means, on the one hand, that energetic laser-generated electron beams could be used to rapidly heat substantial volumes of cold dense matter to produce WDM directly, and, on the other hand, it also allows the electrons to transfer their energy into energetic ions and x-rays, which are also potential routes to generating WDM.

The generation of highly energetic electrons at these intensities is not entirely unexpected \cite{Beg1997,robinson_breaking_2017,robinson_interaction_2018,arefiev_beyond_2016}.  For single electron motion in a strong laser field, one expects that ${\bf p}_\perp = a_0m_ec$ (${\bf p}_\perp$ being the momentum transverse to the wave propagation), so relativistic electrons can be expected for $a_0 > 1$.  On the other hand, single electron theory does not predict the electron energy spectrum, angular distribution of fast electrons, or the power absorption fraction into fast electrons.  A large number of mechanisms have been proposed for how absorption into relativistic electrons actually occurs.  For this reason, numerical simulations are a very important tool. 

The common computational workhorse of laser-plasma interactions is the Particle-in-Cell (PIC) method \cite{birdsall1991}, with popular codes including EPOCH~\cite{2015_Arber_et_al}, Warp-X~\cite{WarpX}, SMILEI~\cite{smileipic}, OSIRIS~\cite{osirispic}, Virtual Laser Plasma Laboratory \cite{vlpl}, ALaDyn~\cite{aladyn}, PIConGPU~\cite{2010_Burau_et_al}, and LSP~\cite{lsp}. 
For relativistic interactions, this fully self-consistent model is particularly important as any model would have to incorporate a fully kinetic, fully relativistic treatment that self-consistently solves Maxwell's equations thus accounting for both the laser field and for fields that are generated due to charge separation and plasma currents.  In addition, modern PIC codes can also incorporate treatments of collisions and ionization.  

A significant review of fast electron generation, in the context of Fast Ignition ICF, was done by Kemp et al. \cite{ajkemp_laser-plasma_2014}.  In summary: there can be good conversion of laser energy into fast electrons, but the fast electrons appear to often have a large angular spread. For Fast Ignition this angular spread was an acute problem, and is one that WDM generation will have to contend with (see also \cite{adebayle_feg_2010}).

PIC codes have also been used to model ion acceleration.  This has led to the identification of a number of acceleration mechanisms, although TNSA \cite{mora_plasma_2003} is probably still the most common mechanism.  This may change with the development of multi-PW lasers \cite{robinson_relativistically_2009}.  PIC codes are reasonably well suited to studying ion acceleration: in the TNSA mechanism, ion acceleration is achieved by electric fields set up the laser-generated fast (multi-MeV) electrons which can make a relatively large excursion into the vacuum and establish a restraining potential on the same scale as their energy.  Thus ions can achieve energies proportional to the characteristic fast electron energy.  PIC simulations largely affirm the core predictions of TNSA theory \cite{mora_plasma_2003}. Thus these need to be regarded as constraining factors when contemplating using such proton/ion beams for WDM generation.

The physics and modelling of the transport of the fast electron beams was reviewed by Robinson et al. \cite{aplrobinson_fetrev_2014} in the context of the Fast Ignition approach to ICF.   The key physics of fast electron transport can be summarized as follows: fast electrons are subject to drag by the background (cold) electrons, scattering from the background ions, and to electric and magnetic fields that arise because of the large current density that such beams carry.  These EM fields arise because currents will be set up in the background material to try to cancel the fast electron current density to a good approximation \cite{bell_resistive_2003,bell_fast_2006}, i.e. ${\bf j}_f + {\bf j}_c \simeq 0$.  Since these materials possess a finite resistivity this implies the existence of an electric field, i.e. ${\bf E} = -\eta{\bf j_f}$, and this also implies magnetic field generation.  As $|j_f|$ could, in principle, be on the order of 10$^{16}\,$Am$^{-2}$ or higher, generating magnetic fields in the 100-1000\,T range is possible.  This also implies that Ohmic heating of the background could be dominant.  This heating will affect the ionization and resistivity of the background in turn.

There are at least two major classes of numerical tools to tackle the problem of fast electron transport: the collisional PIC code, and the `hybrid' code.  Collisional PIC codes are robust tools that treat the plasma fully kinetically with a full set of Maxwell's Equations.  They can therefore treat both transport and absorption in one code.  The inclusion of collisions means that collisional transport phenomena should be modelled as well, however, as Arber points out \cite{2015_Arber_et_al}, one can require many particles per cell in order to accurately reproduce certain phenomena such as the coefficient of thermal conduction.  In addition to this, PIC codes are relatively computationally expensive compared to alternatives such as the hybrid code.  The term `hybrid' code in this context means the type of code that Davies \cite{Davies_PRE_2002} discussed in which the fast electrons are treated kinetically (as particles) and the background electrons and ions are all treated as a fluid (even a static fluid).  This model is computationally cheaper, especially as one does not need to resolve very small time and length-scales.  Partial degeneracy in the background plasma can be included, and one can include a resistivity model beyond binary collision (Spitzer) theory.  The main drawback of hybrid codes is that the standard hybrid code will not self-consistently include laser absorption.  Nonetheless, hybrid codes have been used in a large number of studies and have been able to provide useful insights when the central phenomena are primarily transport phenomena (see \cite{damaclellan_2013,bramakrishna_2010} for instance).

\subsection*{Challenges \& outlook}

Laser excitations with low intensities are mainly applied to the creation and investigation of highly excited solids where the largest challenge is to formulate a consistent quantum theory for the non-equilibrium processes induced. In contrast, highly relativistic laser beams create conditions that approach the realm of traditional plasma physics. For WDM research, these interactions are mainly interesting as sources for secondary X-ray and particle beams. The real challenge for WDM research lies, however, in the regime of intermediate laser intensities. Despite the considerable progress described above, a comprehensive description of situations, where the quiver energy of the electrons in the laser field is comparable with their thermal, Fermi and correlation energies, 
is still elusive. Moreover, there are almost no experimental data available for WDM conditions that could be used to benchmark current theoretical approaches. 

\newpage 
\clearpage

\section{Machine Learning}\label{section_24}
\author{Sebastien Hamel$^{1}$, Sam Vinko$^{2}$, Lei Wang$^{3}$}
\address{
$^1$Lawrence Livermore National Laboratory, Livermore, CA 94550, United States \\
$^2$Department of Physics, Clarendon Laboratory, University of Oxford, Parks Road, Oxford OX1 3PU, UK \\
$^3$Institute of Physics, Chinese Academy of Sciences, Beijing 100190, China
}
\definecolor{psychedelicpurple}{rgb}{0.87, 0.0, 1.0}
\newcommand{\sh}[1]{{\color{purple} #1 }}
\newcommand{\sv}[1]{{\color{blue} #1 }}
\newcommand{\lw}[1]{{\color{psychedelicpurple} #1 }}
\newcommand{\task}[1]{{\color{red}{#1}}}

\subsection*{Introduction}

Machine learning is increasingly transforming computational approaches across plasma physics and high energy density (HED) science, offering new strategies to accelerate simulations, expand modeling capabilities, and integrate data-driven insights. This section reviews recent developments in machine-learned interatomic potentials (MLIPs), which have enabled large-scale atomistic simulations with near first-principles accuracy, including strategies to incorporate electronic temperature effects for warm dense matter applications. It then introduces a variational free energy framework that bypasses data-driven approaches, offering a fundamentally different route to modeling quantum systems at finite temperature. The chapter continues with a survey of surrogate models, highlighting their role in replacing expensive components of complex simulations with efficient, trainable approximations, particularly in inertial confinement fusion and transport modeling. 
We also discuss machine learning approaches applied directly to electronic structure, including the learning of Hamiltonians, density matrices, and spectroscopic observables, which significantly broaden the scope of data-driven modeling in HED physics.
Finally, we turn our discussion to the emerging use of automatic differentiation and hybrid simulation architectures, which are beginning to bridge machine learning and physics-based modeling in a more seamless and differentiable way. 

\subsection*{Machine Learned Interatomic Potentials}

Neural-network (NN) representations of forces and energies from electronic structure calculations were introduced in 2007 by J\"{o}rg Behler and Michele Parrinello~\cite{Behler_PRL_2007}. This approach was a paradigm shift in the simulation of material properties, overcoming time and length scale limitations of electron structure calculations while keeping its accuracy for use in molecular dynamics simulations. These NN are high-dimensional representations of atomic environments where parameters are fitted iteratively (trained) on large sets of curated electronic structure calculation results (training sets). They also incorporated the principle of near-sightedness (where the energy and forces of a configuration of atoms are assumed to be dominated by the contributions of the atoms that are closest) by using descriptors of the local atomic structure in the form of short-ranged atom-centered symmetry functions (ACSF). 

The development of Gaussian Approximation Potentials (GAP)~\cite{Bartok_PRL_2010} was the next major step forward expressing the local energy in terms of a Gaussian process kernel and using the bispectrum (a three-point correlation function) components as descriptors of the neighborhood of each atom. Spectral neighbor analysis potentials (SNAP)~\cite{THOMPSON2015316} used the same bispectrum components as descriptors of the atomic environment but with the energy expressed as a linear combination instead of a Gaussian process kernel resulting in a more scalable computational approach (to billions of atoms)~\cite{Nguyen-Cong2021}. 

The atomic cluster expansion (ACE)~\cite{Drautz_PRB_2019} as a general framework introduced a systematically improvable set of correlation functions for the representation of atomic environments. The development of equivariant Message-Passing neural networks (MPNN) such as MACE~\cite{Batatia_NIPS_2022} or NequIP~\cite{Batzner_NatCommu_2022} was an important milestone. These models represent structures as graphs where atoms are nodes and message-passing operations between nodes results in a more information-rich representation of atomic environments. 
Recent iterations of these models allow for the accurate simulation of millions of atoms~\cite{Musaelian_NatCommu_2023}. 

The combination of these computationally efficient MLIP and High Performance Computing resources enables the calculation of properties such as the Equation-of-State, structure factors, diffusion coefficients, viscosity, as well as constitutive properties such as material strength. They allow for the simulation of materials over timescales comparable to the experimental timescale (10s of ns) in the case of dynamic compression experiments, providing invaluable information to the study of the kinetics of phase transitions~\cite{Nguyen-Cong2024} in the WDM regime.  

By fitting MLIPs to electronic structure energies and forces, a massive speedup compared to AIMD is achieved, but in a sense the electrons are "lost". When the temperature is sufficiently high, a significant fraction of the electrons are thermally excited. In Mermin finite-temperature DFT, this effect is approximated using a Fermi distribution of the occupation numbers of Kohn-Sham orbitals and this electronic entropy has an impact on the forces. For MLIPs to adequately reproduce forces and free energies over a large temperature range, it is important to 
include this electronic entropy. In simulations of the {\em equilibrium state}, this can be addressed by fitting a different MLIP for different temperatures. 
This approach is taken by Kumar {\em et al.} who used a GAP potential (trained on-the-fly) to accelerate DFT-MD calculations of Hugoniot curves~\cite{Kumar_PoP_2024a} and transport properties~\cite{Kumar_PoP_2024b}. Another approach is to augment the structure-based descriptors used in these ML schemes with an electronic temperature based descriptor, see e.g., Zhang {\em et al.}~\cite{20PP-Yuzhi} where a temperature (corresponding to the Fermi smearing used) is included as another feature for a deep neural network (DNN) potential. Ben Mahmoud {\em et al.}~\cite{BenMahmoud_PRB_2022} and Fiedler {\em et al.}~\cite{fiedler_machine_2023} learn the local electronic density of state to augment the structural information and thus render models that are transferable over a certain temperature range.

\subsection*{Challenges \& outlook for MLIPs}

One of the main challenges in constructing a MLIP remains the preparation of adequate training sets. 
One would prefer to operate under interpolation rather than extrapolation, and concepts such as D-optimality~\cite{Podryabinkin_CMS_2017} and information entropy~\cite{schwalbekoda_NatCommun_2025} offer avenues worth investigating to better ascertain this. Uncertainty quantification in general will become very important when using these complex ML models to make predictions of material properties at scales larger than that at which ground-truth electronic structure simulations can be performed.

Inclusion of the electronic temperature in the MLIP in some way, opens the door to the development of Two-Temperature models (TTM) to enable the study of certain non-equilibrium scenarios, see Sec.~\ref{section_23}. In addition, methods to tackle large-scale, spatial temperature gradients still need to be developed for direct simulations of shocks crossing from condensed matter through WDM into the plasma regime.   

Note that the vast majority of MLIPs have been trained on results from DFT-based electronic structure calculations. Since we do not yet have the "exact" exchange-correlation functional there is a fundamental uncertainty that remains and in certain cases a more accurate ground-truth will need to be considered. 

%

\subsection*{Variational free energy approach}

For the purpose of reaching higher accuracy in the simulation in the sign-problematic WDM region, a promising routine is the deep generative model-based variational free energy approach~\cite{SciPostPhys.14.6.154}. Here, one minimizes the variational free energy 
\begin{equation}
F[\rho] = \mathrm{Tr}(H \rho) + k_B T  \mathrm{Tr}(\rho \ln \rho), 
\label{eq:vfe}
\end{equation}
where $\rho$ is the variational density matrix which should be hermitian, positive and normalized. Ref.~\cite{SciPostPhys.14.6.154} expresses the variational density matrix as 
$\rho = \sum _{n} U |n \rangle p_n \langle n| U^\dagger $, where $|n\rangle$ is the many-body base state such as the plane wave states, $p_n$ is a probabilistic model for the state occupation probability, and finally the unitary transformation $U$ is implemented with a flow model~\cite{papamakarios2021normalizing}. 
Therefore, one minimizes the variational free energy by jointly optimizing the two models expressed as generative neural networks. Different from commonly used machine learning approaches, the variational free energy approach does not require training data from another solver. Moreover, the approximation involved in the approach is variational bias due to expressibility and optimization, rather than generalization error. 
Overall, as a finite temperature formulation of the variational Monte Carlo method, the approach scales quartically with the number of electrons in the simulation making physically interesting parameter regions accessible. 
Further development of the approach calls for extension to spinful cases and adequate treatment of the electron-electron cusp condition. One has also explored extensions of the approach in dense hydrogen systems~\cite{Xie_PRL_2023, li2025deep} related to problems in Sec.~\ref{section_16}.

\subsection*{Surrogates and reduced models}

One of the earliest and most accessible applications of machine learning in computational HED physics is the construction of surrogate or reduced models. These aim to replace parts -- or all -— of a costly simulation with a simpler, data-trained approximation. While the training process can be demanding, once built, such models are fast and inexpensive to query. Their usefulness depends largely on the complexity of the underlying physics: simple systems require less data to model accurately, while complex systems—where small input changes produce large output variations—demand much more extensive and carefully curated training sets.

The use of reduced models in simulations is not new: early examples include data tables, spline fits, and interpolation schemes that are still common in large-scale plasma codes. What has changed is the widespread availability of standardized machine learning libraries, such as Scikit-learn, SciPy, NumPy, and Pandas, which now enable more advanced and accessible model-building strategies~\cite{osti_1971305}. In parallel, the rise of deep learning frameworks like PyTorch, TensorFlow, Keras, and JAX has expanded the toolkit for building more powerful surrogates. Although these models are more computationally intensive, they offer greater representational capacity and generalize more effectively across diverse input types, from waveforms to multidimensional tensors. As a result, surrogate models built with these tools can now approach, and in some cases rival, the accuracy of full-scale simulations~\cite{Fiuza:2022}.

The use case for surrogates is broad and they have been successfully deployed in several areas, with a particular emphasis on inertial fusion applications where the need is most urgent, given the huge cost for at-scale computational models~\cite{Spears:2020}.
Building on this foundation, transfer learning techniques were used to integrate both simulation and experimental data into unified predictive models~\cite{Humbird_2021}, allowing for more accurate performance. These models have since been extended to address target design and optimization tasks~\cite{Gammel:2024}, as well as uncertainty quantification and ignition prediction~\cite{Gaffney:2024}, providing actionable insights for experiment planning. 

One of the computational bottlenecks in ICF simulations lies in the modeling of non-local thermodynamic equilibrium (NLTE) processes. To mitigate this, neural network surrogates have been proposed for NLTE spectral opacity calculations, significantly accelerating radiative transfer modeling while maintaining accuracy~\cite{Kluth:2020,Vander:2023}. 
More recently, surrogate models based on graph neural networks have been developed to emulate kinetic plasma solvers, replacing traditional particle-in-cell simulations with fast, differentiable approximations~\cite{Carvalho:2024}. In addition to task-specific models, general-purpose DNN surrogates have been applied to a wide range of plasma regimes~\cite{Kasim:2022}, including the emulation of equations of state for ICF simulations~\cite{Mentzer:2023}. To further improve performance on sparse and high-dimensional data, novel transformer-based architectures incorporating graph-informed hyperparameter optimization have also been proposed~\cite{Olson:2024}, pointing to a growing trend in hybrid models that balance interpretability, speed, and physical fidelity.

Surrogate models generally tend to perform well when used as interpolators of the physics on which they were trained. In contrast, when used for extrapolating they can fail dramatically, and bounding the prediction inaccuracies can be a challenge unless the architecture has a strong and appropriate inductive bias, good regularization, or has been otherwise constrained in training to learn the underlying physics itself. Interestingly, such models do exist and will be touched on briefly in what follows.

\subsection*{Machine learning of electronic structure}

Machine learning has increasingly been applied directly to electronic structure calculations, too. Standard DFT scales cubically with system size, which limits its applicability to large systems. Recent developments aim to bypass the expensive self-consistent field iterations altogether by learning the mapping from atomic structure to electronic structure quantities.

One notable approach is to learn the DFT Hamiltonian matrix directly~\cite{Gong_NatCommun_2023,Zhong_npjCompMat_2023}. These methods use equivariant MPNNs to predict the Hamiltonian in a localized basis, exploiting locality and rotational covariance. Once trained, these models can build up the DFT Hamiltonian matrix directly, enabling electronic structure predictions for systems with thousands of atoms at near-DFT accuracy.

Another strategy focuses on learning the electronic density or the one-electron reduced density matrix. The Materials Learning Algorithms (MALA) framework~\cite{fiedler_machine_2023,fiedler_npj_23,Rackers_MachLearnSciTech_2023,Fiedler_CPC_2025} predicts the local density of states from atomic descriptors, from which all ground-state properties can be derived. Importantly, MALA has demonstrated transferability across phase boundaries and electronic temperatures, making it particularly relevant for WDM applications. Related work has shown that machine learning models based on the one-electron reduced density matrix can generate surrogate electronic structure methods for a range of theories from DFT to full configuration interaction~\cite{Shao_NatCommun_2023}.

Machine learning has also been applied to predict spectroscopic properties. Graph-based neural networks have been used to predict X-ray absorption near-edge structure (XANES) spectra to quantitative accuracy~\cite{Carbone_PRL_2020}, enabling high-throughput spectral sampling across material configuration space. Such approaches are valuable for interpreting experimental diagnostics in WDM research.

\subsection*{Automatic differentiation and hybrid simulations}

The advent of software packages that natively support automatic differentiation (AD) marks a significant development in computational HED science. 
Unlike symbolic or numerical differentiation, AD computes exact derivatives up to machine precision and scales well with complex functions and high-dimensional inputs. Essential for training neural networks, AD is now increasingly used in scientific computing, from the use in libraries for differentiable functionals~\cite{kasim2020xitorch} to powering plasma physics frameworks built in Julia~\cite{juliaplasma2024} and JAX~\cite{torax2024arxiv}.

ADs real impact lies in enabling backpropagation, which allows neural networks to be embedded and trained within larger simulation frameworks~\cite{PhysRevLett.127.126403}. This capability supports  experimental diagnostics such as robust deconvolution of noisy experimental data~\cite{Forte:2024}, and facilitates more efficient implementations of optimization, control~\cite{torax2024arxiv}, uncertainty quantification~\cite{SHIN2023112183}, and inverse problem solving~\cite{PhysRevApplied.20.044017, Kasim_POP_2019, Karim:2024}. It also supports the development of differentiable solvers for transport and other plasma physics applications~\cite{JoglekarThomas2022}. Though still emerging, the integration of AD into scientific computing holds significant promise for expanding the power and flexibility of future simulations.


\newpage 
\clearpage

\section{Databases and open source software}\label{section_25}
\author{Michael Bussmann$^{1}$, Jean Cl\'erouin$^{2}$, Michael S. Murillo$^{3}$, Fran\c{c}ois Soubiran$^{2}$}
\address{
$^1$Helmholtz-Zentrum Dresden-Rossendorf (HZDR), D-01328 Dresden, Germany \\
$^2$Universit\'e Paris-Saclay, CEA, Laboratoire Mati\`ere sous conditions extr\^emes, 91680 Bruy\`eres-le-Ch\^atel, France
$^3$Michigan State University, East Lansing, MI 48824, United States 
}

\subsection*{Introduction}

No single experiment can capture the entire complexity of WDM processes, which span from atomic processes to hydrodynamics of ICF capsules, and from femtosecond laser interactions to nanosecond implosions or planetary interiors. This necessitates multiple codes benchmarked against various experiments, with clear understanding of their capabilities and their limitations. This multi-scale, multi-physics nature presents unique challenges for data science~\cite{2023_bremer_et_al}: a smaller research community, costly extreme-condition experiments, and the need to integrate simulations across different physical regimes. Nevertheless, modern data science approaches offer significant opportunities to accelerate WDM research, especially with increasing compute power and higher-repetition laser systems. Systematic data management enables better leveraging of limited experimental data, more robust model validation, and knowledge transfer between disciplines.

The FAIR principles—Findability, Accessibility, Interoperability, and Reusability—~\cite{Wilkinson2016}, provide a framework for maximizing research data value. For WDM, these principles translate to:
\begin{itemize}
\item {\em {\bf F}indability}: Documenting experimental conditions (temperature, density, composition) and computational parameters with rich metadata. Hugoniot datasets need standardized schemes for searchability.
\item {\em {\bf A}ccessibility}: Providing programmatic access through APIs alongside human interfaces, evolving beyond basic online availability.
\item {\em {\bf I}nteroperability}: Addressing format fragmentation between equation of state data, transport properties, and spectroscopic measurements.
\item {\em {\bf R}eusability}: Ensuring clear licensing and provenance, especially for computational results where methodological choices strongly impact outcomes.
\end{itemize}

While the scientific literature contains rich experimental data, these are rarely compiled into unified datasets. Modern materials database architectures have evolved beyond file collections~\cite{Chen2014} to incorporate programmatic interfaces, federated storage systems, and knowledge graphs with machine learning. The Materials Project~\cite{Jain2013}, AFLOW~\cite{AFLOW2017}, and NOMAD~\cite{Draxl2019} exemplify these advances, combining web accessibility with programmatic interfaces. 

Metadata standards and ontologies~\cite{Ashino2010}, though not yet widely adopted in WDM research, offer significant benefits for addressing the field's complex data challenges. For WDM, where small variations in conditions significantly affect results, comprehensive metadata is essential for reproducibility. Computational codes require precisely documented inputs: DFT-based codes need exchange-correlation functionals and pseudopotentials, pair-potential MD codes rely on interatomic potentials~\cite{Stanek2021,DharmaWardana_PRE_2022}, and hydrocodes require accurate equations of state~\cite{Gaffney_HEDP_2018} and transport properties~\cite{Stanek_PoP_2024,Murillo_FrontiersinPhysics_2022, Johnson_PoP_2024}. WDM data standards face unique challenges requiring extensions beyond conventional materials science to capture:
\begin{itemize}
\item extreme states including non-equilibrium conditions and ultrafast dynamics, 
\item provenance tracking for processed data (e.g., X-ray Thomson scattering measurements), 
\item uncertainty representation for measurements made under challenging conditions. 
\end{itemize}
Developing WDM-specific standards would enable more effective data sharing, improve reproducibility, and facilitate advanced data science applications to the field's unique challenges.

\subsection*{State of the art}

\subsubsection*{Historical Database Development}

The early WDM databases emerged from American and Russian national laboratory experiments (1950s-1980s), available in the American report~\cite{Marsh_LASL_1980} and the Russian {\sc Rusbank}~\cite{Bushman_Rusbank_2000}. Based on shock experiments and Hugoniot relations, these data provide pressure as a function of compression, with pressures up to 10$^5$ GPa from nuclear experiments, now extending to analytical models reaching 10$^6$ GPa~\cite{Burakovski_JAP_2022}. Complementary ambient-temperature data comes from static compression experiments, with unified formats for sharing crystal structures through databases like COD~\cite{COD}. These historical datasets provide invaluable benchmarks despite their limited scope.

\subsubsection*{Equation of State Databases}

Theoretical approaches combining Thomas-Fermi electronic models with ionic models have yielded more complete high-pressure matter descriptions, including QEOS (Livermore~\cite{More_PhysFluids_1988}), SESAME (Los Alamos~\cite{Lyon_Report_1990}), and REOS (Rostock~\cite{Becker_AJSS_2014}). The recent first-principles EOS database by Militzer \textit{et al.}~\cite{FPEOS}, based on PIMC and DFT calculations, covers elements up to Ar with accompanying analysis tools. Developments should focus on database interoperability and uncertainty quantification.

\subsubsection*{Transport Properties and Specialized Databases}

Beyond EOS, electronic transport properties (ionization, conductivities) have been measured through pulsed discharge experiments~\cite{Benage_PRL_1999,Desilva_PRE_2011}, creating databases for expanded plasmas~\cite{Murillo_FrontiersinPhysics_2022}. Notable is a database from isochoric pulsed discharge experiments for Ni, Au, Al, Ag, Ti, Cu, Si at few-eV temperatures~\cite{Clerouin_PoP_2012}, covering expanded plasma conditions where theoretical evaluations diverge.

\subsubsection*{XFEL Revolution in WDM Data}

X-ray free-electron lasers have transformed WDM probing through XRTS spectra analysis (see Sec.\ref{section_13}), thus providing access to electron density and temperatures. These facilities create data management challenges:
\begin{itemize}
\item greater volume and complexity of experimental data,
\item need for sophisticated computational workflows,
\item complex integration with existing databases requiring careful metadata standardization.
\end{itemize}

\subsubsection*{Open Source Development Models}

WDM code development demands synergy between experimental, numerical, and theoretical communities. Open-source approaches serve as both quality assurance and innovation catalyst—as demonstrated by Abinit~\cite{Gonze2020}, whose community-driven development has yielded remarkable capabilities over time.

While proprietary codes remain common (VASP \cite{Kresse_1993}, OFMD~\cite{Lambert_2013}, DRAGON~\cite{MIHAYLOV2024108931}), open-source alternatives are gaining traction: Abinit~\cite{Gonze2020}, Quantum Espresso~\cite{Giannozzi_2009}, and SHRED~\cite{Sharma_PRE_2023} for ab initio calculations; Atomec~\cite{callow_2023_10090435} for average-atom models; LAMMPS~\cite{thompson2022} for precomputed-potential-based simulations; and MULTI2D~\cite{RAMIS2009977} and FLASH~\cite{Plewa2003} for hydrodynamics. Open-source particle-in-cell simulations~\cite{2010_Burau_et_al,2018_Derouillat_et_al,2015_Arber_et_al} increasingly bridge molecular dynamics and fluid simulations~\cite{2024_randolph_et_al,2023_Kluge_et_al}, despite microphysics implementation challenges~\cite{2024_Banjafar_et_al}. Future development demands unified, community-wide approaches that transcend private implementations.

The {\tt libxc} library~\cite{libxc_2018} exemplifies successful standardization, now integrated across most KS-DFT codes, while openPMD~\cite{2024_williams_et_al} establishes crucial data standards for analytics, visualization, and code coupling.

\subsubsection*{Code Coupling and Interoperability}

Recent breakthroughs in coupling large-scale plasma simulations with machine learning on Exascale machines~\cite{2025_kelling_et_al} are driving multi-scale, multi-physics workflows with unprecedented capabilities.

Temperature scalability remains a critical challenge for WDM simulations. The extended method bypasses the \emph{orbital wall}~\cite{Blanchet_pop_2020} by replacing high-temperature orbitals with plane waves~\cite{Zhang_POP_2016,Blanchet_cpc_2022}, while density matrix approaches exploit temperature-induced localization~\cite{Suryanarayana_ChemPhysLet_2017} in codes like SQDFT, PIMC, and SHRED. However, these methods' inefficiency at low temperatures necessitates coupling with KS-DFT~\cite{Bethkenhagen_2020}, potentially causing database discontinuities. The fundamental challenge lies in creating seamless temperature transitions throughout the WDM regime. These interoperability challenges demand:
\begin{itemize}
\item Integration of diverse simulation results into unified databases.
\item Clear metadata indicating simulation methodologies.
\item Rigorous validation across method transitions.
\item Surrogate models connecting disparate scales in multi-physics simulations.
\end{itemize}

\subsubsection*{Workflow Management and Reproducibility}

Modern computational research increasingly relies on sophisticated workflow management to ensure reproducibility and facilitate collaboration. While the WDM community is beginning to adopt these approaches, further development is needed:
\begin{itemize}
\item {\em Containerization technologies} enable consistent environments for simulation and analysis.
\item {\em Workflow managers} can automate complex simulation campaigns orchestrating the use of various codes.
\item {\em Workflow versioning and sharing} support increase in workflow quality, reuse and reproducibility.
\item {\em Provenance tracking} systems document the full history of how results were generated.
\item {\em Version control} for both code and data ensures transparency and reproducibility as well as connects continuous development and integration with machine learning operations.
\item {\em In-memory coupling of simulation codes} enables complex multi-scale, multi-physics simulations.
\end{itemize}
These tools are particularly valuable for WDM research, where complex multi-step simulations are common and reproducibility challenges can arise from the variety of computational approaches employed.

\subsection*{Challenges \& outlook}

\subsubsection*{Machine Learning Applications in WDM}

Machine learning is clearly a promising way for the simulation of WDM systems (see Sec.\ref{section_24}). In particular, ML can be used to manage, analyze, and predict properties from diverse datasets of different origins. For example, multifidelity regression methods and neural network approaches have already been successfully applied to interpolate transport data and predict conductivities \cite{Stanek_PRE_2021}. The application of machine learning to WDM presents unique opportunities:
\begin{itemize}
\item {\em Interpolation in sparse data regions} where experimental access is limited \cite{Stanek_PRE_2021}.
\item {\em Compression} of high-dimensional data.
\item {\em Surrogate models} to reduce the computational expense of DFT-MD and PIMC calculations.
\item {\em Physics-informed neural networks and operators} that respect underlying physical constraints.
\item {\em Uncertainty quantification, anomaly detection and out of distribution detection} in regions far from available training data.
\item {\em Foundation models and knowledge graphs} that connect data from both simulation and experiments in a multi-scale, multi-physics view.
\item {\em Closed-loop experiments} that are informed by models for intelligent coverage of parameter spaces and that feed back into databases, closing the loop between systematic experimental campaigns and data-augmented modeling.
\end{itemize}
As the field develops, maintaining and curating well-structured databases with clear metadata and connecting these data bases becomes increasingly important to enable effective machine learning applications.

\subsubsection*{High-Throughput Analysis Frameworks}

It is evident that large facilities such as XFELs will play a major role in the production of data in the WDM regime. But there is still some room for experiments at much smaller scale that can produce data at high rate and for a small cost. A miniaturization of the aforementioned pulsed approach, inspired by Russian work \cite{Clerouin_PRB_2008}, was recently developed \cite{Jodar_RSI_2024}, which should allow for the creation of a much more complete database in the expanded regime in the near future. The targeted temperatures are $\sim10$'seV for pressures $\sim10$'sGPa for a much more efficient production of data. Solutions like openPMD already provide complex in-memory workflows and code coupling for simulations and experiments~\cite{2022_poeschel_et_al}.

In light of the high-throughput of experimental campaigns at EuXFEL, ESRF and ELI, the development of shared scripts, analysis tools and workflows that are emerging to handle large volume of data is of particular importance. The development of these high-throughput analysis frameworks represents a critical frontier for WDM data science:
\begin{itemize}
    \item {\em Real-time analysis pipelines} enable feedback during experimental campaigns.
\item {\em Standardized analysis tools} ensure consistency across different facilities.
\item {\em Community-developed libraries} leverage expertise from the broader scientific computing community.
\item {\em Scalable computing resources} through cloud or HPC integration support processing of increasingly large datasets.
\end{itemize}

\subsubsection*{Toward Integrated WDM Data Infrastructure}

Looking forward, the WDM community has an opportunity to develop more integrated data infrastructure that connects experimental results, simulation outputs, and analytical tools. The outline of how such a facility could look like was recently presented in a whitepaper of US and European research facilities~\cite{2023_bremer_et_al}. Key components of this vision include:
\begin{itemize}
    \item {\em Community database standards} that enable seamless data exchange.
\item {\em Integrated analysis environments} that combine data access with visualization and processing tools.
\item {\em Cross-validation frameworks} that systematically compare different computational approaches with experimental benchmarks.
\item {\em Collaborative platforms} that facilitate sharing of both data and analysis methods.
\end{itemize}

Laser and photon science facilities currently develop and provide data lifecycle management pipelines and automate experiments and beamlines. Here we foresee future opportunities of accelerating warm dense matter science.

By balancing standardization with innovation, the WDM field can develop data infrastructure that accelerates discovery while respecting the unique challenges of research at extreme conditions.

\subsubsection*{A Community Vision for WDM Data Science}

The warm dense matter community stands at a pivotal juncture, with unprecedented potential for advancement through XFEL facilities, improved computational methods, and emerging data science approaches. Future progress hinges on three key priorities:

\begin{itemize}
\item {\em Standardization}: Adopting FAIR principles and developing WDM-specific metadata standards that address extreme conditions research.
\item {\em Openness}: Expanding open-source software development to enhance transparency and reproducibility.
\item {\em Integration}: Combining physics-based models with machine learning approaches while building consensus around data practices.
\end{itemize}
By merging experimental innovation with modern data science, the WDM community can accelerate discovery despite the inherent challenges of extreme conditions research.

\newpage 
\clearpage
\section{Future facilities}\label{section_26}
\author{Siegfried Glenzer$^{1}$, Nicholas J. Hartley$^{1}$, Bob Nagler$^{1}$, Paul Neumayer$^{2}$, Ulf Zastrau$^{3}$}
\address{
$^1$SLAC National Accelerator Laboratory, Menlo Park, CA, United States \\
$^2$GSI Helmholtzzentrum f\"ur Schwerionenforschung GmbH, 64291 Darmstadt, Germany \\
$^3$European XFEL, Holzkoppel 4, 22869 Schenefeld, Germany
}

\subsection*{Introduction}

The need for high energy input to create WDM and the need for energetic requirements to diagnose WDM states have, over time, led to large scale facilities of drivers (lasers, gas guns, pinches) and diagnostics (XFELS, synchrotrons, long and short-pulse optical lasers for x-ray probing). It was soon realized that lasers for creating warm dense matter need different specifications in energy, pulse duration, or wavelength than lasers for diagnostics. The pioneering first XRTS experiments at the OMEGA~\cite{glenzer2003,glenzer_PRL_2007,Lee_PRL_2009,Kritcher_PRL_2011}, the Titan~\cite{Kritcher_science_2008}, and  VULCAN laser facilities~\cite{GarciaSaiz2008} were thus soon augmented by dedicated experimental facilities at x-ray free electron lasers where high-energy lasers were installed additionally.

Due to the short lived nature of WDM states in the laboratory, experimentalists are faced with a number of difficulties like timing of pump and probe on femto- to picosecond timescales or the limited signal-to-noise ratio of a single x-ray scattering event. This increases the requirements on the facilities in terms of spatio-temporal stability of the delivered (x-ray) laser light, repetition rates, detector (temporal) resolution, as well as data intake and storage.

Modern High Energy Density (HED) endstations at XFEL facilities perform better than 1\,Hz pump-probe experiments, accumulating data from 1000s of shots for the final spectrum. However, the highest pressure conditions are currently only available at Sandia’s Z-pinch~\cite{Matzen2005}, the Omega laser~\cite{Gopalaswamy2019}, and the National Ignition Facility~\cite{Moses_NIF} approaching 100s of GBar. Here, laser-produced X-ray sources with 2-18\,keV photon energies have been extensively used to probe WDM conditions. Relatively soft x-rays from the Janus laser backlighting a Si$_3$N$_4$ foil successfully probed the dissociation-induced metallization of dynamically compressed deuterium with x-ray scattering observing the Compton and plasmon scattering features at temperatures of 0.15\,eV and densities exceeding $2\times10^{20}$~cm$^{-3}$~\cite{Davis2016}. At the Omega laser, WDM experiments developed the 9\,keV Zn x-ray probe delivering x-rays for the in-flight capsule adiabat measurements in laser-driven plastic and beryllium implosions validating radiation-hydrodynamic simulations for capsule adiabats of $1<\alpha<10$ and electron densities exceeding $10^{24}$~cm$^{-3}$ for the first time~\cite{Kritcher_PRL_2011}. The same 9\,keV x-ray source was subsequently used to measure the adiabatic index in Be compressed by counter-propagating shocks demonstrating convergence to the ideal gas limit at $6\times$ compressed Be densities~\cite{Fortmann_PRL_2012}. The highest x-ray energies of 17.9\,keV were also employed at the Omega laser using laser-produced molybdenum x-rays to resolve strong ion-ion correlations in shock-compressed aluminum~\cite{Ma_PRL_2011}. However, the reduced spectral resolution limited the analysis of the scattering spectra to intensity ratios of elastic to inelastic features. Finally, the 9\,keV laser-produced zinc x-ray source was subsequently introduced at NIF and employed in the ignition program~\cite{Glenzer_science_2010}. In 2012, the NIF Discovery Science program was launched. Since then, laser-produced 9,10\,keV x-rays have been employed as backlight and scattering probes of the equation of state at gigabar pressures and the thermodynamic properties of extreme WDM conditions, e.g., in hohlraums~\cite{Kritcher2020} or with the new CPS platform~\cite{MacDonald_POP_2023}. Recent measurements show that the intensity of the elastic scattering feature of x-ray Thomson scattering spectra observe the onset of pressure-driven K-shell delocalization~\cite{Doeppner_nature_2023}.

With the success of ignition and demonstration of burning fusion plasmas on the NIF~\cite{ICF_PRL_2024,Zylstra_nature_2022}, present and future x-ray laser facilities will play an important role to study WDM with the goal to support advances towards higher fusion yield. Here, high repetition rates will allow accurate tests of the hydrodynamic evolution at foot-pressure conditions of fusion capsule implosions including measurements of ion and electron temperatures, densities and ionization states. It is then only natural to seek to further develop and upgrade these facilities to meet the need for high-signal-to noise data acquisition and to push boundaries in achievable pressures and temperatures. Last but not least, a goal of such upgrades should always be to make experiments at WDM conditions more affordable, either by higher reliability and higher repetition rates that require less time per experiment or by downsizing lasers and x-ray sources (at constant power, energies, and achievable conditions) so that more WDM experimental setups can be built at smaller facilities.

\subsection*{State of the art}

In the fall of 2009, the Linac Coherent Light Source (LCLS), the first X-ray free electron laser, started operation~\cite{emma2010}. Since then, free electron lasers have truly become game changers in HED science. The X-ray beam’s unparalleled brilliance, tunable photon energy, and high photon flux allow for unique investigations of matter through interactions that require precise energy resolution or have inherently low cross-sections. Furthermore, the intense, ultrafast X-ray pulses can isochorically heat solid-density materials on micron scales, driving them to plasma temperatures spanning from eVs to keVs. The Matter in Extreme Conditions (MEC) instrument was developed to harness the distinctive capabilities of LCLS for HED science\,\cite{nagler2015matter, Glenzer_JPhysB_2016}.  Since its commissioning in 2012, MEC has facilitated pioneering research leading to key scientific breakthroughs, specifically in dynamic shock compression studies~\cite{Wehrenberg2017}, planetary and earth sciences~\cite{Gleason2015}, relativistic electron flows in solid density matter~\cite{Kluge2018} and WDM research~\cite{Fletcher_NP_2015,mabey_2017}.

The HED scientific instrument~\cite{Zastrau_JSyncRad_2021}, complemented by the HIBEF user consortium, located at the European X-ray Free-Electron Laser Facility GmbH (European XFEL)~\cite{Tschentscher_2017, decking2020mhz} is a unique platform for experiments combining hard X-ray free-electron laser radiation with the capability to generate matter at extreme conditions of pressure, temperature, or electromagnetic fields. The instrument started user operation in May 2019 and has augmented its capabilities within the last years. In general, experiments at HED-HIBEF combine hard X-ray FEL radiation and the capability to generate matter under extreme conditions of pressure, temperature or electric field using the FEL, high energy optical lasers, and diamond anvil cells. Scientific applications comprise studies of matter occurring inside exoplanets, of new extreme-pressure phases and solid-density plasmas, and of structural phase transitions of complex solids in high magnetic fields. The first user experiment took place in May 2019. Several additional capabilities have been meanwhile commissioned such as focusing, spectrometers, monochromators, sample environments. The HIBEF user consortium has contributed a second target chamber, a diamond anvil cell platform, the RE.LA.X and DiPOLE laser systems, and a laser-shock platform.

\subsection*{Challenges \& outlook}

\subsubsection*{MEC-U at the LCLS}

A major upgrade of the MEC instrument at LCLS is being planned. The project has not been formally baselined and remains in the design stage. An overview of the plans for the MEC-U project follows. Further details are available in Ref.~\cite{dyer2023mec}. The high repetition rate laser upgrade will feature a repetition rate ranging from shot-on-demand to $10\,$Hz. In petawatt output mode, the laser will operate at a central wavelength of approximately $800\,$nm, delivering $150\,$J per pulse with a pulse duration of $150\,$fs. The focusing capabilities will achieve spot size of $5\,\mu$m with peak intensity exceeding $5\times10^{20}$\,W/cm$^2$. Additionally, the system will support a nanosecond output mode at $527\,$nm ($2\omega$) with an energy of $150\,$J per pulse. In this mode, the laser pulse shape will be programmable with durations ranging from 3 to $35\,$ns. Focusing will be facilitated by a phase plate, producing a flat-top profile with a diameter between $150\,\mu$m and $1\,$mm.

A second laser system will consist of a high energy long pulse laser, with a shot rate of one shot per 30 min. The central wavelength will be $527\,$nm ($2\omega$), with a potential consideration for operation at approximately $351\,$nm ($3\omega$). The system will deliver a pulse energy of $1\,$kJ for a $5\,$ns pulse, with a pulse profile that can be shaped between 3 and $30\,$ns. Focusing will be achieved using a phase plate, yielding a spot size ranging from $300\,\mu$m to 
$2\,$mm. A potential later upgrade would increase the energy to $5\,$kJ in four beams at $3\omega$.

The experimental systems will include a $4.5\,$m partial sphere chamber, load-locked diagnostic inserters, and a load-lock target inserter. At a later stage, an expansion of the suite of diagnostics with up to 18 deployment locations is envisaged. Additionally, the facility aims to support simultaneous multi-view imaging for dynamic single-shot tomography, as well as the inclusion of small downstream interaction chambers, such as those used for high explosives or dynamic diamond anvil cell (DAC) experiments.

\subsubsection*{Facility for Antiproton \& Ion Research (FAIR)}

FAIR, an international large-scale ion accelerator facility that is currently under construction alongside the GSI Helmholtzzentrum für Schwerionenforschung near Darmstadt/Germany, is envisioned to serve various scientific fields, ranging from the study of compressed baryonic matter over radioactive isotopes far off stability to atomic physics at extreme fields, radiation biology, materials science and HED plasma physics. Concerning HED research, intense heavy-ion pulses up to $5\times10^{11}$ uranium ions per bunch, delivered by the new superconducting heavy-ion synchrotron SIS-100, will be compressed to below $100\,$ns and focused to mm spot sizes. This will yield a volumetric energy deposition of $\sim100\,$kJ/g, with resulting temperatures of $\sim10\,$eV, offering a complementary approach to WDM generation, with unique features such as mm$^3$ sample sizes and homogeneous near-LTE conditions. Various schemes have been devised to produce expanded high-entropy matter states, or high-pressure states by near-isentropic cylindrical compression. Planning is ongoing for the addition of a multi-kJ laser facility that would enable laser-driven x-ray backlighting and to provide state-of-the-art x-ray diagnostics. The international HED@FAIR collaboration is defining the scientific program and is preparing the experimental infrastructure and diagnostic equipment to allow exploiting the unique possibilities for HED research~\cite{Schoenberg_PoP_2020}. Note that with \textit{HIAF}, a similar heavy-ion facility is being constructed in China~\cite{Yang_NIMB_2013}, which also plans an extensive HED program~\cite{Cheng_MRE_2018}.

Intense pulses of relativistic protons and light ions from the FAIR synchrotron will also provide new opportunities for diagnosing dense samples at extreme conditions. The PRIOR-II proton microscope for FAIR has recently been commissioned using protons from the SIS-18 accelerator~\cite{Schanz_RSI_2024}. Using a system of magnetic lenses, PRIOR-II provides imaging with a spatial resolution down to $10\,\mu$m, at exposure times down to $10\,$ns. The density contrast results from both direct nuclear collisions, and by introducing a collimator in the Fourier plane to suppress large scattering angles due to multiple Coulomb collisions. Particle radiography with sub-percent density resolution and centimeter field-of-view will be available. In order to produce such large samples at extreme conditions, a confinement chamber is currently under construction, which will allow the use of up to $250\,$g TNT (equivalent), resulting in pressures up to $\sim1\,$Mbar in cubic centimeter volumes. This can produce strongly-coupled plasmas by shock-compression of high-Z gases, facilitate studies of the EOS of porous or granular materials (relevant for meteorite impacts and asteroid deflection schemes), measurements of material strength and failure at extreme conditions (to test space craft shielding, or better understand explosive forming techniques), or provisions of in-situ diagnostics for the bulk synthesis of novel materials.

\subsubsection*{Chinese HED XFEL facilities}

Two x-ray free electron laser facilities are under construction. The Shenzhen Soft X-ray FEL (S3FEL) facility was approved in 2023 and will operate at the soft x-ray and VUV spectrum ($\sim1-30~$nm). Currently, there are no plans for an HED-like endstation. The Shanghai X-ray Free Electron Laser (SHINE), launched in 2018, is based on an 8~GeV CW superconducting linac and plans to have 3 undulator lines and 10 experimental stations in phase-I, covering the photon energy range of $0.4–25\,$keV.

The SHINE scientific instruments will be build in three FEL phases. The portfolio of FEL-I end-stations includes the "Station of Extreme Light (SEL)".  SEL constitutes an important part of this project, in which the powerful light of the 100 PW-class laser (SEL-100 PW) will combine with the X-ray free laser (XFEL) for research on strong-field QED physics~\cite{wang202213}. In a future upgrade, FEL-III, will host the HED: High Energy Density science end-station. As of late-2024, the design parameters of the beamlines are frozen, the technical and engineering design of FEL-I/II/III and three endstations (AMO, SES, CDS) has been done, the engineering design for five endstations (SSS, HSS, HXS, SFX, CDE) has almost been completed. FEL-II has highest priority, and now the first experiments and science cases for the FEL-II endstations are in preparation. Among these 10 end-stations, the Station of Extreme Light (SEL), which combines the hard X-ray FEL with a 100 PW laser, aims at pioneering cutting-edge researches on strong field QED physics.

\subsubsection*{Further facilities}

The SPring-8 Angstrom Compact free electron LAser (SACLA), is an X-FEL located in Harima Science Garden City (Japan). For WDM research, it offers endstations with long pulse~\cite{Inubushi_2020} or short pulse~\cite{Yabuuchi_2019} laser, as well as a variety of special X-ray modes, such as seeding~\cite{Inoue_2019} and nanometer-scale KB mirror focusing~\cite{Yamada_2022}. Of particular interest is the two-color mode~\cite{Hara_2013}, which allows melting and disordering induced by an XFEL pulse to be probed by the following pulse~\cite{Inoue_2016} -- the probed intensities and delays cannot currently be delivered at other facilities.

The NIF laser system has reached its end of life and many parts should be replaced and upgraded. Dedicated programs to increase the laser energy output to $\sim 3$~MJ are under way~\cite{MACLAREN_2024,dinicola2024}. This includes replacement and upgrades to the power amplifiers and frequency conversion crystals. Overall, the lasers in the facility need to be hardened for the enhanced radiation passing through.

The Laser Megajoule (LMJ) in France is a multi-beam facility designed for indirect drive inertial confinement fusion experiments and research~\cite{Neauport_2024}. It had its first experimental campaign in 2014, its first deuterium shot in 2019 and is currently operating at a hundred kJ level in the frequency-tripled ultraviolet spectral range. It uses diffraction gratings to focus the laser beams onto the target, limiting the amount of the first and second harmonic light. It is designed to reach 660\,kJ in 2027 and 1.3\,MJ later. Finally, the PETAL stage will provide a short-pulse (0.5-10\,ps) 800\,J petawatt laser beam. For diagnostics, X-ray spectrometers and imagers with a spatial resolution $>15\,\mu$m are available. Optical and neutron diagnostics are also available.

In the mid- to far-future, laser-plasma acceleration of electrons and ions is expected to contribute to the creation and diagnostics of WDM~\cite{Albert_2021,Galletti2024}. Such accelerators are driven by high-power lasers that are smaller than their high-energy counterparts but can still deliver, e.g., ion beams capable of isochorically heating the target to several eVs. In addition, they are capable of delivering GeV electrons on cm-length scales (instead of the kilometer-long linear accelerators) and may be capable of shrinking the necessary undulator for a plasma-based XFEL~\cite{Wang2021,Pompili2022,Labat2023}.

\newpage 
\section*{Acknowledgments}

\subsubsection*{Frank Graziani} is supported by the Lawrence Livermore National Laboratory.  Portions of this work were performed under the auspices of the U.S. Department of Energy by Lawrence Livermore National Laboratory under contract DE-AC52-07NA27344. Lawrence Livermore National Security, LLC.\vspace{-0.7em}

\subsubsection*{David Riley} is supported by UK Engineering and Physical Sciences Research Council grants EP/N009487, EP/K009591 and EP/I031464.\vspace{-0.7em}

\subsubsection*{Andrew D. Baczewski} is partly supported by the US Department of Energy Office of Fusion Energy Sciences and Sandia National Laboratories' Laboratory Directed Research and Development (LDRD) Project No.\ 233196.
This work was performed, in part, at the Center for Integrated Nanotechnologies, an Office of Science User Facility operated for the U.S. Department of Energy (DOE) Office of Science. This article has been co-authored by employees of National Technology \& Engineering Solutions of Sandia, LLC under Contract No. DE-NA0003525 with the U.S. Department of Energy (DOE). The authors own all right, title and interest in and to the article and are solely responsible for its contents. The United States Government retains and the publisher, by accepting the article for publication, acknowledges that the United States Government retains a non-exclusive, paid-up, irrevocable, world-wide license to publish or reproduce the published form of this article or allow others to do so, for United States Government purposes. The DOE will provide public access to these results of federally sponsored research in accordance with the DOE Public Access Plan \url{https://www.energy.gov/downloads/doe-public-access-plan}.\vspace{-0.7em}

\subsubsection*{Isabelle Baraffe} is partly supported by STFC grant ST/V000721/1.\vspace{-0.7em} 

\subsubsection*{Mandy Bethkenhagen} is partly supported by the ANR-DFG project HENEHOP (Grant No. ANR-24-CE92-0060) and gratefully acknowledges computational time through the GENCI project GEN15731 (Grant No. A0170915731) hosted by the French computational centers TGCC and CINES as well as the German NHR projects mvk00106 and mvp00026.\vspace{-0.7em} 

\subsubsection*{Michael Bonitz} is supported by the DFG via grants BO 1366-13/2 and BO 1366-16.\vspace{-0.7em} 

\subsubsection*{Peter Celliers} is supported by Lawrence Livermore National Laboratory.  Portions of this work were performed under the auspices of the U.S. Department of Energy by Lawrence Livermore National Laboratory under contract DE-AC52-07NA27344. Lawrence Livermore National Security, LLC.\vspace{-0.7em}

\subsubsection*{Gilles Chabrier} is partly supported by STFC grant ST/V000721/1.\vspace{-0.7em} 

\subsubsection*{Nicolas Chamel} is financially supported by Fonds de la Recherche Scientifique-FNRS (Belgium) under Grant Number IISN 4.4502.19. He is a member of the Brussels Laboratory of the Universe (BLU-ULB), Belgium.\vspace{-0.7em}

\subsubsection*{Gilbert W. Collins} is supported by the Department of Energy [National Nuclear Security Administration] University of Rochester “National Inertial Confinement Fusion Program” under Award Number DE-NA0004144 and U.S. National Science Foundation under Award No. 2205521, Center for Matter at Atomic Pressures (CMAP), a National
Science Foundation (NSF) Physics Frontiers Center,
under Award PHY2020249, and DOE-FES Award DE-SC-0020340.
\vspace{-0.7em}

\subsubsection*{Federica Coppari} is supported by Lawrence Livermore National Laboratory and the DOE Early Career Research Program.  Portions of this work were performed under the auspices of the U.S. Department of Energy by Lawrence Livermore National Laboratory under contract DE-AC52-07NA27344. Lawrence Livermore National Security, LLC.\vspace{-0.7em}

\subsubsection*{Tilo D\"oppner} is partly supported by the U.S. Department of Energy by Lawrence Livermore National Laboratory under Contract No. DE-AC52-07NA27344. Lawrence Livermore National Security, LLC.\vspace{-0.7em}

\subsubsection*{Tobias Dornheim} gratefully acknowledges funding from the Deutsche Forschungsgemeinschaft (DFG) via project DO 2670/1-1.
Tobias Dornheim has been partly supported by the Center for Advanced Systems Understanding (CASUS) financed by Germany’s Federal Ministry of Education and Research (BMBF) and the Saxon state government out of the State budget approved by the Saxon State Parliament. Tobias Dornheim's work is supported by the European Union's Just Transition Fund (JTF) within the project \emph{R\"ontgenlaser-Optimierung der Laserfusion} (ROLF), contract number 5086999001, co-financed by the Saxon state government out of the State budget approved by the Saxon State Parliament. Tobias Dornheim's work is also supported by the European Research Council (ERC) under the European Union’s Horizon 2022 research and innovation programme (Grant agreement No. 101076233, "PREXTREME"). Views and opinions expressed are however those of the authors only and do not necessarily reflect those of the European Union or the European Research Council Executive Agency. Neither the European Union nor the granting authority can be held responsible for them. Tobias Dornheim also acknowledges compute time at Norddeutscher Verbund f\"ur Hoch- und H\"ochstleistungsrechnen (HLRN) under grant mvp00024.\vspace{-0.7em}

\subsubsection*{Luke B. Fletcher} is supported by the Department of Energy Office of Science, Fusion Energy Sciences FWP (Field Work Proposal) No.100866, and by the Department of Energy, Laboratory Directed Research and Development program at SLAC National Accelerator Laboratory, under contract DE-AC02-76SF00515.\vspace{-0.7em}

\subsubsection*{Siegfried Glenzer} is supported by the Department of Energy Office of Science, Fusion Energy Sciences FWP (Field Work Proposal) No. 100182, No.100866, and by the National Science Foundation under NSF Grant No. 1632708 and PHY-2308860.\vspace{-0.7em}

\subsubsection*{Sebastien Hamel} is partly supported by the U.S. DOE by LLNL under Contract No. DEAC52-07NA27344.\vspace{-0.7em}

\subsubsection*{Stephanie B. Hansen} is supported by National Technology \& Engineering Solutions of Sandia, LLC under Contract No. DE-NA0003525 with the U.S. Department of Energy (DOE). The authors own all right, title and interest in and to the article and are solely responsible for its contents. The United States Government retains and the publisher, by
accepting the article for publication, acknowledges that the United States Government retains a non-exclusive, paid-up, irrevocable, world-wide license to publish or reproduce the published form of this article or allow others to do so, for United States Government purposes. The DOE will provide public access to these results of federally sponsored research in accordance with the DOE Public Access Plan \url{https://www.energy.gov/downloads/doe-public-access-plan}.\vspace{-0.7em}

\subsubsection*{Nicholas J. Hartley} is supported by the Department of Energy Office of Science, Fusion Energy Sciences FWP (Field Work Proposal) Nos. 100182 \& 100866.\vspace{-0.7em}

\subsubsection*{Suxing Hu} is supported by the Department of Energy [National Nuclear Security Administration] University of Rochester “National Inertial Confinement Fusion Program” under Award Number DE-NA0004144 and U.S. National Science Foundation under Awards No. 2205521 and No. 2020249.
\vspace{-0.7em}

\subsubsection*{Valentin Karasiev} is supported by the Department of Energy [National Nuclear Security Administration] University of Rochester “National Inertial Confinement Fusion Program” under Award Number DE-NA0004144 and U.S. National Science Foundation under Award No. 2205521.
\vspace{-0.7em}

\subsubsection*{Thomas Kluge} is supported by HIBEF (www.hibef.eu) and partially by the European Commission via H2020 Laserlab Europe V (PRISES) contract no. 871124, and by the German Federal Ministry of Education and Research (BMBF) under contract number 03Z1O511. 
This research used resources of the National Energy Research Scientific Computing Center (NERSC), a U.S. Department of Energy Office of Science User Facility located at Lawrence Berkeley National Laboratory, operated under Contract No. DE-AC02-05CH11231 using NERSC awards FES-ERCAP0023216, ERCAP0031552 and ERCAP0032005.\vspace{-0.7em}

\subsubsection*{Marcus D. Knudson} is supported by National Technology \& Engineering Solutions of Sandia, LLC under Contract No. DE-NA0003525 with the U.S. Department of Energy (DOE). The authors own all right, title and interest in and to the article and are solely responsible for its contents. The United States Government retains and the publisher, by
accepting the article for publication, acknowledges that the United States Government retains a non-exclusive, paid-up, irrevocable, world-wide license to publish or reproduce the published form of this article or allow others to do so, for United States Government purposes. The DOE will provide public access to these results of federally sponsored research in accordance with the DOE Public Access Plan \url{https://www.energy.gov/downloads/doe-public-access-plan}.\vspace{-0.7em}

\subsubsection*{Alina Kononov} is partly supported by the US Department of Energy Science Campaign 1 and Sandia National Laboratories' Laboratory Directed Research and Development (LDRD) Project No.\ 233196. This work was performed, in part, at the Center for Integrated Nanotechnologies, an Office of Science User Facility operated for the U.S. Department of Energy (DOE) Office of Science. This article has been co-authored by employees of National Technology \& Engineering Solutions of Sandia, LLC under Contract No. DE-NA0003525 with the U.S. Department of Energy (DOE). The authors own all right, title and interest in and to the article and are solely responsible for its contents. The United States Government retains and the publisher, by accepting the article for publication, acknowledges that the United States Government retains a non-exclusive, paid-up, irrevocable, world-wide license to publish or reproduce the published form of this article or allow others to do so, for United States Government purposes. The DOE will provide public access to these results of federally sponsored research in accordance with the DOE Public Access Plan \url{https://www.energy.gov/downloads/doe-public-access-plan}.\vspace{-0.7em}

\subsubsection*{Zuzana Kon\^{o}pkov\'{a}} acknowledges European XFEL in Schenefeld, Germany, for provision of X-ray free-electron laser beam time at Scientific Instrument HED (High Energy Density Science), HIBEF user consortium and DESY (Hamburg, Germany), a member of the Helmholtz Association HGF.\vspace{-0.7em}

\subsubsection*{Burkhard Militzer} was supported by the U.S. Department of Energy under award DE-NA0004147.\vspace{-0.7em}

\subsubsection*{Zhandos Moldabekov} is partly supported by the Center for Advanced Systems Understanding (CASUS) financed by Germany’s Federal Ministry of Education and Research (BMBF) and the Saxon state government out of the State budget approved by the Saxon State Parliament. Zhandos Moldabekov acknowledges funding by the European Research Council (ERC) under the European Union’s Horizon 2022 research and innovation programme (Grant agreement No. 101076233, "PREXTREME"). Views and opinions expressed are however those of the authors only and do not necessarily reflect those of the European Union or the European Research Council Executive Agency. Neither the European Union nor the granting authority can be held responsible for them.\vspace{-0.7em}

\subsubsection*{Nadine Nettelmann} acknowledges support through the DFG-grant NE 1734/3-1.\vspace{-0.7em}

\subsubsection*{Benjamin K. Ofori-Okai} is supported by the Department of Energy Office of Science, Fusion Energy Sciences FWP (Field Work Proposal) No. 100866, and by the Department of Energy, Laboratory Directed Research and Development program at SLAC National Accelerator Laboratory, under contract DE-AC02-76SF00515 as part of the Panofsky Fellowship.\vspace{-0.7em}

\subsubsection*{Ivan Oleynik} is supported by the U.S. Department of Energy (DOE) National Nuclear Security Administration (Award Nos. DE-NA-0003910 and DE-NA-0004089), the DOE Fusion Energy Science (Award No. DE-SC0024640), the National Science Foundation (Award No. 2421937), and Academic Collaborative Team (ACT) award by Lawrence Livermore National Laboratory. The computations were performed using leadership-class HPC systems: OLCF Frontier at Oak Ridge National Laboratory (ALCC and INCITE Awards Nos. MAT198 and MAT261), ALCF Aurora at Argonne National Laboratory (ALCC Award “FusAblator”), Perlmutter at National Energy Research Scientific Computing Center (NERSC award FES-ERCAPm3993) and TACC Frontera at the University of Texas at Austin (LRAC Award No. DMR21006).\vspace{-0.7em}

\subsubsection*{Aurora Pribram-Jones} is supported by the U.S. Department of Energy Office of Science under Award DE-SC0024476.\vspace{-0.7em}

\subsubsection*{Alessandra Ravasio} is  supported by the Centre National de la Recherche Scientifique (CNRS) and  gratefully acknowledges fundings from the  ANR-DFG project PROPICE (Grant No. ANR-22-CE92-0031) and ANR project MinDIXI (Grant No. ANR-22-CE49-0006).\vspace{-0.7em}

\subsubsection*{Ronald Redmer} is supported by the Deutsche For\-schungsgemeinschaft (DFG, German Science Foundation) with project 521548786 in the priority program SPP2404 (DeepDyn), project 362460292 in the priority program SPP1992 (The Diversity of Exoplanets), and project 280637173 in the Research Unit FOR2440 (Matter under Planetary Interior Conditions).\vspace{-0.7em}

\subsubsection*{Baerbel Rethfeld} acknowledges the support of the Deutsche Forschungsgemeinschaft (DFG, German
Research Foundation) through the SFB/TRR- 173-268565370 “Spin + X”
(project A08).\vspace{-0.7em}

\subsubsection*{Gerd Steinle-Neumann} is supported by the Deutsche Forschungsgemeinschaft (DFG, German Science Foundation) with project 521548786 in the priority program SPP2404 (DeepDyn).\vspace{-0.7em}

\subsubsection*{Aidan P. Thompson} is supported by the U.S. Department of Energy, Office of Fusion Energy Sciences (OFES) under Field Work Proposal Number 20-023149 and the Plasma Surface Interaction project of the Scientific Discovery through Advanced Computing (SciDAC) program, which is jointly sponsored by the Fusion Energy Sciences (FES) and the Advanced Scientific Computing Research (ASCR) programs within the U.S. Department of Energy Office of Science.
This article has been co-authored by employees of National Technology \& Engineering Solutions of Sandia, LLC under Contract No. DE-NA0003525 with the U.S. Department of Energy (DOE). The authors own all right, title and interest in and to the article and are solely responsible for its contents. The United States Government retains and the publisher, by accepting the article for publication, acknowledges that the United States Government retains a non-exclusive, paid-up, irrevocable, world-wide license to publish or reproduce the published form of this article or allow others to do so, for United States Government purposes. The DOE will provide public access to these results of federally sponsored research in accordance with the DOE Public Access Plan \url{https://www.energy.gov/downloads/doe-public-access-plan}.\vspace{-0.7em}

\subsubsection*{Sam M. Vinko} acknowledges support from the UK EPSRC grants EP/W010097/1 and EP/X025373/1.\vspace{-0.7em}

\subsubsection*{Lei Wang} acknowledges the support of the National Natural Science Foundation of China under grant nos. 92270107 and T2225018.\vspace{-0.7em}

\subsubsection*{Alexander J. White} is supported by the U.S. Department of Energy through the Los Alamos National Laboratory (LANL). Research presented in this article was supported by the Laboratory Directed Research and Development program, Projects Nos. 20230322ER and 20230323ER. Los Alamos National Laboratory is operated by Triad National Security, LLC, for the National Nuclear Security Administration of U.S. Department of Energy (Contract No. 89233218CNA000001).\vspace{-0.7em}

\subsubsection*{Thomas G. White} is supported by the National Science Foundation under Grant Nos. PHY-2045718 and PHY-2409354, and by the U.S. Department of Energy, National Nuclear Security Administration under Award No. DE-NA0004039.\vspace{-0.7em}

\subsubsection*{Ulf Zastrau} acknowledges European XFEL in Schenefeld, Germany, for provision of X-ray free-electron laser beam time at Scientific Instrument HED (High Energy Density Science) and the HIBEF user consortium.\vspace{-0.7em}

\subsubsection*{Eva Zurek} acknowledges the Center for Matter at Atomic Pressures (CMAP), a National Science Foundation (NSF) Physics Frontier Center, under Award PHY-2020249, and the U.S. Department of Energy, Office of Science, Fusion Energy Sciences funding the award entitled High Energy Density Quantum Matter, under Award No.\ DE-SC0020340. This material is based upon work supported by the U.S. National Science Foundation under Cooperative Agreement No. (PHY-2329970).\vspace{-0.7em}

\section*{Acknowledgments for the reviewers}
The guest editors (Panagiotis Tolias, Jan Vorberger, Frank Graziani and David Riley) and the authors would like to express their gratitude to Raymond Jeanloz, Yang Zhao, Dieter Hoffmann, Attila Cangi, and Hanno Kählert for their exceptional effort in refereeing this Roadmap article, and for generously sharing their insight into the topic. The identities of the reviewers are disclosed with the explicit consent of both the reviewers and the journal editors.

\newpage 
\onecolumn
\section*{Appendix: List of acronyms}
\begin{longtblr}[
  caption = {List of acronyms together with the main section of reference.},
  label = {tab:acronyms},
]{c c c}    
    \hline \hline
    AA & Average Atom (model) & Section \ref{section_08} \\
    ACE & atomic cluster expansion & Section \ref{section_24} \\
    AD & Automatic Differentiation & Section \ref{section_24} \\
    AIMD & \emph{Ab Initio} MD (= BOMD = DFT-MD) & Section \ref{section_03} \\
    APS & Advanced Photon Source & Section \ref{section_19} \\
    BD & Brown Dwarfs & Section \ref{section_20} \\
    BOA & break out after burner & Section \ref{section_11} \\
    BOMD & Born-Oppenheimer molecular dynamics & Section \ref{section_03} \\
    BSE & Bethe-Salpeter Equation & Section \ref{section_06} \\
    CEIMC & Coupled Electron-Ion MC & Section \ref{section_16} \\
    CM & Chemical Model & Section \ref{section_08} \\
    CMD & Classical MD & Section \ref{section_07}\\
    CPIMC & Configuration PIMC & Section \ref{section_02} \\
    CSP & Crystal Structure Prediction & Section \ref{section_22} \\
    DAC & Diamond Anvil Cell & Section \ref{section_09} \\
    DC & direct current & General \\
    DCS & Dynamic Compression Sector & Section \ref{section_19} \\
    DFT & Density Functional Theory & Section \ref{section_03} \\
    DiPOLE & diode pumped optical laser for experiments & Section  \ref{section_19}\\
    DNN & Deep Neural Network & Section \ref{section_24} \\
    DOS & Density Of States & General \\
    DPMD & Deep Potential MD & Section \ref{section_03} \\
    DSF & Dynamic Structure Factor & General \\
    DSL & Dynamically Screened Ladder (approximation) & Section \ref{section_06} \\
    ELF & Electron Localization Function & Section \ref{section_22} \\
    EOS & Equation Of State & General \\
    ESA & Effective Static Approximation & Section \ref{section_15} \\
    ESRF & European Synchrotron Radiation Facility & Section \ref{section_19} \\
    EXAFS & Extended X-ray Absorption Fine Structure & Section \ref{section_13b} \\
    ext-FPMD & extended First-Principles MD & Section \ref{section_03} \\
    FAIR (principle) & Findability, Accessibility, Interoperability, Reusability & Section \ref{section_25} \\
    FAIR (facility) & Facility for Antiproton and Ion Research & Sections \ref{section_22}, \ref{section_26} \\
    FEL & Free Electron Laser & General \\
    FSP & Fermion Sign Problem & Section \ref{section_02} \\
    FPEOS & First Principles EOS & Section \ref{section_02} \\
    FT & Finite Temperature & General \\
    FVT & Fluid Variational Theory & Section \ref{section_08} \\
    G1-G2 & G1-G2 scheme &Section \ref{section_06} \\
    GAP & Gaussian Approximation Potentials & Section \ref{section_24} \\
    GDSMFB & Finite-T LDA by S.~Groth \textit{et al.} & Section \ref{section_15}\\
    GFs & Green's Functions  & Section \ref{section_06} \\
    GGA & Generalized Gradient Approximation  & Section \ref{section_03} \\
    GKBA & Generalized Kadanoff-Baym ansatz  & Section \ref{section_06} \\
    gLRT & generalized LRT  & Section \ref{section_06} \\
    GP & Giant Planets  & Section \ref{section_17} \\
    GSA & Ground State Approximation  & Section \ref{section_03} \\
    GSAXS & Grazing-incidence SAXS  & Section \ref{section_13} \\
    GSI & Gesellschaft f\"ur Schwerionenforschung & Section \ref{section_22} \\
    GW & GW approximation of the self energy & Section \ref{section_06} \\
    HDC & High-Density Carbon  & Section \ref{section_21} \\
    HED & High Energy Density & General \\
    HED-HIBEF & High Energy Density-Helmholtz International Beamline for Extreme Fields & Section \ref{section_26} \\
    HK & Hohenberg \& Kohn & Section \ref{section_03} \\
    HNC & HyperNetted Chain & Section \ref{section_15} \\
    HPLF & High Power Laser Facility & Section \ref{section_19} \\
    I:R & Ice-to-Rock (ratio) & Section \ref{section_18} \\
    IDD & InDirect-Drive & Section \ref{section_21} \\
    IFE & Inertial Fusion Energy & Section \ref{section_21} \\
    ICF & Inertial Confinement Fusion & Section \ref{section_21} \\
    IET & Integral Equation Theory & Section \ref{section_15} \\
    IPD & Ionization Potential Depression & General\\
    ITCF & Imaginary Time Correlation Function & Sections \ref{section_05}, \ref{section_15} \\
    JWST & James Webb space telescope & Section \ref{section_20}\\
    KBE & (Keldysh-)Kadanoff-Baym Equations & Section \ref{section_06} \\
    KG & Kubo-Greenwood (formula) & Sections \ref{section_04}, \ref{section_05} \\
    KS-DFT & Kohn-Sham DFT & Section \ref{section_03}  \\
    KSDT & Finite-T LDA by V.~Karasiev  \textit{et al.} & Section \ref{section_15} \\
    KTMD & Kinetic theory MD & Section \ref{section_07} \\
    LCLS & Linac Coherent Light Source & Section \ref{section_26} \\
    LDA & Local Density Approximation & Section \ref{section_03} \\
    LFC & Local Field Correction & Sections \ref{section_05}, \ref{section_15} \\
    LLPT & Liquid-Liquid Phase Transition & Section \ref{section_16} \\
    LMJ & Laser Megajoule & Section \ref{section_26} \\
    LR-TD-DFT & Linear Response TD-DFT & Section \ref{section_04} \\
    LRT & Linear Response Theory & Section \ref{section_05} \\
    LTE & Local Thermodynamic Equilibrium & General \\
    MACE & see MPNN & Section \ref{section_24}\\
    MagLIF & Magnetized Liner Inertial Fusion & Section \ref{section_21} \\
    MC & Monte Carlo & General \\ 
    MD & Molecular Dynamics & General \\
    mDFT & mixed DFT & Section \ref{section_03} \\
    MEC & Matter in Extreme Conditions & Section \ref{section_26} \\
    MGF & Matsubara GF & Section \ref{section_06} \\
    MLIP & Machine Learning Interatomic Potentials & Section \ref{section_24} \\
    MPNN & Message-Passing Neural-Network & Section \ref{section_24} \\
    NEGF & Non-Equilibrium GF & Section \ref{section_06} \\
    NequIP & see MPNN & Section \ref{section_05} \\
    NIC & National Ignition Campaign & Section \ref{section_21}  \\
    NIF & National Ignition Facility & Section \ref{section_21} \\
    NIR & Near-InfraRed & General \\
    NLTE & Non-Local Thermodynamic Equilibrium & General \\ 
    NN & Neural Network & Section \ref{section_24} \\ 
    NS & Neutron Stars & Section \ref{section_20} \\
    NSF & National Science Foundation & Section \ref{section_22} \\
    NSF-OPAL & NSF Optical Parametric Amplifier Line & Section \ref{section_22} \\
    NSO & Nonequilibrium Statistical Operator & Section \ref{section_06} \\
    OCP & One-Component Plasma & Section \ref{section_15} \\
    OF-DFT & Orbital-free DFT & Section \ref{section_03}  \\
    PAMD & Pseudo-Atom MD & Section \ref{section_08} \\
    PBE & Perdew-Burke-Ernzerhof (functional) & Section \ref{section_03} \\
    PB-PIMC & Permutation Blocking PIMC & Section \ref{section_02} \\
    PIC & Particle-In-Cell & Section \ref{section_23} \\
    PIMC & Path Integral MC  & Section \ref{section_02} \\
    PIMD & Path Integral MD & Section \ref{section_16} \\
    PPT & Plasma Phase Transition & Section \ref{section_08}  \\
    PXRDIP & Powder X-Ray Diffraction Image Plate & Section \ref{section_22} \\
    QHD & Quantum Hydrodynamics & Section \ref{section_07} \\
    QMC & Quantum Monte Carlo & Section \ref{section_02} \\
    QMD & Quantum molecular dynamics (= BOMD = DFT-MD) & Section \ref{section_03} \\
    RH & Radiation-Hydrodynamics & Section \ref{section_07} \\
    RIF & Reaction-In-Flight & Section \ref{section_11} \\
    RPA & Random Phase Approximation & General \\
    RPA & radiation pressure acceleration & Section \ref{section_11} \\ 
    RPIMC & Restricted PIMC & Section \ref{section_02} \\
    RT & Rayleigh-Taylor (instability) & Section \ref{section_21} \\
    RT-TD-DFT & Real-Time TD-DFT & Section \ref{section_04}  \\
    S3FEL & Shenzhen Soft X-ray FEL & Section \ref{section_26} \\
    SACLA & SPring-8 Angstrom Compact free electron LAser & Section \ref{section_26} \\
    SAXS & Small-Angle X-ray Scattering & Section \ref{section_13} \\
    sDFT & stochastic DFT & Section \ref{section_03} \\
    SEL & Station of Extreme Light & Section \ref{section_26} \\
    SHINE & Shanghai X-ray Free Electron Laser & Section \ref{section_26} \\
    SI & SuperIonic & Section \ref{section_09} \\
    SNAP & Spectral Neighbor Analysis Potentials & Section \ref{section_24}  \\
    SOP & Streaked Optical Pyrometry & Section \ref{section_10} \\
    SP & Stewart \& Pyatt & Section \ref{section_08} \\
    SQ & Spectral Quadrature & Section \ref{section_03} \\
    SSF & Static Structure Factor & General \\
    STLS & Singwi-Tosi-Land-Sj\"olander (dielectric scheme) & Section \ref{section_15} \\
    TARDIS & TARget Diffraction In-Situ & Section \ref{section_22} \\
    TBL & Thermal Boundary Layer & Section \ref{section_18} \\
    TD-DFT & Time-Dependent DFT & Section \ref{section_04} \\
    TD-KS & Time-Dependent Kohn-Sham & Section \ref{section_04} \\
    TFD & Thomas-Fermi-Dirac (model) & Sections \ref{section_08}, \ref{section_22} \\
    THz & Terahertz & Section \ref{section_14} \\
    TNSA & Target-Normal Sheath Acceleration & Sections \ref{section_10}, \ref{section_11}, \ref{section_23} \\
    TTM & Two Temperature Models & Section \ref{section_24} \\
    UEG & Uniform Electron Gas (= jellium = homogeneous electron gas - HEG) & Section \ref{section_15} \\
    VASP & Vienna \emph{Ab Initio} Simulation Package & Section \ref{section_25} \\
    VISAR & velocity interferometer system for any reflector & Section \ref{section_15} \\ 
    VS & Vashishta-Singwi (dielectric scheme) & Section \ref{section_15} \\
    WD & White Dwarfs & Section \ref{section_20} \\
    WDM & Warm Dense Matter & General \\
    XANES & X-ray Absorption Near-Edge Structure & Section \ref{section_13b} \\
    XAS & X-ray Absorption Spectroscopy & Section \ref{section_13b} \\
    XC & eXchange-Correlation & General \\
    XES & X-ray Emission Spectroscopy & Section \ref{section_13} \\
    XFEL & X-ray Free Electron Laser & General \\
    XRD & X-ray Diffraction & Section \ref{section_13} \\
    XRTS & X-ray Thomson Scattering & Section \ref{section_13} \\
    XUV & extreme ultra violet & Section \ref{section_11}
\end{longtblr}
\twocolumn

\newpage 
\onecolumn
\section*{Appendix: List of symbols}
\begin{longtblr}[
  caption = {List of symbols ordered by sections},
  label = {tab:symbols},
]{l l }    
    \hline \hline
     \SetCell[c=2]{c} \rule[1.5pt]{0.35\textwidth}{1.pt} Preamble \& General \rule[1.pt]{0.35\textwidth}{1.pt} &\\
    $T$ & temperature  \\
    $k_B$ & Boltzmann constant  \\
    $a_B$ & Bohr radius  \\
    $Z$ & charge state  \\
    $c$ & speed of light  \\
    $\hat {H}$ & Hamilton operator  \\
    $m_a$ & mass of a particle of species $a$  \\
    $m_i$ & mass of the $i$-th particle  \\
    $\varepsilon_0$ & vacuum permittivity  \\
    $e_i$ & electrical charge of the $i$-th particle  \\
    $e_a$ & electrical charge of a particle of species $a$ \\
    $\Gamma_{ab}$ & Coulomb coupling parameter between species $a$ and $b$  \\
    $\langle V_{ab}\rangle$ & Mean potential energy between species $a$ and $b$  \\
    $\langle K_{a}\rangle$ & Mean kinetic energy of species $a$ \\
    $d$ & mean particle distance\\
    $n$ & mean density\\
    $n_a$ & mean density of species $a$\\
    $r_s$ & Brueckner parameter\\
    $\Theta_a$ & degeneracy parameter of species $a$\\
    $\chi_a$ & degeneracy parameter of species $a$\\
    $q_F$ & Fermi momentum \\
    $T_F$ & Fermi temperature \\
    $E_F$ & Fermi energy \\
    $\kappa_a$ & screening parameter of species $a$ \\
    $\lambda_a=1/\kappa_a$ & Debye length of species $a$\\
    $\omega_{pl}$ & plasma frequency \\
    $\alpha$ & ionization degree \\
    $E_i^{eff}$ & effective ionization energy\\
    \SetCell[c=2]{c} \rule[1.5pt]{0.4\textwidth}{1.pt} Section \ref{section_02} \rule[1.pt]{0.4\textwidth}{1.pt} &\\
    $\hat{Z}$ & partition function \\
    $S_N$ & permutation group\\
    $\hat{\Pi}_{\sigma}$ & pair exchange operator\\
    $\beta=1/k_BT$ & inverse temperature \\
    $\hat{\rho}$ & density operator \\
    $\hat{K}$ & kinetic energy operator \\
    $\hat{V}$ & potential energy operator \\
    $P$ & number of high temperature beads\\
    $\epsilon=\beta/P$ & fictitious inverse temperature in PIMC\\
    $W(\mathbf{X})$ & statistical weight of configuration $\mathbf{X}$ \\
    $\mathbf{X}$ & configuration \\
    $\rho_T$ & trial density matrix\\
    \SetCell[c=2]{c} \rule[1.5pt]{0.4\textwidth}{1.pt} Section \ref{section_03} \rule[1.pt]{0.4\textwidth}{1.pt} &\\
    $n(\mathbf{r})$ & density at location $\mathbf{r}$\\
    $V_{eff}(\mathbf{r})$ & effective potential at location $\mathbf{r}$\\
    $s$ & density gradient\\
    $q$ & reduced density Laplacian\\
    $T_s[n]$ & kinetic energy functional \\
    $E_{xc}[n]$ & exchange-correlation energy functional\\
    \SetCell[c=2]{c} \rule[1.5pt]{0.4\textwidth}{1.pt} Section \ref{section_04} \rule[1.pt]{0.4\textwidth}{1.pt} &\\
    $\phi(\mathbf{r},t)$ & Kohn-Sham orbitals\\
    $f_k(\mu,T)$ & occupation numbers of eigenstate $k$\\
    \SetCell[c=2]{c} \rule[1.5pt]{0.4\textwidth}{1.pt} Section \ref{section_05} \rule[1.pt]{0.4\textwidth}{1.pt} &\\
    $\delta n(\mathbf{r},t)$ & density fluctuation \\
    $V_{ext} (\mathbf{r},t)$ & external (perturbing) potential\\
    $\hat{\rho}(\mathbf{r})$ & density operator in position space\\
    $\omega$ & frequency, usually given in energy units\\
    $G(q,\omega$ & dynamic local field correction\\
    $G(q,z_l)$ & dynamic local field correction at imaginary Matsubara frequencies\\
    $\xi$ & continuous variable describing the statistics of the system ($\xi=-1$ for Fermions, $\xi=1$ for Bosons)\\
    \SetCell[c=2]{c} \rule[1.5pt]{0.4\textwidth}{1.pt} Section \ref{section_06} \rule[1.pt]{0.4\textwidth}{1.pt} &\\
    $G^{M}$ & Matsubara Green's function \\
    $A(p,\omega)$ & spectral function \\
    $G^{<}$, $G^{>}$ & correlation functions \\
    $\Sigma$ & self energy \\
    $a_i^{\dagger}$, $a_i$ & creation and annihilation operators\\
    $C(\tau)$ & cumulant function\\
    $\epsilon=E/N$ & specific energy\\
    $f(\omega)$ & Fermi function \\
    $f(n,T)$ & free energy per particle\\
    $\rho_{rel}$ & relevant statistical operator\\
    ${\bf P}_n$ & moment of the distribution function\\
    $\sigma(n,T)$ & electrical conductivity\\
    \SetCell[c=2]{c} \rule[1.5pt]{0.4\textwidth}{1.pt} Section \ref{section_07} \rule[1.pt]{0.4\textwidth}{1.pt} &\\
    $n(\mathbf{r},t)$ & density at $\mathbf{r}$ and $t$\\
    $V(\mathbf{r},t)$ & potential at $\mathbf{r}$ and $t$\\
    $\mu(\mathbf{r},t)$ & generalized potential energy\\
    $\nabla\mu(\mathbf{r},t)$ & generalized force field\\
    $F[n]$ & free energy functional\\
    $\mu_0$ & local chemical potential\\
    $\phi(\mathbf{r},t)$ & Hartree mean field potential\\
    \SetCell[c=2]{c} \rule[1.5pt]{0.4\textwidth}{1.pt} Section \ref{section_08} \rule[1.pt]{0.4\textwidth}{1.pt} &\\    
    $n_j$ & density of charge states $j$\\
    $g_j$ & statistical weight of charge state $j$\\
    \SetCell[c=2]{c} \rule[1.5pt]{0.4\textwidth}{1.pt} Section \ref{section_09} \rule[1.pt]{0.4\textwidth}{1.pt} &\\    
    $T_C$ & temperature of onset of high absorption\\
    $E_g$ & energy of band gap\\
    $\mu$ & chemical potential\\
    \SetCell[c=2]{c} \rule[1.5pt]{0.4\textwidth}{1.pt} Section \ref{section_11} \rule[1.pt]{0.4\textwidth}{1.pt} &\\    
    $E_0$ & initial particle energy\\
    $v_{th}$ & thermal velocity\\
    $v_p$ & particle velocity\\
    \SetCell[c=2]{c} \rule[1.5pt]{0.4\textwidth}{1.pt} Section \ref{section_12} \rule[1.pt]{0.4\textwidth}{1.pt} &\\    
    $u_s$ & shock velocity \\
    $u_p$ & particle velocity \\
    \SetCell[c=2]{c} \rule[1.5pt]{0.4\textwidth}{1.pt} Section \ref{section_13} \rule[1.pt]{0.4\textwidth}{1.pt} &\\    
    $S(k,\omega)$ & dynamic structure factor \\
    $F(k,\tau)$ & imaginary time correlation function \\
    $\tau=i\hbar\beta$ & imaginary time \\
    $G(\theta,\phi)$ & geometric form factor\\
    $I_0$ &  initial intensity of the radiation\\
    $r_c$ & classical electron radius \\
    $S(k)$ & static structure factor \\
    $S_{ii}(k)$ & static ion-ion structure factor \\
    $I(k)$ & scattered intensity \\
    $f(k)$ &  atomic/ionic form factor\\
    $q(k)$ &  ionic screening cloud\\
    $d$ &  distance between scattering crystal planes\\
    $\theta$ & scattering angle \\
    \SetCell[c=2]{c} \rule[1.5pt]{0.4\textwidth}{1.pt} Section \ref{section_13b} \rule[1.pt]{0.4\textwidth}{1.pt} &\\    
    $\chi(e)$ & normalized absorption profile\\
    $\mu(E)$ & absorption coefficient\\
    $\mu_0(e)$ & atomic background absorption\\
    $\Delta\mu_0(e)$ & jump in absorption at edge\\
    \SetCell[c=2]{c} \rule[1.5pt]{0.4\textwidth}{1.pt} Section \ref{section_14} \rule[1.pt]{0.4\textwidth}{1.pt} &\\    
    $\bar{\sigma}$ & Drude frequency dependent conductivity\\
    $q_e$ & electron charge \\
    $\sigma_0$ & DC conductivity\\
    $\tau=1/\nu_e$ & electron scattering time \\
    $\nu_e$ & electron collision frequency \\
    $\tilde{t}$ & transmission coefficient\\
    $Z_0$ & impedance of free space \\
    $d$ & film thickness\\
    \SetCell[c=2]{c} \rule[1.5pt]{0.4\textwidth}{1.pt} Section \ref{section_15} \rule[1.pt]{0.4\textwidth}{1.pt} &\\    
    $\zeta$ & spin polarisation\\
    $f_{xc}$ & specific exchange-correlation free energy \\
    $G(r_s,\theta,q)$ & static local field correction \\
    \SetCell[c=2]{c} \rule[1.5pt]{0.4\textwidth}{1.pt} Section \ref{section_16} \rule[1.pt]{0.4\textwidth}{1.pt} &\\    
    $\chi(q,\omega)$ & density response function \\
    $\chi(q)$ & static ($\omega=0$) response function\\
    $K_{xc}$ & exchange-correlation Kernel\\
    \SetCell[c=2]{c} \rule[1.5pt]{0.4\textwidth}{1.pt} Section \ref{section_17} \rule[1.pt]{0.4\textwidth}{1.pt} &\\    
    $\kappa_T$ & isothermal compressibility \\
    $P$ & pressure \\
    $u$ & internal energy \\
    $\alpha$ & thermal expansion coefficient\\
    $C_P$ & heat capacity at constant pressure\\
    $c_S$ & sound velocity \\
    $\gamma$ & Gr\"uneisen parameter \\
    $R_S$ & radius of the planet Saturn\\
    \SetCell[c=2]{c} \rule[1.5pt]{0.4\textwidth}{1.pt} Section \ref{section_18} \rule[1.pt]{0.4\textwidth}{1.pt} &\\    
    $R_P$  & planetary radius \\
    \SetCell[c=2]{c} \rule[1.5pt]{0.4\textwidth}{1.pt} Section \ref{section_20} \rule[1.pt]{0.4\textwidth}{1.pt} &\\    
    $L_{nuc}$ & nuclear luminosity due to proton fusion \\
    $Y_{\odot}$ & hydrogen-helium abundance ratio\\
    $Z_{\odot}$ & fraction of heavy elements\\
    $M_{\odot}$ & solar mass \\
    $B_{at}$ & atomic magnetic field strength \\
    \SetCell[c=2]{c} \rule[1.5pt]{0.4\textwidth}{1.pt} Section \ref{section_23} \rule[1.pt]{0.4\textwidth}{1.pt} &\\    
    $\gamma$ & Keldysh parameter \\
    $\lambda$ & wave length \\
    $I_0$ & band gap/ionization energy\\
    $p_{\perp}$ & momentum transverse to the wave propagation\\
    $a_0$ & normalized laser strength parameter\\
    $\eta$ & resistivity \\
    $\mathbf{j}_f$ & fast electron current component\\
    $\mathbf{j}_c$ & cold electron current component\\
    $\mathbf{E}$ & electric field\\
    \SetCell[c=2]{c} \rule[1.5pt]{0.4\textwidth}{1.pt} Section \ref{section_24} \rule[1.pt]{0.4\textwidth}{1.pt} &\\    
    $U$ & unitary transformation within a flow model\\
    $p_n$ & probability for state occupations\\
    $|n\rangle$ & many body base state \\
    \SetCell[c=2]{c} \rule[1.5pt]{0.4\textwidth}{1.pt} Section \ref{section_26} \rule[1.pt]{0.4\textwidth}{1.pt} &\\    
    $\alpha$ & adiabat
    \end{longtblr}
\twocolumn

\bibliographystyle{iop-art}
\bibliography{bibliography_sorted_cleared}

\end{document}